\documentclass[12pt]{nmsuth01jcw}

\usepackage{amsmath,amsthm}

\usepackage{graphicx}

\usepackage{axodraw}

\usepackage{rotating}

\usepackage{lscape}

\usepackage{longtable}

\usepackage{epsf}

\usepackage[section]{placeins}

\def\deg{\hbox{$^\circ$}}

\def\pphi{\hbox{$\phi_{p}$ }}

\newcommand{\prl}[3]{Phys. Rev. Lett. {\bf #1}, #2 (#3)}
\newcommand{\prd}[3]{Phys. Rev. D {\bf #1}, #2 (#3)}
\newcommand{\plb}[3]{Phys. Lett.  B {\bf #1}, #2 (#3)}
\newcommand{\nima}[3]{Nucl. Inst. Methods A {\bf #1}, #2 (#3)}
\newcommand{\listserver}[1]{E866 Internal Memo, #1, { \em unpublished }}

\newcommand{\npb}[3]{Nucl. Phys. B {\bf #1}, #2, (#3)}
\newcommand{\zpc}[3]{Z. Phys. C {\bf #1}, #2, (#3)}
\newcommand{\epjc}[3]{European Phys. J. C {\bf #1}, #2, (#3)}
\newcommand{\ijmpa}[3]{Int. J. Mod. Phys. A {\bf #1}, #2, (#3)}

\newlength{\singlespace}
\newlength{\doublespace}
\begin{document}

\setlength{\baselineskip}{\doublespace}



\pagenumbering{roman}
\pagestyle{empty}
\renewcommand{\baselinestretch}{2}
\begin{center}
MEASUREMENT OF CONTINUUM DIMUON PRODUCTION IN 800-GEV/C PROTON-NUCLEON COLLISIONS\\
\vspace{0.1in}
BY\\
\vspace{0.1in}
JASON C. WEBB, B.S.
\end{center}
\vspace{1.0in}
\begin{center}
A dissertation submitted to the Graduate School\\
\vspace{0.1in}
in partial fulfillment of the requirements\\
\vspace{0.1in}
for the degree \\
\vspace{0.1in}
Doctor of Philosophy
\end{center}
\vfill
\begin{center}
Major Subject: Physics
\end{center}
\vspace{1.0in}
\begin{center}
New Mexico State University\\
\vspace{0.1in}
Las Cruces, New Mexico\\
\vspace{0.1in}
December 2002
\end{center}

\pagestyle{plain}
\noindent
``Measurement of Continuum Dimuon Production in 800-GeV/c Proton-Nucleon Collisions,'' 
a dissertation prepared by
Jason C. Webb 
in partial fulfillment of the requirements for the degree, 
Doctor of Philosophy,
has been approved and accepted by the following:

\setlength{\baselineskip}{\singlespace}
\vspace{0.1in}
\begin{flushleft}
\ \ 

\hrulefill
\newline
Linda Lacey
\newline
Dean of the Graduate School
\vspace{0.5in}

\hrulefill
\newline
Vassili Papavassiliou
\newline
Chair of the Examining Committee
\vspace{0.5in}

\hrulefill
\newline
Date
\vspace{0.5in}
\newline
Committee in charge:
\end{flushleft}

\setlength{\baselineskip}{\doublespace}
Dr. Vassili Papavassiliou, Chair

Dr. William R. Gibbs

Dr. Vincent P. Gutschick

Dr. Gary S. Kyle

\begin{center}
DEDICATION
\end{center}

For Mom.

\begin{center}
ACKNOWLEDGMENTS
\end{center}

This dissertation would not have been possible without the hard work and
support of many people.  I would like to begin by thanking my adviser for 
his guidance of my research, for always having time to answer my questions
about most any subject, for teaching me everything I know about
particle physics (albeit only a fraction of what he tried to), and for
putting up with numerous drafts of this thesis.  I would also like to 
express my gratitude for the hard work and dedication of the FNAL 
E866/NuSea collaboration.  I have enjoyed being a part of this experiment,
and I couldn't have gotten this done without you all.

Over the past few years, I have leaned heavily upon the advice and help
of several members of the collaboration, and I would like to thank them 
each individually.  First is Paul Reimer, whose work on the Monte Carlo 
was invaluable to virtually every aspect of the data analysis, and whose 
help with the theoretical calculations was so important to the conclusions 
drawn from the results.  It is difficult to imagine making this measurement
without his help.  It is also difficult to imagine doing this experiment 
without the efforts of Eric Hawker and Rusty Towell, whose analysis of the 
data preceded mine.  Their work paved the way for me, making my job a 
significantly easier one.  FNAL E866 addressed many physics topics, and 
several analyses were performed in parallel with my own.  I thank 
Ting Chang, Bill Lee, Mike Leitch, Bryon Mueller, Chuck Brown, 
Donald Koetke, and Maxim Vasiliev for their hard work on these projects, 
which often yielded useful results for my purposes.  In particular I'd 
like to thank Maxim for his assistance with the combinatoric background.  
Finally, Pat McGaughey and Carl Gagliardi have provided many suggestions
over the years which helped me overcome obstacles in the analysis (and
occassionally pointed out ones which I had missed).  I appreciate all of
their efforts which have made this thesis immeasurably better.

I would not be here today without a solid education, and so I would like to 
thank the faculty of my alma mater, Valparaiso University, for the foundation 
that they laid which enabled me to succeed in graduate school.  In particular 
I would like to thank Don Koetke, whose nuclear physics course got me started 
out on the road I'm currently on.   I owe them all my thanks (and a 
colloquium).  I also owe the faculty here at New Mexico State University
my thanks for the many interesting and thought provoking courses, seminars and
discussions over the years.  I would especially like to thank Gary Kyle, for 
sending me off to Fermilab to work on E866 my first summer as a graduate 
student and for supporting me ever since.

I've been fortuante to make several friends here in Las Cruces, without 
whom life here in Las Cruces would not have been very enjoyable.  Friday
nights will be sorely missed.

Finally, I have been blessed in life by two wonderful parents who truly
taught me the most important lessons I have ever learned in life.  Mom
passed away just as I started working towards this thesis, and so I 
dedicate it to her.  But I probably couldn't have finished it without 
the support and encouragement of my Dad.

Dad, I got it done.



\vspace{0.5in}
\begin{center}
{\normalsize VITA}
\end{center}
\ \ 

\begin{flushleft}
\begin{tabular}{ll}
July 27, 1973 & Born in Danville, Illinois \\
 & \\
May 25, 1995 & B.S. in Physics, Valparaiso University, Valparaiso, Indiana \\
 & \\
\end{tabular}
\end{flushleft}
\vspace{0.1in}
\begin{center}
PROFESSIONAL  AND HONORARY SOCIETIES
\end{center}
\begin{flushleft}
American Physical Society

Sigma Pi Sigma
\end{flushleft}
\vspace{0.1in}
\begin{center}PUBLICATIONS
\end{center}
\begin{flushleft}
\setlength{\baselineskip}{\singlespace}
\ \\
E.~A.~Hawker { et al.}  [FNAL E866/NuSea Collaboration],
``Measurement of the Light Antiquark Flavor Asymmetry in the Nucleon Sea,''
Phys.\ Rev.\ Lett.\  {\bf 80}, 3715 (1998) \\ \

J.~C.~Peng { et al.}  [FNAL E866/NuSea Collaboration],
``$\bar{d}/\bar{u}$ Asymmetry and the Origin of the Nucleon Sea,''
Phys.\ Rev.\ D {\bf 58}, 092004 (1998) \\ \

M.~A.~Vasilev { et al.}  [FNAL E866 Collaboration],
``Parton Energy Loss Limits and Shadowing in Drell-Yan Dimuon Production,''
Phys.\ Rev.\ Lett.\  {\bf 83}, 2304 (1999) \\ \

M.~J.~Leitch { et al.}  [FNAL E866/NuSea Collaboration],
``Measurement of $J/\psi$ and $\psi^\prime$ Suppression in p A Collisions at 800-GeV/c,''
Phys.\ Rev.\ Lett.\  {\bf 84}, 3256 (2000) \\ \

C.~N.~Brown { et al.}  [FNAL E866 Collaboration],
``Observation of Polarization in bottomonium production at $\sqrt{s} =  38.8$-GeV,''
Phys.\ Rev.\ Lett.\  {\bf 86}, 2529 (2001) \\ \

R.~S.~Towell { et al.}  [FNAL E866/NuSea Collaboration],
``Improved Measurement of the $\bar{d}/\bar{u}$ Asymmetry in the Nucleon Sea,''
Phys.\ Rev.\ D {\bf 64}, 052002 (2001) \\ \

J.~C.~Webb { et al.}  [FNAL E866/NuSea Collaboration], ``Measurement of Continuum Dimuon Production in 800-GeV/c Proton-Nucleon Interactions'', in preparation.

\setlength{\baselineskip}{\doublespace}

\end{flushleft}
\vspace{0.1in}
\begin{center}
FIELD OF STUDY
\end{center}
\begin{flushleft}
Major Field: Physics
\end{flushleft}


\begin{center}
ABSTRACT
\end{center}
\vspace{0.3in}
\begin{center}
MEASUREMENT OF CONTINUUM DIMUON PRODUCTION IN 800-GEV/C PROTON-NUCLEON COLLISIONS
\\
BY
\\
JASON C. WEBB, B.S.
\end{center}
\vspace{0.3in}
\begin{center}
Doctor of Philosophy

New Mexico State University

Las Cruces, New Mexico, 2002

Dr. Vassili Papavassiliou, Chair
\end{center}
\vspace{0.3in}
\hspace{\parindent}
Fermilab Experiment 866 has performed an absolute measurement of continuum
dimuon (Drell-Yan) cross sections in 800-GeV/c $pp$ and $pd$ interactions.
Results differential in the mass, Feynman-$x$ ($x_F$) and transverse 
momenta ($p_T$) of the dimuon pairs are reported.  These results represent
the most extensive study of the Drell-Yan process to date, based on a data
sample of 175,000 dimuon events covering the widest range in kinematics
yet achieved ($4.2 \leq M \leq 16.85$ GeV and $-0.05 \leq x_F \leq 0.8$)
with this level of precision.
The cross sections are primarily sensitive to the magnitude and shape of the
light antiquark distributions ($\bar{d}(x)$ and $\bar{u}(x)$) in the nucleon,
but also provide important information on the valence quarks as well as the
gluons.
They are in good agreement with other existing proton-induced Drell-Yan
experiments.  There is also general agreement between the data and
next-to-leading-order calculations based on various sets of parton 
distribution functions.  Differences between data and theory are examined,
and the potential impact of these data on future parameterizations of the
parton distributions are discussed.


\tableofcontents

\newpage
\addcontentsline{toc}{section}{LIST OF TABLES}

\section*{
  \listtablename
  }
  \contentsline {table}{\numberline {1.1}{\ignorespaces Properties of the light quarks.}}{3}
\contentsline {table}{\numberline {1.2}{\ignorespaces Experimental $K$-factors.}}{14}
\contentsline {table}{\numberline {2.1}{\ignorespaces Specifications of the hodoscope planes.}}{29}
\contentsline {table}{\numberline {2.2}{\ignorespaces Specifications of the drift chamber planes.}}{31}
\contentsline {table}{\numberline {2.3}{\ignorespaces Specifications of the proportinal tube planes.}}{32}
\contentsline {table}{\numberline {2.4}{\ignorespaces Trigger configuration (PhysA).}}{37}
\contentsline {table}{\numberline {2.5}{\ignorespaces Trigger configuration (PhysB).}}{38}
\contentsline {table}{\numberline {3.1}{\ignorespaces Definition of the data sets.}}{41}
\contentsline {table}{\numberline {3.2}{\ignorespaces Field strengths for each data set. }}{59}
\contentsline {table}{\numberline {4.1}{\ignorespaces Number of events in each data set which survived all data cuts. }}{66}
\contentsline {table}{\numberline {4.2}{\ignorespaces Spill quality cuts.}}{74}
\contentsline {table}{\numberline {4.3}{\ignorespaces \baselineskip \singlespace \relax SEM calibration measurements on the Meson East beam-line. }}{76}
\contentsline {table}{\numberline {4.4}{\ignorespaces SEM offsets.}}{78}
\contentsline {table}{\numberline {4.5}{\ignorespaces \baselineskip \singlespace \relax Percent molecular and atomic abundance of second deuterium fill. }}{79}
\contentsline {table}{\numberline {4.6}{\ignorespaces \baselineskip \singlespace \relax Target pressures (in psi) for selected data sets. }}{80}
\contentsline {table}{\numberline {4.7}{\ignorespaces \baselineskip \singlespace \relax Integrated luminosities in the E866 data sets. }}{82}
\contentsline {table}{\numberline {4.8}{\ignorespaces Rate-dependent corrections to the hydrogen and deuterium yields.}}{84}
\contentsline {table}{\numberline {5.1}{Scaling form $M^3d^2\sigma /dMdx_F$ for the hydrogen cross section.}}{103}
\contentsline {table}{\numberline {5.2}{Scaling form $M^3d^2\sigma /dMdx_F$ for the deuterium cross section.}}{116}
\%contentsline {table}{\numberline {5.2}{(continued)}}{126}
\ 
\baselineskip \singlespace \relax 
\contentsline {table}{\numberline {5.3}{Invariant form $E d^3\sigma /dp^3$ for the hydrogen cross section over the range -0.05 $\leq x_F \leq $ 0.15.}}{146}
\baselineskip \doublespace \relax 
\ 
\baselineskip \singlespace \relax 
\contentsline {table}{\numberline {5.4}{Invariant form $E d^3\sigma /dp^3$ for the hydrogen cross section over the range 0.15 $\leq x_F \leq $ 0.35.}}{150}
\baselineskip \doublespace \relax 
\ 
\baselineskip \singlespace \relax 
\contentsline {table}{\numberline {5.5}{Invariant form $E d^3\sigma /dp^3$ for the hydrogen cross section over the range 0.35 $\leq x_F \leq $ 0.55.}}{155}
\baselineskip \doublespace \relax 
\ 
\baselineskip \singlespace \relax 
\contentsline {table}{\numberline {5.6}{Invariant form $E d^3\sigma /dp^3$ for the hydrogen cross section over the range 0.55 $\leq x_F \leq $ 0.8.}}{161}
\baselineskip \doublespace \relax 
\ 
\baselineskip \singlespace \relax 
\contentsline {table}{\numberline {5.7}{Invariant form $E d^3\sigma /dp^3$ for the deuterium cross section over the range -0.05 $\leq x_F \leq $ 0.15.}}{166}
\baselineskip \doublespace \relax 
\ 
\baselineskip \singlespace \relax 
\contentsline {table}{\numberline {5.8}{Invariant form $E d^3\sigma /dp^3$ for the deuterium cross section over the range 0.15 $\leq x_F \leq $ 0.35.}}{170}
\baselineskip \doublespace \relax 
\ 
\baselineskip \singlespace \relax 
\contentsline {table}{\numberline {5.9}{Invariant form $E d^3\sigma /dp^3$ for the deuterium cross section over the range 0.35 $\leq x_F \leq $ 0.55.}}{176}
\baselineskip \doublespace \relax 
\ 
\baselineskip \singlespace \relax 
\contentsline {table}{\numberline {5.10}{Invariant form $E d^3\sigma /dp^3$ for the deuterium cross section over the range 0.55 $\leq x_F \leq $ 0.8}}{182}
\baselineskip \doublespace \relax 
\ 
\baselineskip \singlespace \relax 
\contentsline {table}{\numberline {6.1}{\ignorespaces $K^\prime $-factors, where $K^\prime = \frac {\sigma ^{\text {exp}}}{\sigma ^{\text {NLO}}}$, for 800-GeV $pp$ and $pd$ dimuon production.}}{203}
\baselineskip \doublespace \relax

\newpage
\addcontentsline{toc}{section}{LIST OF FIGURES}
\listoffigures

\newpage
\pagenumbering{arabic}

\def\nevents{175,000}
\def\ndevents{120,000}
\def\nhevents{55,000}

\cleardoublepage

\section{INTRODUCTION}

Investigation into the field of lepton-pair production in hadronic
interactions began with an 
experiment carried out at Brookhaven National Laboratory in 1970 
\cite{bib:BNL-DY}.  This experiment studied the reaction 
$p$U$ \; \rightarrow \mu^+ \mu^- X$
for proton energies between 22 and 29 GeV, resulting in pair masses between
1 and 6.7 GeV\footnote{
\setlength{\baselineskip}{\singlespace}
Throughout this thesis, we will utilize ``God-given'' units, setting
$\hbar = c = 1$.}.  The data above $\sim 3$ GeV (where the $J/\psi$ resonance
family would later be found) exhibited a rapidly decreasing continuum of 
muon pairs.
The steeply falling nature of the cross section was explained later that year
by Drell and Yan \cite{bib:DY}, who were interested in dilepton production
as a possible application of the quark-parton model of hadron structure 
outside of deeply-inelastic
scattering experiments.  Their model of quark-antiquark annihilation
through the electromagnetic interaction, which
has become known as the Drell-Yan process, was generally successful in 
describing the data and would only improve as our understanding of the 
strong interaction improved.

The Drell-Yan process still remains an active area of experimental and 
theoretical 
research some thirty years later.  It has played a key role in developing 
the mathematical technology of perturbative Quantum Chromodynamics (QCD), 
being one of the first
processes to be calculated to next-to-leading order
${\cal O}(\alpha_s)$,
and remains one of the few processes to be calculated to 
next-to-next-to-leading ${\cal O}(\alpha_s^2)$.  Experimentally it has 
provided a wealth of 
information about nucleon structure; its confirmation of the quark-parton
model and its verification of the quark charge assignments being two 
notable early applications.  In this thesis we describe and report the
results of an absolute measurement of the dimuon cross sections from the
interaction of an 800-GeV proton beam with hydrogen and deuterium targets.
As will be shown, these data provide important information about the 
distributions of antiquarks in the nucleon sea.

\subsection{Background}

The disciplines of high-energy and particle physics consist of the search for
the ultimate constituents of matter and the study of the fundamental 
interactions between them.  With the introduction of Quantum Chromodynamics 
in the mid-1970s, a consistent treatment of the weak, electromagnetic 
and strong interactions was finally realized.  The gauge field theories of 
Quantum Electrodynamics (QED) and the Weinberg-Glashow-Salam model had long 
treated the electromagnetic and weak interactions felt by the leptons -- the 
electron ($e$), muon ($\mu$), tau ($\tau$), and associated neutrinos 
($\nu_e$, $\nu_\mu$, $\nu_\tau$) -- as the exchange of gauge field bosons 
(the photon ($\gamma$) and the $W^\pm$ and $Z^0$ bosons).  Attempts to 
formulate similar theories for the strongly interacting hadrons, however,
initially met with only  limited success.  The realization 
by Gell-Mann and (independently) Zweig in 1964 that the hadron spectrum could 
be explained in terms of three fractionally charged fermionic constituents, 
called quarks, set the stage for the development of QCD and its inclusion in 
what has become known as the Standard Model.

Although Gell-Mann and Zweig's static quark model needed only three
different quark flavors, experiment confirms the existence of six, with
larger numbers being ruled out in the framework of the Standard Model
by recent experiments.
The six quarks are conventionally known as up ($u$), down 
($d$), strange ($s$), charm ($c$), bottom ($b$) and top ($t$) and are
found only in bound states called hadrons: 
mesons ($\pi^\pm$, $\pi^0$, $J/\psi$, etc...) are quark-antiquark ($q\bar{q}$)
bound states, while baryons ($p$, $n$, $\Delta^{++}$, etc...) are the bound
states of three quarks ($qqq$).
Table \ref{table:stquarkmodel} shows the quantum 
numbers of the three quarks of the first three quarks.
In order to avoid 
violating the Pauli principle in baryons such as the $\Delta^{++}$ ($uuu$), 
the 
static quark model had to add an additional quantum number called color.  Each
quark carries one of three colors (red, green or blue).  An object in which
all three colors (or one color and its anti-color) are present is a 
color-neutral (color-singlet) state.

\begin{table}
\centering
\caption[Properties of the light quarks.]{
   \label{table:stquarkmodel}
   \setlength{\baselineskip}{\singlespace}
   Properties of the light quarks.  Shown are the isospin ($I$) and its 
   third component ($I_3$), strangeness ($S$), baryon number ($B$) and 
   charge ($Q/e$).}
  \vspace{10pt}
  \begin{tabular}{|c|c|c|c|c|c|}
  \hline
  flavor         & $I$ & $I_3$ & $S$ & $B$ & $Q/e$ \\
  \hline\hline
  u & $1/2$ &  $1/2$ & $0$  & $1/3$ &  $2/3$ \\
  d & $1/2$ & $-1/2$ & $0$  & $1/3$ & $-1/3$ \\
  s &   $0$ &    $0$ & $-1$ & $1/3$ & $-1/3$ \\
  \hline
\end{tabular}

\end{table}

Color was an {\it ad hoc} addition to the static quark model.  The role 
which it plays in QCD is more fundamental.  QCD is a non-Abelian color gauge 
field theory, the three colors representing the fundamental basis of the
$SU(3)$ symmetry group.  Color plays essentially the same role in the
strong interaction as electric charge in electromagnetism, the principal
difference being the number of ``charge'' states -- two electric charge states
($+$ and $-$) versus six color states ($r$, $g$, $b$, $\bar{r}$, $\bar{g}$,
$\bar{b}$).\footnote{Red, green, blue and their anti-colors.}
The quarks interact through the exchange of QCD's gauge field bosons, the
gluons ($g$).  Unlike the photons in QED, which are electrically neutral, the 
gluons carry color charge.  This leads to some important differences.  
A well-known property of QED is the running of the electromagnetic coupling 
constant $\alpha \approx \frac{1}{137}$.  A test charge polarizes the
virtual $e^+e^-$ pairs surrounding it, leading to screening of the test
charge at larger distances, resulting in an effective reduction in $\alpha$.
Probing the test charge at smaller distance scales, i.e. larger
momentum transfers ($q^2$), results in an observed increase in the 
coupling.  The same effect occurs in QCD, but the color-charged gluons
are self interacting.  This leads to an antiscreening of the color test 
charge, resulting in a decrease in the strong coupling constant ($\alpha_s$) 
with increasing $q^2$.  
This property of the strong interaction is known as 
asymptotic freedom.


Owing to the mathematical complexity of the $SU(3)\otimes SU(2)\otimes U(1)$ 
Standard Model, exact solutions are often not possible for physical 
interactions.  We must therefore rely upon perturbation theory to obtain 
meaningful predictions from the model.  This has never presented a problem
in either the weak or electromagnetic interactions, where one can simply 
carry out the perturbative expansion in the coupling constant to whatever 
level of precision is desired.  
However, the strong interaction derives its name from the fact
that it is a {\it strong} interaction.  For momentum scales on the order of the
proton mass, the size of the strong coupling constant 
($\alpha_s(M_p^2) \sim 1$) renders fixed-order perturbation theory
meaningless -- all orders in the perturbative expansion are significant.
Fortunately, the asymptotic freedom exhibited by QCD results in a strong 
coupling constant which decreases with increasing $q^2$.  At momentum-transfers
typical of high-energy fixed-target experiments, $\alpha_s$ becomes 
sufficiently 
small that it may be reliably used as an expansion parameter.
Nevertheless, it remains sufficiently large that terms next-to-leading order
in $\alpha_s$ may remain important.

An additional complication in hadronic interactions is that we cannot
be certain which of the interacting partons (the quark, antiquark and gluon 
constituents) are participating in the interaction.  We can at best speak
of the probability of finding a given parton carrying a fraction $x$ 
(called Bj\"orken-$x$) of the interacting hadron's momentum, which we in
general denote by $f_{q/A}(x)$ for a parton of type $q$ in a hadron of type
$A$.  Calculations of physical processes in the Standard Model are thus 
predicated on prior knowledge of these parton distribution functions (PDF's).

\begin{figure}
\centering
\includegraphics[width=0.9\linewidth]{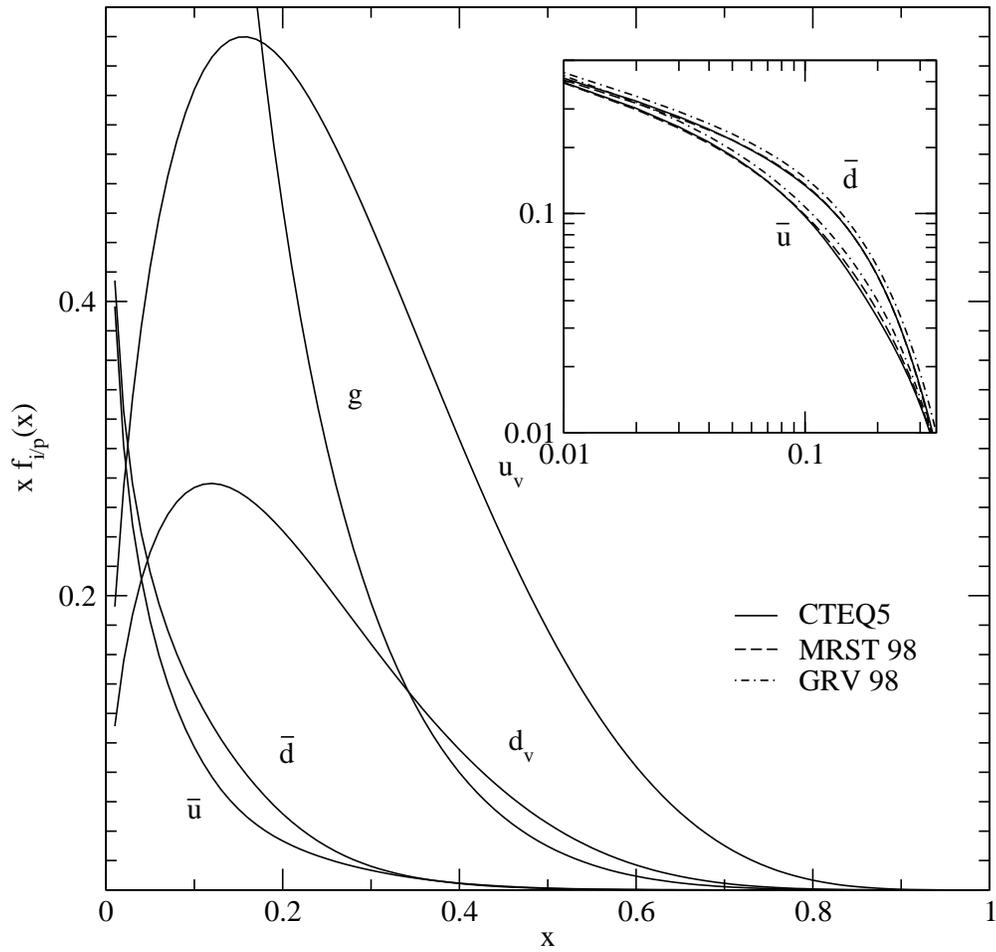}
\caption[CTEQ5 parton distributions.]{
	\setlength{\baselineskip}{\singlespace}
	\label{fig:pdfs}
	CTEQ5 parton distributions.  Shown are the valence quark and light anti-quark 
	distributions, and the gluons.  The inset compares $\bar{d}(x)$
	and $\bar{u}(x)$ from the CTEQ5 \protect\cite{bib:CTEQ}, MRST 98 
	\cite{bib:MRST} and GRV 98 \protect\cite{bib:GRV} fits.
	}
\end{figure}

Of particular importance to many experiments are the parton distributions 
in the nucleon (i.e. $u(x) = f_{u/p}(x)$, $\bar{u}(x) = f_{\bar{u}/p}(x)$, 
etc...).  Early parameterizations of the PDF's relied on fits to the
structure functions measured in deeply-inelastic lepton-hadron scattering
experiments (DIS), which 
are primarily sensitive to the
light quark distributions ($u(x)$ and $d(x)$).
More modern parameterizations, such as those performed
by the CTEQ, MRST and GRV collaborations \cite{bib:MRST,bib:CTEQ,bib:GRV} 
use several different physical processes to extract complementary
information about the parton distributions.  
The lepton-charge asymmetry observed
in $W^\pm$ production provides additional 
information about the light quark distributions, while jet production 
and prompt photon measurements are used to constrain the gluon distributions.
The Drell-Yan process for dilepton production, which involves quark-antiquark
annihilation, provides constraints on the light antiquark distributions 
($\bar{u}(x)$ and $\bar{d}(x)$) in the nucleon sea.
The CTEQ 5 set of parton distributions is shown in figure \ref{fig:pdfs},
along with the light antiquark distributions from the MRST 98 and GRV 98
distributions.

%

\subsection{The Drell-Yan Process}

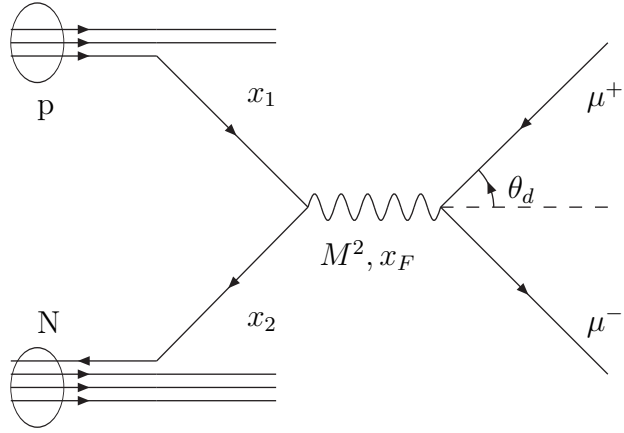
\begin{figure}[h]
  \begin{center}
  \begin{picture}(230,150)(0,50)

    \ArrowLine(0,170)(55,170)  \ArrowLine(55,170)(112,113)
    \ArrowLine(0,175)(55,175)  \Line(55,175)(100,175)
    \ArrowLine(0,180)(55,180)  \Line(55,180)(100,180)
    \Oval(10,175)(15,10)(0)

    \ArrowLine(0,50)(55,50)    \Line(55,50)(100,50)
    \ArrowLine(0,45)(55,45)    \Line(55,45)(100,45)
    \ArrowLine(0,40)(55,40)    \Line(55,40)(100,40)
    \ArrowLine(112,113)(55,55)\ArrowLine(55,55)(0,55)
    \Oval(10,45)(15,10)(0)
    \Photon(112,113)(162,113){5}{5}
    \ArrowLine(225,175)(162,113)
    \ArrowLine(162,113)(225,50)
    \DashLine(162,113)(225,113){5}
    \ArrowArc(162,113)(20,0,45)

    \Text(10,150)[l]{p}
    \Text(10,70)[l]{N}    
    \Text(95,155)[]{{$x_{1}$}}
    \Text(95,70)[]{{$x_{2}$}}    
    \Text(135,95)[]{{$M^{2},x_{F}$}}    
    \Text(225,155)[]{{$\mu^{+}$}}
    \Text(225,70)[]{{$\mu^{-}$}}    
    \Text(188,120)[l]{$\theta_{d}$}
  \end{picture}
  \end{center}
  \caption{
	\label{fig:dylo}
	The Drell-Yan process for dimuon production.
	}
\end{figure}

Prior to the introduction of QCD, hadronic interactions were calculated with
some success using the quark-parton model, whereby the interacting partons
(quarks and antiquarks) were treated as 
being free of the strong interactions
on the timescale of a hard 
interaction (the impulse approximation).  Cross sections involving hadrons
were then reduced to the comparatively simple problem of calculating the 
parton-level cross sections and folding in the probability to find the 
particular
parton configuration.
Figure \ref{fig:dylo} shows the Feynman diagram for the Drell-Yan process 
for lepton pair production in proton-nucleon collisions in the dimuon channel 
studied by E866.  In the initial state, a quark (antiquark) carrying a 
fraction $x_1$ of the beam's momentum annihilates an antiquark (quark) 
carrying a fraction $x_2$ of the target's momentum.
This results in a time-like ($q^2>0$) intermediate photon, which decays 
into the final state $\mu^+\mu^-$ pair.  

\subsubsection{Kinematics}

In the lab, we measure the
invariant mass of the muon pair, and their total longitudinal momentum.
The mass of the dimuon is related to the momentum fractions of the interacting
partons by
\begin{equation}
q^2 = M^2 = s \, x_1 x_2
\label{eqn:dimu-mass}
\end{equation}
\noindent where $s$ is the total four-momentum squared of the interacting
hadrons.  The longitudinal momentum of the dimuon, as a fraction of its
maximum possible value ($p_L^{\text{max}}\approx\sqrt{s}/2$), 
is referred to as Feynman-$x$ ($x_F$).  It is related
to the initial state momentum fractions by 
\begin{equation}        
x_F = \frac{p_L}{p_L^{\text{max}}} = x_1 - x_2.
\label{eqn:dimu-xf}
\end{equation}
\noindent Thus, measurement of the invariant mass and $x_F$ of the muon
pair allows us to determine the momentum fractions carried by the interacting
partons.  Solving for $x_1$ and $x_2$ in equations \ref{eqn:dimu-mass} and
\ref{eqn:dimu-xf} we find
\begin{equation}
        \label{eqn:kinvert}
        x_{1,2} = \pm \frac{1}{2} ( x_F \pm \sqrt{x_F^2 + 4\tau} )
\end{equation}
\noindent where we have taken $\tau = x_1 x_2 = M^2 / s$.

In order to completely specify the final state of the interaction, we
need four more variables.  One of these is the momentum of the muon pair
transverse to the beam, denoted by $p_T$.  This is expected to be small
in the Drell-Yan process: $<p_T>$ $\approx$ 0.3 GeV, non-zero only due to 
Fermi smearing of the momenta
of the interacting partons.  The larger values of $p_T$ which are observed
in dilepton production require additional partons in the final state, which
are provided by the 
higher-order QCD processes which we discuss later.

\begin{figure}
\centering
\begin{picture}(300,100)(0,0)
\Text(0,90)[l]{$\mu^+\mu^-$ C.M.}
\Text(15,10)[l]{$pN$ plane}
\Line(0,0)(100,100)
\Line(100,100)(300,100)
\Line(300,100)(200,0)
\Line(200,0)(0,0)

\ArrowLine(60,50)(150,50)
\Text(70,60)[l]{$p$}

\ArrowLine(200,15)(150,50)
\Text(180,15)[l]{N}

\Line(150,50)(150,95)
\Text(130,90)[l]{$-\hat{y}$}

\DashLine(150,50)(240,50){5}

\Line(150,50)(200,35)
\Text(205,40)[l]{$\hat{z}$}

\LongArrow(150,50)(175,90) \Text(180,90)[l]{$\mu^+$}
\CArc(150,50)(15,-17,57)
\Text(167,58)[l]{$\theta_d$}

\end{picture}

\caption{
	\label{fig:CSFrame}
	Definition of the Collin-Soper frame.
	}
\end{figure}
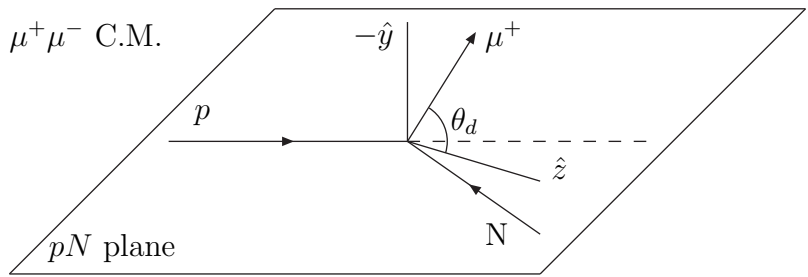

The three other variables which characterize the process are the polar
decay angle $\theta_d$, and the azimuthal production and decay angles
$\phi_p$ and $\phi_d$ respectively.  These variables are most naturally defined
with respect to the $q\bar{q}$ annihilation axis.  However, when $p_T$ is
non-zero, this axis becomes difficult to measure.  The angular variables 
are therefore measured with respect to the Collins-Soper frame 
\cite{bib:csframe}, which is shown schematically in figure \ref{fig:CSFrame}.
In this frame, the $z$-axis is taken to be  parallel to the bisector of the 
angle between the interacting hadrons in the rest frame of the virtual photon.


\subsubsection{Cross Section}

The cross section for the Drell-Yan process is easily constructed in the 
quark-parton model.  The assumption that the interacting quark and antiquark 
are free on the timescale of the hard interaction 
allows us to factor the cross section into a short distance $q\bar{q}$ 
annihilation term ($\sigma_{q\bar{q}}$), and a term expressing the probability
of finding the particular partonic configuration.  The $q\bar{q}$ annihilation
cross section is simply given by \cite{bib:DY}
\begin{equation}
\label{eqn:dyc-qqbar}
\sigma_{q\bar{q}} = \frac{1}{3}\frac{4\pi\alpha^2}{3M^2}e_q^2 
\end{equation}
\noindent which is just the cross section for the annihilation of two fermions
in QED, multiplied by a factor of $1/3$ required by color conservation
and a factor
of $e_q^2$ to account for the fractional charge of the quarks.  Multiplying
by the probability of a quark (antiquark) in the beam annihilating an 
antiquark (quark) in the target and summing over all parton species we have 
an expression for the Drell-Yan cross section differential in $x_1$ and $x_2$
\begin{equation}
\label{eqn:dyc-x1x2}
\frac{d^2\sigma_{AB}}{dx_1dx_2} =  \frac{1}{3}\frac{4\pi\alpha^2}{3M^2}
\sum_{q\in u,d,s,...} e_q^2 \left[  f_{q/A}(x_1)f_{\bar{q}/B}(x_2) + f_{\bar{q}/A}(x_1)f_{q/B}(x_2)  \right].
\end{equation}
\noindent Expressed in terms of quantities measured in the lab we obtain
\begin{equation}
\label{eqn:dyc-mxf}
M^2\frac{d^2\sigma_{AB}}{dM^2dx_F} =  \frac{1}{3} \frac{4\pi\alpha^2}{3M^2}
\frac{x_1x_2}{x_1+x_2}
\sum_{q} e_q^2 \left[  f_{q/A}(x_1)f_{\bar{q}/B}(x_2) + f_{\bar{q}/A}(x_1)f_{q/B}(x_2)  \right]
\end{equation}
\noindent where 
equation \ref{eqn:kinvert} is used to evaluate $x_1$ and $x_2$.
Alternatively, we can express the cross section in terms of the dimensionless
variables $\tau$ and $y = \frac{1}{2} \ln \frac{x_1}{x_2}$ (the rapidity)
\begin{equation}
  \label{eqn:dyscaling}
  s\frac{d^2\sigma_{AB}}{d\tau dy} = \frac{1}{3}\frac{4\pi\alpha^2}{3\tau} 
  \sum_{q} e_q^2 \left(  f_{q/A}(x_1)f_{\bar{q}/B}(x_2) + f_{\bar{q}/A}(x_1)f_{q/B}(x_2)  \right).
\end{equation}
\noindent This is known as the scaling form of the cross section.

\subsubsection{Predictions}


The Drell-Yan model makes some important predictions about the dilepton 
cross sections.  The left-hand side of equation \ref{eqn:dyscaling} is a
function of the variables $\tau$, $y$ and $s$, while the right-hand side is 
only a function of $\tau$ and $y$.  Thus, the model predicts that the 
cross sections multiplied by $s$, and measured at different 
beam energies, but the same kinematics,
should be the same -- the cross sections should scale with $\frac{1}{s}$,
and this is generally observed 
\cite{bib:FNAL-E288-SCALING,bib:FNAL-E615,bib:CERN-R-209}.


Another important prediction is the shape of the angular distributions.
The angular distribution of the $\mu^+$ in the pair rest frame can be 
expressed as \cite{bib:DY-THEORY-ANG}
\begin{equation}
\label{eqn:dyangfull}
\frac{d^2\sigma}{d\phi_d \;d\cos\theta_d} \propto 1 + \lambda \; {\cos}^2 \, \theta_d + \mu \; \sin \, 2\theta_d \; \text{cos} \, \phi_d + \nu / 2 \; {\sin}^2 \, \theta_d \; \cos \, 2\phi_d
\end{equation}
\noindent where $\lambda$, $\mu$ and $\nu$ are functions of the other kinematic
variables.  For the Drell-Yan process, the assumption of massless quarks
implies that the virtual photon is transversely polarized ($\lambda = 1$, 
$\mu = \nu = 0$), resulting in the prediction
\begin{equation}
\label{eqn:dyang}
\frac{d^2\sigma}{d\cos\theta_d} \propto 1 + {\cos}^2 \, \theta_d
\end{equation}
\noindent when $\theta_d$ is measured with respect to the $q\bar{q}$ 
annihilation axis.  
Several experiments 
\cite{bib:FNAL-E615,bib:E772-DY-POL,bib:E866-JPSI-POL,bib:E866-UP-POL} have 
confirmed that the dileptons are produced transversely polarized.

Despite these successes in describing experimental data, the Drell-Yan model
also had some initial failures.  Because there are no particles in the final state
off which to recoil, any transverse momentum in the dilepton must 
originate with the initial state.  The mechanism for this is most likely Fermi
smearing of the parton distributions, and the expected size of this 
intrinsic $k_T$ is small ($\sim 300$ MeV).  The observed ($s$-dependent) 
values of $\langle p_T \rangle \sim 1.2$ GeV cannot be reproduced by this mechanism
in the Drell-Yan process.

A second problem came about when absolute measurements of the dilepton
cross sections were compared to Drell-Yan calculations based on the
early PDF fits to DIS structure functions.  Table \ref{table:dykfactor}
shows the $K$-factors, defined as the ratio of measured to calculated
cross sections ($K = \sigma_{\text{meas.}} / \sigma_{\text{DY}}$),
from several different experiments using different beams and targets.
Typically, experiments measured a cross section which was a (nearly constant)
factor of two larger than the Drell-Yan model and DIS structure function 
measurements implied.  Along with the discrepancy in the $p_T$ distribution, 
this was taken as a sign that QCD would play an important role in dilepton 
production.

\begin{table}[!htb]

  \begin{center}
  \caption[Experimental $K$-factors.]{
    \label{table:dykfactor}
    \setlength{\baselineskip}{\singlespace}
    Experimental $K$-factors.}
    \vspace{10pt}
    \begin{tabular}{|lc|c|c|c|}
      \hline 
      \multicolumn{2}{|c|}{Experiment}  & Interaction & Beam Momentum          & $K = \sigma_{\text{meas.}} / \sigma_{\text{DY}}$ \\
      \hline\hline
      E288      & \protect\cite{bib:FNAL-E288} & $p$ $Pt$              & 300/400 GeV &  $\sim 1.7$ \\
      \hline
      WA39      & \protect\cite{bib:CERN-WA39} & $\pi^{\pm}$ $W$       & 39.5 GeV    &  $\sim 2.5$ \\
      \hline
      E439      & \protect\cite{bib:FNAL-E439} & $p$ $W$               & 400 GeV     &  $1.6\pm 0.3$ \\
      \hline
                &                      & ($\bar{p}$ - $p$)$Pt$ & 150 GeV     &  $2.3\pm 0.4$ \\ 
                &                      & $p$ $Pt$              & 400 GeV     &  $3.1 \pm 0.5 \pm 0.3$ \\
      NA3       & \protect\cite{bib:CERN-NA3}  & $\pi^{\pm}$ $Pt$      & 200 GeV     &  $2.3 \pm 0.5$ \\
                &                      & $\pi^{-}$   $Pt$      & 150 GeV     &  $2.49 \pm 0.37$ \\
                &                      & $\pi^{-}$   $Pt$      & 280 GeV     &  $2.22 \pm 0.33$ \\
      \hline
      NA10      & \protect\cite{bib:CERN-NA10} & $\pi^-$ $W$           & 194 GeV     &  $\sim 2.77\pm 0.12$ \\
      \hline
      E326      & \protect\cite{bib:FNAL-E326} & $\pi^-$ $W$           & 225 GeV     &  $2.70\pm 0.08\pm 0.40$ \\      
      \hline
      E537      & \protect\cite{bib:FNAL-E537} & $\bar{p}$ $W$         & 125 GeV     &  $2.45\pm 0.12 \pm 0.20$ \\
      \hline
      E615      & \protect\cite{bib:FNAL-E615} & $\pi^-$ $W$           & 252 GeV     &  $1.78\pm 0.06$ \\ 
      \hline
    \end{tabular}
  \end{center}
  
\end{table}


\subsection{QCD Modifications}
\label{sec:qcdmod}

\begin{figure}

\centering

Gluon Bremsstrahlung ($q\bar{q}g$)

\vspace{0.25in}

\begin{minipage}[c]{0.46\linewidth}
\centering
\begin{picture}(100,100)(0,0)
\ArrowLine(0,0)(50,25)
\ArrowLine(50,25)(50,75)
\ArrowLine(50,75)(0,100)
\Gluon(50,75)(100,100){5}{5}
\Photon(50,25)(100,0){5}{5}
\end{picture}
\end{minipage}
\begin{minipage}[c]{0.46\linewidth}
\centering
\begin{picture}(100,100)(0,0)
\ArrowLine(0,0)(50,25)
\ArrowLine(50,25)(50,75)
\ArrowLine(50,75)(0,100)
\Gluon(50,25)(100,0){5}{5}
\Photon(50,75)(100,100){5}{5}
\end{picture}
\end{minipage}

\vspace{0.25in}

Gluon Compton Scattering ($qg$ and $\bar{q}g$)

\vspace{0.25in}

\begin{minipage}[c]{0.46\linewidth}
\centering
\begin{picture}(100,100)(0,0)
\ArrowLine(0,0)(50,25)
\ArrowLine(50,25)(50,75)
\Gluon(50,75)(0,100){5}{5}
\ArrowLine(50,75)(100,100)
\Photon(50,25)(100,0){5}{5}
\end{picture}
\end{minipage}
\begin{minipage}[c]{0.46\linewidth}
\centering
\begin{picture}(100,100)(0,0)
\Gluon(0,0)(50,25){5}{5}
\ArrowLine(50,25)(50,75)
\ArrowLine(50,75)(0,100)
\ArrowLine(50,25)(100,0)
\Photon(50,75)(100,100){5}{5}
\end{picture}
\end{minipage}

\vspace{0.25in}

\begin{minipage}[c]{0.46\linewidth}
\centering
\begin{picture}(100,100)(0,0)
\Gluon(0,0)(25,50){5}{5}
\ArrowLine(0,100)(25,50)
\ArrowLine(25,50)(75,50)
\ArrowLine(75,50)(100,100)
\Photon(75,50)(100,0){5}{5}
\end{picture}
\end{minipage}
\begin{minipage}[c]{0.46\linewidth}
\centering
\begin{picture}(100,100)(0,0)
\ArrowLine(25,50)(0,0)
\Gluon(0,100)(25,50){5}{5}
\ArrowLine(75,50)(25,50)
\ArrowLine(100,0)(75,50)
\Photon(75,50)(100,100){5}{5}
\end{picture}
\end{minipage}

\vspace{0.25in}

Vertex correction ($q\bar{q}$)

\vspace{0.25in}

\begin{minipage}[c]{0.46\linewidth}
\centering
\begin{picture}(100,100)(0,0)
\ArrowLine(0,0)(50,50)
\ArrowLine(50,50)(0,100)
\Photon(50,50)(100,50){5}{5}
\Gluon(35,35)(35,65){3}{4}
\Gluon(20,20)(20,80){3}{9}
\Vertex(2,50){1}
\Vertex(7,50){1}
\Vertex(12,50){1}
\end{picture}
\end{minipage}

\caption[Next-to-leading order contributionsto the 
	dilepton cross section.]{
	\label{fig:DYNLO}
	\setlength{\baselineskip}{\singlespace}
	Next-to-leading order contributions to the 
	dilepton cross section.  The final state muon pair is omitted.
	}

\end{figure}
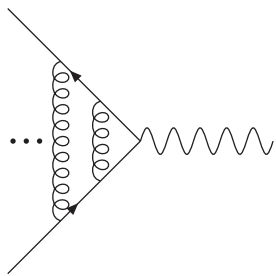

With the introduction of QCD, the basic physics behind the Drell-Yan process
was confirmed theoretically.  That quark-antiquark annihilation gives rise
to dilepton pairs is a consequence of the electromagnetic interaction.  
QCD provides substantial, albeit well-known, corrections to the
leading-order (LO) electromagnetic contribution described by 
Drell and Yan.  The diagrams which contribute at next-to-leading order 
(NLO or 
${\cal O}(\alpha_s)$) are shown in figure \ref{fig:DYNLO}.  The 
gluon-bremsstrahlung and vertex-correction diagrams represent corrections to 
the basic $q\bar{q}$ annihilation picture of the Drell-Yan process, increasing
the magnitude of the theoretical cross section and providing a mechanism to 
reach larger values of $p_T$.
Resummation techniques have been developed, which enable the calculation
of such corrections to all orders in $\alpha_s$.  
Taking $K$ equal to the resummed cross section divided by the LO contribution
gives
$K \approx 1.8$ for $\alpha_s$ evaluated at $q^2$ typical of fixed-target 
experiments.  This already accounts for much of the experimental $K$-factor.


The contributions described above are simply modifications to the basic
$q\bar{q}$ picture of the Drell-Yan process -- the main interaction is still
the electromagnetic annihilation of the interacting partons, with the strong
interaction only modifying the initial states of the $q\bar{q}$ pair.
The gluon Compton-scattering diagram, however, is fundamentally different.
Here, the incoming quark (antiquark) scatters off a gluon 
through the strong interaction, emitting the virtual photon which decays in 
the final state of the interaction.  
Theoretical studies of this process by Berger {\it et al.} \cite{bib:Berger} 
have shown that for transverse momenta large compared to the mass of the 
pair ($p_T > M/2$), the gluon Compton scattering diagrams dominate
the cross section (accounting for $\approx 80\%$ of the NLO calculation).
Therefore, with increasing $p_T$, the dilepton cross section
becomes less sensitive to the antiquark distributions, and more sensitive to
the gluons.


\subsection{Motivation}

The primary goal of Fermilab Experiment 866 (FNAL E866/NuSea) was to measure
the ratio of continuum dimuon cross sections in proton-deuteron ($\sigma_{pd}$)
interactions to those in proton-proton ($\sigma_{pp}$) interactions.  These 
measurements are sensitive to the $x$-dependence of the ratio $\bar{d}/\bar{u}$ 
in the nucleon, and were motivated by several experiments 
\cite{bib:NMC-gsr,bib:NA51,bib:E665-gsr} which demonstrated a large and 
previously unanticipated flavor asymmetry $\bar{d}/\bar{u} > 1$.  The E866 
results \cite{bib:E866-Hawker,bib:E866-Towell}, which were based on a large 
sample of dimuon events ($\approx 360,000$) induced by an 800-GeV proton beam 
($\sqrt{s} = 38.8$ GeV) interacting with hydrogen and deuterium targets, 
provided the first determination of the strong $x$-dependence 
of this flavor asymmetry.

Absolute measurements of the hydrogen and deuterium cross sections, the
topic of this thesis, provide additional information on the magnitude and
absolute shape of the light antiquarks.  This can be easily shown by writing 
the leading-order cross section expression in equation \ref{eqn:dyc-mxf}
for various targets.
Evaluating the sum for the hydrogen cross section, and taking the limit of 
large $x_F$ appropriate for much of the E866 data\footnote{\setlength{\baselineskip}{\singlespace}In the large $x_F$ 
limit, $x_1$ is sufficiently large compared to $x_2$ that we can neglect terms
in the cross section with the antiquark in the beam and the quark in the 
target.} we have
\begin{equation}
  \left(M^2\frac{d^2\sigma}{dM^2dx_F}\right)_{x_1 \gg x_2}
  \approx \frac{1}{3}\frac{4\pi\alpha^2}{3M^2} 
  \left(
  \frac{4}{9}u(x_1)\bar{u}(x_2) +
  \frac{1}{9}d(x_1)\bar{d}(x_2)
  \right).
\label{eqn:dylopp}
\end{equation}
\noindent Owing to the larger charge-squared factor in front of the 
$u\bar{u}$ term and the size of the up distribution compared to that
of the down distribution in the proton, about $80\%$ of the proton cross 
section at leading order is due to the $u\bar{u}$ annihilation.  Thus, to the 
extent
that the valence distributions are known (DIS measurements constrain these
well in our kinematic range), the proton cross section provides a direct 
measurement of the $\bar{u}(x)$ distribution.

The deuterium cross section can be similarly written.  
Nuclear effects in deuterium are small, and confined to small $x_2$
\cite{bib:E772-nucdep}.  This allows us to express the deuterium cross 
section as the sum
$\sigma_{pd} = \sigma_{pp} + \sigma_{pn}$.  If we assume that charge
symmetry holds, then the parton distributions in the proton and neutron
are related through $u(x) = d^n(x)$, $\bar{u}(x) = \bar{d}^n(x)$, etc....
This yields an expression for the deuterium cross section (per nucleon) 
\begin{equation}
\label{eqn:dylopd}
  \left(M^2\frac{d^2\sigma}{dM^2dx_F}\right)_{x_1 \gg x_2}
  \approx \frac{1}{3}\frac{4\pi\alpha^2}{3M^2} 
  \left\{
  \frac{4 u(x_1) + d(x_1)}{9}
  \right\}
  \left[
  \bar{d}(x_2) + \bar{u}(x_2)
  \right]
\end{equation}
\noindent where we have again taken the large-$x_F$ limit.
Thus, the deuterium cross section is directly proportional to the sum of the
light antiquarks $\bar{d}(x) + \bar{u}(x)$.


In this thesis we describe and report results on the absolute measurement 
of continuum dimuon production in 800-GeV $pp$ and $pd$ interactions.
The doubly-differential ($M^3d^2\sigma/dMdx_F$) and triply-differential
($Ed^3\sigma/dp^3$) cross sections have been measured over a large 
range in the mass ($4 \leq M \leq 16.85$ GeV), 
$x_F$ ($-0.05 \leq x_F \leq 0.8$) and 
$p_T$ ($0 \leq p_T \leq 7$ GeV) 
of the muon pair.  These data represent
the first such measurement of the $pp$ cross sections over an extended 
range of kinematics, while the $pd$ cross sections have been measured
over a wider kinematic range with greater statistical precision than has 
previously been achieved.

Traditionally, the Drell-Yan process is discussed in the context of
probing the light antiquark distributions in the proton.  And indeed, 
that was the primary motivation of this experiment.  But it is clear from 
equations \ref{eqn:dylopp} and \ref{eqn:dylopd} that dimuon production also 
provides sensitivity to the valence distributions.  While the kinematic range
covered by previous dimuon measurements \cite{bib:FNAL-E605} has been limited 
to regions where the valence quarks are well constrained by DIS measurements, 
the E866 data cover a much wider kinematic range.  Our data probe the valence
out to $x\sim 0.8$, which provides an important measurement of the large-$x$
behavior of the valence quarks complementary to existing DIS measurements
\cite{Zhang:2001hr}.

\newpage
\section{APPARATUS}
   \label{chapter:apparatus}

\begin{figure}[!htb]
  \begin{center}
  \includegraphics[width=0.9\linewidth,clip]{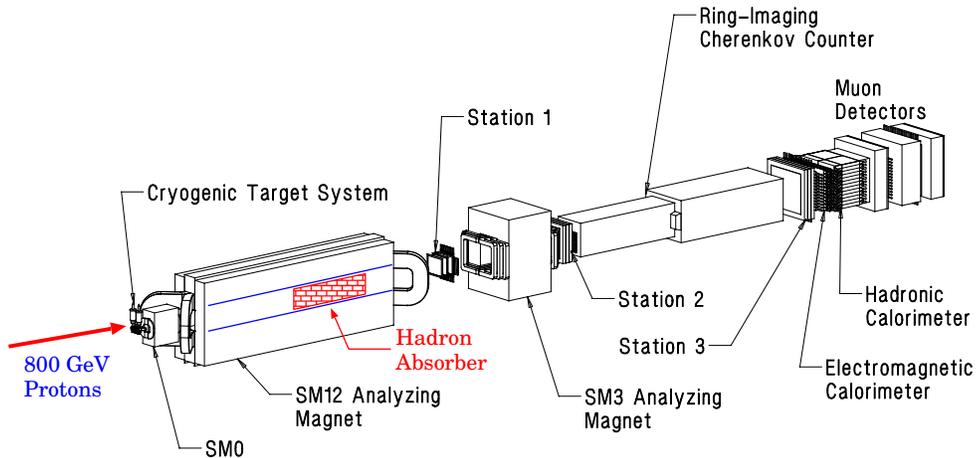}
  \end{center}
  \vspace{-0.5in}
  \caption{  \label{fig:MEAST_SPECTROMETER}
	FNAL E866/NuSea Spectrometer.
	}

\end{figure}

Fermilab Experiment 866 used a modified version of the dimuon spectrometer 
\cite{bib:MEAST-SPECTROMETER} located in the Meson East experimental area at the 
Fermi National Accelerator Laboratory.  Previous experiments which used this 
spectrometer were E605, E772 and E789.  Figure \ref{fig:MEAST_SPECTROMETER}
shows a drawing of the E866 spectrometer, which was designed to detect 
oppositely-charged muon pairs while minimizing sensitivity to everything else. 
The coordinate system of the spectrometer was defined in the following way: 
the $z$-axis of the spectrometer was parallel to the nominal direction of the 
beam; the $x$-axis laid parallel to the floor of the experimental hall, 
pointing to the left facing station-1 from the targets; the $y$-axis 
then formed the right-handed Cartesian coordinate system by pointing up, 
perpendicular to the floor of the experimental hall.  The origin of the 
coordinate system was chosen as the center of the upstream face of the SM12 
magnet.

\subsection{Beam line}
   \label{apparatus_beamline}

An 800-GeV proton beam was extracted from the Fermilab Tevatron, split 
at the switchyard and sent to several fixed target experiments which 
ran concurrently
during the 1996 fixed target run.
The beam was transported down the east beam line of the Meson 
experimental area and arrived at the target in 20-second-long spills, 
with new spills arriving once every minute.  The accelerator's frequency 
was 53 MHz, which meant that during the spill the protons arrived 
grouped into ``buckets'' separated by 19 ns.  The typical 
intensities which were used for E866 were between $1\times 10^{11}$ and 
$2\times 10^{12}$ protons per spill, depending on trigger rates and 
radiation-safety requirements for a particular spectrometer configuration.

The position, size and intensity of the beam was constantly monitored 
at various points along the beam line by several detectors.  The beam's 
position and size were monitored by 
segmented wire ion chambers (SWICs).  
The last SWIC in the beam line was located approximately 188 cm upstream 
of the target.   It had a 2 mm horizontal wire spacing and a 0.5 mm vertical 
wire spacing.  During the experiment, the beam spot on the target was 
approximately 6 mm wide by 1 mm high.

Three different types of detectors monitored the intensity of the beam.  The
primary detector was a secondary-emission monitor (SEM), located at about
100 m upstream of the targets.  In addition to the SEM counter, an ion chamber
(IC3) and a quarter-wave RF cavity were used.  These additional detectors 
allowed us to study the linearity and offsets of the SEM.  


The luminosities of the targets were also
monitored during the experiment with a pair of four element scintillator 
telescopes located about
$85 \deg$ from the beam direction in the horizontal plane.  The data from
these detectors (referred to as AMON and WMON) were used to determine 
beam duty factor, data acquisition live-time and which target was in the
beam during the spill.

\subsection{Targets}
   \label{apparatus_targets}

\begin{figure}

  \begin{center}
     \includegraphics[clip,width=4in,viewport=150 75 500 600,angle=-90]{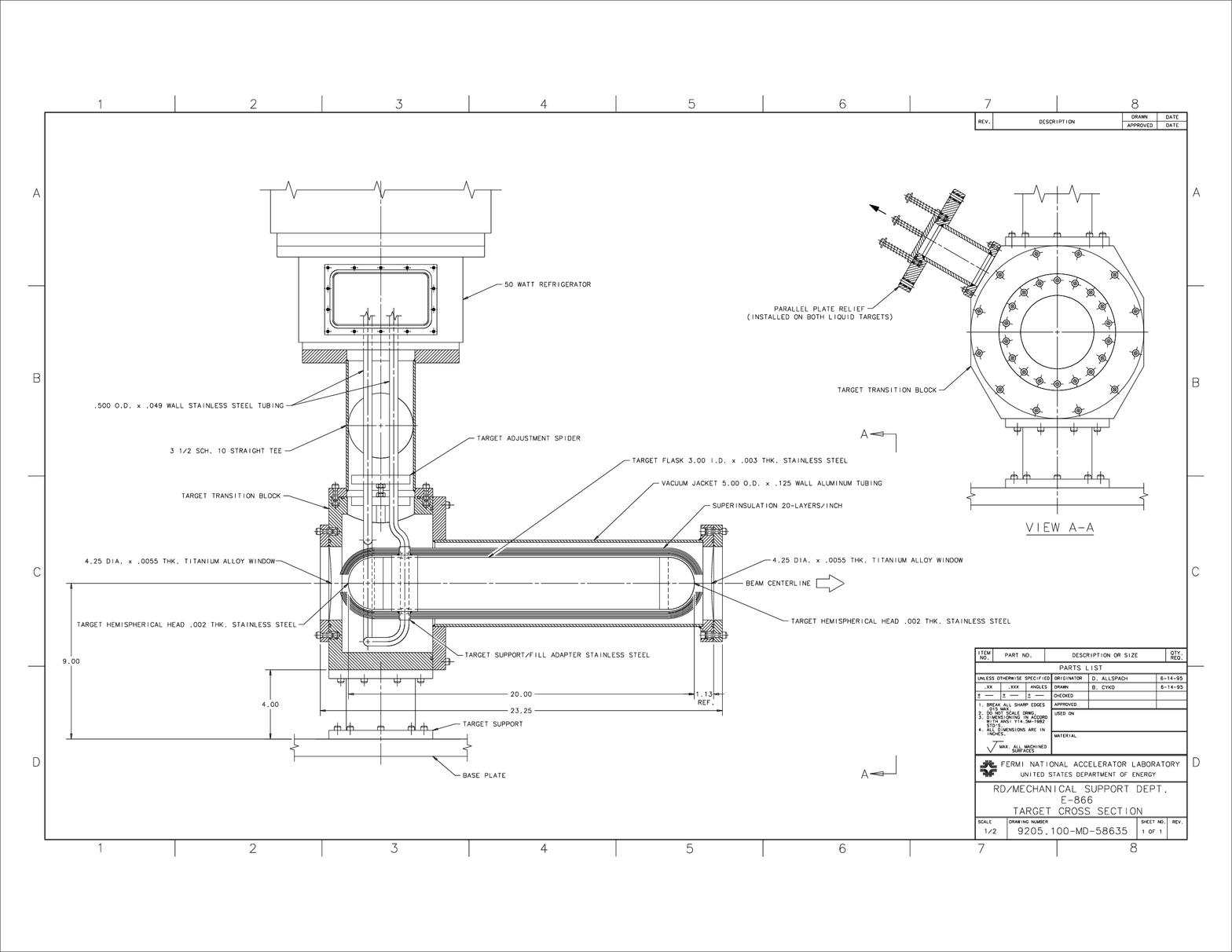}
  \end{center}
  \caption{\label{apparatus_targetflask_figure}One of the three identical target flasks used in E866.}
\end{figure}

Three identical target flasks, one of which is shown in figure 
\ref{apparatus_targetflask_figure}, were mounted on a movable table 
located at $z \approx -400$ cm.  They consisted of a 17-inch long 
cylindrical shell with a 3-inch inside diameter.  Hemispherical end caps
with a 3-inch diameter sealed either end, creating a 20-inch 
(50.8-cm) long
target along the beam axis.\footnote{\setlength{\baselineskip}{\singlespace}
It was confirmed by visual inspection after the experiment that
the $z$-axis of each target was aligned properly with the beam.}  Facing 
the targets from upstream, the leftmost target was filled with liquid hydrogen
and the rightmost target was filled with liquid deuterium.  The center target 
was empty and held at vacuum to  measure the background rates due to the beam 
interacting with the SWIC, the material in the target flasks, and the air and 
other materials downstream of the target.  Immediately downstream of the 
target, filling the aperture of the SM0 magnet, was a bag filled with helium 
to reduce background interactions in the vicinity of the targets.

The targets were cycled periodically to minimize the systematic 
uncertainties between the hydrogen and deuterium measurements.  The programed 
sequence was 5 spills on the deuterium target followed by 1 spill on the empty 
target, 5 spills on the hydrogen target, and one more spill on the empty 
target.  The target control computer counted only ``good'' spills -- spills 
which the DAQ determined had a sufficiently large number of protons.
Upon receipt of the appropriate number of good spill signals, which were sent 
after the end of the spill, the target-control computer initiated the move to 
the next target in the sequence.  The move was completed before the beginning
of the next spill.  

A set of four switches on the target table were engaged by a stationary
roller mounted in front of the table.  The positions of these switches 
corresponded to the beam striking the center of each of the targets, plus a 
fourth position where the targets were clear of the beam.  When engaged, each 
switch completed a circuit which was read out by the DAQ, recording which
target was in the beam.  The switches were also tied into the beam interlock 
system, which required that one of the four switches be engaged for beam to 
be allowed to enter the target area.  This ensured that the beam would not 
be able to hit the sides of the target flasks and cause a radiation
hazard.

\subsection{Spectrometer Magnets}

The E866 spectrometer used three dipole magnets whose magnetic fields were
oriented in the $x$ direction (bending charged particles in the $y$ direction).
The two magnets closest to the targets were SM0 and SM12.  These magnets were 
used to optimize the acceptance of the spectrometer for different ranges of 
dimuon masses.  The SM12 magnet was used to focus dimuon pairs on the 
downstream detector elements.  Access to higher-mass events, which tend
to have larger opening angles, was possible by using larger magnetic fields
in SM12.  This would ensure that such events were kept inside the aperture
of the SM12 magnet and the detector stations.

The SM0 magnet was used to optimize the spectrometer for lower mass
dimuon events in the low-mass data set.  In this data set, the SM0 magnet
was used to defocus (increase the opening angle of) the $\mu^+\mu^-$ pairs 
before they entered SM12.  This increased the probability that
the muons would miss the dump.  More importantly, this increased the
probability that the events would miss the dump cut used
to eliminate beam-like muons, which were indistinguishable from the
high flux of muons produced in the dump.


The SM3 magnet was located between the station-1 and station-2 detectors.
It was used in the experiment to analyze the momentum of the muons by
measuring the deflection it produced in the tracks.  It was operated using 
a single current of $4230$ A throughout the experiment.

%
%
%
%
%

\subsection{Beam Dump and Hadronic Absorbing Wall}

In order to prevent damage to the downstream detectors from the beam 
the target was followed by a large, water-cooled beam dump whose upstream
face was located at $z \approx 173 $ cm.  The beam dump consisted of 255 cm 
of copper, which spanned -15.25 cm $\leq y \leq$ 15.25 cm at its thickest
point.  At its widest, the dump blocked the entire x-aperture of the magnet.
The beam dump was nominally centered on the $z$-axis, but had settled over
the years and developed a small slope relative to SM12 such that it was
approximately $0.2''$ low at $z = 68''$ and $0.4''$ low at $z = 530''$.

\begin{figure}
 
  \begin{center}
     \includegraphics[width=0.9\linewidth]{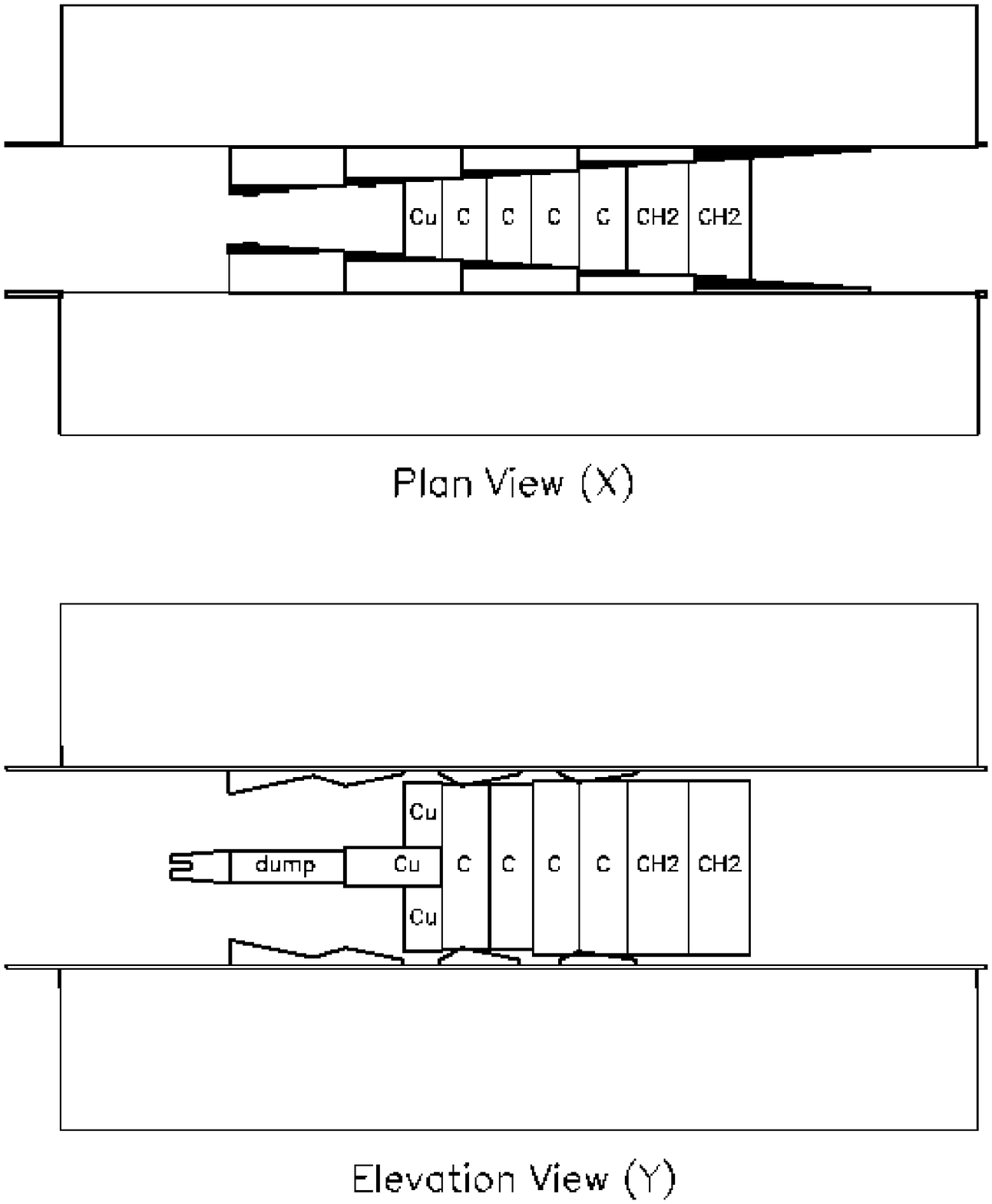}
  \end{center}
  \caption{ \label{apparatus_absorber_figure}Side and elevation views of the absorbing wall and beam dump.}
\end{figure}

A large hadronic 
absorbing wall was located immediately downstream of the beam dump.
The purpose of this wall was to fully attenuate the intense flux of
hadrons produced in the target and the dump.  The absorbing wall consisted of
61 cm of copper\footnote{\setlength{\baselineskip}{\singlespace}
There was a 7.6 cm gap between the copper and
carbon sections of the absorbing wall in the region behind the dump.}
 in the most upstream section, followed by a 205.8 cm section 
of carbon, a 68.6 cm long section containing both carbon and polyethylene, 
and a 183 cm section of polyethylene.  The configuration
of the beam dump and absorbing wall is shown in figure 
\ref{apparatus_absorber_figure}.

\subsection{Detector Stations}
\enlargethispage{\baselineskip}

Four detector stations were responsible for triggering on dimuon events
and tracking the muons which traversed the spectrometer.  The first three 
detector stations consisted of hodoscope- and drift chamber-planes, while 
station-4 consisted of hodoscope- and proportional-tube planes.  The fast 
response time of the hodoscopes was utilized to trigger on events with the 
expected dimuon signature, at which point the slower responding drift chambers
could be read out.  The proportional-tubes and hodoscopes at station-4 were 
used to provide some position information to the trigger, and 
discriminate against any hadrons which made it through the absorbing wall.

\subsubsection{Hodoscope Planes}

Hodoscope planes were located at each of the four detector stations and were
used to trigger on dimuon events.  Each hodoscope plane was split into two 
half-planes of parallel scintillator paddles attached to photomultiplier tubes 
by plexiglass light guides.  Stations 1, 3 and 4 each had two 
hodoscope planes, with their paddles oriented in the X and Y directions 
respectively.  Station-2 had only one hodoscope plane, with its paddles 
oriented parallel to the floor.  A vertical gap between the Y half-planes
and the central X hodoscopes was inserted in order  to avoid triggering on 
the high rate of beam-like muons from the decay of pions which were produced
copiously in the dump.  
The design specifications of the hodoscope planes can be found in table 
\ref{apparatus_hodoscope_layout}.  A more precise alignment of the hodoscopes 
was achieved by examining the distributions of positions of muon tracks 
at each hodoscope plane when a given hodoscope in that plane fired.

\begin{table}
  \caption[Specifications of the hodoscope planes.]{\label{apparatus_hodoscope_layout}
	\setlength{\baselineskip}{\singlespace}
	Specifications of the hodoscope planes.  Distances measured in inches.
        }
  \begin{center}
    \begin{tabular}{|c|c|c|c|c|c|c|}
      \hline
      detector & z position & \# of counters & counter width & gap & $\Delta x$ & $\Delta y$ \\
      \hline\hline
      Y1 &  769.78 & 16 & 2.5   & 0.38 & 47.50  & 40.75 \\
      X1 &  770.72 & 12 & 4.0   & 0.47 & 47.53  & 40.78 \\
      Y2 & 1114.94 & 16 & 3.0   & 0.66 & 64.625 & 48.625 \\
      X3 & 1822.00 & 12 & 8.68  & 1.0  & 105.18 & 92.00 \\
      Y3 & 1832.00 & 13 & 7.5   & 0.0  & 104.00 & 92.00 \\
      Y4 & 2035.50 & 14 & 8.0   & 0.0  & 116.00 & 100.00 \\
      X4 & 2131.12 & 16 & 7.125 & 0.0  & 126.00 & 114.00 \\
      \hline
    \end{tabular}
  \end{center}

\end{table}

\subsubsection{Drift Chambers}
Each of the first three detector stations contained six drift-chamber planes,
arranged in three pairs of planes with parallel wire orientations (referred to
as ``views'').  Wires were oriented horizontally in the ``Y'' view, and at 
an angle of $+14$\deg \ and $-14$\deg \ with respect to the X axis in the ``V'' and 
``U'' views, respectively.  
The wires in the second plane in each pair were offset
by half the cell size of the drift chamber.  The plane in each pair closest 
to the target was denoted as the ``unprimed'' plane, while the plane in the 
pair furthest from the target was denoted as the ``primed'' plane.  These 
planes provide information on the X and Y intercept of the muon tracks at the 
detector station, with redundant information about the Y position.

The drift chambers were operated with a gas mixture of 49.7\% argon, 49.6\%
ethane and 0.7\% ethanol, mixed by volume at a constant temperature of 
25\deg F.  The anode wires at Station-1 were gold-plated tungsten wire.
Stations 2 and 3 used silver-coated beryllium-copper wires as anodes.  All
of the anodes were 25 $\mu$m in diameter.  The cathode wires were all 
62.5-$\mu$m silver-coated beryllium-copper wire.  The drift chambers were 
operated at voltages between 1700 and 2200 volts.  Typical drift
velocities were 
$\approx$ 50 $\mu$m/ns.  
Drift chamber specifications can
be found in table \ref{apparatus_chamber_specifications}.

The signals from the drift chambers were read out by a fast amplifier and
discriminator system.  Single-hit time-to-digital converters (TDCs), which 
only record the first hit on the wire during an event, were used to measure 
the drift time.  The combination of good hits together with their associated 
drift times in all three views result in a ``triplet'' hit in each station.
The bank of the triplets was saved to provide information used later in
reconstructing the muon tracks in the analysis.

\begin{table}
  \caption[Specifications of the drift chamber planes.]{
  \label{apparatus_chamber_specifications}
	\setlength{\baselineskip}{\singlespace}
	Specifications of the drift chamber planes.  Distances measured in inches.
	}
  \vspace{10pt}
  \begin{tabular}{|c|c|c|c|c|c|}

    \hline
    detector & Z-position & \# of wires & cell size & aperture(X$\times$Y) & operating voltage\\
    \hline\hline
    V1           &     724.69  &    200  &        0.25   &    48$\times$40      &    +1700\\
    V1$^\prime$  &     724.94  &    200  &        0.25   &    48$\times$40      &    +1700\\
    Y1           &     740.81  &    160  &        0.25   &    48$\times$40      &    +1700\\
    Y1$^\prime$  &     741.06  &    160  &        0.25   &    48$\times$40      &    +1700\\
    U1           &     755.48  &    200  &        0.25   &    48$\times$40      &    +1700\\
    U1$^\prime$  &     755.73  &    200  &        0.25   &    48$\times$40      &    +1700\\
    \hline
    V2           &    1083.40  &    160  &        0.388  &    66$\times$51.2    &    $-$2000\\
    V2$^\prime$  &    1085.52  &    160  &        0.388  &    66$\times$51.2    &    $-$2000\\
    Y2           &    1093.21  &    128  &        0.40   &    66$\times$51.2    &    $-$2000\\
    Y2$^\prime$  &    1095.33  &    128  &        0.40   &    66$\times$51.2    &    $-$2000\\
    U2           &    1103.25  &    160  &        0.388  &    66$\times$51.2    &    $-$1950\\
    U2$^\prime$  &    1105.37  &    160  &        0.388  &    66$\times$51.2    &    $-$1975\\
    \hline
    V3           &    1790.09  &    144  &        0.796  &    106$\times$95.5   &    $-$2200\\
    V3$^\prime$  &    1792.84  &    144  &        0.796  &    106$\times$95.5   &    $-$2150\\
    Y3           &    1800.20  &    112  &        0.82   &    106$\times$91.8   &    $-$2200\\
    Y3$^\prime$  &    1802.95  &    112  &        0.82   &    106$\times$91.8   &    $-$2200\\
    U3           &    1810.24  &    144  &        0.796  &    106$\times$95.5   &    $-$2200\\
    U3$^\prime$  &    1812.99  &    144  &        0.796  &    106$\times$95.5   &    $-$2200\\
    \hline

  \end{tabular}

\end{table}

\subsubsection{Proportional Tubes}
The detectors at station-4 provided both trigger-information and muon 
discrimination capabilities.  Located downstream of the  
electromagnetic and hadronic 
calorimeters,\footnote{\setlength{\baselineskip}{\singlespace}
The calorimeters were left over 
from a previous experiment.  For E866, they were unnecessary and were only 
used to eliminate any hadrons which made it through the absorbing wall.}
station-4 consisted of two hodoscope 
planes (X4 and Y4) and three proportional-tube planes (PTY1, PTX and PTY2).  
Each of the proportional-tube planes had two layers of $1\times 1$-inch cells. 
Adjacent layers were offset by half a cell to cover the dead region between 
adjacent cells.  The proportional-tubes were operated using the same gas 
mixture used in the drift chambers.   To reduce the probability of any hadrons 
making it through the calorimeters, a 3 foot wall of zinc and a 4 inch wall of
lead were placed between the calorimeters and the station-4 detectors.  
Furthermore, 3 feet of concrete was placed between PTY1 and X4 and between 
PTX and PTY2.  This provided a total of 16.6 hadronic-interaction lengths 
upstream of the Y4 plane.  The only detectable particles which could reach 
the station-4 detectors were muons.  Specifications of the proportional-tubes 
may be found in table \ref{apparatus_proptube_specifications}.

\begin{table}
  \caption[Specifications of the proportinal tube planes.]{
	\label{apparatus_proptube_specifications}
	\setlength{\baselineskip}{\singlespace}
	Specifications of the proportinal tube planes. Distances measured in inches.
	}
  \vspace{10pt}
  \begin{center}
    \begin{tabular}{|c|c|c|c|c|}
      \hline
      detector & Z-position & No.of wires & cell size & aperture X$\times$Y\\
      \hline\hline
      PTY1  &    2041.75  &   120     &     1.0   &     117$\times$120\\
      PTX   &    2135.875 &  135      &     1.0   &     135.4$\times$121.5\\
      PTY2  &    2200.75  &   143     &     1.0   &     141.5$\times$143\\
      \hline
    \end{tabular}
  \end{center}
\end{table}

\subsection{Trigger}
\label{chapter:apparatus:trigger}
Despite the fact that the absorbing wall all but eliminated the rate
of hadrons produced in the dump, the rates of single and dimuon events
from both the target and dump were too large for the drift chambers to
be read out on every muon which was detected.   Therefore the information from
the hodoscopes was used to trigger on events which had the expected signature 
of a dimuon event originating in the target.  The trigger was also 
configured to take prescaled samples of events which would be used later to 
reconstruct the combinatoric background.  

\begin{landscape}

\begin{figure}
  \begin{center}
 \includegraphics[width=0.9\linewidth]{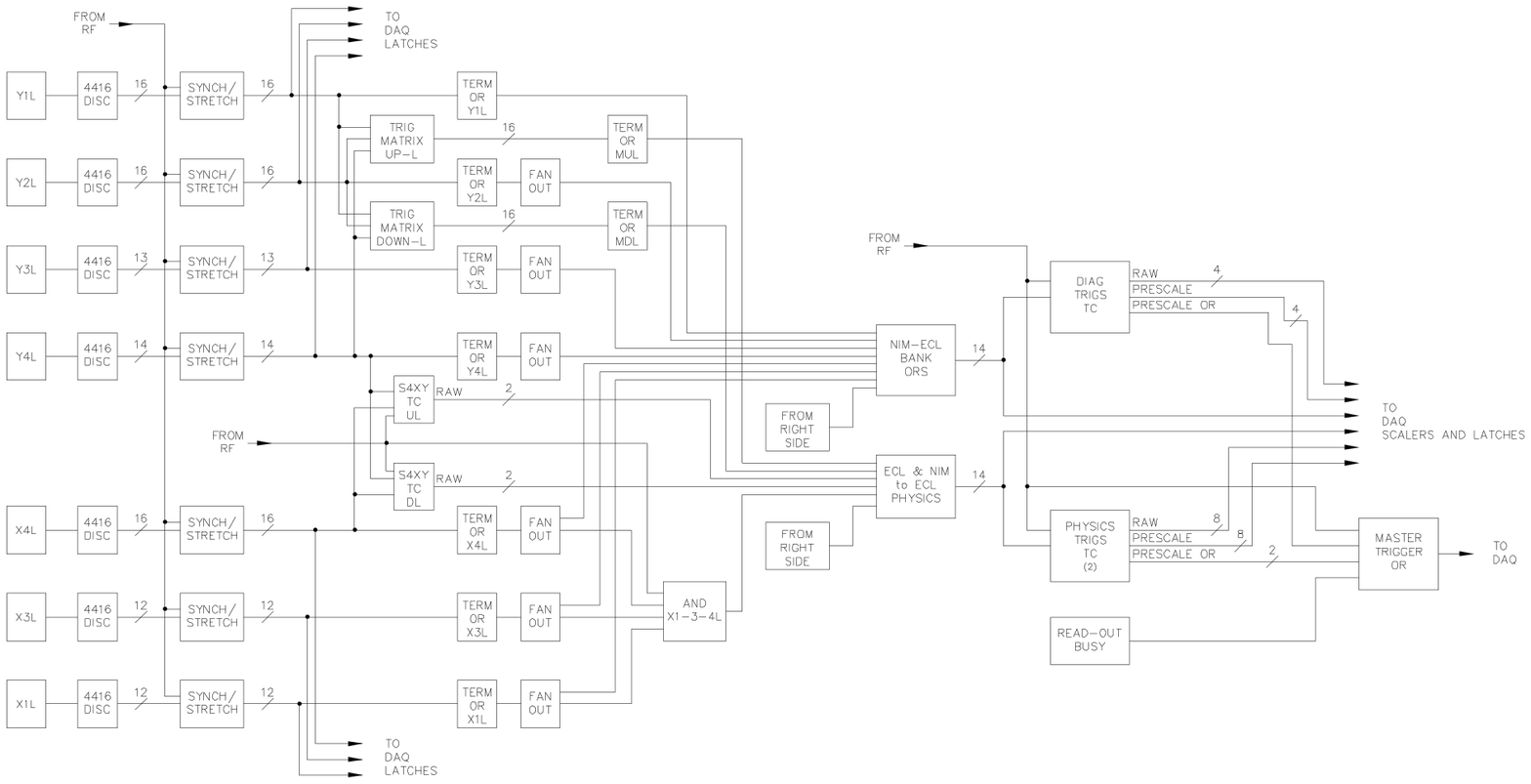}
  \end{center}
  \caption[Block diagram of the E866 trigger.]{  \label{apparatus_trigger_figure}
	\setlength{\baselineskip}{\singlespace}
	Block diagram of the E866 trigger.  Only the inputs and electronics
    	for the left side of the spectrometer are shown.
	}
\end{figure}

\end{landscape}

The trigger system used in E866 was upgraded from that used for previous
experiments \cite{bib:E866-Trigger}.  The purpose of the upgrade was to increase
the flexibility of the trigger, and to improve the acceptance of high
$p_{T}$ dimuon events.  The main elements of the trigger system were the 
Trigger Matrix Modules and the Track Correlators.  Figure 
\ref{apparatus_trigger_figure} shows a diagram of the trigger electronics for
the left hand side of the spectrometer.

The Trigger Matrix Modules were responsible for determining whether a track
was likely to be from a dimuon event which originated at the target.  The 
outputs of the Y hodoscopes at stations 1, 2 and 4 were input into the 
Trigger Matrix Modules.  The hit patterns registered
by the hodoscopes were tested against a preselected set of hit patterns
which corresponded to a muon from a dimuon event in the target
traversing the spectrometer.  These hit patterns where determined from
Monte Carlo studies prior to the experiment.  Some matrix elements had 
considerable rates of muons passing through them which actually originated in 
the dump.  In order to take larger rates of target events, these matrix 
elements were turned off.

Four sets of Trigger Matrix Modules were used in the trigger system.  These
were referred to as matrix-up-left (MUL), matrix-down-left (MDL), 
matrix-up-right (MUR) and matrix-down-right (MDR).  They corresponded to
single muons on the left- or right-hand side of the spectrometer which
went above or below the dump.  These, along with the
outputs of the Terminator/OR's and the Station-4 XY (S4XY) Track Correlators, 
were the inputs to the Track Correlators.  The job of the Track Correlators 
was to filter the inputs and determine if a trigger should be fired and the 
event written to tape. There were three Track Correlators capable of 
triggering an event: Physics A, Physics B and Diagnostic.

Physics A (PhysA) was the main physics Track Correlator.  Like all of the 
Track Correlators, it could handle four sets of trigger conditions.  Table 
\ref{apparatus_trigger_specifications} shows the triggers that were used
in each of the data sets in this analysis. 
The triggers in each of the data sets performed the same functions.
The PhysA1 trigger selected oppositely-charged muons which traversed opposite 
sides of the spectrometer, while the PhysA3 and PhysA4 triggers selected 
oppositely-charged muons which traversed the same side of the spectrometer.  
The PhysA2 trigger was the like-sign trigger.  It selected same-sign muon 
pairs on opposite sides of the spectrometer.

In addition to the requirements on the charge and geometry of the muons
applied by the matrix, the Track Correlators used in the low- and 
intermediate-mass data sets added constraints on the X and Y positions of 
the muons at station-4 (the S4xxx entries in the table.)  These constraints 
were designed to veto events which populated areas where the dump dominated 
the event rate.

The Physics B (PhysB) track correlator was used to sample events which were
used to monitor the efficiencies and systematics of the detector.  An example 
of one of the many B triggers used in the experiment appears in table 
\ref{apparatus_trigger_specificationsb}.  The PhysB1 trigger was designed to 
read out any event with muons 
traveling along paths on opposite sides of the spectrometer, regardless
of their relative sign.  The PhysB2 trigger was often configured to trigger on
events with a single muon traveling along one of the predetermined matrix roads
that a muon in a dimuon event would follow.  These data were used to extract
the combinatoric background, as will be discussed later.

\begin{table}
  \caption[Trigger configuration (PhysA).]{
	\label{apparatus_trigger_specifications}
	\setlength{\baselineskip}{\singlespace}
	Trigger configuration (PhysA).  Configurations shown for the low-, 
	intermediate- and high-mass data.    The symbols ``*'', ``+'' and 
	``!'' correspond to logical AND, OR and NOT, respectively.
	}
  \begin{center}
    \begin{tabular}{|c|c|c|c|}
      \hline
      Mass Setting & Number & Prescale & Requirements \\      
      \hline\hline      
      low & PhysA1 & 1 & (MUL * MDR) + (MUR * MDL)\\
          & PhysA2 & 4 & (MUL * MUR) + (MDL * MDR)\\
          & PhysA3 & 1 & (MUL * MDL) * (S4UL2 * S4DL2)\\
          & PhysA4 & 1 & (MUR * MDR) * (S4UR2 * S4DR2)\\
      \hline\hline
      int & PhysA1 & 1   & ( (MUL * MDR) * ( !S4DL1 + !S4UR1 ) ) + \\
          &        &     & ( (MUR * MDL) * ( !S4DR1 + !S4UL1 ) )   \\
          & PhysA2 & 2,1 & ( (MUL * MUR) * ( !S4DL1 + !S4DR1 ) ) + \\
          &        &     & ( (MDL * MDR) * ( !S4UL1 + !S4UR1 ) )   \\
          & PhysA3 & 1   & ( (MUL * MDL) * \\
          &        &     & ( !S4UL1 + !S4DL1 ) * ( S4UL2 * S4DL2 ) )\\
          & PhysA4 & 1   & ( (MUR * MDR) * \\
          &        &     & ( !S4UR1 + !S4DR1 ) * ( S4UR2 * S4DR2 ) )\\
      \hline\hline
      high & PhysA1 & 1 & (MUL * MDR) + (MUR * MDL)\\
           & PhysA2 & 4 & (MUL * MUR) + (MDL * MDR)\\
           & PhysA3 & 1 & (MUL * MDL)\\
           & PhysA4 & 1 & (MUR * MDR)\\
      \hline 
    \end{tabular}
  \end{center}

\end{table}

\begin{table}
  \caption[Trigger configuration (PhysB).]{
	\label{apparatus_trigger_specificationsb}
	\setlength{\baselineskip}{\singlespace}
	Trigger configuration (PhysB).  An example of a PhysB trigger used
	in the high-mass data.
    	The symbols ``*'', ``+'' and ``!'' correspond to logical AND, OR
    	and NOT, respectively.
	}
  \begin{center}
    \begin{tabular}{|c|c|c|c|}
      \hline
      Mass Setting & Number & Prescale & Requirements \\      
      \hline\hline      
      high & PhysB1 & 800 & X134L * X134R \\
           & PhysB2 & 1000 & ( MUL + MDL + MUR + MDR ) \\
           & PhysB3 &   -  & - \\
           & PhysB4 &   -  & - \\
      \hline 
    \end{tabular}
  \end{center}

\end{table}

\subsection{Data Acquisition System}

The E866 data acquisition system (DAQ) was essentially an upgraded 
version of the system used in E789.  The DAQ had three areas of 
responsibility: event readout, data archiving and online analysis.  
The backbone of the event readout was the Nevis Transport System.  Upon 
receipt of a signal from one of the Track Correlators, a busy signal was
raised which inhibited further triggers from being accepted.  Simultaneously, 
the first word of the event was inserted onto 
the transport bus.   Signals
were sent to the TDC readouts on the drift chambers and the coincidence
registers (CR's) on the hodoscopes.  When signaled, the TDC's and CR's began 
digitizing their signals for insertion onto the bus.  Each 
hodoscope and proportional-tube hit resulted in the insertion of an 
identifying word on the bus.  The TDC's would also start a timer
which would be terminated by the amplified signal from the drift chamber.
This provided a measurement of the drift time in the chamber, which was 
inserted  onto the transport bus along with the number of the struck wire.

At this point, the data stream on the transport bus was fed into the Versa
Module Eurocard (VME) based archiving system.  During the 20s spill cycle, 
data were first transported to a pair of high-speed memory boards.  Once one 
of the boards was full, it was drained into a large memory buffer while the 
other board was filled.  Upon completion of the spill, the data were then 
formatted and sent down the VME pipeline to the tape archiving system.  Here
the data were written out to 8mm Exabyte tapes.  Additionally, a fraction of
the event data and all of the spill information (number of triggers fired, 
SEM counts, target position, etc...) were sent to the online monitoring
systems.  

\enlargethispage{\baselineskip}

The online monitoring system consisted of a
database system interfaced
to several graphical tools, and the E866 online analysis code.  The database
system allowed us to monitor the status of various components of the 
beamline and spectrometer.  Graphical displays of the luminosity, magnet
voltages, livetimes, etc...,
gave an overall indication of the health of the
spectrometer.  The online analysis code analyzed the data that were sent to
it via the VME pipeline.  The histograms generated by the analysis code 
were accessible in real time by making use of PAW (Physics Analysis Workstation) 
\cite{bib:POS} global 
sections.  This allowed
us to monitor the detector planes, watching for and correcting any 
inefficiencies which might develop.


\newpage
\def\ydumpcut{10.2 cm}
\section{EVENT RECONSTRUCTION}
   \label{chapter:analysis}


By the time the experiment ended, approximately $250 \, \text{GB}$ of data
had been recorded to tape.  The data on these tapes represented the hits
on each chamber and hodoscope plane for each event, the general 
characteristics of the beam in any given spill and the overall state of the
spectrometer.  The reconstruction of the kinematics of the events from these
raw data required several steps.  After configuring the analysis for the
specific data set, the tracks were reconstructed from the chamber hits.
Once the tracks had been found, their paths through the SM0 and SM12
magnets had to be reconstructed while accounting for energy loss
and multiple scattering in the absorbing wall and beam dump.
These procedures are outlined in greater detail below.

\subsection{Data Sets}
\label{analysis_data_sets}

During the nine months starting in September 1996 in which FNAL E866/NuSea
took data, over 
360,000 continuum dimuon events 
were written to tape.  
These events were grouped into several sequentially numbered data sets 
which were differentiated by the polarities and currents used to energize
the three spectrometer magnets.  Each data set fell into one of three
categories, defined by the mass range for which the spectrometer was
optimized.  These categories were referred  to as the low-, intermediate- 
and high-mass data, corresponding to
spectrometer settings optimized for masses near the J$/\psi$, between the 
J$/\psi$ and $\Upsilon$ masses, and near the $\Upsilon$ mass, respectively.
Table \ref{table:datasets} lists the data sets used in this analysis and the
mass setting to which they belong.  It also tabulates the currents
used to power the spectrometer magnets, and whether the targets were filled
with either the first or second sample of cryogenic liquids.  

\begin{table}[h] 
   \centering
   \caption[Definition of the data sets.]{
	\label{table:datasets}
	\setlength{\baselineskip}{\singlespace}
	Definition of the data sets.  The SM3 magnet was operated at
	a single current of 4230 A, with the same polarity as the SM12
        current.
	}
   \vspace{10pt}
   \begin{tabular}{|c|c|c|c|c|}
     \hline
     mass setting   & 
     data set       & 
     SM0 current    & 
     SM12 current   & 
     target fill \\
     \hline \hline      
     low          & 5   & $-2100$ A  &  $2800$ A  &  first   \\
                  & 10  &  $2100$ A  & $-2800$ A  &  second  \\
     \hline
     intermediate & 9   &  $0$ A     &  $2800$ A  &  second  \\
     \hline 
     high         & 7   &  $0$ A     &  $4000$ A  &  first   \\
                  & 8   &  $0$ A     &  $4000$ A  &  second  \\
                  & 11  &  $0$ A     & $-4000$ A  &  second  \\
     \hline
   \end{tabular}
\end{table}


\subsection{First Pass}
     \label{analysis_first_pass}

Since only about $\sim 1\%$ of the events recorded to tape were
due to a continuum dimuon event originating from the target, 
multiple analyses of the entire data sample were not feasible.
It was therefore necessary to reduce this data sample to more manageable 
levels.  A first pass analysis was performed which filtered out many of
the bad (non-dimuon) events in the data sample, while preserving almost all 
of the events we were interested in.  The computing resources for this 
analysis were provided by the Fermilab Computing Division.  A number of IBM 
workstations were linked together into four parallel computing clusters called
``farms'', which were able to process the complete data sample in 
approximately two months.


   \subsubsection{Track Reconstruction}
      \label{analysis_first_pass_track_reconstruction}

   The analysis began by reading in the configuration of the 
   spectrometer from various files, some specified by the user and
   some loaded automatically based on the data set being analyzed. 
   Once configured, the analysis code examined the hits in the drift
   chambers at stations 2 and 3 to find a set of candidate tracks.  If
   four of the six planes in a station registered a hit, the position
   of that hit was considered a possible track reconstruction point.
   Points which were inconsistent with the trigger which fired were
   dropped.  The remaining reconstruction points were iteratively combined
   to form track segments, so long as those segments were consistent
   with a single particle forming both points.

   The track segments were then compared with hits in station 1 to 
   reduce further the number of track segments.  Each segment was first
   extended to the SM3 bend plane, located between stations 1 and 2.
   The charge and momentum of the particle was not yet known, so a 
   window-of-interest at station 1 which pointed back to the target in
   \newpage \noindent
   the X direction was searched.  If none of the hits at station 1 met
   the search criteria, the track segment was discarded.

   Once the track segment between stations 2 and 3 was connected with
   the station 1 reconstruction point (or points), the charge and 
   momentum of the track were determined from the direction and size of
   the deflection of the track at SM3.  The track segment was then 
   extended back to station 4 to verify that it was indeed a muon.  
   The concrete and lead shielding present at station 4 absorbed any 
   hadrons and electrons which managed to penetrate the absorbing 
   wall, but also caused any muons to multiple scatter.  Since at
   this point the analysis knew the momentum of the track, a window
   consistent with the expected multiple scattering could be searched
   for hits in the five station 4 detectors.  If the track fired less
   than three of the station 4 detectors, it was rejected on the grounds
   that it was not likely to be a muon.

   \subsubsection{Event Kinematics}
      \label{analysis_first_pass_event_reconstruction}

At this point in the analysis, the charge, momenta and positions 
at the various detectors of the tracks in the event were known.  
It had also been determined at this point that each of the tracks 
was due to a muon.  The tracks were then propagated back through the 
SM12 and SM0 magnetic fields, accounting for the energy lost in the 
hadronic absorbing wall and (if applicable) the dump.  The paths of 
the tracks were reconstructed inside the magnetic field in 46-cm steps, 
using field maps which had been measured prior to the experiment.  
A correction\footnote{\setlength{\baselineskip}{\singlespace}The most probable energy lost in a given
section of absorbing wall or dump was parameterized versus the momentum
of the track.  This energy loss parameterization was used to estimate the
energy lost by the muon.} was also made to compensate for the energy lost by 
the muon where the track passed through either the absorbing wall or the beam 
dump.

   In addition to suffering energy loss inside the absorbing wall and dump, 
   the muons also underwent multiple Coulomb scattering (MCS).  The correction
   for this began by reconstructing the track back to an effective
   scattering bend plane (ZSCPLN) and recording its position and momentum
   there.  The track was then completely analyzed, noting the deviations
   in X and Y from the nominal target center.  A correction to the 
   momentum vector at the ZSCPLN was calculated from these deviations,
   and the reconstruction of the track repeated from that point.  This
   procedure was iterated until the track converged on the nominal center
   of the target.

The production of pions and other hadrons in the target contributed to a 
large rate of single muons in the spectrometer as they decayed between the
target and dump.  A significant fraction of these muons followed a beam-like
trajectory through the dump and absorbing wall.  Because there was no way
to differentiate between one of these single muons and a muon which was
a part of a valid dimuon event, these muons could cause ambiguities in
the analysis that could not be resolved.  To prevent this from happening,
muons which passed through a \ydumpcut \\ horizontal band 
centered on the nominal beam position at the dump were eliminated from
the event sample.



After the kinematics of the individual tracks were reconstructed
at the target, they were combined to form dimuon pairs.   Less than 0.08\% 
of the reconstructed events contained more than two tracks.  From the charge 
of the tracks and their momenta, the kinematics of the event were determined.  
Events which reconstructed to a mass $< 2.0 \, \text{GeV}$ were rejected
on the grounds that the acceptance for such pairs arising from the target 
was zero.\footnote{\setlength{\baselineskip}{\singlespace}
There was a problem in the application of this cut which 
applied to events with more than two tracks.  If one of the pairs which 
could be combined from the tracks in the event was below the mass cut,
the entire event was cut rather than the particular combination of
tracks.  Studies indicate that the error introduced by this is on the
order of 0.1\%, which is negligible compared to other uncertainties in
the experiment.}  Events whose uniterated vertex\footnote{\setlength{\baselineskip}{\singlespace}The position at
which the tracks made their closest approach before the MCS corrections
were made.} indicated they were unlikely to have originated in the target
were also cut.  This cut was realized in terms of the $z$-coordinate of the
uniterated vertex (ZUNIN), which was restricted to values less than 
350 inches from the center of the target.

\subsection{Second Pass}
   
  
  The goal of the first pass analysis was to reduce the data sample to
a more manageable level.  The purpose of the second-pass analysis was to
reconstruct the kinematics of the events as accurately as possible, 
minimizing any systematic errors which would affect the cross sections.
In the remainder of this chapter, we will discuss the differences between
the algorithms the first- and second-pass analyses used to correct for
multiple scattering and energy loss.  We will also outline the procedures used
to determine the strengths of the fields in the spectrometer magnets, and 
the position and angles of the beam at the target.
 
\subsubsection{Multiple Scattering}

The ZSCPLN was used to correct for the multiple scattering that 
the muons underwent in traversing the absorbing wall and beam dump
as described above.  The first pass-analysis used a single position
for the ZSCPLN, which was empirically determined by minimizing the
correlation between the mass of the dimuon pair and the mean ZUNIN.  
This approach ignored the additional scattering that events which 
passed through the dump were subject to.  The second-pass analysis based 
the position of the ZSCPLN on the length of the dump which the muon 
penetrated.  It was found that for every 1 cm the track traveled through 
the dump, the ZSCPLN should be moved upstream 0.7 cm.


The second pass analysis also employed a different technique for finding
the optimal position of the ZSCPLN.  Monte Carlo studies indicated that
the correlation between the mass of the events and ZUNIN was due to energy
loss fluctuations in the beam dump and hadronic absorber, rather than
difficulties in reconstructing the track angles as had previously been
believed.  The ZSCPLN in the second-pass analysis was placed at the position
which optimized the reconstruction of the track angles at the target 
for Monte Carlo events.  This resulted in a ZSCPLN located $12$ m downstream
of the target, near the end of the carbon section of the hadron absorber.
The method used in the first pass analysis placed the ZSCPLN $22$ m 
downstream of the target, well outside of the absorbing wall.


\subsubsection{Energy Loss}
 \label{subsec:energyloss}

As will be discussed below, determination of the strength of the magnetic 
fields from the data relied on an accurate reconstruction of the masses of
the J$/\psi$ and $\Upsilon$ resonances, and the ZUNIN distribution of the
events.  Deviations between the average energy lost by the muons and the
parameterization used to correct for the energy loss in the analysis could
cause systematic shifts in both mass and $z$-vertex reconstruction, and thus 
the measurement of the magnetic fields.

\begin{figure}
  \begin{center}
  \includegraphics[width=0.9\linewidth,clip]{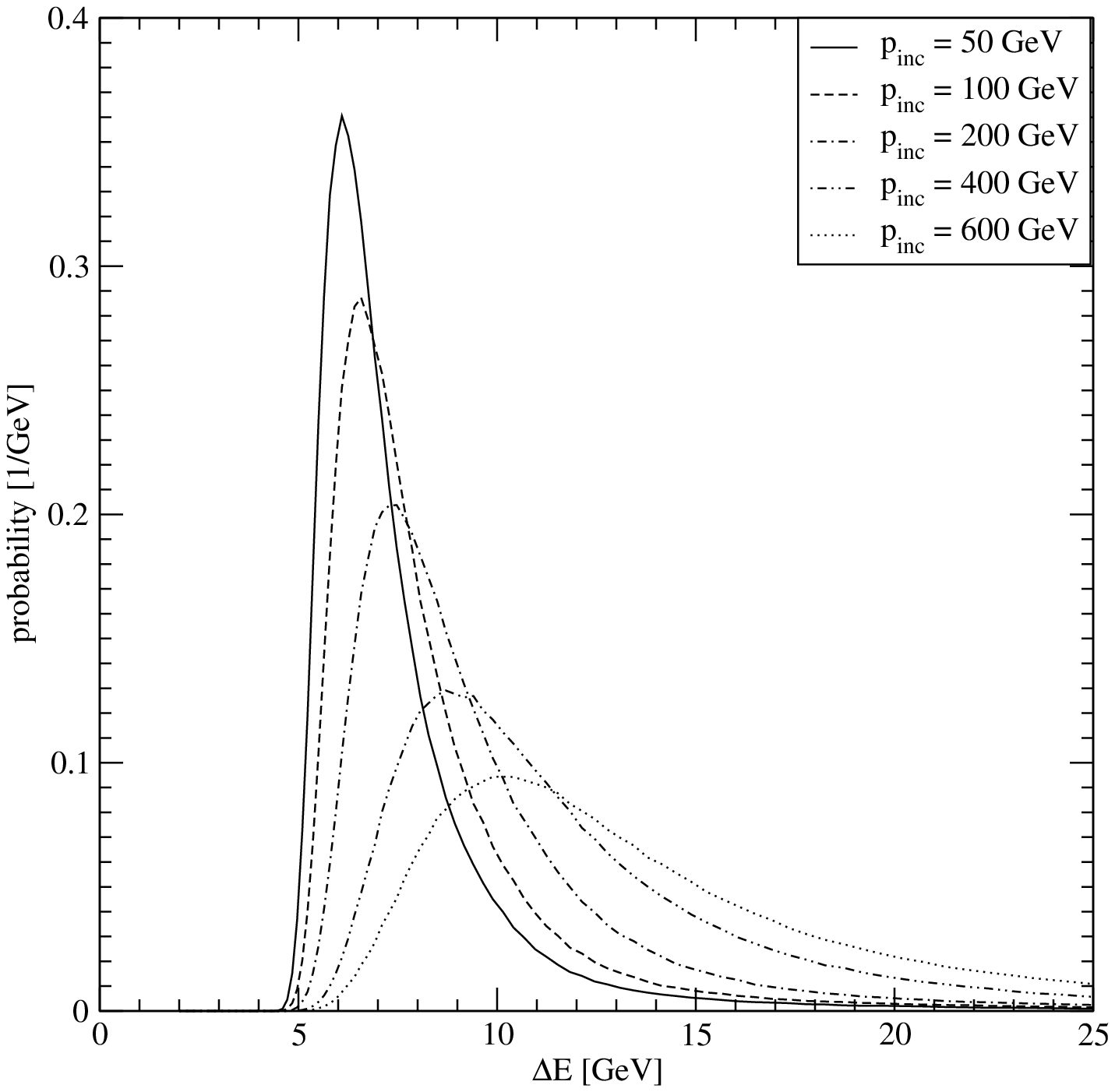}
  \caption[Energy loss  for muons passing through the dump.]{
	\label{fig:dedx-dump}
	\setlength{\baselineskip}{\singlespace}
	Energy loss  for muons passing through the dump (144'' of
  	copper). Various incident energies are shown.
	}
  
  \end{center}
\end{figure}

\begin{figure}
  \centering
  \includegraphics[width=0.9\linewidth]{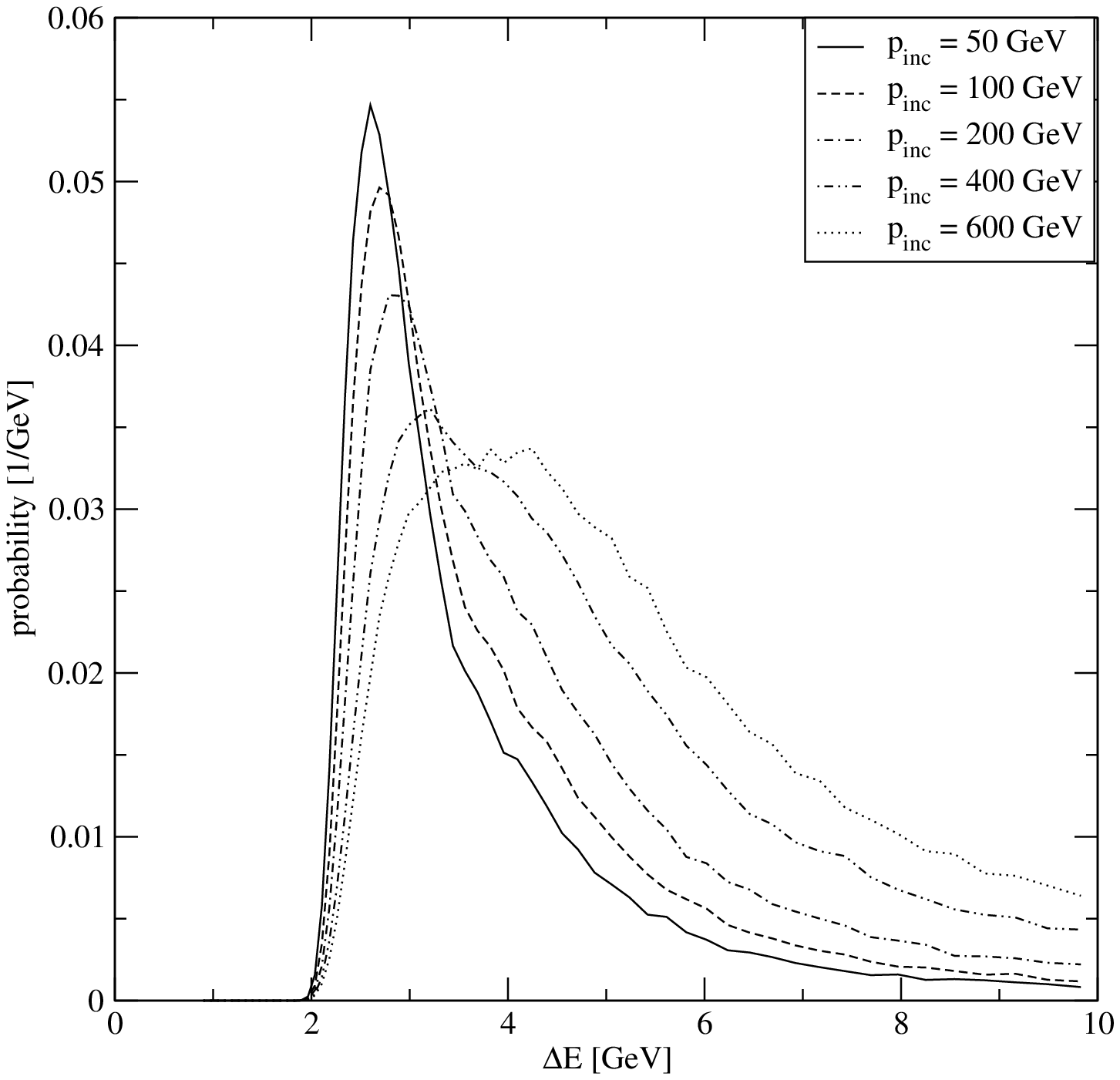}
  \caption[Energy loss  for muons passing through the absorber wall.]{
	\label{fig:dedx-abs}
	\setlength{\baselineskip}{\singlespace}
	Energy loss  for muons passing through the absorber wall.
   	Various incident energies are shown.
	}

\end{figure}

Every muon which passed through the spectrometer was subject to a large
amount of energy loss in the absorbing wall and possibly the beam dump.
A typical $200 \, \text{GeV}$ muon lost on average $\sim$ 2 GeV in the
absorber.  Figures \ref{fig:dedx-dump} and \ref{fig:dedx-abs} show the
energy lost in traversing the beam dump and absorbing wall for muons 
with different incident momenta.  The calculations were performed using
the TRAMU muon transport program \cite{bib:TRAMU}.  The large, non-Gaussian 
tail skews the mean to much larger energy losses than the most probable value. 
It was important to parameterize the energy loss in the reconstruction of the
data with emphasis on accurate (on average) reconstruction of the kinematics 
of the events rather than optimal resolution.  Thus, the mean energy loss was 
used in this analysis.

Given the incident momentum of the muons on a specified material, 
the parameterization of the mean energy loss is a well defined problem.
However, the incident momentum of the muons on the absorbing wall or
dump was the quantity which we wished to estimate from our measurement
of the momentum downstream of SM12 and some parameterization of the
random distribution of energy loss inside the absorbing wall and 
dump.  To further complicate the problem, muons which lost a large
amount of momentum in the dump/absorber could be swept out of the 
acceptance.  Muons with lower incident momenta would be more susceptible 
to this, introducing an acceptance effect into the parameterization of
the energy loss.

\begin{figure}
     \centering
      \includegraphics[width=0.9\linewidth]{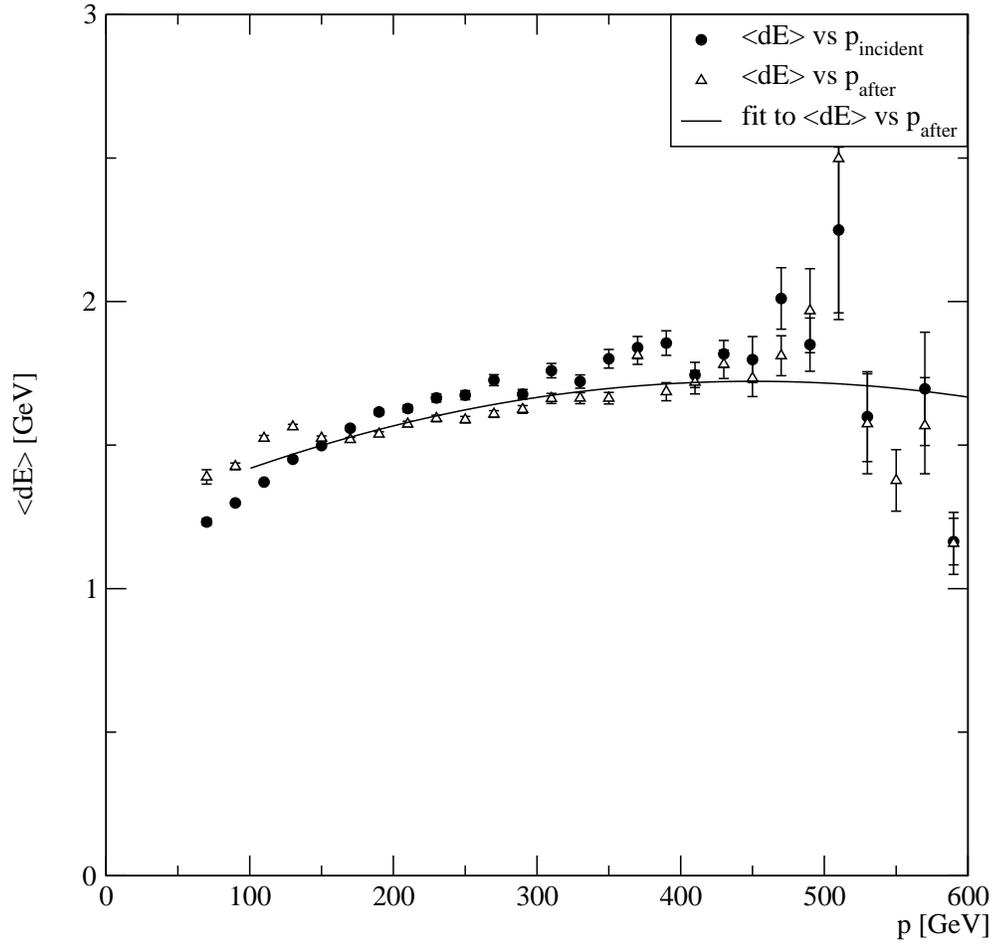}
  \caption[Mean energy lost by muons in copper section of the absorber.]{
	\label{fig:dedxparam}
	\setlength{\baselineskip}{\singlespace}
	Mean energy lost by muons in copper section of the absorber.
        Shown is the energy loss versus the momentum
        of the muon entering the section (circles) and the energy loss
        versus the momentum of the muon exiting the section (triangles).
        The solid line is the fit to the second set of points which was
        used in reconstructing the events.
	}
  
\end{figure}

Figure \ref{fig:dedxparam} shows the mean energy lost by muons in the 
$24''$ section of the absorbing wall according to the Monte Carlo 
(where we knew the energy lost by each muon thrown).  We plot the mean
energy loss versus the momentum incident upon the $24''$ of copper, and
versus the momentum after the $24''$ of copper.  The ``after'' 
parameterization is generally flatter than the ``incident'' parameterization.
This is due to the fact that larger mean energy losses in the ``incident''
parameterization will be shifted to smaller values of $p_{\text{after}}$.
It should be clear from the figure that this effect is important, especially
at lower momenta where we have most of our statistics in the data.

Figure \ref{fig:dedxparam} also shows the fit to the energy loss in the
$24''$ copper section of the absorbing wall.  The parameterizations which
we used for all of the absorber sections in the high-mass data are given
by
\begin{eqnarray}
\begin{aligned}
\nonumber & dE_{Cu}   & = 1.2202 + 0.2228\times 10^{-2} p_\text{after} - 0.2472\times 10^{-5} p_\text{after}^2 \\
\nonumber & dE_{C}    & = 0.2461 + 0.3135\times 10^{-3} p_\text{after} - 0.4362\times 10^{-6} p_\text{after}^2 \\
\nonumber & dE_{CH_2} & = 0.1857 + 0.2073\times 10^{-3} p_\text{after} - 0.3501\times 10^{-6} p_\text{after}^2 \\
\end{aligned}
\end{eqnarray}
\noindent where $p_\text{after}$ is the momentum immediately downstream of
the absorbing wall section being traced through.  Deviations in the shape
of this parameterization, but not intercept, were found for the low- and
intermediate-mass data sets.  When used to reconstruct Monte Carlo events,
systematic shifts in the masses and ZUNIN of the events are held to
acceptable levels.

   \subsubsection{Magnetic Fields}

   \label{chapter:analysis:bfields:tweeks}

The magnetic fields in the three spectrometer magnets were measured
prior to the installation of the absorbing wall.  Measurements were made at 
each of the planned operating currents, except for the 4000 A setting of the 
SM12 magnet.  This setting was not measured due to concerns over the 
structural integrity of certain parts of the magnet under the increased strain 
of higher magnetic forces.  At the time, it was anticipated that the 2800 A 
setting of the SM12 field would be the largest current at which that magnet 
would be operated.  Later during the experiment, the SM12 magnet was 
successfully operated at the 4000 A setting, providing the largest and 
cleanest sample of continuum dimuon events in the E866 data sample.


The magnet map used in the high-mass data utilized a measurement of the SM12
field from a previous experiment.  However, the upstream portion of the 
magnet was reconfigured for E866.  This region of the magnet map relied on
an OPERA \cite{bib:OPERA} calculation of the magnetic field, matched to the 
field map in the
known region of the magnet.  A similar procedure had to be performed to
determine the magnetic fields in the beam dump, where it was not possible
to measure the fields.

\begin{figure}
\centering
\includegraphics[width=0.8\linewidth,clip]{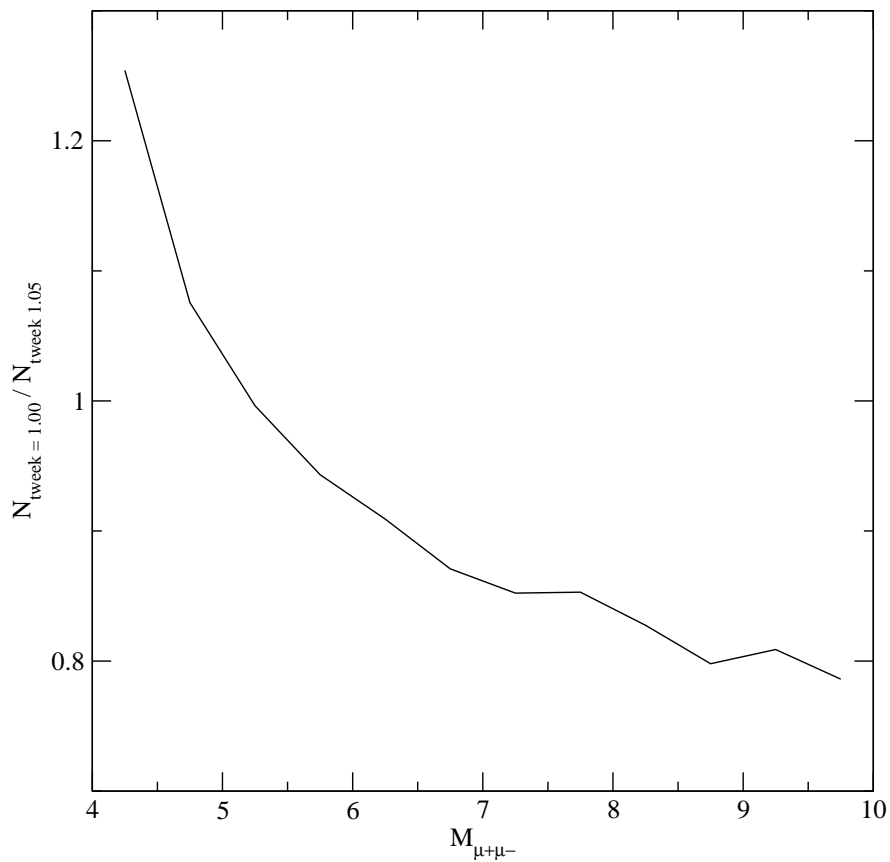}
\caption[Sensitivity of the acceptance to the magnetic fields.]{
	\label{fig:tweeks}
	\setlength{\baselineskip}{\singlespace}
	Sensitivity of the acceptance to the magnetic fields. The ratio
	of Monte Carlo events generated with the nominal SM12 field 
        to those generated with the SM12 field augmented by $5\%$ is
	shown as a function of mass in the range $0.5\leq x_F < 0.55$.
	}
\end{figure}

Although the shapes of the magnetic fields were well determined, the 
absolute strengths of the fields were uncertain at the $\sim$ 1\% level.  
Monte Carlo studies indicated that this level of uncertainty in the SM0 and
SM12 magnets could be a significant source of uncertainty in the acceptance 
of the spectrometer, especially near its low mass edge.  Figure 
\ref{fig:tweeks} shows the ratio of Monte Carlo events generated with
magnetic fields differing by 5\% using a fast version of the E866 Monte Carlo. 
The relative acceptance falls by nearly $40\%$ over the given mass
range.  A systematic uncertainty of $\sim 1\%$ in the SM12 field would
correspond to a $\sim 8\%$ point-to-point systematic uncertainty in the
acceptance, which would have been unacceptable given the precision of our data.
Furthermore, systematic 
shifts in the SM3 field would cause systematic shifts in the reconstructed
kinematics of the events. Such shifts were a problem, considering the 
steeply-falling nature of the cross section we were trying to measure.

The general procedure for determining the absolute field strengths from
the data, once the impact of the energy loss parameterization had been
limited, was as follows.  We began by fitting the masses of any resonances
($J/\psi$, $\Upsilon$)
in the mass spectrum to a Gaussian plus a quadratic background, noting the
reconstructed mass of the peaks.  We also estimated the mean of the 
ZUNIN distribution.  Deviations of these quantities from
the nominal position of the target and the published resonance masses
indicated an incorrect value for the SM12 and SM3 fields respectively.
The data set was then re-analyzed with a multiplicative constant -- referred
to as a tweek ({\em sic.}) -- applied to each
of the magnetic fields to try to compensate.  This was performed
iteratively until the resonance masses converged to their published 
values, and ZUNIN was centered on the nominal target position.

The procedure outlined above was essentially the procedure used to determine
the tweeks on the SM12 and SM3 magnets in the high-mass data.  It is 
a slight simplification, since a shift in the SM3 magnet would also 
have caused a shift in the ZUNIN of the events due to the increased 
momentum and resulting decrease in the deflection of the muon by SM12.  
The one important change to note was that only events which missed the
dump were studied.  Muons which hit the dump went through a somewhat
uncertain amount of material, contributing an uncertainty to the amount
of energy loss which should have been corrected for.  Additionally, the 
ZUNIN of the events with one track in the dump was largely determined by
the other track in the event.  This reduced the sensitivity of ZUNIN
to the magnetic field in SM12.

Additional problems arose in the analysis of the low and intermediate mass
data sets.  In the intermediate-mass data set, we had only a limited number
of events to work with.  The reconstructed $\Upsilon$ mass, the only
resonance available in that data set, was too uncertain to provide an 
accurate assessment of the SM3 tweek.  Therefore, the value found in the
high mass analysis was used for SM3, and the usual procedure of finding
the SM12 tweek using the ZUNIN distribution performed.

A similar problem occurred in the low-mass data set, where we had an extra
magnetic field (SM0) to worry about.  In principle, since we could resolve 
both the J$/\psi$ and $\psi'$ resonances, we had the extra information 
needed to tune the SM0 field.  In practice, the $\psi'$ mass was relatively 
insensitive to the SM3 field, for a reasonable set of tweeks.  Once again,
the SM3 field tweek from the high-mass data was used in this data set.
The SM0 tweek was found from the fits to the $J/\psi$ mass, and the usual
ZUNIN fitting procedure used to extract the SM12 tweek.

The procedures described above provided an estimate of the average field 
strength in a given data set.  It also provided a way to estimate the
systematic uncertainties on the tweeks by studying the impact of a given
change in tweek on the ZUNIN distribution and the resonance
masses.  Uncertainties in the mass and vertex fits could then be translated
into uncertainties in the magnetic fields.  Once the uncertainties in the
magnetic fields were determined, the corresponding acceptance uncertainty
could be calculated by running the Monte Carlo using different input
fields.  We will describe this in greater detail in the next chapter.

\begin{figure}
\includegraphics[clip,width=0.95\linewidth]{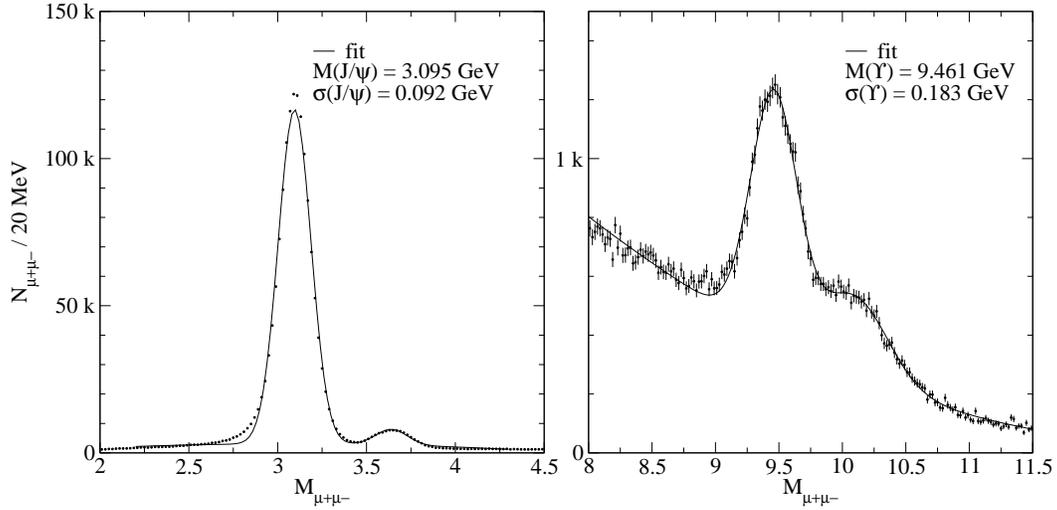}
\caption[Reconstructed $J/\psi$ and $\Upsilon$ resonances.]{
	\setlength{\baselineskip}{\singlespace}
	\label{fig:tuneup}
	Reconstructed $J/\psi$ and $\Upsilon$ resonances.  Left panel
 	shows the low-mass data, right panel the high-mass data.  Data
	is the sum of the hydrogen and deuterium targets.
	}
\end{figure}
\begin{figure}
\centering
\includegraphics[width=0.5\linewidth]{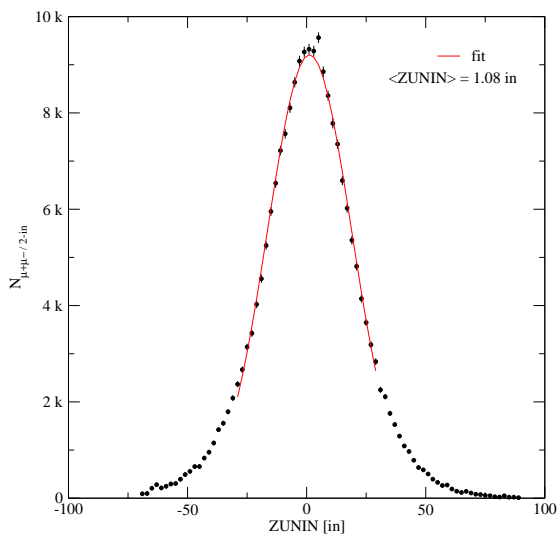}
\caption[Uniterated $z$-vertex distribution.]{
	\label{fig:tuneupz}
	\setlength{\baselineskip}{\singlespace}
	Uniterated $z$-vertex distribution.  Data is the sum of both
	targets, corrected for empty-target background.  The fit is to
	a Gaussian curve.
	}
\end{figure}

Figure \ref{fig:tuneup} shows the $J/\psi$ and $\Upsilon$ mass regions
from the low- and high-mass data, when reconstructed using the optimal
magnetic field tweeks.
The reconstructed $J/\psi$ and $\Upsilon$ masses of 3.095 and 9.461 GeV
compare quite well with the published values of 3.097 and 9.46 GeV 
\cite{bib:PDB}.
The intrinsic widths of the resonances (which are best expressed in keV),
are smeared out by energy loss and multiple scattering in the absorbing
wall and dump -- our mass resolution is 92 MeV at the $J/\psi$ mass and
183 MeV near the $\Upsilon$.
Figure \ref{fig:tuneupz} shows a fit to the ZUNIN distribution of the 
continuum dimuon regions, where the data has been summed over all data sets.
In practice, fits to the ZUNIN distributions and resonance regions were
performed separately in each data set.  The deviations of the fits from
their expected values amount to uncertainties of only $\pm 0.10\%$
and $\pm 0.15\%$ in the SM0/12 and SM3 tweeks, respectively.

One other source of systematic error in the tweeks was uncovered using
the Monte Carlo.  When Monte Carlo events were retracked through the
reconstruction algorithm, the momentum found at the SM3 bend plane was 
$\sim$ 0.06\% larger than that which was actually generated by the
Monte Carlo.  Thus, the correct field tweeks were subject to a small
offset which had to be accounted for to obtain a more accurate simulation
of the spectrometer.  The origin of this effect was unclear. It may have 
been due to slight differences between the actual and simulated geometry of
the apparatus, resolution effects coupled to the single bend plane 
approximation used for SM3, or some combination of these and other
issues.  This additional source of systematic uncertainty in the SM3
field does not contribute significantly to the overall size of the
systematic uncertainties in the acceptance.

In table \ref{magnet_tweeks} we list the tweeks found for each mass
setting using the previously described procedures.  Studies of the stability
of the mean ZUNIN and $\Upsilon$ masses versus data set and run number in 
the high-mass data suggest that a single value for the SM3 field tweek
is reasonable.  These studies also indicated that the SM12 field 
was reliably stable, and a single value was used for each data set
within a given mass setting.

   \begin{table}
     \caption{	 
        \label{magnet_tweeks}
	Field strengths for each data set.
	}    
     \begin{center}
     \begin{tabular}{|c|c|c|c|c|}
       \hline
       Mass setting   & Data set & SM0 tweek  & SM12 tweek & SM3 tweek  \\
       \hline\hline 
       low  & 5    & $1.0140 \pm 0.0010$ & $1.0068 \pm 0.0010$ & $1.0049 \pm 0.0015$ \\
            & 10   & $1.0140 \pm 0.0010$ & $1.0068 \pm 0.0010$ & $1.0049 \pm 0.0015$ \\ 
       \hline
       int  & 9    & N/A                 & $1.0093 \pm 0.0010$ & $1.0049 \pm 0.0015$ \\
       \hline
            & 7    & N/A                 & $0.9865 \pm 0.0010$ & $1.0049 \pm 0.0015$ \\
       high & 8    & N/A                 & $0.9865 \pm 0.0010$ & $1.0049 \pm 0.0015$ \\
            & 11   & N/A                 & $0.9865 \pm 0.0010$ & $1.0049 \pm 0.0015$ \\
       \hline
     \end{tabular}
     \end{center}

   \end{table}

\subsubsection{Beam Position and Angles}
  \label{beam_pos_and_ang}

  The position of the beam was determined by fitting the distribution
  of the $x$ and $y$ vertices of the raw data at the target position.  
  The uniterated $x$ and $y$ positions of the two tracks in the event 
  were averaged together to yield the $x$ and $y$ vertices.  
  Run-to-run variations in beam position  were found to be negligible
  for our purposes.
  %

The beam angle was found through an iterative process involving the Monte 
Carlo.  In principle, the Y component (and similarly the X component) of the 
beam angle could be found from just the data by looking at the muon
momentum vectors
 $\theta_{y}^{dimuon} = ( p_{y}^{+} + p_{y}^{-} ) / ( p^{+} + p^{-} )$.
However, asymmetries in the dump cut in the first pass analysis 
limited the accuracy of this technique.  An independent test of the 
validity of the beam angle extracted from the data was to check the
\pphi distribution in the data against that of the Monte Carlo.  Since we
know that the \pphi distribution must be isotropic, and having generated 
an isotropic distribution in the Monte Carlo, any difference between the
distribution in the data versus the Monte Carlo could only result from a
difference in the beam angle between the experiment and the Monte Carlo.
The beam angles were improved by minimizing these differences,
and we estimate that the uncertainty in the beam angle contributes no
more than 5\% ($\sim 0.7$ GeV) of the transverse momentum of an
average event.

\newpage
\section{ANALYSIS}

\label{chapter:cross}

The differential cross section is defined as the number of interactions
per target particle leading to an event with  kinematics between
$\Omega$ and $\Omega + \delta \Omega$ per number of incident particles per
unit area.  We express this as 
\begin{equation}
\label{eqn:crossdef}
\frac{d\sigma}{d\Omega} = \frac{N/\Delta \Omega}{N_T(\bar{N}_i/A)}.
\end{equation}
\noindent where $N$ is the number of dimuon events from the target, $N_T$ the
number of target particles in the path of the beam, $\bar{N}_i$ the mean number of protons incident
on the target and $A$ the cross sectional area of the beam.

The simple expression in equation \ref{eqn:crossdef} is somewhat misleading,
since we did not measure the number of $pp$ and $pd$ interactions which led 
to a dimuon event directly. Instead, we measured the yields of oppositely 
charged muon pairs from the interaction of a proton beam with target flasks 
containing liquid $H_2$ and $D_2$.  These yields were subject to contamination
by background events originating from the interaction of the beam with the 
SWIC and the front and back faces of the target flasks, and from the random 
coincidence of two uncorrelated muons produced in the same beam bucket.  The 
yields were also subject to losses due to the acceptance of the spectrometer 
and inefficiencies in the detection of the muons.  Equation \ref{eqn:crossdef} 
also relies on quantities such as the number of target particles
and incident protons, which were also not directly measured, and the cross 
sectional area of the beam, which was not measured with any great precision.

  Fortunately, we can rewrite equation \ref{eqn:crossdef} entirely in terms
  of quantities which were directly measured, while eliminating terms which 
  were not measured well.
  We define the number of dimuons, corrected for backgrounds, 
  as $N_{\mu^+\mu^-}$.  
  This can be related to the number of interactions in equation 
  \ref{eqn:crossdef} by
  \begin{equation}
  \label{eqn:dimu-yield}
  N = \frac{ N_{\mu^+\mu^-} }{ \alpha \epsilon } = \frac{ N_{\mu^+\mu^-}^\text{target} - N_{\mu^+\mu^-}^\text{background} }{ \alpha \epsilon }
  \end{equation}
  \noindent where $\alpha$ is the geometric acceptance of the spectrometer, 
  defined as the probability that both tracks in a given event would 
  traverse all of the active detector elements of the spectrometer and
  survive all data cuts,
  $\epsilon$ is the efficiency with which the given event was detected, 
  and the ``target'' and ``background'' superscripts refer to events which 
  originated in the liquid target and background events respectively.  
  The number of target particles can be expressed in the usual way:
  \begin{equation}
  \label{eqn:ntgts}
  N_T = N_A \, \rho \, L \, A
  \end{equation}
  \noindent
  where $N_A$ is Avogadro's number, $\rho$ is the density of cryogenic liquid, 
  $L$ is the length of the target, and $A$ is the cross sectional area of the 
  beam.

To get the mean number of incident protons, we need to average over the
attenuated flux in the target.
The number of protons at any point along the length $z$ of the target can be 
expressed in terms of the incident number of protons and the hadronic 
absorption length $\lambda$, as 
\begin{equation}
N_i(z) = N_p e^{-z/\lambda}.
\end{equation}
\noindent Averaging over the total length of the target $L$ yields
\begin{equation}
\label{eqn:attenfact}
\bar{N_i} = \frac{1}{L} \int_0^L N_i(z) \, dz = \frac{N_p \lambda}{L} ( 1 - e^{-L/\lambda} ).
\end{equation}

Substituting equations \ref{eqn:dimu-yield}-\ref{eqn:attenfact} into equation
\ref{eqn:crossdef} results in an expression for the differential cross section:
\begin{equation}
\label{eqn:crossdefined}
\frac{d\sigma}{d\Omega} = \frac{N_{\mu^+\mu^-}}{ { \mathcal L }  \alpha \epsilon \Delta \Omega},
\end{equation}
\noindent where we have grouped several of the factors in the denominator into 
a single quantity, called the luminosity, given by
\begin{equation}
\label{eqn:luminosity}
{ \mathcal L } = N_A \rho \lambda ( 1 - e^{-L/\lambda} ) N_p.
\end{equation}
\noindent We will now explain in detail how each of the quantities in equation
\ref{eqn:crossdefined} was determined.



\subsection{Event Yields}

The total event sample produced by the second pass analysis contained several
questionable events.  
These included backgrounds from uncorrelated muon pairs and correlated pairs 
originating outside of the target volume, events which were reconstructed by 
the analysis but were not responsible for the trigger which caused the DAQ to 
read out, and events which occupied regions of the spectrometer which 
complicated the reconstruction of their kinematics.  
These events were either filtered out of the second pass event sample, or 
independently binned and subtracted from it, to obtain the event yields used 
to calculate the cross sections.
The same set of cuts were used on the Monte Carlo events used to calculate
the acceptance.
Simulations showed that contributions from the interactions
of secondary hadrons produced in the target were negligible.

\begin{figure}
\centering
\includegraphics[clip,width=0.85\linewidth]{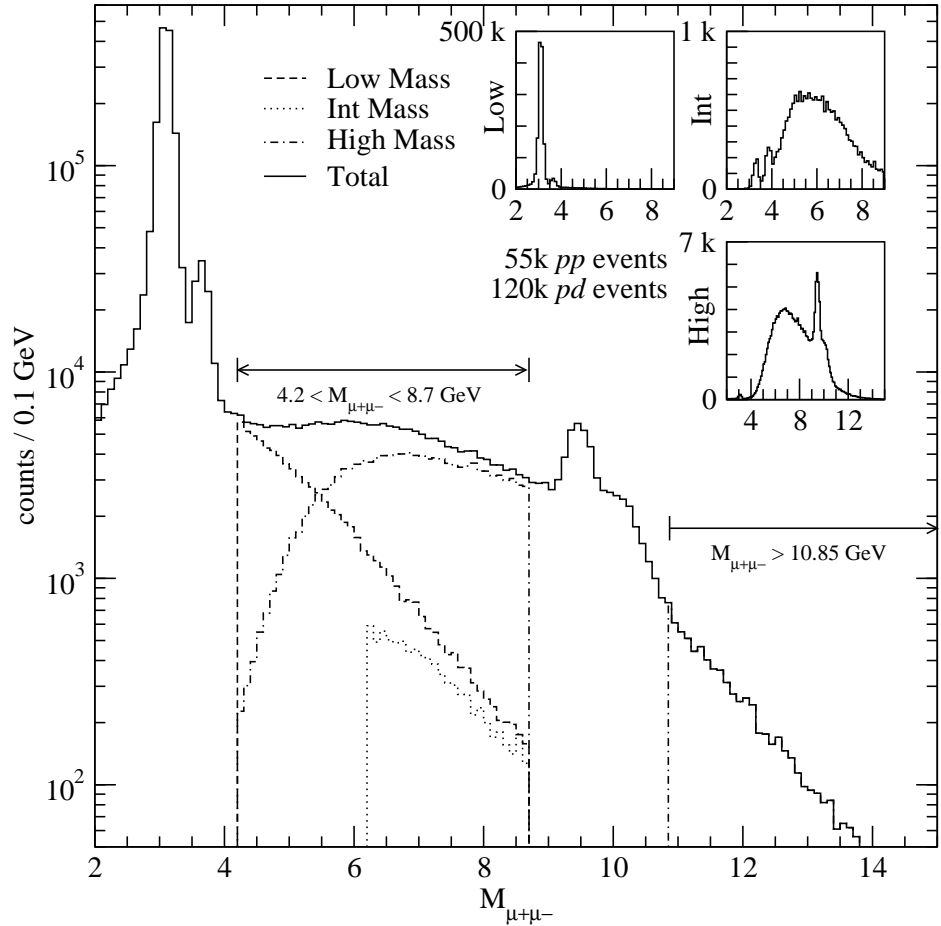}
\caption[Total E866 mass spectrum surviving all cuts.]{
	\label{fig:massplots}
	\setlength{\baselineskip}{\singlespace}
	Total E866 mass spectrum surviving all cuts. The contribution
	of the low, intermediate and high mass data sets superimposed in their
 	respective allowed mass ranges.  Inset figures show the low, 
	intermediate and high mass data sets on a linear scale over their 
	entire mass range.
	}
\end{figure}

In order to avoid the $J/\psi$ and $\Upsilon$ resonances, we selected events
within the mass range $4.2 \leq M_{\mu^+\mu^-} \leq 8.7$ GeV, and events
with masses $M > 10.85$ GeV.  As the acceptance above the $\Upsilon$ resonance
was quite small in the low and intermediate mass data sets, we only used the
high mass data in that region.  Due to uncertainties in the magnetic fields, 
data from the intermediate mass setting below $6.2$ GeV were also discarded.
Figure \ref{fig:massplots} shows the total E866 mass spectrum, with the 
contributions from the low-, intermediate- and high-mass data sets in their
respective allowed mass ranges.  The resonance regions have been retained,
but all other cuts were applied.  The number of events surviving all 
cuts is shown in table \ref{table:survivors}.

\begin{table}\centering
  \caption{
	\label{table:survivors}
	Number of events in each data set which
        survived all data cuts.
	}
  \vspace{10pt}
  \begin{tabular}{|c|c|c|c|}
  \hline
  Set & H$_2$ target & D$_2$ target & Empty target\\
  \hline\hline
  5  & 17217 & 36962 & 699  \\
  7  & 9548  & 21030 & 425  \\
  8  & 20049 & 44129 & 1007 \\ 
  9  & 1768  & 4966  & 63   \\
  10 & 1658  & 3509  & 60   \\
  11 & 4850  & 10545 & 230  \\
  \hline\hline
  total & 55090 & 121241 & 2484 \\
  \hline
  \end{tabular}
\end{table}

\subsubsection{Target Flask Background}

\begin{figure}
  \centering
  \includegraphics[width=0.85\linewidth,clip]{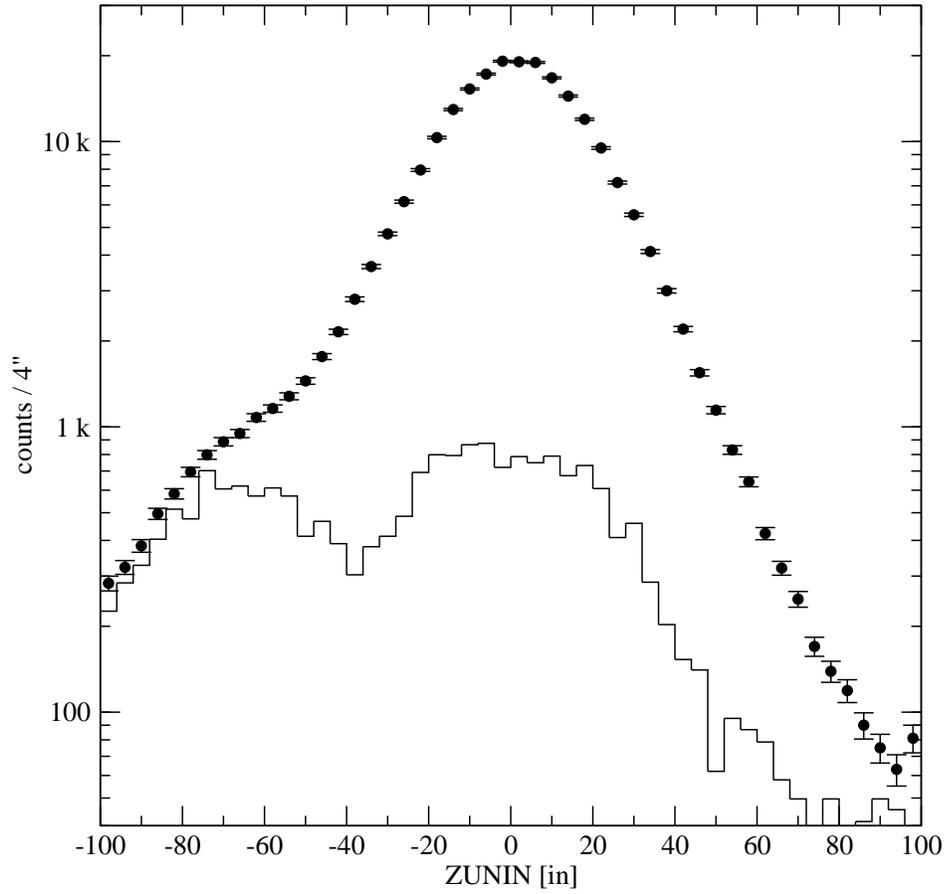}
  \caption[Comparison of target yields to empty-target background.]{
	\setlength{\baselineskip}{\singlespace}
	Comparison of target yields to empty-target background.
	The circles represent the sum of the hydrogen and deuterium yields,
	the histogram represents the normalized empty-target contamination.}
  \label{fig:targetflaskbackground}
\end{figure}

One of the larger sources of contamination was due to the beam interacting
with the SWIC ($\approx$ 188 cm upstream of the target) and the walls
of the target flasks which contained the cryogenic liquids.
Figure \ref{fig:targetflaskbackground} shows the yields from the cryogenic 
liquid and empty target flasks plotted versus ZUNIN.  The SWIC is clearly 
visible in the empty target distributions, and much of the contamination
was removed by simply applying cuts (ZUNIN $> -50''$ in the low mass data, 
and ZUNIN $> -60''$ in the intermediate and high mass data).  Events with
ZUNIN $>60''$ in the low mass data, and ZUNIN $> 90''$ in the intermediate
and high mass data were also rejected, as they were likely to have originated
in material downstream of the target.

The remaining background from the target flask was measured during the
experiment, allowing us to subtract it from our event yields.  A correction
was made to the empty target yields based on the uniterated $z$-vertex of
the event to account for the $\approx 7\%$ and $\approx 14\%$ attenuation
of the beam between the front and back end caps of the flasks containing
liquid hydrogen and liquid deuterium.  This correction was given by
\begin{equation}
N_{\mu^+\mu^-}^{empty} = \left\{ 
   \begin{aligned}
	& N_{empty} \times e^{-L/\lambda}, & \text{if }  \text{ZUNIN} > 0 \\
	& N_{empty}, & \text{if }  \text{ZUNIN} < 0 
   \end{aligned}
            \right.
\end{equation}
\noindent where we corrected events which appeared to come from downstream
of the nominal target position $z>0$ by the attenuation of the cryogenic 
liquid appropriate to the target.

\subsubsection{Combinatoric Background}

Another important background source was the random coincidence of oppositely
charged, uncorrelated muons produced in the target in the same beam bucket.
The numbers and kinematic distributions of these random events were estimated
using event samples triggered on the like-sign (PhysA2) and single-muon 
(PhysB2) triggers.  These random events were typically confined to masses 
below 5 to 6 GeV and were only important in the low- and intermediate-mass data 
sets, the latter being a non-issue due to the $6$-GeV cut in the intermediate
mass.

\begin{figure}
\centering
\includegraphics[width=0.8\linewidth,clip]{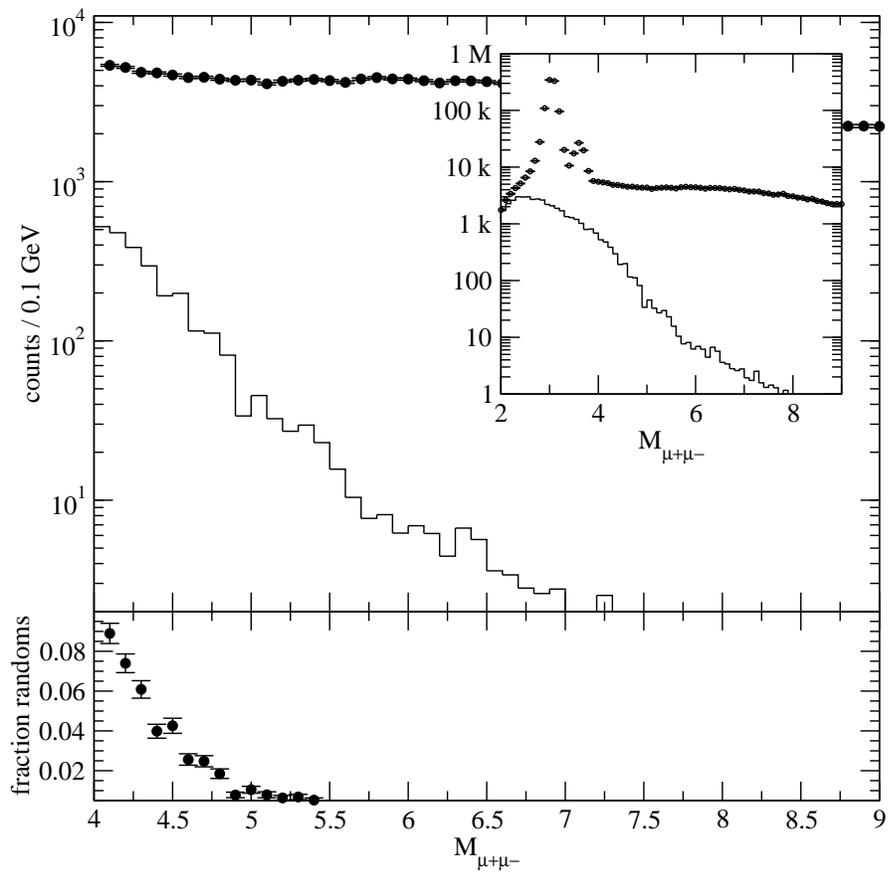}
\caption[Comparison of the target yields to the combinatoric background.]{
	\setlength{\baselineskip}{\singlespace}
	\label{fig:combback}Comparison of the target yields
	(dots) to the combinatoric background (histogram).  The top 
	panel shows the unsubtracted mass distributions, with the inset 
	showing the same distributions down to $2$ GeV.  The bottom panel 
	shows the fractional contribution of randoms vs. mass.
	}

\end{figure}

The single-muon trigger was used to trigger on what appeared to be single
muons from the target traversing one of the matrix elements.  Due to the
high rate of such events from the target, these triggers were prescaled
when in use during normal run conditions.  Single-muon event samples were 
taken either during the normal course of the data collection or in special 
low-intensity runs.  The resulting event samples were subjected to the
same analysis chain as the opposite-sign data.  The single muons were then 
randomly combined with other single muons to form an uncorrelated dimuon event.
This produced a sample of random dimuon events which, when properly normalized,
could be subtracted from the data sample.  Figure \ref{fig:combback} compares 
the total dimuon yield in the data (summed over both targets and the
high and low mass data sets) to the combinatoric background.  The low mass 
data set contributes the largest contamination -- approximately $5\%$ of its
total yield was due to randoms.

\begin{figure}
\centering
\includegraphics[width=0.8\linewidth,clip]{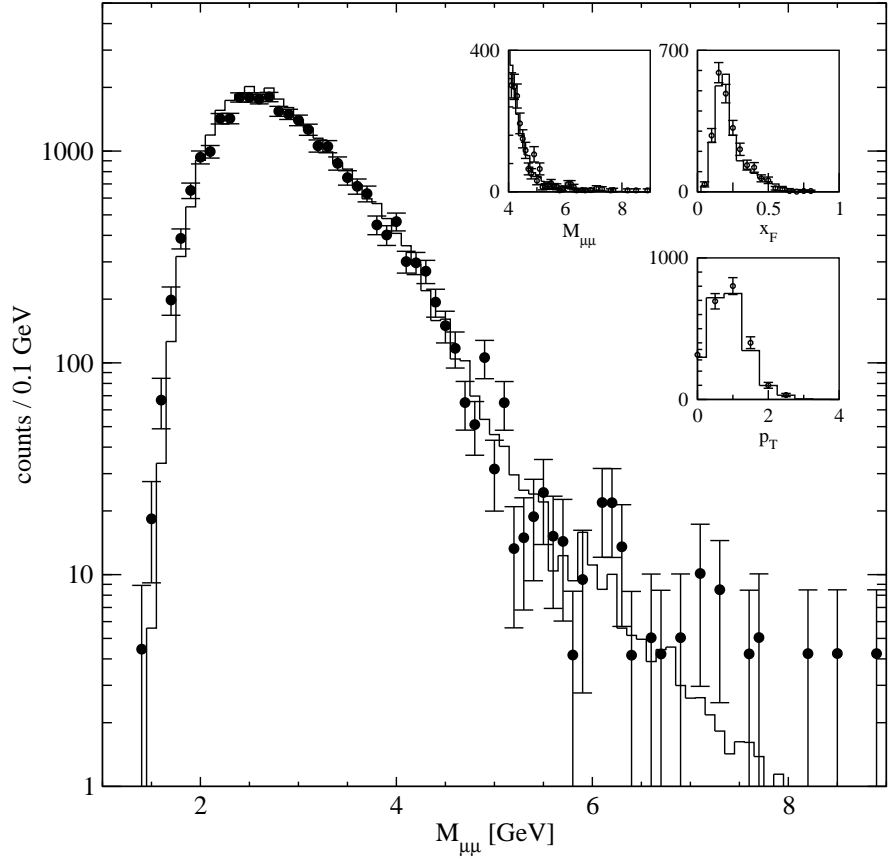}

\caption[Comparison of the data and combinatoric like-sign events.]{
	\setlength{\baselineskip}{\singlespace}
	\label{fig:lsnorm}
	Comparison of the data (circles) and combinatoric (histogram) like-sign events.
	The data and
	combinatoric like-signs are summed over the low and high mass data sets
	and both targets.  
	The inset figures show the mass, $x_F$ and $p_T$
	like-sign distributions over the mass range used in this analysis.
	}

\end{figure}

The normalization of the randoms was accomplished by comparing a subset
of the randoms (those with two muons with the same charge), to events
which satisfied the like-sign trigger in the data.  These like-sign events 
were analyzed in the same way as any other event, except that one of the tracks
was reflected about $y=0$ when the kinematics of the events were calculated.
Figure \ref{fig:lsnorm} shows the real like-sign events compared to the
combinatoric like-sign events, scaled to match the real like-signs.  


\subsubsection{Trigger Cuts}

For any given event, the possibility existed that the muons which fired
the trigger and resulted in the event being read out by the DAQ were not
the same muons which were reconstructed by the analysis.  The easiest way
for this to occur was to have a single, beam-like muon fire a set of 
hodoscopes, one of which was outside of the matrix, while simultaneously a 
second, low-momentum muon fired the missing matrix element.  When combined
with a second, uncorrelated, oppositely charged muon which fired a valid
matrix element, this would appear to be a high $p_T$ continuum dimuon 
event.
We therefore had to check that the muons in each dimuon event satisfied the 
trigger condition which had caused the reading out of that event.  
This was done by comparing the reconstructed 
positions of the muons at each hodoscope plane to the aperture of the 
hodoscopes in the trigger which fired.  If the track was outside that
aperture (which was widened slightly so that we would not lose events
due to chamber resolution effects), it was discarded.  Events which fired
PhysA2 (the like-sign trigger) were 
binned separately from events which fired the other PhysA triggers.

In addition to the trigger cuts outlined above, comparisons of the data and
Monte Carlo suggested that there was a problem with the trigger hardware
corresponding to certain matrix elements.  
To first order, the acceptance
of the spectrometer should be left-right symmetric.
A separate analysis \cite{bib:E866-UP-POL}, which studied 
events from the dump in the high mass data, noticed that several matrix 
elements yielded much more data relative to Monte Carlo on one side of
the spectrometer than the other.  
This discrepancy was much larger than could be explained by left-right
asymmetries in the spectrometer, and thus these matrix elements were
suspect.
Similar studies were performed for the target data, confirming that there was 
a problem with several matrix-elements.  
These matrix elements, which tended to be at the edge of the acceptance 
where the statistics were generally poor.
Events with muons occupying these elements were cut from the final event 
sample.

\subsubsection{Event Geometry}

Several events which passed through all of the second pass cuts occupied
regions of the spectrometer which complicated the reconstruction of their
kinematics, or were judged to raise questions about their origins.  Events
which had a muon which appeared beam-like (tracks with 
$|\theta_y^{target}|<0.006$ or tracks which were on the wrong side of the
dump given their charge) were discarded.  Such tracks could have originated
from the decay of a hadron produced in the dump, and also followed a 
trajectory which skimmed the edge of the dump, calling into question the
amount of energy lost by the track.  

Another problematic class of events was found by studying the Monte Carlo.
The energy lost by tracks in the Monte Carlo was plotted versus the $x$
position of the tracks at SM3.  Tracks near the edge of the acceptance, 
$|x_{SM3}| > 24''$, had significantly more energy loss in them than 
tracks with $|x_{SM3}| < 24''$.  The source of this excess energy loss
was the walls and teeth of the SM12 magnet, indicating that this class
of tracks passed through a region of the magnetic field which was highly
uncertain.

As discussed in the previous chapter, energy-loss fluctuations and multiple
coulomb scattering cause the events to appear to reconstruct away from the 
actual interaction point in the target.  To limit the fluctuation of energy 
loss and the effects of multiple scattering in the data, and to provide
additional discrimination against events which originated outside
the target volume, we required the event to reconstruct to within
$2.5''$ of the nominal transverse position of the beam.

\subsubsection{Spill Quality Cuts}

During each beam spill, a number of quantities were monitored which told
us the overall quality of the data being collected.  These included the
intensity of the beam (SEM and IC3), the luminosity of the target (AMON), 
the live-time of the spectrometer (AMONSB/AMON) and the beam duty
factor (DUTY).  Table \ref{table:spillcuts} lists the spill quality cuts
which were required for each spectrometer setting.

\begin{table}
\centering
\caption[Spill quality cuts.]{
	Spill quality cuts.  Events were kept if they obeyed the inequality.
	}
  \vspace{10pt}
  \label{table:spillcuts}
  \begin{tabular}{|c|ccc|}
  \hline
      & \multicolumn{3}{c|}{mass setting} \\
  \hline
  Cut & low & intermediate & high \\
  \hline\hline
  SEM           &  $>150$  & $>100$  &  $>100$    \\
  IC3           &  $>5000$ & $>5000$ &  $>5000$   \\
  AMONSB / AMON &  $>0.9$  & $>0.9$  &  $>0.9$    \\
  DUTY          &  $>25$   & $>50$   &  $>25$     \\
  \hline
  \end{tabular}
\end{table}

\subsection{Luminosity}

The intensity of the beam in each spill was measured with the SEM6 counter
on the Meson East beam line.  This recorded a SEM count which was proportional
to the number of protons in the spill.  The SEM6 counter was calibrated 
during special calibration runs where a thin copper foil was inserted into
the beam and a SEM count accumulated during a number of spills.  By studying
the rate of the 1368 keV $\gamma$'s emitted by the $^{24}$Na produced in the 
interaction, the number of protons incident on the foil could be deduced,
and the SEM response calibrated.  These calibrations rely upon a measurement
of the $^{24}$Na production cross section \cite{bib:NA24-CROSS} of
$3.90\pm 0.11$ mb\footnote{\setlength{\baselineskip}{\singlespace}Although reference \cite{bib:NA24-CROSS}
measures the $^{24}Na$ cross section using 400-GeV protons, it is generally
believed that the cross section is independent of energy.}.

Table \ref{table:me6semcal} shows the results of several calibration 
measurements dating back to the original E605 experiment.  Based on these 
measurements (taking the more recent measurements with greater weight), we 
estimate a SEM response of $(0.79 \pm 0.051) \times 10^8$ protons / 
SEM\cite{bib:CNB-NORMALIZATION}.
The uncertainty in the normalization was determined by adding the uncertainty 
due to the $^{24}$Na cross section in quadrature with the standard deviation 
of the measurements in table \ref{table:me6semcal}.  



\begin{table}
  \caption{
	\setlength{\baselineskip}{\singlespace}
	SEM calibration measurements on the Meson East beam-line.
	}
  \label{table:me6semcal}
  \begin{center}
    \begin{tabular}{|c|c|c|c|c|c|}
      \hline
      Date & Protons & Protons/SEM $\times 10^8$\\
      \hline \hline
      01/26/84 & 2.8E13 & .83 \\
      02/11/85 & 6.9E12 & .78 \\
      03/25/85 & 2.0E12 & .85 \\
      05/06/85 & 2.0E12 & .90 \\
      07/25/85 & 9.9E13 & .80 \\
      08/29/85 & 9.6E13 & .80 \\
      11/10/87 & 3.4E13 & .79 \\
      11/16/87 & 2.4E13 & .77 \\
      02/11/88 & 3.6E13 & .78 \\      
      10/29/96 & 1.7E12 & .885 \\
      10/29/96 & 8.5E11 & .915 \\
      12/20/96 & 5.1E13 & $.76$ \\
      08/15/97 & 5.0E11 & $.86$ \\
      08/29/97 & 2.0E11 & $.77$ \\
      \hline
    \end{tabular}
  \end{center}

\end{table}

Studies of the linearity of the SEM response were also performed by 
comparing the SEM counts to the luminosity as measured by the AMON and 
WMON detectors, and comparing to a redundant measurement of the beam intensity
using the ion chambers.  These studies showed the SEM response was linear with 
increasing beam intensity, but subject to a constant offset which varied from 
run to run.  These offsets, shown in table 
\ref{table:semoff}, were subtracted from the SEM counts before calculating
how many protons were incident on the target in each spill.

\begin{table}
  \caption[SEM offsets.]{
	\setlength{\baselineskip}{\singlespace}
	SEM offsets.  These were subtracted from each spill.
	$I_{\text{run}}$ is the run number.
	}
  \label{table:semoff}
  \begin{center}
    \begin{tabular}{|c|c|c|}
      \hline
      Starting run & Ending run & SEM offset \\
      \hline \hline
      1551 & 1588 & 27 \\
      1589 & 1620 & $1.479 \times I_{\text{run}} - 2436$ \\
      1631 & 1690 & 28 \\
      1879 & 1884 & 28 \\      
      1886 & 1887 & 155 \\
      1888 & ------ & 275 \\
      1889 & 1944 & 375 \\
      1945 & 2020 & 130 \\
      2021 & 2046 & 0 \\
      2049 & 2053 & 98 \\
      2056 & 2078 & 100 \\
      2079 & 2115 & 30 \\
      \hline
    \end{tabular}
  \end{center}

\end{table}

Both the density and attenuation lengths of the targets depended upon the
exact composition of the targets.  Therefore, assays of the cryogenic 
liquids were performed using samples taken when the target flasks were 
emptied for the Christmas shutdown (end of data set 7) and at the end of 
the planned run (end of data set 11).  The cryogenic liquids used in the 
first fill had negligible amounts of contaminants.  The same was true
for the liquid hydrogen in the second fill.  However, two separate 
analyses of the second fill's liquid deuterium showed a significant
HD contamination.  Table \ref{table:contamination} shows the
measured contamination and the estimated nucleon content in the 
target.  Once the final $pp$ cross sections had been measured, the
deuterium cross sections were corrected for the $\sim 3\%$ contamination.

\begin{table}
 \caption{
	\setlength{\baselineskip}{\singlespace}
	Percent molecular and atomic abundance of second deuterium fill.
	}
  \begin{center}
  \begin{tabular}{|c|c|c|}
    \hline
    Molecule & percent abundance & percent abundance \\
             & (first analysis)  & (second analysis) \\
    \hline\hline
    $D_{2}$ & $94.13 \pm 0.58$ & $92.7 \pm 0.8$ \\
    $HD$    & $ 5.82 \pm 0.58$ & $6.89 \pm 0.69$ \\
    $H_{2}$ & $ 0.05 \pm 0.01$ & $0.147 \pm 0.015$ \\    
    other   & --               & -- \\ 
    \hline
    \hline
    Atom    & \multicolumn{2}{c|}{estimated abundance} \\
    \hline 
    D       & \multicolumn{2}{c|}{$97.00\pm 0.6$} \\
    H       & \multicolumn{2}{c|}{$ 3.00\pm 0.6$} \\
    \hline
  \end{tabular}
  \end{center}
  \label{table:contamination}
 
\end{table}

In order to determine the target densities, the 
vapor pressures of the cryogenic targets were measured and periodically
recorded.  The high circulation rate of the liquids, combined with the
comparatively low beam current ensured that the cryogenic liquids did not 
boil during the spill which would have reduced the densities of the targets.
Table \ref{table:targetpressures} shows the mean target
pressure for selected data sets in the experiment.  

\begin{table}[h]

    \caption{  \label{table:targetpressures}
	\setlength{\baselineskip}{\singlespace}
	Target pressures (in psi) for selected data sets.
	}
  \begin{center}
    \begin{tabular}{|c|c|c|}
      \hline
      Data set & hydrogen & deuterium \\
      \hline \hline
      5        & 14.97    &  14.92 \\
      7        & 15.04    &  14.96 \\
      8        & 15.11    &  15.17 \\
      11       & 15.15    &  15.21 \\
      \hline
    \end{tabular}

  \end{center}
\end{table}

Cryogenic data tables \cite{bib:CRYO-TABLES} were used to calculate the 
target densities from the average vapor pressures.  Taking the pressure
$P$ to be measured in psi, and density $\rho$ in g/cm$^3$, the expression
for the density of hydrogen is given by
\begin{equation}
  \label{eqn:lh2density}
  \frac{1}{\rho_h} = 62.473 \, \{ 0.2115 + 1.171 \times 10^{-3} P - 1.109 \times 10^{-5} P^2 \},
\end{equation}
\noindent and that for deuterium by
\begin{equation}
  \label{eqn:ld2density}
  \rho_d = 4.028 \times 10^{-3} \{ 43.291 - 3.4176 \frac{P}{14.6959} + 0.5783 ( \frac{P}{14.6959} )^2 \}
\end{equation}
\noindent  These result in densities which vary from data set
to data set by no more than $\sim 2\%$, which is negligible
compared to the $6.5\%$ uncertainty in 
the beam normalization.
From the measured pressures, the densities of
the hydrogen and deuterium targets were determined to be
\begin{eqnarray}
\nonumber \rho_h = 0.0706 \, \text{g/cm}^3 \\
\nonumber \rho_d = 0.1630 \, \text{g/cm}^3.
\end{eqnarray}

As reported in the particle data book \cite{bib:PDB}, the hadronic interaction
lengths for hydrogen and deuterium are $50.8 \, \text{g/cm}^2$ and
$54.7 \, \text{g/cm}^2$ respectively.  To express these lengths in terms of 
the physical length, as opposed to the ``thickness'' of the target, we need 
to divide by the density of the targets.  For the cryogenics used in the first
fill, and the liquid hydrogen from the second fill, this is all we needed to 
do.  For the liquid deuterium from the second fill, however, we needed to 
correct for the hydrogen contamination by taking the weighted average.  The 
resulting hadronic absorption lengths are
\begin{eqnarray}
\begin{aligned}
\nonumber & \lambda_{lh2}                & = 719.5 \, \text{cm} \\
\nonumber & \lambda_{ld2}^{(1\text{st})} & = 335.6 \, \text{cm} \\
\nonumber & \lambda_{ld2}^{(2\text{nd})} & = 340.6 \, \text{cm}
\end{aligned}
\end{eqnarray}
\noindent where the superscripts indicate either the first or second fill.
The total integrated luminosities for each data set and the entire data sample
are shown in table \ref{table:luminosity}.

\begin{table}
  \centering
  \caption{
	\setlength{\baselineskip}{\singlespace}
	Integrated luminosities in the E866 data sets.
	}
  \vspace{10pt}
  \label{table:luminosity}
  \begin{tabular}{|c|c|c|}
     \hline
      Data Set & Integrated $H_2$        & Integrated $D_2$        \\
               & Luminosity (nucleon/nb) & Luminosity (nucleon/nb) \\
     \hline\hline
     5  & $4.74 \times 10^6$ & $1.04 \times 10^7$ \\ 
     7  & $1.97 \times 10^7$ & $4.34 \times 10^7$ \\
     8  & $4.04 \times 10^7$ & $8.95 \times 10^7$ \\
     9  & $2.95 \times 10^6$ & $6.65 \times 10^6$ \\
     10 & $4.72 \times 10^5$ & $1.03 \times 10^6$ \\
     11 & $1.03 \times 10^7$ & $2.28 \times 10^7$ \\
     \hline\hline
     Total & $7.86 \times 10^7$ & $1.74 \times 10^8$ \\ 
     \hline
  \end{tabular}
\end{table}

\subsection{Trigger Efficiencies}

In order to properly measure the efficiencies of the hodoscopes, it was
necessary to use a sample of events which were triggered independently
of the hodoscopes being measured.  Because we only use the information
from the Y hodoscopes in the experiment, events which triggered on the
X hodoscopes could provide this independent measurement.  Unfortunately,
the rate of events which satisfied the $\text{X134L}\cdot\text{X134R}$ trigger 
(a muon on either side of the spectrometer) was prohibitively high
to take along with the physics data.  
Therefore, several specialized runs were taken periodically to determine the 
efficiencies of the Y hodoscopes.  A specialized trigger required that 
five hodoscope planes on one side of the spectrometer were hit.  Reconstructed
single muon tracks which fired at least six hodoscopes (one in each plane) 
were then studied to determine the efficiency of the hodoscope in the seventh 
plane.  The typical efficiencies measured were $\sim 97\%$.  However, these 
specialized runs did not measure the efficiencies near the edge of the 
acceptance with sufficient precision.  To increase the statistics in these 
regions, we used events from the full data set.  Although these events 
introduced a bias from the trigger to the efficiency measurement, they did
probe the edge of the acceptance with far more statistics.  The efficiencies
measured here were checked against those of the specialized runs and found
to be largely consistent. 


\subsection{Chamber Efficiencies}

In order to study the efficiency of each detector plane, we began with a
sample of tracks which would have been reconstructed regardless of whether
or not the particular detector plane being studied had fired or not.  This
was possible due to the redundant tracking information at each station.
Each plane was divided into quadrants, and the ratio of the number of tracks
which fired in a given quadrant to the total number of tracks passing
through the quadrant was used to estimate the efficiency of the drift
chamber.

The use of single-hit TDC's in the DAQ electronics introduced an 
additional source of inefficiency.  A second charged particle which 
passed through a drift cell following the first particle in an event
would not have its position recorded to tape, resulting in the event
being more difficult to reconstruct.  This effect would have a definite
dependence on the event rate, and therefore the beam intensity.

\begin{figure}
  \centering

  \includegraphics[width=0.8\linewidth,clip]{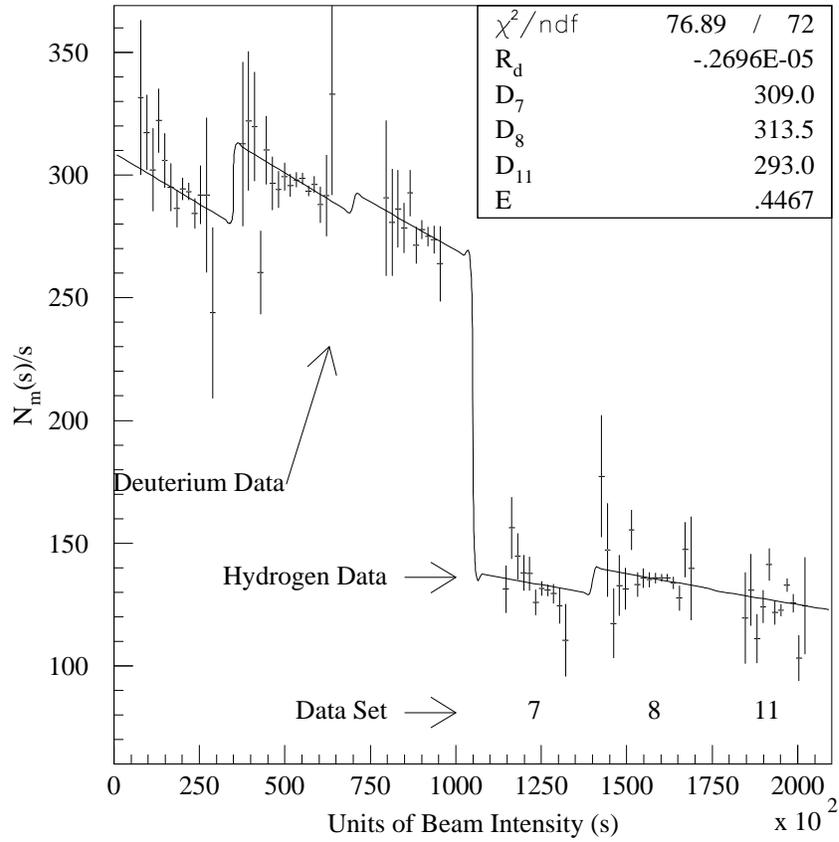}

  \caption[Rate dependent inefficiency.]{
	\setlength{\baselineskip}{\singlespace}	
	\label{fig:highrate}Rate dependent inefficiency.
	Dimuon yield ($N_m$) per unit beam intensity 
	($s$) versus beam intensity for the high mass data, as given in 
	reference \protect\cite{bib:THESIS-RUSTY}.
	}

\end{figure}

To correct for the rate dependence, the number of reconstructed events 
per unit beam intensity (measured in terms of SEM counts) was plotted
versus the beam intensity and fit to a straight line.  The fits for hydrogen
and deuterium were constrained by the relative number of extra drift chamber
hits in an average event $F=\frac{N^{\text{extra}}_d}{N^{\text{extra}}_h}$
for each target \cite{bib:THESIS-RUSTY}.  This meant 
that only one parameter, the slope $R_d$ of the efficiency-versus-intensity 
plot for deuterium, needed to be determined for each mass setting.  
Figure \ref{fig:highrate} shows the fit for the high mass data, and table 
\ref{table:ratedep} shows the corrections obtained by reference 
\cite{bib:THESIS-RUSTY}.

\begin{table}
  
  \caption[Rate-dependent corrections to the hydrogen and deuterium yields.]{
	\label{table:ratedep}
	\setlength{\baselineskip}{\singlespace}
	Rate-dependent corrections to the hydrogen and deuterium yields.  Each event
    	is weighted by $1 - SEM \times R_i$, where $i$ is the target.
	}
  \begin{center}
    \begin{tabular}{|c|c|c|c|}
      \hline
      Mass Setting & $R_d$ & F & $R_h = R_d / F$ \\
      \hline \hline
      Low          & $-0.2035 \times 10^{-4}$ & 1.51 & $-.1348 \times 10^{-4}$ \\
      Intermediate & $-0.4976 \times 10^{-5}$ & 1.18 & $-.4152 \times 10^{-5}$ \\
      High         & $-0.2696 \times 10^{-5}$ & 1.78 & $-.1515 \times 10^{-5}$ \\
      \hline
    \end{tabular}
  \end{center}

\end{table}

\subsection{Monte Carlo and Acceptance}

\label{chapter:montecarlo}

The acceptance of the spectrometer was a complicated function involving the
detector geometry, the magnitudes of the magnetic fields in each of the
three spectrometer magnets, the efficiencies of the detectors and trigger,
and the various physical processes which muons traversing the spectrometer
were subject to.   This level of complexity required the use of a Monte Carlo
simulation to determine the acceptance of the spectrometer.

The E866 Monte Carlo consisted of custom routines designed to simulate the
production of dimuon events, the passage of the resulting muons through the 
spectrometer, and the detector response to their passage.  The output of the
Monte Carlo was a file whose format was identical to that produced by the DAQ
during the experiment.  This enabled us to analyze the Monte Carlo events with
the exact same analysis chain as the real data, minimizing any systematic 
differences between the data and Monte Carlo.

To further minimize these differences, the Monte Carlo was configured to 
simulate as closely as possible the state of the spectrometer while it was
taking data, the physical processes the muons were subject to in passing
through the spectrometer, and the physical processes giving rise to 
the dimuon events.  In chapter \ref{chapter:analysis} we discussed how
we obtain information about the state of the spectrometer (magnetic fields,
detector alignments and efficiencies, etc...).  
We will discuss how events were generated in the Monte Carlo,
the simulation of the physical processes the resulting muons were subject to,
and the final set of calculations needed to compute the acceptance of the
spectrometer.

\subsubsection{Event Generation}

In order to generate a continuum dimuon event, we had to sample the six
kinematic variables $M_{\mu^+\mu^-}$, $x_F$, $p_T$, $\phi_p$, $\phi_d$, and 
$\theta_d$ from distributions which closely approximate the actual differential
cross section $d^6\sigma/dMdx_Fdp_Td\theta_dd\phi_pd\phi_d$. {\it A priori} we 
know the cross section differential in the angular variables goes like,
\begin{equation}
	\frac{d^3\sigma}{d\theta_dd\phi_dd\phi_p} \propto 
	1 + \cos^2(\theta_d)
\end{equation}
\noindent since there is no preferred direction in the experiment to give
a $\phi$ dependence, and the virtual photon in a $q\bar{q}$ annihilation
is produced transversely polarized as predicted by QED and confirmed by
experiment 
\cite{bib:FNAL-E615,bib:E772-DY-POL,bib:E866-JPSI-POL,bib:E866-UP-POL}.

Next-to-leading order calculations of the doubly-differential cross section 
$d^2\sigma/dMdx_F$ agree well with existing measurements of the deuterium and 
nuclear cross sections, but problems arise in QCD in calculating the full 
triple differential cross section $d^3\sigma/dMdx_Fdp_T$.  Therefore, an
empirical form \cite{bib:FNAL-E288} for the $p_T$ dependence of the cross 
section was used
\begin{equation}
	\label{eqn:d3sample}
	\frac{d^3\sigma_{\text{empirical}}}{dMdx_Fdp_T} 
	\propto
	\frac{1}{p_T^0}
	\frac{10 p_T / p_T^0}{(1+(p_T/p_T^0)^2)^6} \times
	\frac{d^2\sigma_{\text{NLO}}}{dMdx_F}
\end{equation}
where $p_T^0 = 2.8$ GeV was fit to both the hydrogen and deuterium data, and 
the MRST 98 \cite{bib:MRST98} partons were used in the NLO calculations of 
$d^2\sigma/dMdx_F$.

\subsubsection{Physics Simulation and Detector Response}

Once an event had been sampled from the above distributions, the kinematics of
each muon were calculated and boosted into the lab frame, and the event vertex
sampled from 
\begin{equation}
	P(z) = \frac{1}{\lambda} \frac{e^{-z/\lambda}}{1 - e^{-L/\lambda}}.
\end{equation}
\noindent Inside the target, the very small amount of multiple scattering 
the muons were subject to was simulated.  This multiple scattering was found 
to be negligible.

After the effects on the muons in the target had been simulated, the muon's
positions were extrapolated to the point at which the SM0/SM12 field began.
The same magnetic field maps which were used in the analysis as well as 
the overall field strengths as computed from the data were used here.
The muons were propagated through the magnetic field one at a time in 5 cm
steps.  At each step in the field, the magnetic field kick was calculated, 
using 
\begin{eqnarray}
      dp_x = -q \; dz \, B_y \\
      dp_y =  q \; dz \, B_x
\end{eqnarray}
which neglected only the small $z$ component of the force.\footnote{\setlength{\baselineskip}{\singlespace}To be
precise, we computed the changes in the $x$ and $y$ components of the muon's
momentum, then applied momentum conservation to calculate the final $z$
component of the momentum.  It can be shown that this causes a 
systematic error in the momentum at each step which is negligible over
the entire magnetic field.}

In addition to the simulation of the magnetic fields, the effects of the beam 
dump and absorbing wall had to be included.  Multiple scattering was simulated
in each 5 cm step of the dump and absorbing wall, using the {\it gmols} routine
from the standard GEANT \cite{bib:GEANT} simulation package.  This was the 
only case where a GEANT routine was used in the Monte Carlo.  Multiple 
scattering was also simulated in the materials found in the helium bags which 
filled SM0 and SM12 downstream of the absorber, the air and helium the muons 
had to pass through, the hodoscope and drift chamber planes, and the bulk 
material at station 4.

  The energy lost by muons traversing the spectrometer was sampled from a 
  set of fast lookup tables generated using Von Ginniken's TRAMU program 
  \cite{bib:TRAMU}.
  Twelve tables for muons with incident momenta between 10 and 600 GeV were
  interpolated to obtain the energy loss for muons of arbitraty momenta.
  This number was then scaled from the length of material assumed in the 
  table to the step size of 5 cm, and subtracted from the muon's momenta
  in each step.
  A new energy loss was sampled once the muon had traversed the length of
  material specified by the energy loss routines.
  This greatly simplified the energy-loss simulation in the dump, where the
  total length of material the muon would pass through was not known 
  beforehand.
  The only other places where energy loss was simulated was in the teeth and
  walls of the SM12 magnet, and the hadronic and EM calorimeters at station 4.
  A precise simulation of the energy loss was unnecessary here, so a constant
  energy loss was subtracted in these materials.

Equation \ref{eqn:crossdefined} indicates separate corrections for the 
acceptance of the spectrometer and detector efficiencies.  In reality,
the Monte Carlo simulated detector inefficiencies based on the
measurements described above.  Thus, the acceptance calculations contained
the efficiency corrections.  Only the rate-dependent inefficiencies
were corrected for separately.

\subsubsection{Acceptance Calculation}

Once a sufficient number of Monte Carlo events had been recorded, they were
subjected to the same analysis chain and cuts described in chapters 
\ref{chapter:analysis} and \ref{chapter:cross}.  The only difference was
that spill quality cuts were not applied to the Monte Carlo.  Once we
had filtered the Monte Carlo event sample, the acceptance in any given bin 
could  be calculated as the ratio of the number of surviving Monte Carlo 
events in that bin divided by the total number of events generated in that bin
\begin{equation}
        \label{eqn:alpha}
	\alpha(M,x_F,p_T) = \frac{N_{\text{MC}}^{\text{acc}}(M,x_F,p_T)}{N_{\text{MC}}^{\text{gen}}(M,x_F,p_T)}.
\end{equation}


Calculating the number of accepted and generated Monte Carlo events in each 
bin would have been a straightforward counting exercise, save for two reasons.
First, comparisons of the data to the Monte Carlo revealed differences on the 
order of $10$ to $15\%$ which we wished to eliminate.\footnote{\setlength{\baselineskip}{\singlespace}These differences
were consistent with the observed differences between the measured cross
sections, and the NLO calculations upon which the Monte Carlo generator
was based.}  Second, in order
to eliminate the systematic error between the hydrogen and deuterium cross 
sections introduced by finite Monte Carlo statistics in the acceptance 
calculation, the same sample of Monte Carlo events were used to calculate the 
acceptance for hydrogen and deuterium.  Since the event sample had been 
generated using either a hydrogen or deuterium target, the events had to
be reweighted by the cross section appropriate to the target being 
calculated.  

Differences between the data and Monte Carlo were reduced by 
reweighting the Monte Carlo events as a function of $M$, $x_F$ and $p_T$.
These functions were determined by polynomial fits to the ratios of 
the data divided by the Monte Carlo.  Furthermore, the $p_T$ cross section 
was seen to vary as a function of $M$ and $x_F$.  This correlation was dealt 
with by fitting $p_T^0$ in each bin, resulting in the empirical form 
\begin{equation}
  \label{pt0func}
  p_T^0(M,x_F) = 1.21  + 0.350 M - 0.0182 M^2 + 1.37 x_F + 2.60 x_F^2 
\end{equation}
which was the same for both targets, within the uncertainties in the fit 
parameters.  The resulting weighting functions for the Monte Carlo were 
consistent with the differences observed between the measured cross sections 
and the double-differential cross section tables used to generate the Monte 
Carlo.

Monte Carlo events which had been generated using hydrogen as the target were 
reweighted with the deuterium cross section and vertex distribution when 
calculating the deuterium cross section.  Thus, the deuterium yield calculated 
from a hydrogen Monte Carlo event sample was given by
\begin{equation}
   N_{\text{D2MC}} = \sum_{i=1}^{N_\text{H2MC}} \frac{d^3\sigma_{pd}(M^i,x_F^i,p_T^i)}{d^3\sigma_{pp}(M^i,x_F^i,p_T^i)} \times   \left( \frac{e^{-z/\lambda_{\text{D2}}}}{e^{-z/\lambda_{\text{H2}}}} \right)
\end{equation}
Hydrogen yields were calculated from deuterium Monte Carlo samples in a 
similar manner:
\begin{equation}
   N_{\text{H2MC}} = \sum_{i=1}^{N_\text{D2MC}} \frac{d^3\sigma_{pp}(M^i,x_F^i,p_T^i)}{d^3\sigma_{pd}(M^i,x_F^i,p_T^i)} \times   \left( \frac{e^{-z/\lambda_{\text{H2}}}}{e^{-z/\lambda_{\text{D2}}}} \right). 
\end{equation}
\noindent This procedure was carried out for both the accepted and 
generated events, maintaining the relative normalization of the thrown and
accepted Monte Carlo, and equation \ref{eqn:alpha} was used to compute the 
acceptance.

Figures \ref{fig:dvmc_pair} and \ref{fig:dvmc_detector} compare the data and
Monte Carlo for several reconstructed pair- and detector-quantities, summed 
over both targets and averaged over all data sets.  The agreement between the 
data and Monte Carlo is quite good for most physics and detector distributions.
The discrepancies in the single track momentum distributions, which occur at
larger momenta where the statistics are poor, are thought to be due to 
beam-alignment problems.  The other major discrepancy is in the ZUNIN distributions.
This is partly due to a slightly wider ZUNIN distribution in the Monte Carlo
than is observed in the data, but mainly due to the centroids of the data
and Monte Carlo distributions not being precisely aligned.\footnote{\setlength{\baselineskip}{\singlespace}This is
a major component of the estimated systematic errors discussed below.}

\begin{figure}
\centering
\includegraphics[clip,width=0.85\linewidth]{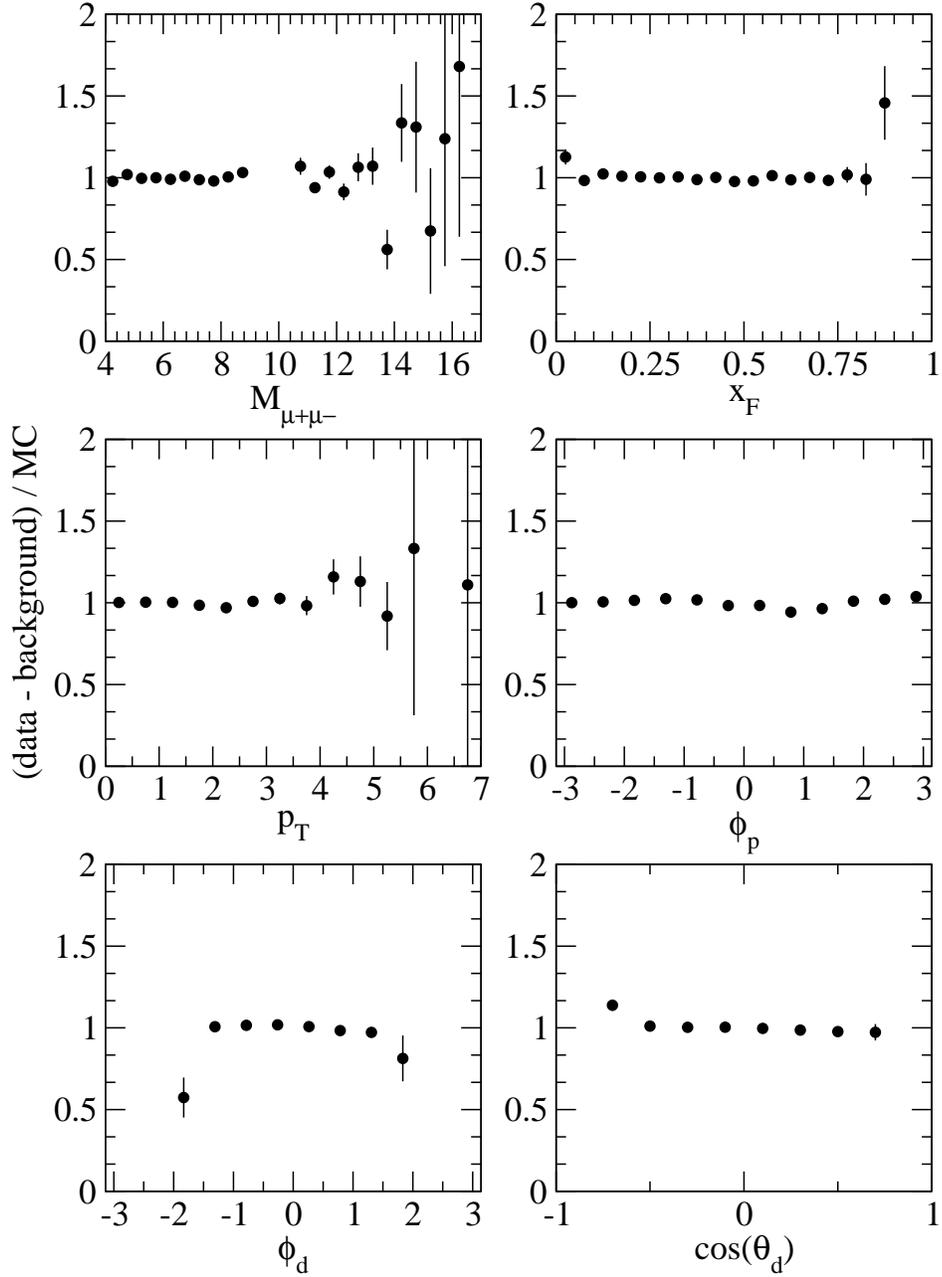}
\caption[Comparison between data and Monte Carlo yields.]{
	\setlength{\baselineskip}{\singlespace}
	\label{fig:dvmc_pair}Comparison between data and Monte Carlo yields.
	The ratios of data to Monte Carlo yields versus
	several ``pair'' quantities (mass, $x_F$, $p_T$, $\phi_d$, 
	$\phi_p$, $cos(\theta_d)$ and ZUNIN) and the positive and negative track momenta.
	Ratios
	averaged over all targets and data sets.
	}
\end{figure}

\begin{figure}
\centering
\includegraphics[clip,width=0.85\linewidth]{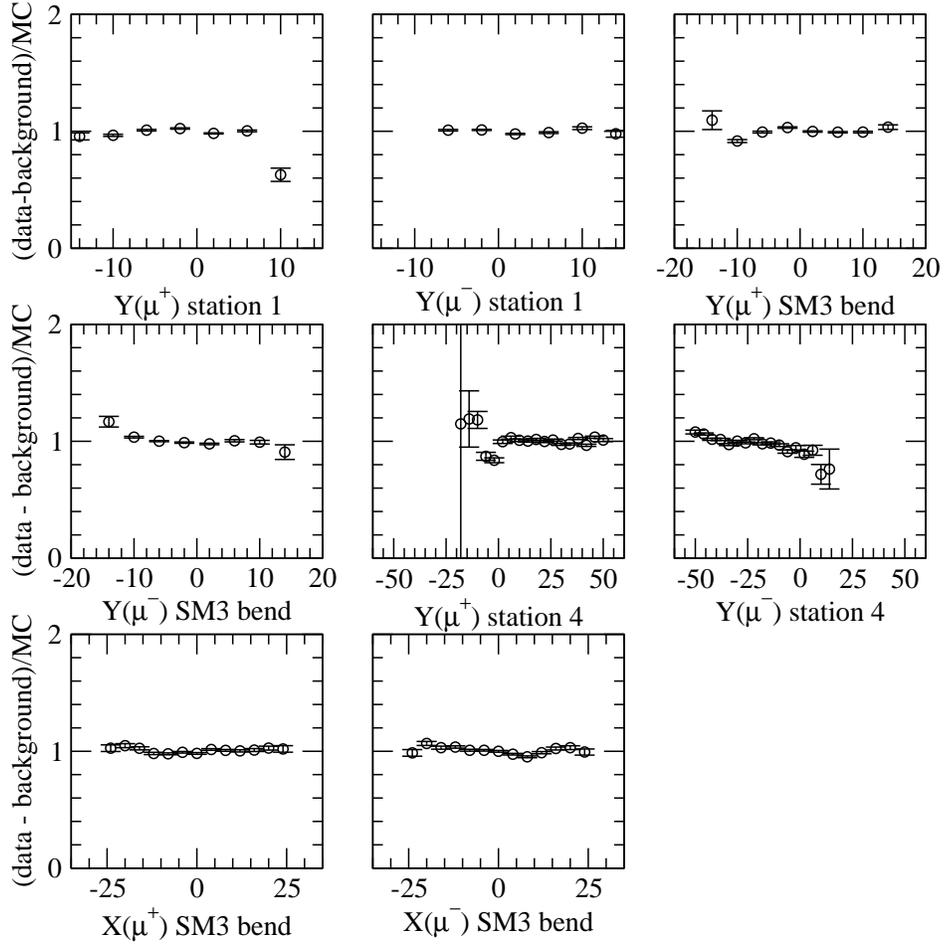}
\caption[Comparison between data and Monte Carlo yields.]{
	\setlength{\baselineskip}{\singlespace}
	\label{fig:dvmc_detector}Comparison between data and Monte Carlo yields.
	The ratios of data to Monte Carlo yields versus the positions of the
	positive (negative) track projected onto the station 1 hodoscope-,
	SM3 bend- and station 4 hodoscope-planes.
	Ratios	averaged over all targets and data sets.
	}
\end{figure}

\subsection{Combining the Data}

Once all of the data were binned and subjected to all of the corrections 
described above, we had a measurement of the triple-differential cross sections
$d^3\sigma(pp)/dMdx_Fdp_T$ and $d^3\sigma(pd)/dMdx_Fdp_T$ for each of the
six data sets used in the analysis.  For the purposes of the following 
discussion, we denote the measured cross section in a given $M$, $x_F$, 
$p_T$ bin as $\sigma_i \pm \delta \sigma_i \pm \Delta \sigma_i$, where $i$ 
represents the data set, $\delta \sigma_i$ is the statistical uncertainty 
and $\Delta \sigma_i$ is the systematic uncertainty.

The results from the six data sets were first combined into three results
corresponding to each mass setting.  We used the weighted average
\begin{equation}
\label{eqn:weighted_average}
\sigma \pm \delta \sigma = 
	\frac{{\sum_{i}} \sigma_i \omega_{i}}{\sum_i \omega_i} \pm
	\sqrt{\frac{1}{\sum_i \omega_i}}
\end{equation}
where the weights were given by 
$\omega_i = \frac{1}{\left(\delta \sigma_i\right)^2}$.  After the average
cross section for each mass setting was calculated, the three mass settings
were combined into a single result, again using equation 
\ref{eqn:weighted_average}.
This procedure neglected the
systematic uncertainties between data sets and mass settings, which were
sufficiently small so that the average was not affected much by the omission.
Also, it would not have been appropriate to use the systematic
uncertainties in computing the averages, as most of the systematic uncertainty
in the cross sections were (by design) common to the hydrogen and deuterium
measurements.

\subsubsection{Systematic Uncertainties}

Computing the systematic uncertainty in the combined cross section was a 
fairly straightforward exercise in error propagation.  The
standard error propagation formula for an expression $f(x_1,x_2,...,x_N)$
is given by
\begin{equation}
\label{eqn:errprop}
\left( \Delta f \right)^2 = \sum_{i=1}^N \left( \frac{\delta f}{\delta x_i} \right)^2 \left( \Delta x_i \right)^2.
\end{equation}

\noindent From this formula, it can be shown that the systematic uncertainty 
in equation \ref{eqn:weighted_average} is given by

\begin{equation}
\label{eqn:syst_unc_in_weighted_average}
\left( \Delta \sigma \right)^2 = \frac{\sum_i \omega_i^2 \left( \Delta \sigma_i \right)^2}{\left(\sum_i \omega_i\right)^2}
\end{equation}

\noindent where the $\omega_i$ are the statistical weights used in the weighted
average.  One need only compute the systematic uncertainty in each data set
(mass setting) $\Delta \sigma_i$.

Recalling equation \ref{eqn:crossdefined}, the systematic uncertainty on a
given cross section measurement will, by equation \ref{eqn:errprop}, be
given by
\begin{equation}\label{eqn:errprop_cross}
\left(\frac{\Delta \sigma_i}{\sigma_i}\right)^2 =
\left(\frac{\Delta \alpha_i}{\alpha_i}\right)^2 +
\left(\frac{\Delta \epsilon_i}{\epsilon_i}\right)^2 +
\left(\frac{\Delta {\cal L}_i}{{\cal L}_i}\right)^2 +
\left(\frac{\Delta N_{\mu^+\mu^-}}{N_{\mu^+\mu^-}}\right)^2
\end{equation}

\noindent We must be careful here, however.  The uncertainty in the
luminosity $\Delta {\cal L}$, for instance, is dominated by a $\pm 6.5\%$
uncertainty in the beam normalization.  This uncertainty is common to 
all data sets and mass settings.  It would not be appropriate to treat
it as an independent
$6.5\%$ uncertainty in each measurement.  We also assume that the 
uncertainties in detector efficiencies $\Delta \epsilon$ are common to
each data set in a given mass setting, since events with similar kinematics
will occupy similar regions of the spectrometer.  We discuss how the 
systematic uncertainties were treated below.

\subsubsection{Uncorrelated Systematic Uncertainties}

The systematic uncertainties, not including the $\pm 6.5\%$ normalization
uncertainty, fell into three categories: uncorrelated uncertainties, 
uncorrelated uncertainties which are common to a given mass setting, and 
correlated uncertainties which are common to a given mass setting.
In general, we treated these systematic uncertainties separately.  In the
end they were combined 
in quadrature to estimate the total systematic uncertainty
on the averaged cross sections.

The only uncorrelated systematic uncertainty which was not shared between
the data sets in a given mass setting was the uncertainty in the acceptance
due to the statistical uncertainty in the Monte Carlo.  The uncertainty in each
mass setting due to this statistical uncertainty was therefore estimated 
to be, according to the prescription of equation \ref{eqn:syst_unc_in_weighted_average}
\begin{equation}
\label{eqn:est_acc_syst_statistical}
\left(\Delta \sigma_{\alpha,{\text stat}}\right)^2 =
\frac{
	\sum_i \omega_i^2 
      	\left( \Delta \sigma^{\alpha, \text{stat}}_i \right)^2
      }
      {
	\left(\sum_i\omega_i\right)^2
      }
\end{equation}

\noindent where $\Delta \sigma^{\alpha, \text{stat}}_i = \frac{\Delta \alpha_i^{\text{stat}}}{\alpha_i} \sigma_i$
is the uncertainty in the $i^{\text{th}}$ measurement due to the statistical
uncertainty in the acceptance $\Delta \alpha_i$.

The next step combined all of the uncorrelated uncertainties which were 
common to a given mass setting.  These included the uncertainties in the
acceptance in each mass setting, calculated above, and the uncertainties
in the detector efficiencies.  Again we use equation \ref{eqn:syst_unc_in_weighted_average}
\begin{equation}
\label{eqn:est_eff_syst}
\left({\Delta \sigma_{\text{uncorrelated}}}\right)^2 = 
\frac{
	\sum_i \omega_i^2 
	\left\{ 
	\left(\Delta \sigma_{\alpha}\right)^2 +
	\left(\Delta \epsilon_{\text{hodo}}\right)^2 +
	\left(\Delta \epsilon_{\text{chamber}}\right)^2 +
	\left(\Delta \epsilon_{\text{trig}}\right)^2
	\right\}
     }
     {
	\left(\sum_i \omega_i\right)^2
     }
\end{equation}
\noindent where the sum is now over all mass settings instead of all data sets 
in a given mass setting.

The remaining source of systematic uncertainty was the uncertainty in the
acceptance due to uncertainties in the magnetic-field tweeks
as measured in chapter \ref{chapter:analysis}.  Unlike the previous sources
of uncertainties we have discussed, we expect this uncertainty to be at 
least partially correlated from mass setting to mass setting -- the SM3 
tweek was wholly determined by measurements from the high mass data.

An estimation of the size of this uncertainty in any given mass setting
was performed using a Monte Carlo based on the code described above,
modified to propagate a given event through the spectrometer multiple
times with different settings of the magnetic fields.  In this modified
Monte Carlo, all random processes, such as energy loss, multiple scattering
and detector resolutions and efficiencies, were turned off.  The result
was a highly correlated sample of Monte Carlo events where any differences
in reconstructed yields were due solely to acceptance effects.

\subsubsection{Correlated Systematic Uncertainties}

The systematic uncertainty in the acceptance due to the uncertainties in
the magnetic field in any given mass setting could then be estimated by
\begin{equation}
\label{eqn:est_acc_syst_tweek}
\left( \frac{\Delta \alpha_{\text{tweek}}}{\alpha }\right ) 
\approx 2 \;
|\frac{N _{\text{nominal}+\delta}- N_{\text{nominal}}}{N_{\text{nominal}+\delta}  + N_{\text{nominal}}}|.
\end{equation}
\noindent Here, $N _{\text{nominal}}$ represented the number of Monte Carlo
events subjected to the nominal magnetic field tweeks in traversing the 
spectrometer which survived the standard analysis chain.  Similarly, 
$N _{\text{nominal}+\delta}$ were the number of Monte Carlo events subjected 
to a set of augmented magnetic fields in traversing the 
spectrometer, which were again subjected to the standard analysis chain.
The field tweeks in this case were all augmented by the estimates of their
uncertainties,\footnote{
\setlength{\baselineskip}{\singlespace}
In the case of the low mass data, the SM12 and SM3 
field tweeks were augmented by their systematic uncertainties while the
SM0 field tweek was reduced by its uncertainty.  This resulted in the
maximum possible systematic uncertainty in the acceptance for that 
spectrometer configuration.} which should 
result in the maximum possible uncertainty.

This results in a systematic uncertainty in the cross section for each
mass setting given by
\begin{equation}\label{eqn:est_sigma_syst_tweek}
\Delta \sigma_i^{\alpha,\text{tweek}} = \left( \frac{\Delta \alpha_{\text{tweek}}}{\alpha }\right ) \sigma_i.
\end{equation}
\noindent To estimate the uncertainty in the total cross section, we depart
from the prescription given by equation \ref{eqn:syst_unc_in_weighted_average}.
We do this, because equation \ref{eqn:syst_unc_in_weighted_average} is based
on the assumption that the systematic uncertainties obey Gaussian statistics.
This is a good assumption for many of the sources of uncorrelated systematic
uncertainties.  Here it is not.  Therefore we treat the systematic uncertainty
$\Delta \sigma_i^{\alpha,\text{tweek}}$ as a potentially ``missing'' or
``excess''  cross section for each mass setting, for which we wish to obtain an
average estimate for the total cross section.  We take
\begin{equation}\label{eqn:corr_syst}
\Delta \sigma_{\text{correlated}} =
\frac{
  \sum_i \omega_i \Delta \sigma_i^{\alpha,\text{tweek}}
     }
     {
  \sum_i \omega_i
     }
\end{equation}
\noindent where once again the $\omega_i$ are the statistical weights of
the given mass setting.  Finally, we add the correlated and uncorrelated
systematic uncertainties in quadrature to obtain an estimate for the total 
systematic uncertainty
\begin{equation}
\label{eqn:total_syst_uncertainty}
\Delta \sigma = \sqrt{ \left( \Delta \sigma_{\text{uncorrelated}} \right)^2 +
                       \left( \Delta \sigma_{\text{correlated}}   \right)^2
                     }.
\end{equation}

\subsubsection{Integrating the Data}

After the data were combined using equation \ref{eqn:weighted_average}, we
integrated the triple-differential cross sections over $p_T$ to obtain the 
double-differential cross sections $d^2\sigma(pp)/dMdx_F$ and
$d^2\sigma(pd)/dMdx_F$.  The integrated cross sections were given
by
\begin{equation}
M^3\frac{d^2\sigma}{dMdx_F} = M^3\sum_i^N \left( \frac{d^3\sigma}{dMdx_Fdp_T} \right)_i \Delta p_{T,i} + f(M,x_F;\alpha)
\end{equation}
\noindent where the sum is over all $N$ $p_T$ bins whose widths are denoted 
$\Delta p_{T,i}$ and $f(M,xF;\alpha)$ represents a correction explained
below.  The statistical and systematic uncertainties\footnote{\setlength{\baselineskip}{\singlespace}Correlated and
uncorrelated systematic uncertainties were first treated separately, then 
combined in quadrature to obtain the total systematic uncertainty.} on the integrated cross 
sections were calculated using the usual error propagation in equation
\ref{eqn:errprop_cross} -- in other words, the uncertainties (multiplied
by $\Delta p_T$) were added in quadrature.

The correction in the integral $f(M,x_F;\alpha)$ accounted for the fact that
the acceptance of the spectrometer could not reach the full kinematic limit
in $p_T$.  For each bin in mass and $x_F$, the triple-differential cross
sections were integrated only over the range in $p_T$ where the acceptance
was nonzero.  Where the acceptance was zero fir all data sets,
we made an estimate of the missing cross section $f(M,x_F;\alpha)$.
The estimate was based on the empirical $p_T$ distribution given in equation 
\ref{eqn:d3sample} and the parameterization of $p_T^0$ given in equation 
\ref{pt0func}, extrapolated from the largest nonzero $p_T$ bin to infinity.

The resulting corrections were well below $1\%$ over much of the kinematic
range covered.  Only at larger masses did the correction to the cross
sections approach $5\%$.  While the empirical form for the $p_T$ dependence
of the cross section does a fairly good job of reproducing the data,
it tends to underestimate the tail of the distribution.  Given this, and the
uncertainties in the parameterization of $p_T^0$, we estimate that the
correction $f(M,x_F;\alpha)$ is uncertain by about $\pm 15\%$.  This
uncertainty was added in quadrature to the systematic uncertainties in
the final result.

\newpage
\section{RESULTS}

Fermilab Experiment 866 has measured continuum dimuon production cross 
sections in 800-GeV $pp$ and $pd$ interactions.  These data, which are
presented below, represent the most extenive study of the differential $pp$ 
cross to date, while the $pd$ data offer better precision over a broader
range of $x_F$ than previous measurements have achieved.  In this chapter
we tabulate and plot the double- and triple-differential cross sections,
while deferring a discussion of the results until the next chapter.

\subsection{FNAL E866/NuSea Results -- $M^3 d^2\sigma / dMdx_F$}

The scaling form $M^3d^2\sigma/dMdx_F$ for the hydrogen and deuterium 
cross sections are tabulated below in tables \ref{table:ppd2sigma} and 
\ref{table:pdd2sigma}.  The data are tabulated in mass and 
$x_F \equiv \frac{2p_L}{\sqrt{s}}$ bins.  The mass and $x_F$ ranges are 
shown in the tables, as are the mean mass, $x_F$ and $p_T$ in each bin.
The data in the tables are plotted in figures
\ref{fig:d2sigma_first} through \ref{fig:d2sigma_last}, and compared to
NLO calculations of the cross sections using different sets of parton 
distributions.

\newcommand{\spanner}[1]{
  \multicolumn{7}{|c|}{} \\ \nopagebreak
  \multicolumn{7}{|c|}{#1} \\ \nopagebreak
  \multicolumn{7}{|c|}{} \\ \nopagebreak
}

\newcommand{\myheading}[0]{
  \hline 
  \multicolumn{7}{|c|}{} \\
  Mass [GeV] & $<M>$ & $<x_F>$ & $<p_T>$ & $M^3d^2\sigma/dMdx_F$ & Stat. Error & Syst. Error \\
  \multicolumn{7}{|c|}{} \\
  \hline
  \multicolumn{7}{c}{} \\
}

\newcommand{\xfbin}[2]{
    \hline \nopagebreak 
    \spanner{#1} \nopagebreak
    #2 \nopagebreak
    \multicolumn{7}{|c|}{}\\ \nopagebreak
    \hline \nopagebreak
  \multicolumn{7}{c}{}\\ \nopagebreak
}

\begin{landscape}

\begin{longtable}{|*{7}{c}|} 

\caption[Scaling form $M^3d^2\sigma/dMdx_F$ for the hydrogen cross section.]
	{\label{table:ppd2sigma}
	Scaling form $M^3d^2\sigma/dMdx_F$ ( in units of nb GeV$^2$ / nucleon )
        for the hydrogen cross section.  Statistical and point-to-point 
	systematic uncertainties are shown separately.  The normalization 
	is subject to an uncertainty of $\pm 6.5\%$.
	} \\

\endfirsthead

\caption{(continued)}\\

\myheading

\endhead

\myheading

\xfbin{$ -0.05 \leq x_F < 0.05 $}{

  4.20 - 4.70 & 4.64 & 0.046 & 0.65 & 7.92E$+0$1 & 5.60E$+0$1 & 1.06E$+0$1 \\ 
  4.70 - 5.20 & 5.01 & 0.044 & 0.36 & 1.23E$+0$1 & 6.75E$+0$0 & 1.11E$+0$0 \\ 
  5.20 - 5.70 & 5.45 & 0.039 & 0.48 & 3.01E$+0$1 & 1.81E$+0$1 & 3.84E$+0$0 \\ 
  5.70 - 6.20 & 6.07 & 0.036 & 1.10 & 4.02E$+0$1 & 1.47E$+0$1 & 3.06E$+0$0 \\ 
  6.20 - 6.70 & 6.46 & 0.029 & 0.93 & 1.08E$+0$1 & 3.52E$+0$0 & 5.38E$-0$1 \\ 
  6.70 - 7.20 & 6.89 & 0.019 & 0.68 & 7.46E$+0$0 & 2.07E$+0$0 & 2.78E$-0$1 \\ 
  7.20 - 7.70 & 7.54 & 0.024 & 0.84 & 6.17E$+0$0 & 2.56E$+0$0 & 4.24E$-0$1 \\ 
  7.70 - 8.20 & 8.04 & 0.028 & 0.94 & 5.50E$+0$0 & 9.03E$-0$1 & 1.91E$-0$1 \\ 
  8.20 - 8.70 & 8.46 & 0.016 & 0.85 & 3.70E$+0$0 & 4.69E$-0$1 & 1.07E$-0$1 \\ 
  10.85 - 11.85 & 10.86 & 0.047 & 1.43 & 8.02E$+0$0 & 8.02E$+0$0 & 8.85E$-0$1 \\ 

} 

\xfbin{$ 0.05 \leq x_F < 0.10 $}{

  4.20 - 4.70 & 4.46 & 0.083 & 0.77 & 2.67E$+0$1 & 2.76E$+0$0 & 7.73E$-0$1 \\ 
  4.70 - 5.20 & 4.90 & 0.080 & 0.76 & 1.54E$+0$1 & 2.28E$+0$0 & 7.09E$-0$1 \\ 
  5.20 - 5.70 & 5.43 & 0.083 & 0.86 & 1.06E$+0$1 & 2.83E$+0$0 & 1.03E$+0$0 \\ 
  5.70 - 6.20 & 5.95 & 0.083 & 0.94 & 1.24E$+0$1 & 2.59E$+0$0 & 7.74E$-0$1 \\ 
  6.20 - 6.70 & 6.48 & 0.078 & 0.98 & 6.37E$+0$0 & 1.82E$+0$0 & 4.82E$-0$1 \\ 
  6.70 - 7.20 & 6.93 & 0.076 & 0.86 & 6.13E$+0$0 & 2.08E$+0$0 & 2.29E$-0$1 \\ 
  7.20 - 7.70 & 7.50 & 0.080 & 0.98 & 6.60E$+0$0 & 1.95E$+0$0 & 4.68E$-0$1 \\ 
  7.70 - 8.20 & 7.95 & 0.078 & 1.04 & 4.17E$+0$0 & 5.43E$-0$1 & 2.23E$-0$1 \\ 
  8.20 - 8.70 & 8.44 & 0.077 & 1.03 & 3.71E$+0$0 & 4.83E$-0$1 & 1.75E$-0$1 \\ 
  10.85 - 11.85 & 11.17 & 0.081 & 1.15 & 3.46E$+0$0 & 1.51E$+0$0 & 7.66E$-0$2 \\ 

}

\xfbin{$ 0.10 \leq x_F < 0.15 $}{

  4.20 - 4.70 & 4.43 & 0.129 & 0.91 & 2.34E$+0$1 & 1.38E$+0$0 & 5.31E$-0$1 \\ 
  4.70 - 5.20 & 4.92 & 0.127 & 0.90 & 1.59E$+0$1 & 1.26E$+0$0 & 4.18E$-0$1 \\ 
  5.20 - 5.70 & 5.41 & 0.130 & 0.92 & 1.54E$+0$1 & 1.58E$+0$0 & 5.82E$-0$1 \\ 
  5.70 - 6.20 & 5.95 & 0.128 & 0.91 & 1.26E$+0$1 & 1.45E$+0$0 & 3.60E$-0$1 \\ 
  6.20 - 6.70 & 6.44 & 0.130 & 1.04 & 9.98E$+0$0 & 1.22E$+0$0 & 2.94E$-0$1 \\ 
  6.70 - 7.20 & 6.97 & 0.130 & 0.99 & 6.86E$+0$0 & 1.05E$+0$0 & 3.05E$-0$1 \\ 
  7.20 - 7.70 & 7.46 & 0.128 & 1.10 & 5.06E$+0$0 & 4.80E$-0$1 & 1.84E$-0$1 \\ 
  7.70 - 8.20 & 7.95 & 0.127 & 1.10 & 4.56E$+0$0 & 3.67E$-0$1 & 1.47E$-0$1 \\ 
  8.20 - 8.70 & 8.44 & 0.127 & 1.12 & 3.11E$+0$0 & 2.61E$-0$1 & 9.49E$-0$2 \\ 
  10.85 - 11.85 & 11.22 & 0.129 & 1.10 & 1.36E$+0$0 & 3.42E$-0$1 & 5.81E$-0$2 \\ 

}

\xfbin{$ 0.15 \leq x_F < 0.20 $}{

  4.20 - 4.70 & 4.43 & 0.177 & 1.00 & 1.92E$+0$1 & 9.66E$-0$1 & 1.21E$+0$0 \\ 
  4.70 - 5.20 & 4.94 & 0.177 & 0.96 & 1.45E$+0$1 & 7.20E$-0$1 & 2.55E$-0$1 \\ 
  5.20 - 5.70 & 5.43 & 0.177 & 1.02 & 1.20E$+0$1 & 9.89E$-0$1 & 2.71E$-0$1 \\ 
  5.70 - 6.20 & 5.91 & 0.178 & 1.02 & 8.96E$+0$0 & 7.88E$-0$1 & 1.99E$-0$1 \\ 
  6.20 - 6.70 & 6.45 & 0.179 & 1.02 & 8.04E$+0$0 & 8.68E$-0$1 & 3.04E$-0$1 \\ 
  6.70 - 7.20 & 6.96 & 0.179 & 1.06 & 6.22E$+0$0 & 4.59E$-0$1 & 1.98E$-0$1 \\ 
  7.20 - 7.70 & 7.46 & 0.177 & 1.12 & 4.82E$+0$0 & 2.98E$-0$1 & 1.18E$-0$1 \\ 
  7.70 - 8.20 & 7.94 & 0.176 & 1.12 & 3.61E$+0$0 & 2.45E$-0$1 & 9.41E$-0$2 \\ 
  8.20 - 8.70 & 8.44 & 0.176 & 1.19 & 3.00E$+0$0 & 1.98E$-0$1 & 6.28E$-0$2 \\ 
  10.85 - 11.85 & 11.18 & 0.178 & 1.09 & 1.08E$+0$0 & 2.41E$-0$1 & 7.37E$-0$2 \\ 
  11.85 - 12.85 & 12.23 & 0.177 & 1.37 & 1.25E$+0$0 & 5.62E$-0$1 & 1.54E$-0$1 \\ 
  12.85 - 14.85 & 13.31 & 0.185 & 1.24 & 1.17E$+0$1 & 1.03E$+0$1 & 3.13E$-0$1 \\ 

}\pagebreak

\xfbin{$ 0.20 \leq x_F < 0.25 $}{

  4.20 - 4.70 & 4.43 & 0.226 & 1.02 & 1.68E$+0$1 & 6.71E$-0$1 & 2.97E$-0$1 \\ 
  4.70 - 5.20 & 4.93 & 0.225 & 1.02 & 1.26E$+0$1 & 6.54E$-0$1 & 2.35E$-0$1 \\ 
  5.20 - 5.70 & 5.43 & 0.225 & 1.00 & 9.38E$+0$0 & 5.79E$-0$1 & 1.77E$-0$1 \\ 
  5.70 - 6.20 & 5.95 & 0.228 & 1.05 & 8.45E$+0$0 & 5.19E$-0$1 & 2.48E$-0$1 \\ 
  6.20 - 6.70 & 6.46 & 0.228 & 0.99 & 7.22E$+0$0 & 3.73E$-0$1 & 1.88E$-0$1 \\ 
  6.70 - 7.20 & 6.95 & 0.227 & 1.09 & 5.18E$+0$0 & 2.72E$-0$1 & 1.09E$-0$1 \\ 
  7.20 - 7.70 & 7.45 & 0.227 & 1.12 & 3.91E$+0$0 & 2.28E$-0$1 & 1.13E$-0$1 \\ 
  7.70 - 8.20 & 7.94 & 0.225 & 1.16 & 3.73E$+0$0 & 1.97E$-0$1 & 8.98E$-0$2 \\ 
  8.20 - 8.70 & 8.44 & 0.224 & 1.19 & 2.76E$+0$0 & 1.89E$-0$1 & 6.10E$-0$2 \\ 
  10.85 - 11.85 & 11.22 & 0.231 & 1.45 & 7.48E$-0$1 & 1.34E$-0$1 & 4.23E$-0$2 \\ 
  11.85 - 12.85 & 12.33 & 0.226 & 1.33 & 2.02E$+0$0 & 7.89E$-0$1 & 6.25E$-0$1 \\ 
  12.85 - 14.85 & 13.22 & 0.226 & 1.04 & 2.36E$+0$0 & 1.82E$+0$0 & 3.90E$-0$1 \\ 

}

\xfbin{$ 0.25 \leq x_F < 0.30 $}{

  4.20 - 4.70 & 4.44 & 0.274 & 1.04 & 1.23E$+0$1 & 5.17E$-0$1 & 2.79E$-0$1 \\ 
  4.70 - 5.20 & 4.95 & 0.276 & 1.04 & 1.01E$+0$1 & 4.91E$-0$1 & 1.74E$-0$1 \\ 
  5.20 - 5.70 & 5.44 & 0.276 & 1.04 & 7.25E$+0$0 & 4.24E$-0$1 & 2.03E$-0$1 \\ 
  5.70 - 6.20 & 5.94 & 0.277 & 1.01 & 7.53E$+0$0 & 3.04E$-0$1 & 1.93E$-0$1 \\ 
  6.20 - 6.70 & 6.45 & 0.276 & 1.04 & 5.64E$+0$0 & 2.62E$-0$1 & 1.49E$-0$1 \\ 
  6.70 - 7.20 & 6.94 & 0.276 & 1.09 & 4.46E$+0$0 & 2.09E$-0$1 & 8.33E$-0$2 \\ 
  7.20 - 7.70 & 7.43 & 0.275 & 1.12 & 3.31E$+0$0 & 1.77E$-0$1 & 7.80E$-0$2 \\ 
  7.70 - 8.20 & 7.94 & 0.276 & 1.17 & 2.93E$+0$0 & 1.67E$-0$1 & 5.92E$-0$2 \\ 
  8.20 - 8.70 & 8.44 & 0.275 & 1.26 & 2.69E$+0$0 & 1.53E$-0$1 & 5.86E$-0$2 \\ 
  10.85 - 11.85 & 11.23 & 0.278 & 1.19 & 6.93E$-0$1 & 1.09E$-0$1 & 5.55E$-0$2 \\ 
  11.85 - 12.85 & 12.29 & 0.282 & 1.31 & 4.83E$-0$1 & 1.71E$-0$1 & 5.21E$-0$2 \\ 
  12.85 - 14.85 & 13.31 & 0.278 & 1.57 & 4.68E$-0$1 & 3.33E$-0$1 & 5.77E$-0$2 \\ 

}

\xfbin{$ 0.30 \leq x_F < 0.35 $}{

  4.20 - 4.70 & 4.45 & 0.324 & 1.03 & 1.00E$+0$1 & 4.63E$-0$1 & 2.09E$-0$1 \\ 
  4.70 - 5.20 & 4.95 & 0.325 & 1.01 & 8.85E$+0$0 & 4.15E$-0$1 & 2.02E$-0$1 \\ 
  5.20 - 5.70 & 5.45 & 0.326 & 1.02 & 6.91E$+0$0 & 3.03E$-0$1 & 1.89E$-0$1 \\ 
  5.70 - 6.20 & 5.95 & 0.325 & 1.02 & 5.86E$+0$0 & 2.31E$-0$1 & 1.21E$-0$1 \\ 
  6.20 - 6.70 & 6.44 & 0.324 & 1.03 & 4.72E$+0$0 & 1.98E$-0$1 & 8.29E$-0$2 \\ 
  6.70 - 7.20 & 6.94 & 0.325 & 1.11 & 4.03E$+0$0 & 1.78E$-0$1 & 1.01E$-0$1 \\ 
  7.20 - 7.70 & 7.45 & 0.325 & 1.15 & 3.37E$+0$0 & 1.50E$-0$1 & 7.64E$-0$2 \\ 
  7.70 - 8.20 & 7.93 & 0.325 & 1.17 & 2.51E$+0$0 & 1.41E$-0$1 & 4.85E$-0$2 \\ 
  8.20 - 8.70 & 8.43 & 0.325 & 1.19 & 2.22E$+0$0 & 1.33E$-0$1 & 4.33E$-0$2 \\ 
  10.85 - 11.85 & 11.32 & 0.326 & 1.27 & 6.56E$-0$1 & 8.22E$-0$2 & 5.27E$-0$2 \\ 
  11.85 - 12.85 & 12.26 & 0.328 & 1.12 & 9.65E$-0$1 & 4.94E$-0$1 & 4.78E$-0$2 \\ 
  12.85 - 14.85 & 13.30 & 0.329 & 1.31 & 4.92E$-0$1 & 1.78E$-0$1 & 4.26E$-0$2 \\ 

}

\xfbin{$ 0.35 \leq x_F < 0.40 $}{

  4.20 - 4.70 & 4.46 & 0.374 & 1.02 & 7.82E$+0$0 & 4.14E$-0$1 & 2.43E$-0$1 \\ 
  4.70 - 5.20 & 4.95 & 0.376 & 1.04 & 6.54E$+0$0 & 3.20E$-0$1 & 2.26E$-0$1 \\ 
  5.20 - 5.70 & 5.45 & 0.375 & 1.02 & 5.87E$+0$0 & 3.87E$-0$1 & 4.67E$-0$1 \\ 
  5.70 - 6.20 & 5.96 & 0.375 & 0.99 & 4.78E$+0$0 & 2.64E$-0$1 & 1.63E$-0$1 \\ 
  6.20 - 6.70 & 6.45 & 0.375 & 1.05 & 3.86E$+0$0 & 1.62E$-0$1 & 6.75E$-0$2 \\ 
  6.70 - 7.20 & 6.94 & 0.375 & 1.10 & 3.30E$+0$0 & 1.48E$-0$1 & 9.22E$-0$2 \\ 
  7.20 - 7.70 & 7.44 & 0.376 & 1.17 & 2.67E$+0$0 & 1.21E$-0$1 & 8.27E$-0$2 \\ 
  7.70 - 8.20 & 7.95 & 0.375 & 1.13 & 2.21E$+0$0 & 1.24E$-0$1 & 6.03E$-0$2 \\ 
  8.20 - 8.70 & 8.44 & 0.374 & 1.16 & 1.90E$+0$0 & 1.09E$-0$1 & 4.06E$-0$2 \\ 
  10.85 - 11.85 & 11.32 & 0.375 & 1.15 & 4.27E$-0$1 & 7.23E$-0$2 & 5.12E$-0$3 \\ 
  11.85 - 12.85 & 12.26 & 0.374 & 1.29 & 3.45E$-0$1 & 6.23E$-0$2 & 1.31E$-0$2 \\ 
  12.85 - 14.85 & 13.43 & 0.377 & 1.31 & 3.77E$-0$1 & 1.91E$-0$1 & 1.30E$-0$2 \\ 

}

\xfbin{$ 0.40 \leq x_F < 0.45 $}{

  4.20 - 4.70 & 4.46 & 0.424 & 0.99 & 6.45E$+0$0 & 3.68E$-0$1 & 2.25E$-0$1 \\ 
  4.70 - 5.20 & 4.97 & 0.423 & 1.01 & 5.32E$+0$0 & 2.50E$-0$1 & 1.52E$-0$1 \\ 
  5.20 - 5.70 & 5.46 & 0.424 & 0.97 & 4.13E$+0$0 & 1.93E$-0$1 & 1.28E$-0$1 \\ 
  5.70 - 6.20 & 5.95 & 0.425 & 1.01 & 3.76E$+0$0 & 1.53E$-0$1 & 8.14E$-0$2 \\ 
  6.20 - 6.70 & 6.45 & 0.424 & 1.02 & 3.20E$+0$0 & 1.24E$-0$1 & 8.05E$-0$2 \\ 
  6.70 - 7.20 & 6.95 & 0.424 & 1.12 & 2.50E$+0$0 & 1.16E$-0$1 & 6.28E$-0$2 \\ 
  7.20 - 7.70 & 7.45 & 0.424 & 1.12 & 2.04E$+0$0 & 1.02E$-0$1 & 5.64E$-0$2 \\ 
  7.70 - 8.20 & 7.94 & 0.424 & 1.15 & 1.64E$+0$0 & 9.63E$-0$2 & 5.20E$-0$2 \\ 
  8.20 - 8.70 & 8.44 & 0.424 & 1.20 & 1.50E$+0$0 & 9.25E$-0$2 & 5.29E$-0$2 \\ 
  10.85 - 11.85 & 11.28 & 0.423 & 1.36 & 3.63E$-0$1 & 4.95E$-0$2 & 8.00E$-0$3 \\ 
  11.85 - 12.85 & 12.24 & 0.424 & 1.29 & 2.46E$-0$1 & 5.32E$-0$2 & 2.24E$-0$2 \\ 
  12.85 - 14.85 & 13.46 & 0.429 & 1.41 & 6.71E$-0$1 & 5.45E$-0$1 & 2.18E$-0$2 \\ 

}\pagebreak

\xfbin{$ 0.45 \leq x_F < 0.50 $}{

  4.20 - 4.70 & 4.46 & 0.473 & 0.95 & 4.20E$+0$0 & 2.80E$-0$1 & 1.27E$-0$1 \\ 
  4.70 - 5.20 & 4.95 & 0.475 & 0.99 & 3.55E$+0$0 & 2.12E$-0$1 & 9.38E$-0$2 \\ 
  5.20 - 5.70 & 5.46 & 0.475 & 0.99 & 3.22E$+0$0 & 1.54E$-0$1 & 6.59E$-0$2 \\ 
  5.70 - 6.20 & 5.95 & 0.474 & 0.97 & 2.54E$+0$0 & 1.20E$-0$1 & 5.07E$-0$2 \\ 
  6.20 - 6.70 & 6.44 & 0.474 & 1.03 & 1.89E$+0$0 & 9.63E$-0$2 & 3.76E$-0$2 \\ 
  6.70 - 7.20 & 6.94 & 0.474 & 1.09 & 1.88E$+0$0 & 9.42E$-0$2 & 5.12E$-0$2 \\ 
  7.20 - 7.70 & 7.43 & 0.474 & 1.06 & 1.50E$+0$0 & 8.66E$-0$2 & 5.23E$-0$2 \\ 
  7.70 - 8.20 & 7.94 & 0.473 & 1.08 & 1.21E$+0$0 & 8.03E$-0$2 & 2.65E$-0$2 \\ 
  8.20 - 8.70 & 8.44 & 0.474 & 1.24 & 8.83E$-0$1 & 7.12E$-0$2 & 2.05E$-0$2 \\ 
  10.85 - 11.85 & 11.29 & 0.472 & 1.24 & 2.50E$-0$1 & 4.24E$-0$2 & 8.17E$-0$3 \\ 
  11.85 - 12.85 & 12.33 & 0.471 & 1.18 & 1.74E$-0$1 & 4.25E$-0$2 & 6.59E$-0$3 \\ 
  12.85 - 14.85 & 13.49 & 0.474 & 0.90 & 9.36E$-0$2 & 3.83E$-0$2 & 8.91E$-0$3 \\ 

}

\xfbin{$ 0.50 \leq x_F < 0.55 $}{

  4.20 - 4.70 & 4.48 & 0.524 & 1.04 & 2.81E$+0$0 & 2.42E$-0$1 & 1.13E$-0$1 \\ 
  4.70 - 5.20 & 4.97 & 0.524 & 1.01 & 2.56E$+0$0 & 1.76E$-0$1 & 1.14E$-0$1 \\ 
  5.20 - 5.70 & 5.46 & 0.523 & 1.01 & 2.28E$+0$0 & 1.23E$-0$1 & 6.86E$-0$2 \\ 
  5.70 - 6.20 & 5.94 & 0.524 & 1.00 & 1.83E$+0$0 & 9.65E$-0$2 & 6.27E$-0$2 \\ 
  6.20 - 6.70 & 6.45 & 0.525 & 1.03 & 1.51E$+0$0 & 7.93E$-0$2 & 3.54E$-0$2 \\ 
  6.70 - 7.20 & 6.93 & 0.523 & 1.07 & 1.27E$+0$0 & 7.58E$-0$2 & 3.37E$-0$2 \\ 
  7.20 - 7.70 & 7.44 & 0.523 & 1.07 & 9.84E$-0$1 & 7.11E$-0$2 & 2.33E$-0$2 \\ 
  7.70 - 8.20 & 7.94 & 0.523 & 1.14 & 9.85E$-0$1 & 7.06E$-0$2 & 2.27E$-0$2 \\ 
  8.20 - 8.70 & 8.45 & 0.523 & 1.18 & 8.52E$-0$1 & 6.21E$-0$2 & 1.93E$-0$2 \\ 
  10.85 - 11.85 & 11.23 & 0.527 & 1.14 & 1.84E$-0$1 & 3.27E$-0$2 & 1.28E$-0$3 \\ 
  11.85 - 12.85 & 12.16 & 0.525 & 1.24 & 7.93E$-0$2 & 2.51E$-0$2 & 2.80E$-0$3 \\ 
  12.85 - 14.85 & 13.46 & 0.524 & 1.12 & 1.69E$-0$1 & 7.29E$-0$2 & 1.30E$-0$2 \\ 
  14.85 - 16.85 & 14.87 & 0.531 & 0.77 & 1.80E$+0$0 & 1.80E$+0$0 & 6.50E$-0$1 \\ 

}

\xfbin{$ 0.55 \leq x_F < 0.60 $}{

  4.20 - 4.70 & 4.46 & 0.571 & 0.97 & 1.45E$+0$0 & 1.61E$-0$1 & 6.48E$-0$2 \\ 
  4.70 - 5.20 & 4.96 & 0.575 & 0.97 & 1.57E$+0$0 & 1.20E$-0$1 & 6.31E$-0$2 \\ 
  5.20 - 5.70 & 5.45 & 0.573 & 0.91 & 1.44E$+0$0 & 9.82E$-0$2 & 3.99E$-0$2 \\ 
  5.70 - 6.20 & 5.96 & 0.573 & 0.89 & 1.16E$+0$0 & 7.25E$-0$2 & 3.03E$-0$2 \\ 
  6.20 - 6.70 & 6.44 & 0.573 & 1.00 & 1.15E$+0$0 & 7.83E$-0$2 & 2.50E$-0$2 \\ 
  6.70 - 7.20 & 6.94 & 0.572 & 1.06 & 8.69E$-0$1 & 6.05E$-0$2 & 1.99E$-0$2 \\ 
  7.20 - 7.70 & 7.42 & 0.573 & 1.09 & 7.86E$-0$1 & 5.48E$-0$2 & 1.84E$-0$2 \\ 
  7.70 - 8.20 & 7.94 & 0.572 & 1.16 & 6.01E$-0$1 & 5.13E$-0$2 & 2.25E$-0$2 \\ 
  8.20 - 8.70 & 8.44 & 0.573 & 1.03 & 5.52E$-0$1 & 5.05E$-0$2 & 1.82E$-0$2 \\ 
  10.85 - 11.85 & 11.26 & 0.570 & 1.16 & 1.16E$-0$1 & 2.27E$-0$2 & 4.33E$-0$3 \\ 
  11.85 - 12.85 & 12.21 & 0.576 & 1.03 & 9.16E$-0$2 & 2.83E$-0$2 & 6.40E$-0$3 \\ 
  12.85 - 14.85 & 13.28 & 0.570 & 1.14 & 2.50E$-0$1 & 2.02E$-0$1 & 5.16E$-0$3 \\ 

}

\xfbin{$ 0.60 \leq x_F < 0.65 $}{

  4.20 - 4.70 & 4.49 & 0.622 & 0.95 & 1.41E$+0$0 & 1.74E$-0$1 & 8.92E$-0$2 \\ 
  4.70 - 5.20 & 4.96 & 0.623 & 0.94 & 9.66E$-0$1 & 1.14E$-0$1 & 5.80E$-0$2 \\ 
  5.20 - 5.70 & 5.46 & 0.622 & 0.93 & 8.05E$-0$1 & 8.03E$-0$2 & 3.26E$-0$2 \\ 
  5.70 - 6.20 & 5.96 & 0.623 & 0.95 & 7.15E$-0$1 & 5.35E$-0$2 & 2.29E$-0$2 \\ 
  6.20 - 6.70 & 6.43 & 0.623 & 1.02 & 6.39E$-0$1 & 5.76E$-0$2 & 2.27E$-0$2 \\ 
  6.70 - 7.20 & 6.93 & 0.624 & 1.02 & 6.79E$-0$1 & 5.35E$-0$2 & 2.52E$-0$2 \\ 
  7.20 - 7.70 & 7.44 & 0.623 & 1.02 & 5.44E$-0$1 & 6.51E$-0$2 & 3.05E$-0$2 \\ 
  7.70 - 8.20 & 7.93 & 0.624 & 1.03 & 3.85E$-0$1 & 4.04E$-0$2 & 1.19E$-0$2 \\ 
  8.20 - 8.70 & 8.45 & 0.624 & 1.08 & 3.08E$-0$1 & 3.93E$-0$2 & 1.04E$-0$2 \\ 
  10.85 - 11.85 & 11.24 & 0.625 & 1.13 & 1.09E$-0$1 & 3.33E$-0$2 & 6.55E$-0$3 \\ 
  11.85 - 12.85 & 12.18 & 0.620 & 1.34 & 6.14E$-0$2 & 2.93E$-0$2 & 5.11E$-0$3 \\ 
  12.85 - 14.85 & 13.16 & 0.609 & 0.67 & 2.48E$-0$2 & 1.13E$-0$2 & 2.18E$-0$3 \\ 

}

\xfbin{$ 0.65 \leq x_F < 0.70 $}{

  4.20 - 4.70 & 4.47 & 0.673 & 0.97 & 6.90E$-0$1 & 1.18E$-0$1 & 6.38E$-0$2 \\ 
  4.70 - 5.20 & 4.97 & 0.670 & 0.90 & 5.91E$-0$1 & 7.45E$-0$2 & 3.13E$-0$2 \\ 
  5.20 - 5.70 & 5.45 & 0.670 & 0.88 & 6.39E$-0$1 & 7.28E$-0$2 & 3.90E$-0$2 \\ 
  5.70 - 6.20 & 5.95 & 0.673 & 0.89 & 4.48E$-0$1 & 6.71E$-0$2 & 1.74E$-0$2 \\ 
  6.20 - 6.70 & 6.46 & 0.671 & 0.92 & 3.80E$-0$1 & 5.04E$-0$2 & 1.68E$-0$2 \\ 
  6.70 - 7.20 & 6.94 & 0.669 & 1.03 & 2.48E$-0$1 & 3.91E$-0$2 & 1.40E$-0$2 \\ 
  7.20 - 7.70 & 7.43 & 0.673 & 1.06 & 1.97E$-0$1 & 4.09E$-0$2 & 1.65E$-0$2 \\ 
  7.70 - 8.20 & 7.95 & 0.670 & 1.08 & 2.21E$-0$1 & 3.71E$-0$2 & 1.27E$-0$2 \\ 
  8.20 - 8.70 & 8.43 & 0.669 & 1.07 & 1.04E$-0$1 & 2.42E$-0$2 & 5.03E$-0$3 \\ 
  10.85 - 11.85 & 11.30 & 0.674 & 1.22 & 5.25E$-0$2 & 1.52E$-0$2 & 2.79E$-0$3 \\ 
  11.85 - 12.85 & 12.42 & 0.679 & 1.27 & 3.87E$-0$2 & 2.09E$-0$2 & 3.70E$-0$3 \\ 
  12.85 - 14.85 & 13.77 & 0.677 & 0.83 & 9.03E$-0$2 & 4.41E$-0$2 & 3.04E$-0$3 \\ 

}\pagebreak

\xfbin{$ 0.70 \leq x_F < 0.75 $}{

  4.20 - 4.70 & 4.53 & 0.720 & 0.80 & 5.31E$-0$1 & 1.26E$-0$1 & 4.75E$-0$2 \\ 
  4.70 - 5.20 & 5.01 & 0.720 & 0.90 & 4.21E$-0$1 & 8.97E$-0$2 & 2.60E$-0$2 \\ 
  5.20 - 5.70 & 5.45 & 0.719 & 0.94 & 2.44E$-0$1 & 3.87E$-0$2 & 1.16E$-0$2 \\ 
  5.70 - 6.20 & 5.94 & 0.723 & 1.04 & 2.68E$-0$1 & 4.36E$-0$2 & 2.31E$-0$2 \\ 
  6.20 - 6.70 & 6.45 & 0.720 & 1.00 & 2.04E$-0$1 & 4.72E$-0$2 & 1.62E$-0$2 \\ 
  6.70 - 7.20 & 6.93 & 0.720 & 0.86 & 1.62E$-0$1 & 2.65E$-0$2 & 1.24E$-0$2 \\ 
  7.20 - 7.70 & 7.41 & 0.726 & 1.00 & 1.78E$-0$1 & 3.30E$-0$2 & 1.06E$-0$2 \\ 
  7.70 - 8.20 & 7.95 & 0.719 & 1.07 & 3.54E$-0$1 & 1.61E$-0$1 & 1.21E$-0$2 \\ 
  8.20 - 8.70 & 8.42 & 0.717 & 1.13 & 1.05E$-0$1 & 2.43E$-0$2 & 1.05E$-0$2 \\ 
  10.85 - 11.85 & 11.28 & 0.728 & 1.30 & 3.61E$-0$2 & 2.34E$-0$2 & 4.30E$-0$4 \\ 
  12.85 - 14.85 & 13.14 & 0.723 & 1.33 & 3.83E$-0$1 & 2.72E$-0$1 & 4.21E$-0$2 \\ 

}

\xfbin{$ 0.75 \leq x_F < 0.80$}{

  4.20 - 4.70 & 4.48 & 0.770 & 1.01 & 3.20E$-0$1 & 1.11E$-0$1 & 5.33E$-0$2 \\ 
  4.70 - 5.20 & 4.96 & 0.767 & 0.84 & 4.98E$-0$1 & 2.48E$-0$1 & 3.98E$-0$2 \\ 
  5.20 - 5.70 & 5.45 & 0.768 & 0.65 & 1.28E$-0$1 & 2.67E$-0$2 & 9.63E$-0$3 \\ 
  5.70 - 6.20 & 5.95 & 0.771 & 0.96 & 1.76E$-0$1 & 4.18E$-0$2 & 1.37E$-0$2 \\ 
  6.20 - 6.70 & 6.43 & 0.769 & 0.82 & 9.89E$-0$2 & 2.22E$-0$2 & 5.33E$-0$3 \\ 
  6.70 - 7.20 & 6.94 & 0.767 & 0.83 & 8.01E$-0$2 & 2.43E$-0$2 & 5.63E$-0$3 \\ 
  7.20 - 7.70 & 7.45 & 0.773 & 0.97 & 1.09E$-0$1 & 4.14E$-0$2 & 1.91E$-0$2 \\ 
  7.70 - 8.20 & 7.93 & 0.772 & 1.42 & 1.04E$-0$1 & 3.55E$-0$2 & 9.37E$-0$3 \\ 
  8.20 - 8.70 & 8.47 & 0.776 & 0.80 & 7.31E$-0$2 & 2.44E$-0$2 & 6.20E$-0$3 \\ 
  10.85 - 11.85 & 11.39 & 0.765 & 1.52 & 6.65E$-0$2 & 4.97E$-0$2 & 5.74E$-0$3 \\ 

}

\end{longtable} 
\pagebreak

\begin{longtable}{|ccccccc|} 

\caption[Scaling form $M^3d^2\sigma/dMdx_F$ for the deuterium cross section.]{\label{table:pdd2sigma}
	Scaling form $M^3d^2\sigma/dMdx_F$ ( in units of nb GeV$^2$ / nucleon )
        for the deuterium cross section.  Statistical and point-to-point 
	systematic uncertainties are shown separately.  The normalization 
	is subject to an uncertainty of $\pm 6.5\%$.
	} \\

\endfirsthead

\caption{(continued)}\\

\myheading

\endhead

\myheading

\xfbin{$ -0.05 \leq x_F < 0.05 $}{

  4.70 - 5.20 & 4.94 & 0.044 & 0.96 & 4.50E$+0$1 & 2.77E$+0$1 & 5.70E$-0$1 \\ 
  5.20 - 5.70 & 5.40 & 0.039 & 0.55 & 1.81E$+0$1 & 8.30E$+0$0 & 2.00E$+0$0 \\ 
  5.70 - 6.20 & 6.10 & 0.037 & 1.04 & 1.21E$+0$1 & 4.70E$+0$0 & 1.46E$+0$0 \\ 
  6.20 - 6.70 & 6.48 & 0.022 & 1.03 & 1.25E$+0$1 & 2.75E$+0$0 & 9.14E$-0$1 \\ 
  6.70 - 7.20 & 6.95 & 0.025 & 0.90 & 9.76E$+0$0 & 1.73E$+0$0 & 4.41E$-0$1 \\ 
  7.20 - 7.70 & 7.49 & 0.027 & 0.97 & 7.15E$+0$0 & 2.47E$+0$0 & 1.12E$+0$0 \\ 
  7.70 - 8.20 & 8.01 & 0.024 & 0.99 & 5.62E$+0$0 & 6.72E$-0$1 & 2.47E$-0$1 \\ 
  8.20 - 8.70 & 8.45 & 0.017 & 0.97 & 4.32E$+0$0 & 4.32E$-0$1 & 2.15E$-0$1 \\ 
  10.85 - 11.85 & 11.12 & 0.012 & 0.44 & 8.71E$-0$1 & 3.58E$-0$1 & 1.75E$-0$2 \\ 

}

\xfbin{$ 0.05 \leq x_F < 0.10 $}{

  4.20 - 4.70 & 4.48 & 0.084 & 0.78 & 2.78E$+0$1 & 2.08E$+0$0 & 7.52E$-0$1 \\ 
  4.70 - 5.20 & 4.92 & 0.081 & 0.83 & 2.18E$+0$1 & 1.73E$+0$0 & 7.37E$-0$1 \\ 
  5.20 - 5.70 & 5.43 & 0.084 & 0.77 & 1.73E$+0$1 & 2.55E$+0$0 & 1.11E$+0$0 \\ 
  5.70 - 6.20 & 5.97 & 0.080 & 0.98 & 1.30E$+0$1 & 1.81E$+0$0 & 6.84E$-0$1 \\ 
  6.20 - 6.70 & 6.46 & 0.079 & 0.94 & 7.82E$+0$0 & 1.18E$+0$0 & 4.34E$-0$1 \\ 
  6.70 - 7.20 & 6.90 & 0.080 & 0.90 & 6.23E$+0$0 & 1.32E$+0$0 & 4.38E$-0$1 \\ 
  7.20 - 7.70 & 7.50 & 0.082 & 0.91 & 6.18E$+0$0 & 1.21E$+0$0 & 5.15E$-0$1 \\ 
  7.70 - 8.20 & 7.97 & 0.078 & 1.03 & 4.18E$+0$0 & 3.05E$-0$1 & 2.02E$-0$1 \\ 
  8.20 - 8.70 & 8.45 & 0.076 & 1.03 & 3.58E$+0$0 & 2.51E$-0$1 & 1.58E$-0$1 \\ 
  10.85 - 11.85 & 11.21 & 0.085 & 0.98 & 1.84E$+0$0 & 6.56E$-0$1 & 7.84E$-0$3 \\ 

}

\xfbin{$ 0.10 \leq x_F < 0.15 $}{

  4.20 - 4.70 & 4.44 & 0.130 & 0.91 & 2.64E$+0$1 & 9.63E$-0$1 & 4.98E$-0$1 \\ 
  4.70 - 5.20 & 4.93 & 0.128 & 0.86 & 1.72E$+0$1 & 7.96E$-0$1 & 3.55E$-0$1 \\ 
  5.20 - 5.70 & 5.42 & 0.129 & 0.92 & 1.45E$+0$1 & 1.01E$+0$0 & 4.16E$-0$1 \\ 
  5.70 - 6.20 & 5.93 & 0.130 & 0.98 & 1.02E$+0$1 & 7.86E$-0$1 & 2.91E$-0$1 \\ 
  6.20 - 6.70 & 6.45 & 0.129 & 0.93 & 8.98E$+0$0 & 8.55E$-0$1 & 2.84E$-0$1 \\ 
  6.70 - 7.20 & 6.97 & 0.130 & 1.04 & 7.18E$+0$0 & 6.49E$-0$1 & 2.73E$-0$1 \\ 
  7.20 - 7.70 & 7.49 & 0.128 & 1.07 & 5.91E$+0$0 & 4.46E$-0$1 & 1.97E$-0$1 \\ 
  7.70 - 8.20 & 7.95 & 0.129 & 1.13 & 4.92E$+0$0 & 2.49E$-0$1 & 1.82E$-0$1 \\ 
  8.20 - 8.70 & 8.45 & 0.127 & 1.13 & 3.86E$+0$0 & 1.87E$-0$1 & 1.02E$-0$1 \\ 
  10.85 - 11.85 & 11.20 & 0.126 & 1.17 & 1.44E$+0$0 & 3.19E$-0$1 & 1.55E$-0$1 \\ 
  11.85 - 12.85 & 12.03 & 0.135 & 1.72 & 2.12E$+0$0 & 1.04E$+0$0 & 9.62E$-0$2 \\ 

}

\xfbin{$ 0.15 \leq x_F < 0.20 $}{

  4.20 - 4.70 & 4.43 & 0.177 & 1.03 & 2.17E$+0$1 & 6.01E$-0$1 & 4.30E$-0$1 \\ 
  4.70 - 5.20 & 4.93 & 0.177 & 1.00 & 1.68E$+0$1 & 5.27E$-0$1 & 3.45E$-0$1 \\ 
  5.20 - 5.70 & 5.43 & 0.177 & 0.97 & 1.17E$+0$1 & 6.14E$-0$1 & 2.86E$-0$1 \\ 
  5.70 - 6.20 & 5.92 & 0.178 & 1.00 & 1.05E$+0$1 & 5.32E$-0$1 & 2.30E$-0$1 \\ 
  6.20 - 6.70 & 6.46 & 0.179 & 1.08 & 7.81E$+0$0 & 4.77E$-0$1 & 2.93E$-0$1 \\ 
  6.70 - 7.20 & 6.98 & 0.179 & 1.08 & 6.90E$+0$0 & 3.01E$-0$1 & 2.73E$-0$1 \\ 
  7.20 - 7.70 & 7.44 & 0.176 & 1.11 & 5.35E$+0$0 & 2.25E$-0$1 & 1.52E$-0$1 \\ 
  7.70 - 8.20 & 7.95 & 0.176 & 1.15 & 4.46E$+0$0 & 1.73E$-0$1 & 1.29E$-0$1 \\ 
  8.20 - 8.70 & 8.43 & 0.176 & 1.16 & 3.63E$+0$0 & 1.41E$-0$1 & 7.58E$-0$2 \\ 
  10.85 - 11.85 & 11.24 & 0.178 & 1.11 & 1.17E$+0$0 & 2.06E$-0$1 & 5.42E$-0$2 \\ 
  11.85 - 12.85 & 12.06 & 0.166 & 0.91 & 5.16E$-0$1 & 1.91E$-0$1 & 5.58E$-0$2 \\ 
  12.85 - 14.85 & 13.13 & 0.195 & 1.25 & 1.82E$+0$0 & 1.79E$+0$0 & 5.40E$-0$2 \\ 

}\pagebreak

\xfbin{$ 0.20 \leq x_F < 0.25 $}{

  4.20 - 4.70 & 4.44 & 0.225 & 1.04 & 1.82E$+0$1 & 4.23E$-0$1 & 3.32E$-0$1 \\ 
  4.70 - 5.20 & 4.93 & 0.225 & 1.01 & 1.35E$+0$1 & 4.00E$-0$1 & 2.44E$-0$1 \\ 
  5.20 - 5.70 & 5.43 & 0.226 & 1.00 & 1.03E$+0$1 & 3.93E$-0$1 & 2.06E$-0$1 \\ 
  5.70 - 6.20 & 5.94 & 0.227 & 1.03 & 8.80E$+0$0 & 3.40E$-0$1 & 2.30E$-0$1 \\ 
  6.20 - 6.70 & 6.46 & 0.229 & 1.05 & 7.77E$+0$0 & 2.64E$-0$1 & 1.91E$-0$1 \\ 
  6.70 - 7.20 & 6.95 & 0.227 & 1.09 & 5.87E$+0$0 & 1.87E$-0$1 & 1.15E$-0$1 \\ 
  7.20 - 7.70 & 7.43 & 0.226 & 1.11 & 4.73E$+0$0 & 1.55E$-0$1 & 1.16E$-0$1 \\ 
  7.70 - 8.20 & 7.94 & 0.226 & 1.18 & 4.06E$+0$0 & 1.35E$-0$1 & 1.02E$-0$1 \\ 
  8.20 - 8.70 & 8.45 & 0.225 & 1.17 & 3.35E$+0$0 & 1.24E$-0$1 & 6.75E$-0$2 \\ 
  10.85 - 11.85 & 11.17 & 0.227 & 1.25 & 7.07E$-0$1 & 7.43E$-0$2 & 1.97E$-0$2 \\ 
  11.85 - 12.85 & 12.31 & 0.226 & 1.23 & 5.63E$-0$1 & 1.38E$-0$1 & 4.95E$-0$2 \\ 
  12.85 - 14.85 & 13.19 & 0.229 & 0.85 & 3.61E$-0$1 & 1.48E$-0$1 & 2.88E$-0$2 \\ 

}

\xfbin{$ 0.25 \leq x_F < 0.30 $}{

  4.20 - 4.70 & 4.44 & 0.275 & 1.01 & 1.46E$+0$1 & 3.45E$-0$1 & 2.39E$-0$1 \\ 
  4.70 - 5.20 & 4.94 & 0.275 & 1.03 & 1.15E$+0$1 & 3.20E$-0$1 & 1.97E$-0$1 \\ 
  5.20 - 5.70 & 5.45 & 0.276 & 1.02 & 9.32E$+0$0 & 2.96E$-0$1 & 2.30E$-0$1 \\ 
  5.70 - 6.20 & 5.95 & 0.277 & 1.01 & 7.72E$+0$0 & 1.98E$-0$1 & 2.19E$-0$1 \\ 
  6.20 - 6.70 & 6.45 & 0.276 & 1.02 & 6.47E$+0$0 & 1.72E$-0$1 & 1.68E$-0$1 \\ 
  6.70 - 7.20 & 6.95 & 0.276 & 1.08 & 5.15E$+0$0 & 1.39E$-0$1 & 8.95E$-0$2 \\ 
  7.20 - 7.70 & 7.44 & 0.275 & 1.15 & 4.11E$+0$0 & 1.21E$-0$1 & 9.33E$-0$2 \\ 
  7.70 - 8.20 & 7.94 & 0.276 & 1.19 & 3.29E$+0$0 & 1.08E$-0$1 & 6.60E$-0$2 \\ 
  8.20 - 8.70 & 8.45 & 0.275 & 1.20 & 2.85E$+0$0 & 1.01E$-0$1 & 5.41E$-0$2 \\ 
  10.85 - 11.85 & 11.23 & 0.275 & 1.19 & 8.09E$-0$1 & 6.51E$-0$2 & 2.10E$-0$2 \\ 
  11.85 - 12.85 & 12.23 & 0.278 & 1.04 & 3.41E$-0$1 & 8.92E$-0$2 & 2.38E$-0$2 \\ 
  12.85 - 14.85 & 13.42 & 0.270 & 1.29 & 4.11E$-0$1 & 2.70E$-0$1 & 2.52E$-0$2 \\ 

}

\xfbin{$ 0.30 \leq x_F < 0.35 $}{

  4.20 - 4.70 & 4.44 & 0.323 & 1.04 & 1.13E$+0$1 & 3.26E$-0$1 & 2.56E$-0$1 \\ 
  4.70 - 5.20 & 4.95 & 0.325 & 1.05 & 8.45E$+0$0 & 2.57E$-0$1 & 1.86E$-0$1 \\ 
  5.20 - 5.70 & 5.45 & 0.325 & 1.00 & 8.01E$+0$0 & 2.07E$-0$1 & 1.77E$-0$1 \\ 
  5.70 - 6.20 & 5.95 & 0.325 & 1.02 & 6.21E$+0$0 & 1.54E$-0$1 & 1.53E$-0$1 \\ 
  6.20 - 6.70 & 6.45 & 0.325 & 1.07 & 5.23E$+0$0 & 1.30E$-0$1 & 8.99E$-0$2 \\ 
  6.70 - 7.20 & 6.95 & 0.325 & 1.10 & 4.45E$+0$0 & 1.16E$-0$1 & 8.46E$-0$2 \\ 
  7.20 - 7.70 & 7.44 & 0.325 & 1.15 & 3.57E$+0$0 & 9.93E$-0$2 & 7.96E$-0$2 \\ 
  7.70 - 8.20 & 7.93 & 0.325 & 1.22 & 2.98E$+0$0 & 9.51E$-0$2 & 5.55E$-0$2 \\ 
  8.20 - 8.70 & 8.44 & 0.325 & 1.19 & 2.48E$+0$0 & 8.81E$-0$2 & 4.77E$-0$2 \\ 
  10.85 - 11.85 & 11.29 & 0.326 & 1.11 & 5.01E$-0$1 & 4.47E$-0$2 & 2.97E$-0$2 \\ 
  11.85 - 12.85 & 12.17 & 0.326 & 1.18 & 5.05E$-0$1 & 1.10E$-0$1 & 3.74E$-0$2 \\ 
  12.85 - 14.85 & 13.41 & 0.328 & 1.16 & 3.18E$-0$1 & 1.06E$-0$1 & 1.92E$-0$2 \\ 

}

\xfbin{$ 0.35 \leq x_F < 0.40 $}{

  4.20 - 4.70 & 4.45 & 0.375 & 1.00 & 7.97E$+0$0 & 2.55E$-0$1 & 1.80E$-0$1 \\ 
  4.70 - 5.20 & 4.96 & 0.374 & 1.01 & 6.47E$+0$0 & 1.93E$-0$1 & 1.81E$-0$1 \\ 
  5.20 - 5.70 & 5.46 & 0.375 & 0.99 & 5.90E$+0$0 & 1.58E$-0$1 & 9.73E$-0$2 \\ 
  5.70 - 6.20 & 5.95 & 0.374 & 1.02 & 4.91E$+0$0 & 1.20E$-0$1 & 8.34E$-0$2 \\ 
  6.20 - 6.70 & 6.44 & 0.374 & 1.06 & 4.08E$+0$0 & 1.03E$-0$1 & 6.99E$-0$2 \\ 
  6.70 - 7.20 & 6.94 & 0.374 & 1.10 & 3.50E$+0$0 & 9.44E$-0$2 & 9.94E$-0$2 \\ 
  7.20 - 7.70 & 7.44 & 0.374 & 1.16 & 2.92E$+0$0 & 8.24E$-0$2 & 1.01E$-0$1 \\ 
  7.70 - 8.20 & 7.94 & 0.375 & 1.15 & 2.51E$+0$0 & 8.00E$-0$2 & 1.10E$-0$1 \\ 
  8.20 - 8.70 & 8.44 & 0.374 & 1.18 & 2.08E$+0$0 & 7.96E$-0$2 & 8.28E$-0$2 \\ 
  10.85 - 11.85 & 11.27 & 0.375 & 1.16 & 5.13E$-0$1 & 4.04E$-0$2 & 4.33E$-0$3 \\ 
  11.85 - 12.85 & 12.29 & 0.376 & 1.17 & 2.05E$-0$1 & 3.65E$-0$2 & 1.12E$-0$2 \\ 
  12.85 - 14.85 & 13.31 & 0.378 & 1.08 & 1.94E$-0$1 & 4.76E$-0$2 & 1.20E$-0$2 \\ 
  14.85 - 16.85 & 15.48 & 0.390 & 1.64 & 2.40E$-0$1 & 1.77E$-0$1 & 1.76E$-0$1 \\ 

}

\xfbin{$ 0.40 \leq x_F < 0.45 $}{

  4.20 - 4.70 & 4.45 & 0.423 & 1.03 & 6.09E$+0$0 & 2.22E$-0$1 & 1.68E$-0$1 \\ 
  4.70 - 5.20 & 4.96 & 0.424 & 0.99 & 5.09E$+0$0 & 1.58E$-0$1 & 1.59E$-0$1 \\ 
  5.20 - 5.70 & 5.46 & 0.424 & 0.99 & 4.28E$+0$0 & 1.22E$-0$1 & 1.29E$-0$1 \\ 
  5.70 - 6.20 & 5.95 & 0.424 & 1.00 & 3.78E$+0$0 & 9.71E$-0$2 & 1.10E$-0$1 \\ 
  6.20 - 6.70 & 6.45 & 0.424 & 1.02 & 3.12E$+0$0 & 8.00E$-0$2 & 7.37E$-0$2 \\ 
  6.70 - 7.20 & 6.93 & 0.424 & 1.08 & 2.60E$+0$0 & 7.43E$-0$2 & 6.73E$-0$2 \\ 
  7.20 - 7.70 & 7.44 & 0.424 & 1.12 & 2.10E$+0$0 & 6.57E$-0$2 & 4.68E$-0$2 \\ 
  7.70 - 8.20 & 7.94 & 0.425 & 1.13 & 1.86E$+0$0 & 6.46E$-0$2 & 6.32E$-0$2 \\ 
  8.20 - 8.70 & 8.44 & 0.424 & 1.18 & 1.63E$+0$0 & 6.21E$-0$2 & 4.68E$-0$2 \\ 
  10.85 - 11.85 & 11.26 & 0.425 & 1.17 & 4.33E$-0$1 & 3.40E$-0$2 & 1.10E$-0$2 \\ 
  11.85 - 12.85 & 12.23 & 0.427 & 1.27 & 2.70E$-0$1 & 3.32E$-0$2 & 1.10E$-0$2 \\ 
  12.85 - 14.85 & 13.56 & 0.426 & 1.15 & 1.39E$-0$1 & 2.81E$-0$2 & 6.47E$-0$3 \\ 
  14.85 - 16.85 & 15.59 & 0.434 & 1.29 & 4.24E$-0$1 & 2.41E$-0$1 & 1.20E$-0$1 \\ 

}\pagebreak

\xfbin{$ 0.45 \leq x_F < 0.50 $}{

  4.20 - 4.70 & 4.46 & 0.473 & 1.00 & 4.49E$+0$0 & 2.03E$-0$1 & 3.45E$-0$1 \\ 
  4.70 - 5.20 & 4.96 & 0.474 & 0.97 & 3.68E$+0$0 & 1.31E$-0$1 & 1.16E$-0$1 \\ 
  5.20 - 5.70 & 5.46 & 0.474 & 1.01 & 3.07E$+0$0 & 9.86E$-0$2 & 6.13E$-0$2 \\ 
  5.70 - 6.20 & 5.95 & 0.474 & 0.99 & 2.75E$+0$0 & 7.76E$-0$2 & 5.63E$-0$2 \\ 
  6.20 - 6.70 & 6.44 & 0.475 & 1.02 & 2.29E$+0$0 & 6.88E$-0$2 & 5.10E$-0$2 \\ 
  6.70 - 7.20 & 6.94 & 0.474 & 1.07 & 1.91E$+0$0 & 6.02E$-0$2 & 4.62E$-0$2 \\ 
  7.20 - 7.70 & 7.44 & 0.473 & 1.16 & 1.57E$+0$0 & 5.62E$-0$2 & 3.89E$-0$2 \\ 
  7.70 - 8.20 & 7.93 & 0.474 & 1.13 & 1.33E$+0$0 & 5.24E$-0$2 & 2.79E$-0$2 \\ 
  8.20 - 8.70 & 8.44 & 0.474 & 1.08 & 1.21E$+0$0 & 7.74E$-0$2 & 1.68E$-0$2 \\ 
  10.85 - 11.85 & 11.31 & 0.476 & 1.19 & 3.62E$-0$1 & 3.11E$-0$2 & 1.06E$-0$2 \\ 
  11.85 - 12.85 & 12.26 & 0.475 & 1.09 & 2.01E$-0$1 & 2.96E$-0$2 & 7.79E$-0$3 \\ 
  12.85 - 14.85 & 13.43 & 0.471 & 1.08 & 1.01E$-0$1 & 2.71E$-0$2 & 4.90E$-0$3 \\ 
  14.85 - 16.85 & 16.62 & 0.466 & 0.96 & 7.46E$-0$2 & 7.46E$-0$2 & 2.04E$-0$2 \\ 

}

\xfbin{$ 0.50 \leq x_F < 0.55 $}{

  4.20 - 4.70 & 4.46 & 0.523 & 0.97 & 3.08E$+0$0 & 1.60E$-0$1 & 1.55E$-0$1 \\ 
  4.70 - 5.20 & 4.97 & 0.525 & 0.99 & 2.57E$+0$0 & 1.11E$-0$1 & 1.79E$-0$1 \\ 
  5.20 - 5.70 & 5.46 & 0.524 & 0.95 & 2.08E$+0$0 & 7.41E$-0$2 & 7.49E$-0$2 \\ 
  5.70 - 6.20 & 5.95 & 0.524 & 0.97 & 1.95E$+0$0 & 6.28E$-0$2 & 5.96E$-0$2 \\ 
  6.20 - 6.70 & 6.45 & 0.524 & 0.97 & 1.66E$+0$0 & 5.34E$-0$2 & 3.90E$-0$2 \\ 
  6.70 - 7.20 & 6.95 & 0.524 & 1.07 & 1.29E$+0$0 & 4.83E$-0$2 & 3.00E$-0$2 \\ 
  7.20 - 7.70 & 7.44 & 0.523 & 1.05 & 1.16E$+0$0 & 4.65E$-0$2 & 2.67E$-0$2 \\ 
  7.70 - 8.20 & 7.93 & 0.523 & 1.11 & 9.72E$-0$1 & 4.40E$-0$2 & 3.70E$-0$2 \\ 
  8.20 - 8.70 & 8.44 & 0.524 & 1.16 & 8.48E$-0$1 & 4.03E$-0$2 & 2.75E$-0$2 \\ 
  10.85 - 11.85 & 11.30 & 0.522 & 1.14 & 2.00E$-0$1 & 2.14E$-0$2 & 8.15E$-0$3 \\ 
  11.85 - 12.85 & 12.27 & 0.527 & 1.03 & 1.23E$-0$1 & 2.46E$-0$2 & 8.66E$-0$3 \\ 
  12.85 - 14.85 & 13.55 & 0.524 & 1.19 & 8.12E$-0$2 & 1.94E$-0$2 & 3.91E$-0$3 \\ 

}

\xfbin{$ 0.55 \leq x_F < 0.60 $}{

  4.20 - 4.70 & 4.47 & 0.572 & 1.04 & 2.17E$+0$0 & 1.37E$-0$1 & 9.50E$-0$2 \\ 
  4.70 - 5.20 & 4.97 & 0.574 & 0.96 & 1.75E$+0$0 & 8.45E$-0$2 & 6.55E$-0$2 \\ 
  5.20 - 5.70 & 5.45 & 0.574 & 1.00 & 1.51E$+0$0 & 6.37E$-0$2 & 4.53E$-0$2 \\ 
  5.70 - 6.20 & 5.95 & 0.574 & 0.96 & 1.29E$+0$0 & 4.89E$-0$2 & 2.80E$-0$2 \\ 
  6.20 - 6.70 & 6.44 & 0.573 & 1.02 & 1.07E$+0$0 & 4.29E$-0$2 & 2.27E$-0$2 \\ 
  6.70 - 7.20 & 6.94 & 0.572 & 1.02 & 9.56E$-0$1 & 3.99E$-0$2 & 2.22E$-0$2 \\ 
  7.20 - 7.70 & 7.44 & 0.573 & 1.08 & 7.75E$-0$1 & 3.52E$-0$2 & 1.78E$-0$2 \\ 
  7.70 - 8.20 & 7.93 & 0.573 & 1.04 & 6.78E$-0$1 & 3.54E$-0$2 & 2.12E$-0$2 \\ 
  8.20 - 8.70 & 8.44 & 0.573 & 1.10 & 6.17E$-0$1 & 3.44E$-0$2 & 2.01E$-0$2 \\ 
  10.85 - 11.85 & 11.26 & 0.574 & 1.12 & 1.48E$-0$1 & 1.71E$-0$2 & 2.14E$-0$3 \\ 
  11.85 - 12.85 & 12.35 & 0.574 & 1.23 & 1.11E$-0$1 & 1.96E$-0$2 & 5.64E$-0$3 \\ 
  12.85 - 14.85 & 13.51 & 0.572 & 1.11 & 6.24E$-0$2 & 1.77E$-0$2 & 8.75E$-0$3 \\ 
  14.85 - 16.85 & 15.19 & 0.597 & 0.56 & 6.89E$-0$3 & 6.89E$-0$3 & 2.44E$-0$3 \\ 

}

\xfbin{$ 0.60 \leq x_F < 0.65 $}{

  4.20 - 4.70 & 4.48 & 0.624 & 1.01 & 1.13E$+0$0 & 9.93E$-0$2 & 5.28E$-0$2 \\ 
  4.70 - 5.20 & 4.96 & 0.623 & 0.92 & 1.07E$+0$0 & 6.80E$-0$2 & 4.01E$-0$2 \\ 
  5.20 - 5.70 & 5.47 & 0.624 & 0.90 & 8.64E$-0$1 & 4.54E$-0$2 & 3.58E$-0$2 \\ 
  5.70 - 6.20 & 5.96 & 0.622 & 0.94 & 7.77E$-0$1 & 4.28E$-0$2 & 2.64E$-0$2 \\ 
  6.20 - 6.70 & 6.45 & 0.623 & 0.92 & 6.47E$-0$1 & 3.39E$-0$2 & 2.70E$-0$2 \\ 
  6.70 - 7.20 & 6.94 & 0.622 & 0.99 & 5.81E$-0$1 & 3.25E$-0$2 & 1.73E$-0$2 \\ 
  7.20 - 7.70 & 7.44 & 0.624 & 1.08 & 4.47E$-0$1 & 3.15E$-0$2 & 1.69E$-0$2 \\ 
  7.70 - 8.20 & 7.96 & 0.622 & 1.03 & 4.30E$-0$1 & 3.25E$-0$2 & 1.30E$-0$2 \\ 
  8.20 - 8.70 & 8.45 & 0.625 & 1.09 & 3.49E$-0$1 & 2.76E$-0$2 & 9.92E$-0$3 \\ 
  10.85 - 11.85 & 11.28 & 0.621 & 1.19 & 9.51E$-0$2 & 1.42E$-0$2 & 3.28E$-0$3 \\ 
  11.85 - 12.85 & 12.33 & 0.621 & 1.21 & 8.06E$-0$2 & 1.61E$-0$2 & 6.38E$-0$3 \\ 
  12.85 - 14.85 & 13.38 & 0.622 & 1.25 & 2.90E$-0$2 & 7.26E$-0$3 & 1.73E$-0$3 \\ 
  14.85 - 16.85 & 15.89 & 0.623 & 3.48 & 4.67E$-0$2 & 4.67E$-0$2 & 1.78E$-0$1 \\ 

}

\xfbin{$ 0.65 \leq x_F < 0.70 $}{

  4.20 - 4.70 & 4.48 & 0.669 & 0.95 & 5.85E$-0$1 & 7.05E$-0$2 & 5.31E$-0$2 \\ 
  4.70 - 5.20 & 4.96 & 0.670 & 0.93 & 6.19E$-0$1 & 4.74E$-0$2 & 3.13E$-0$2 \\ 
  5.20 - 5.70 & 5.47 & 0.673 & 0.90 & 5.69E$-0$1 & 3.99E$-0$2 & 2.79E$-0$2 \\ 
  5.70 - 6.20 & 5.96 & 0.673 & 0.88 & 4.49E$-0$1 & 2.94E$-0$2 & 1.59E$-0$2 \\ 
  6.20 - 6.70 & 6.45 & 0.673 & 1.02 & 4.01E$-0$1 & 2.91E$-0$2 & 1.33E$-0$2 \\ 
  6.70 - 7.20 & 6.95 & 0.673 & 0.99 & 3.65E$-0$1 & 2.89E$-0$2 & 1.63E$-0$2 \\ 
  7.20 - 7.70 & 7.44 & 0.671 & 1.11 & 2.81E$-0$1 & 2.54E$-0$2 & 1.25E$-0$2 \\ 
  7.70 - 8.20 & 7.93 & 0.671 & 1.10 & 2.44E$-0$1 & 3.18E$-0$2 & 1.53E$-0$2 \\ 
  8.20 - 8.70 & 8.43 & 0.670 & 1.06 & 1.91E$-0$1 & 2.09E$-0$2 & 9.33E$-0$3 \\ 
  10.85 - 11.85 & 11.33 & 0.674 & 0.97 & 5.01E$-0$2 & 1.46E$-0$2 & 3.46E$-0$3 \\ 
  11.85 - 12.85 & 12.26 & 0.675 & 0.95 & 4.39E$-0$2 & 1.34E$-0$2 & 4.85E$-0$3 \\ 
  12.85 - 14.85 & 13.52 & 0.679 & 0.95 & 2.01E$-0$2 & 8.56E$-0$3 & 5.89E$-0$3 \\ 

}\pagebreak

\xfbin{$ 0.70 \leq x_F < 0.75 $}{

  4.20 - 4.70 & 4.50 & 0.721 & 1.03 & 3.58E$-0$1 & 6.46E$-0$2 & 3.68E$-0$2 \\ 
  4.70 - 5.20 & 4.96 & 0.722 & 0.91 & 3.17E$-0$1 & 3.93E$-0$2 & 2.03E$-0$2 \\ 
  5.20 - 5.70 & 5.47 & 0.722 & 0.87 & 2.08E$-0$1 & 2.19E$-0$2 & 1.06E$-0$2 \\ 
  5.70 - 6.20 & 5.94 & 0.723 & 0.82 & 2.31E$-0$1 & 3.15E$-0$2 & 1.44E$-0$2 \\ 
  6.20 - 6.70 & 6.43 & 0.723 & 0.90 & 1.88E$-0$1 & 2.83E$-0$2 & 1.72E$-0$2 \\ 
  6.70 - 7.20 & 6.95 & 0.722 & 0.98 & 1.69E$-0$1 & 1.82E$-0$2 & 8.32E$-0$3 \\ 
  7.20 - 7.70 & 7.45 & 0.722 & 1.00 & 1.35E$-0$1 & 1.79E$-0$2 & 7.68E$-0$3 \\ 
  7.70 - 8.20 & 7.93 & 0.721 & 0.98 & 1.00E$-0$1 & 1.42E$-0$2 & 5.09E$-0$3 \\ 
  8.20 - 8.70 & 8.42 & 0.722 & 0.98 & 8.28E$-0$2 & 1.37E$-0$2 & 4.85E$-0$3 \\ 
  10.85 - 11.85 & 11.28 & 0.719 & 1.16 & 3.11E$-0$2 & 7.74E$-0$3 & 3.80E$-0$3 \\ 
  11.85 - 12.85 & 12.25 & 0.733 & 0.74 & 3.29E$-0$2 & 1.42E$-0$2 & 3.08E$-0$3 \\ 
  12.85 - 14.85 & 13.68 & 0.717 & 1.73 & 4.94E$-0$3 & 4.94E$-0$3 & 1.78E$-0$3 \\ 

}

\xfbin{$ 0.75 \leq x_F < 0.80 $}{

  4.20 - 4.70 & 4.47 & 0.777 & 1.20 & 1.89E$-0$1 & 6.62E$-0$2 & 2.48E$-0$2 \\ 
  4.70 - 5.20 & 4.95 & 0.773 & 0.90 & 1.52E$-0$1 & 2.82E$-0$2 & 1.37E$-0$2 \\ 
  5.20 - 5.70 & 5.44 & 0.769 & 0.75 & 1.02E$-0$1 & 1.62E$-0$2 & 6.22E$-0$3 \\ 
  5.70 - 6.20 & 5.98 & 0.770 & 0.83 & 1.19E$-0$1 & 2.45E$-0$2 & 8.09E$-0$3 \\ 
  6.20 - 6.70 & 6.45 & 0.772 & 0.90 & 1.27E$-0$1 & 5.28E$-0$2 & 8.69E$-0$3 \\ 
  6.70 - 7.20 & 6.92 & 0.769 & 0.93 & 8.04E$-0$2 & 1.20E$-0$2 & 6.24E$-0$3 \\ 
  7.20 - 7.70 & 7.47 & 0.769 & 0.83 & 4.95E$-0$2 & 1.09E$-0$2 & 5.26E$-0$3 \\ 
  7.70 - 8.20 & 7.88 & 0.769 & 1.06 & 6.40E$-0$2 & 1.42E$-0$2 & 6.45E$-0$3 \\ 
  8.20 - 8.70 & 8.44 & 0.775 & 0.93 & 3.33E$-0$2 & 1.06E$-0$2 & 1.48E$-0$3 \\ 
  10.85 - 11.85 & 11.30 & 0.770 & 1.01 & 1.36E$-0$2 & 6.22E$-0$3 & 2.71E$-0$3 \\ 
  11.85 - 12.85 & 12.30 & 0.765 & 1.35 & 2.08E$-0$2 & 1.20E$-0$2 & 3.21E$-0$3 \\ 
  12.85 - 14.85 & 13.59 & 0.773 & 0.61 & 1.09E$-0$2 & 1.09E$-0$2 & 4.06E$-0$3 \\ 

}

\end{longtable} 

\end{landscape} 

\begin{landscape}\centering

\newlength{\mycaplen}
\setlength{\mycaplen}{0.9\linewidth}

\begin{figure}\centering

\begin{minipage}[tb]{0.45\linewidth}
\includegraphics[width=\linewidth,clip]{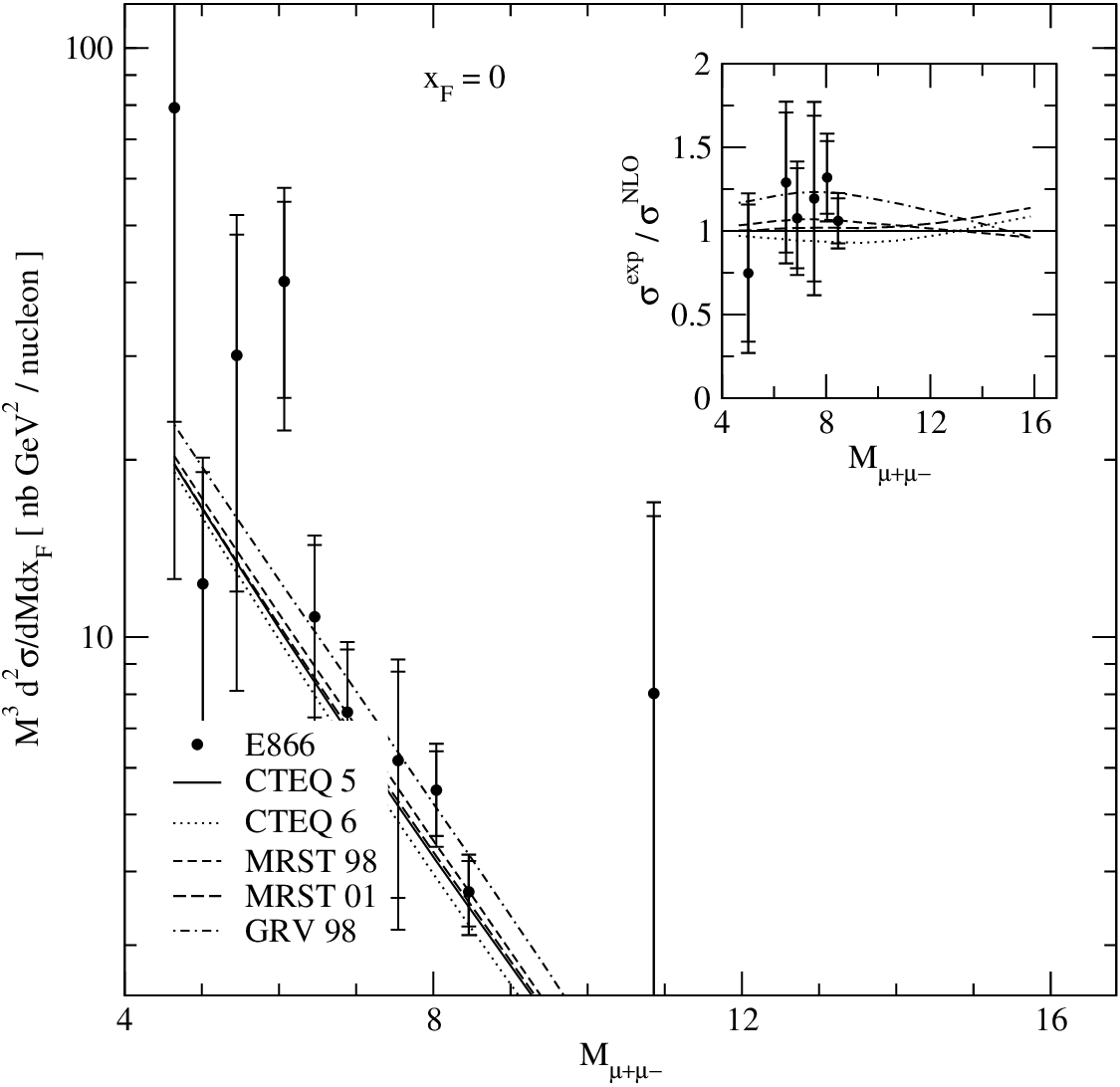}
\end{minipage}
\begin{minipage}[tb]{0.45\linewidth}
\includegraphics[width=\linewidth,clip]{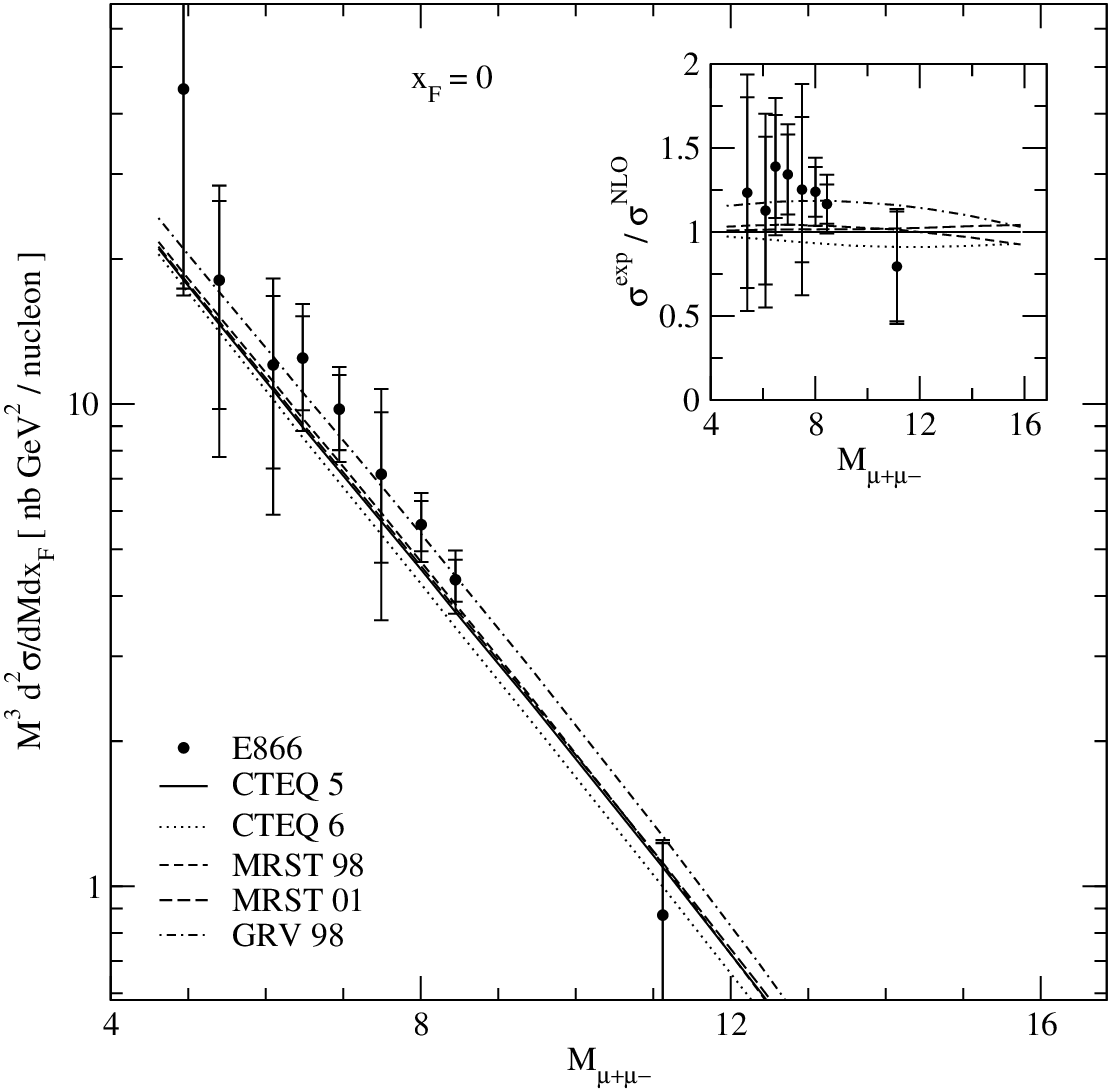}
\end{minipage}

\addtocontents{lof}{\  }
\addtocontents{lof}{\setlength{\baselineskip}{\singlespace}}
\begin{minipage}[tb]{\mycaplen}
\caption[ 
	Scaling form $M^3d^2\sigma/dMdx_F$ of the 
	hydrogen and deuterium cross sections for 
	$-0.05 \leq x_F < 0.05$.]{
	\setlength{\baselineskip}{\singlespace}
	\label{fig:d2sigma_first}Scaling form $M^3d^2\sigma/dMdx_F$ of the 
	hydrogen (left) and deuterium (right) cross sections for 
	$-0.05 \leq x_F < 0.05$.  The data are compared to NLO calculations 
	based on several different sets of parton distributions.  The inset 
	shows the ratio of the experimentally measured cross sections to NLO 
	calculations based on the CTEQ 5 partons.
	}
\end{minipage}
\addtocontents{lof}{\setlength{\baselineskip}{\doublespace}}

\end{figure}

\begin{figure}\centering

\begin{minipage}[tb]{0.45\linewidth}
\includegraphics[width=\linewidth,clip]{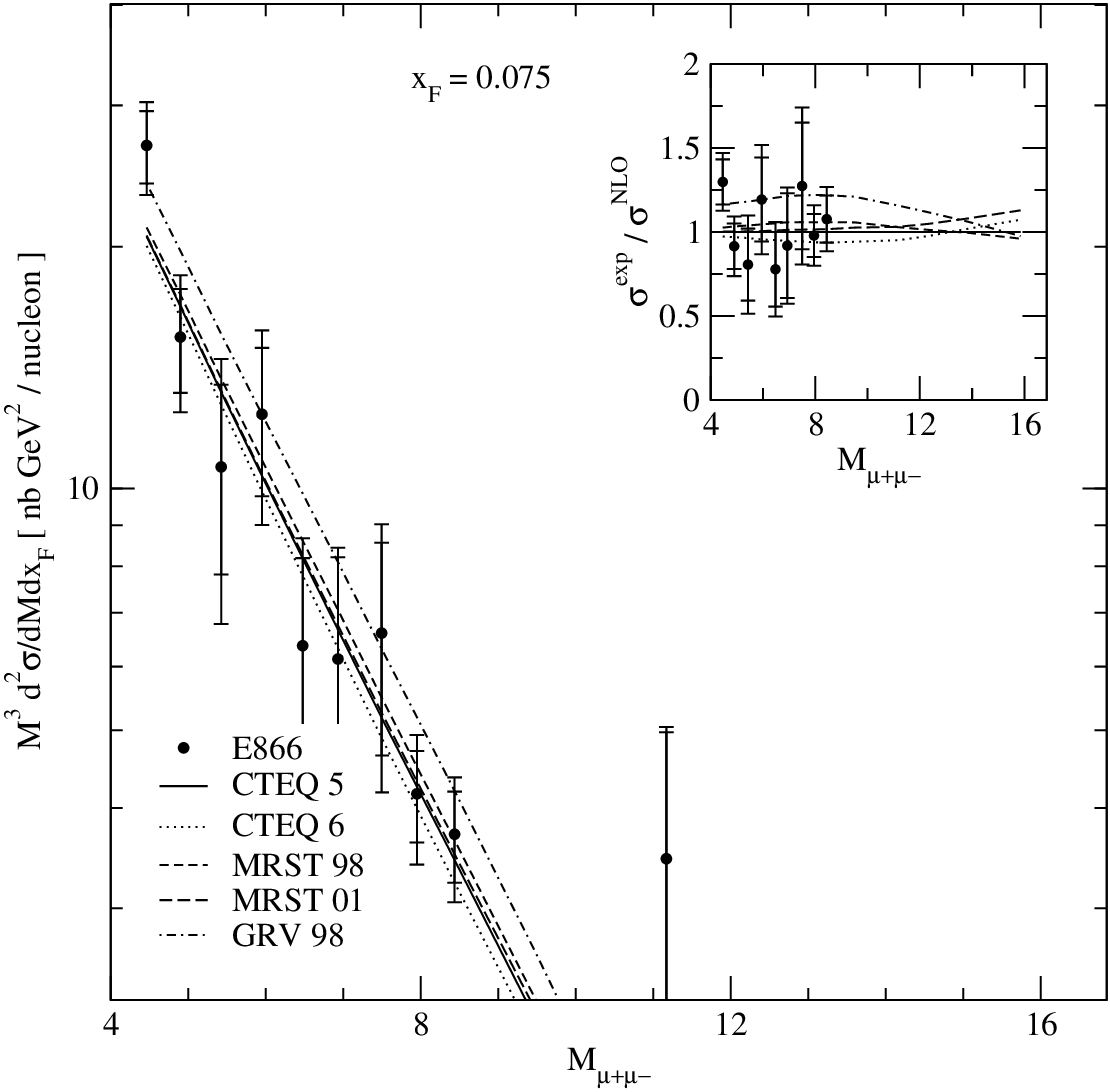}
\end{minipage}
\begin{minipage}[tb]{0.45\linewidth}
\includegraphics[width=\linewidth,clip]{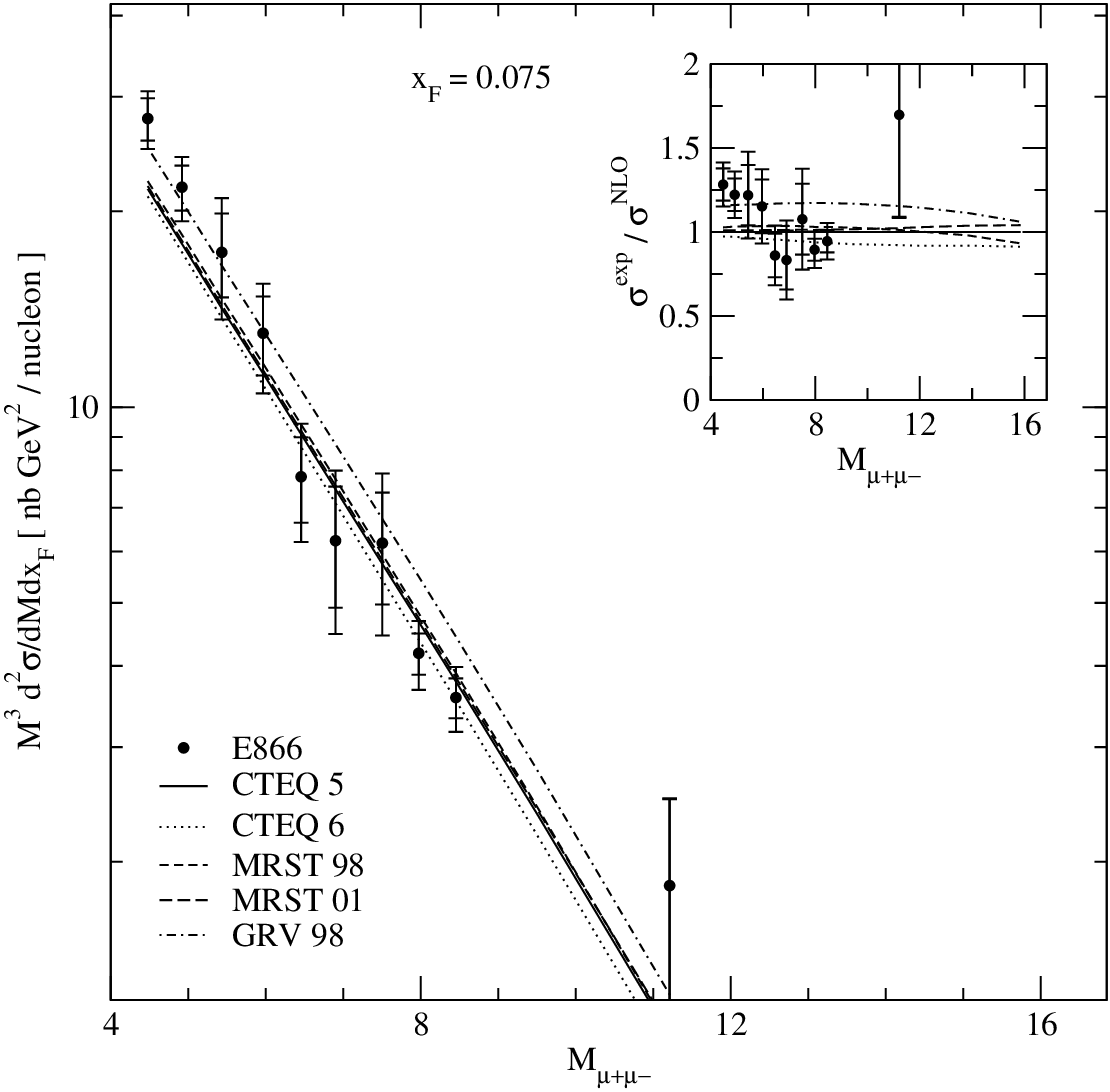}
\end{minipage}

\addtocontents{lof}{\  }
\addtocontents{lof}{\setlength{\baselineskip}{\singlespace}}
\begin{minipage}[tb]{\mycaplen}
\caption[Scaling form $M^3d^2\sigma/dMdx_F$ of the hydrogen and 
	deuterium cross sections for $0.05 \leq x_F < 0.1$.]{
	\setlength{\baselineskip}{\singlespace}
	Scaling form $M^3d^2\sigma/dMdx_F$ of the hydrogen (left) and 
	deuterium (right) cross sections for $0.05 \leq x_F < 0.1$.  The data 
	are compared to NLO calculations based on several different sets of 
	parton distributions.  The inset shows the ratio of the experimentally
	measured cross sections to NLO calculations based on the CTEQ 5 
	partons.
	}
\end{minipage}
\addtocontents{lof}{\setlength{\baselineskip}{\doublespace}}
\end{figure}

\begin{figure}\centering

\begin{minipage}[tb]{0.45\linewidth}
\includegraphics[width=\linewidth,clip]{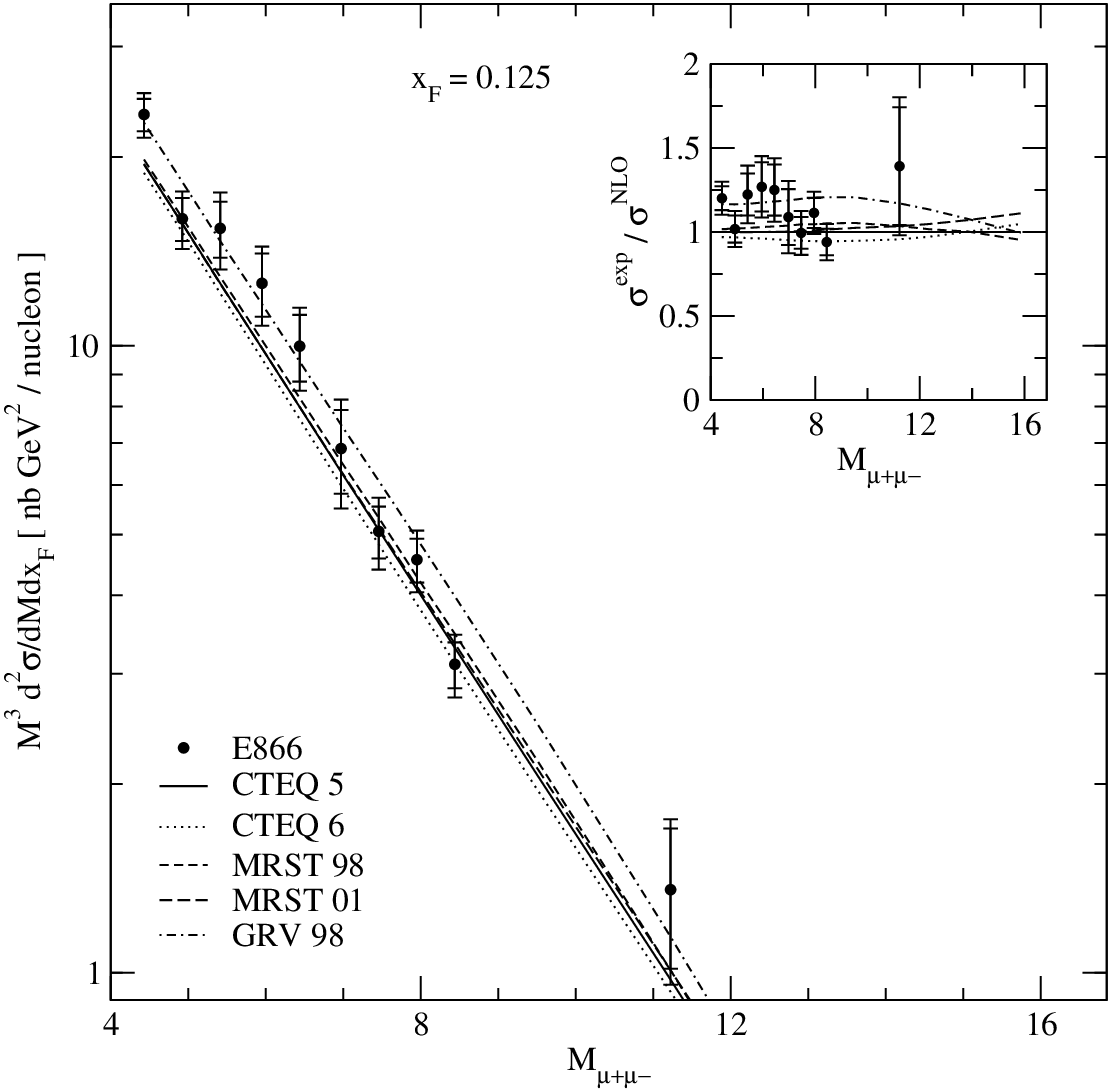}
\end{minipage}
\begin{minipage}[tb]{0.45\linewidth}
\includegraphics[width=\linewidth,clip]{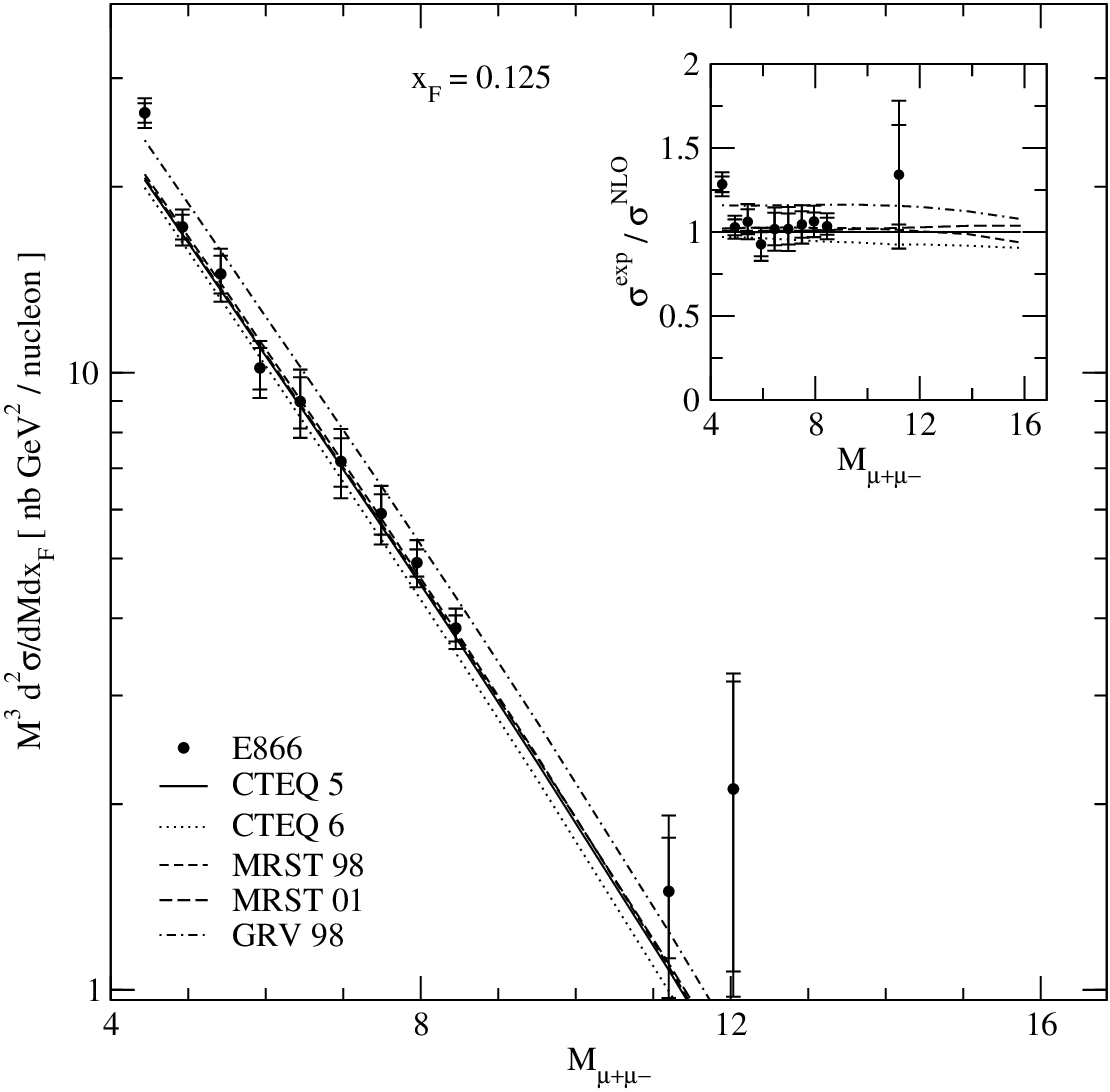}
\end{minipage}

\addtocontents{lof}{\  }
\addtocontents{lof}{\setlength{\baselineskip}{\singlespace}}
\begin{minipage}[tb]{\mycaplen}
\caption[Scaling form $M^3d^2\sigma/dMdx_F$ of the hydrogen and 
	deuterium cross sections for $0.1 \leq x_F < 0.15$.]{
	\setlength{\baselineskip}{\singlespace}
	Scaling form $M^3d^2\sigma/dMdx_F$ of the hydrogen (left) and 
	deuterium (right) cross sections for $0.1 \leq x_F < 0.15$.  The data are compared to 
	NLO calculations based on several different sets of parton distributions.
	The inset shows the ratio of the experimentally measured cross sections to
	NLO calculations based on the CTEQ 5 partons.
	}
\end{minipage}
\addtocontents{lof}{\setlength{\baselineskip}{\doublespace}}
\end{figure}

\begin{figure}\centering

\begin{minipage}[tb]{0.45\linewidth}
\includegraphics[width=\linewidth,clip]{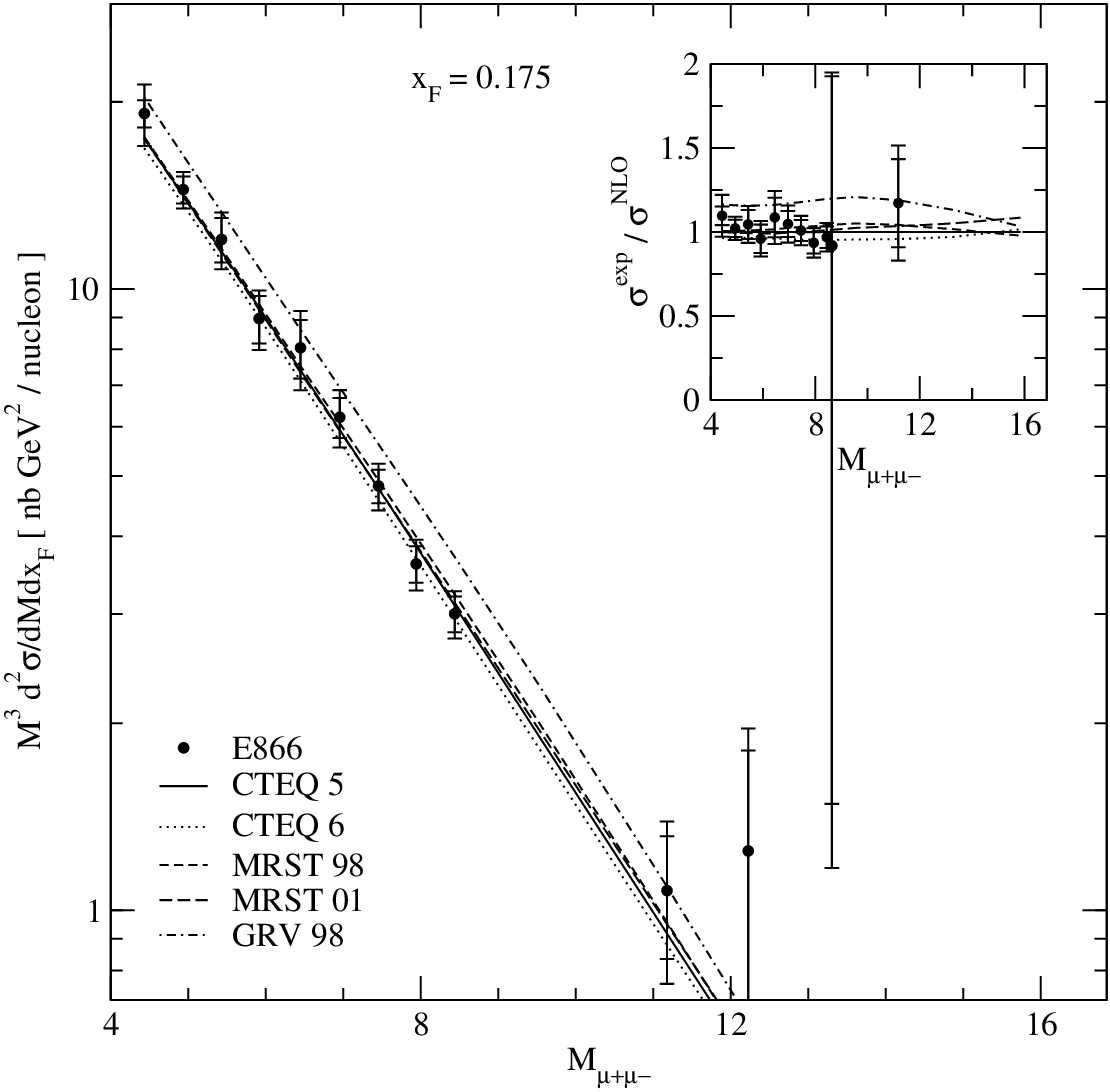}
\end{minipage}
\begin{minipage}[tb]{0.45\linewidth}
\includegraphics[width=\linewidth,clip]{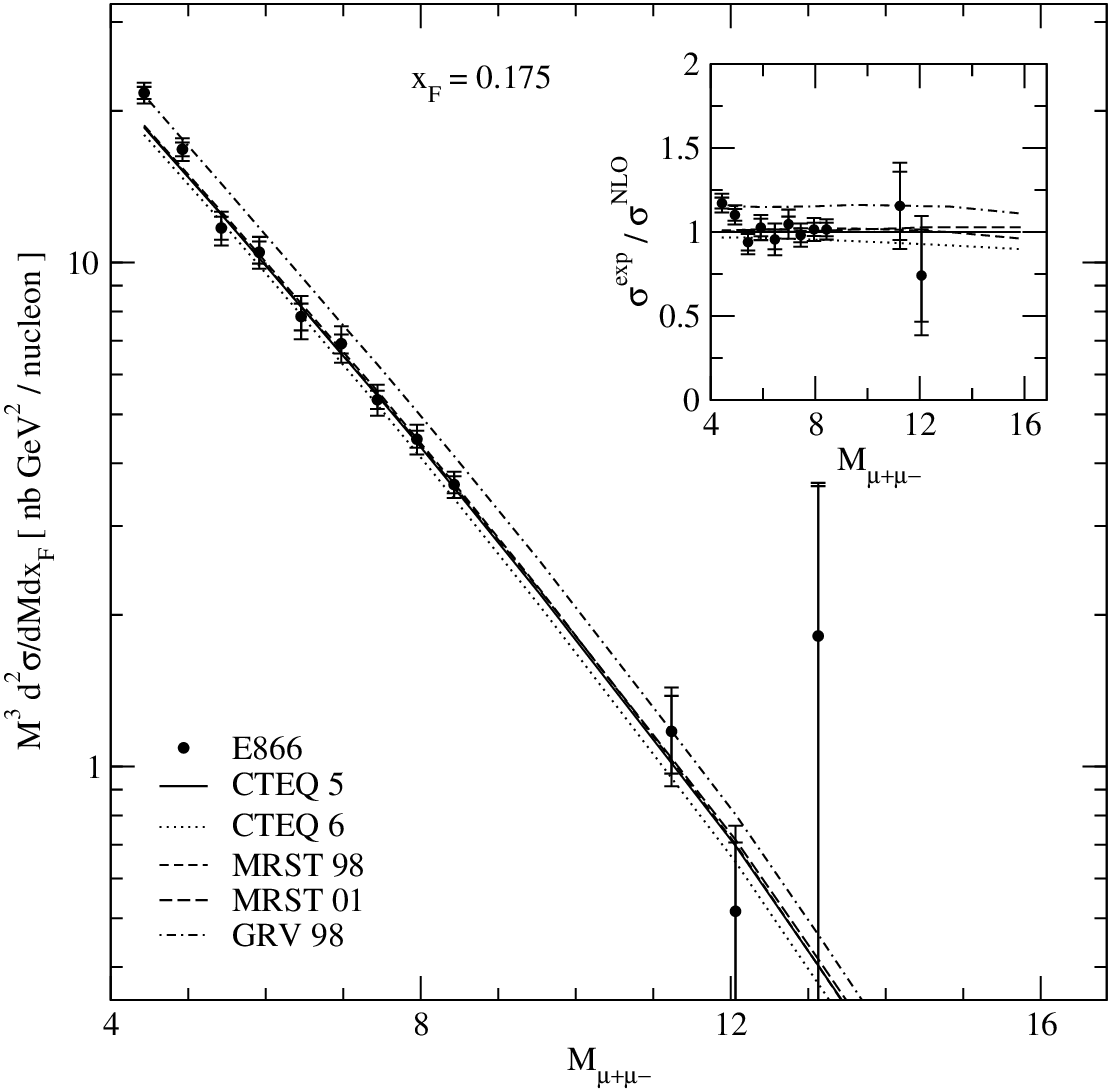}
\end{minipage}

\addtocontents{lof}{\  }
\addtocontents{lof}{\setlength{\baselineskip}{\singlespace}}
\begin{minipage}[tb]{\mycaplen}
\caption[Scaling form $M^3d^2\sigma/dMdx_F$ of the hydrogen and 
	deuterium cross sections for $0.15 \leq x_F < 0.2$.]{
	\setlength{\baselineskip}{\singlespace}
	Scaling form $M^3d^2\sigma/dMdx_F$ of the hydrogen (left) and 
	deuterium (right) cross sections for $0.15 \leq x_F < 0.2$.  The data are compared to 
	NLO calculations based on several different sets of parton distributions.
	The inset shows the ratio of the experimentally measured cross sections to
	NLO calculations based on the CTEQ 5 partons.
	}
\end{minipage}
\addtocontents{lof}{\setlength{\baselineskip}{\doublespace}}
\end{figure}

\begin{figure}\centering

\begin{minipage}[tb]{0.45\linewidth}
\includegraphics[width=\linewidth,clip]{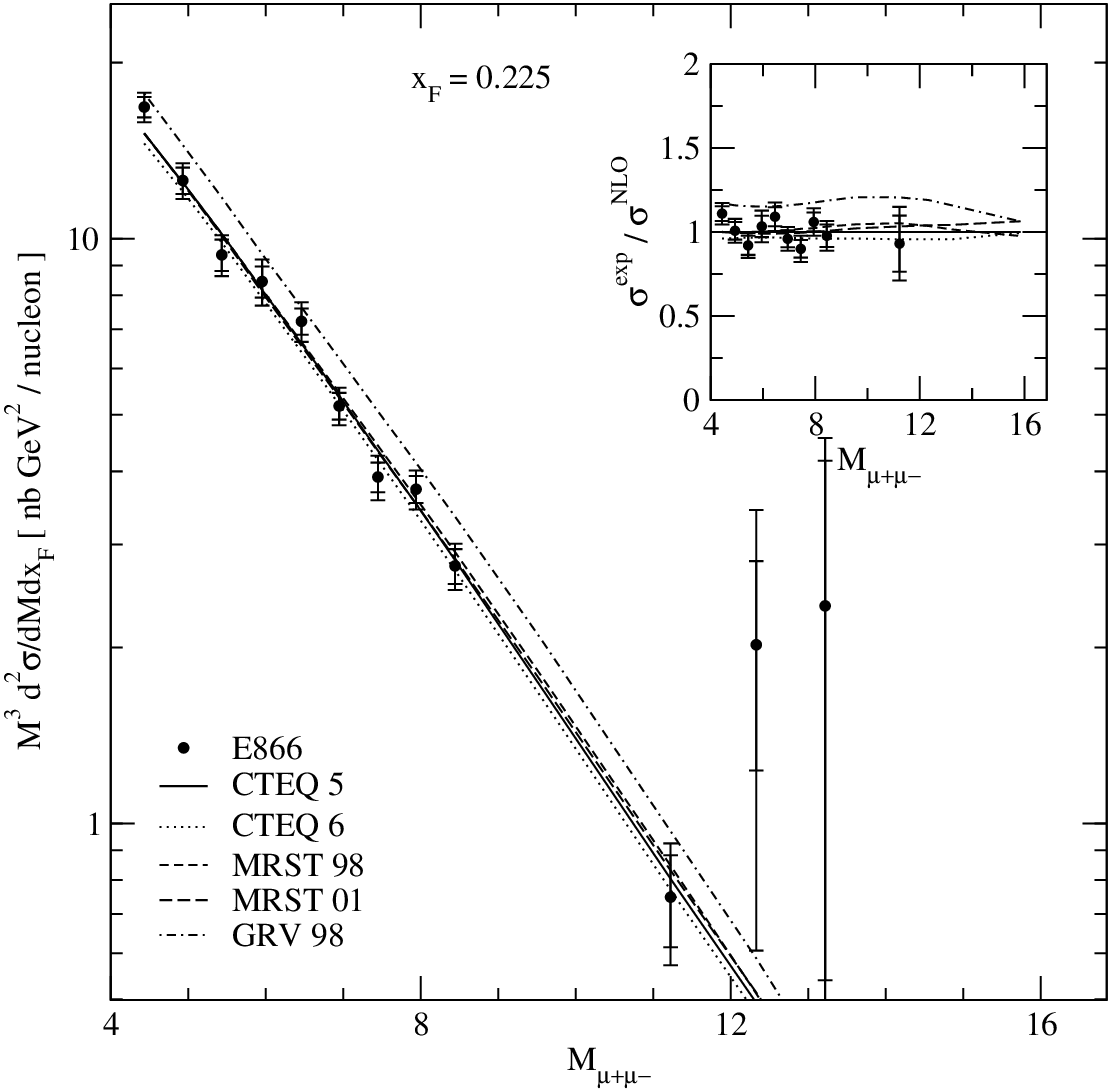}
\end{minipage}
\begin{minipage}[tb]{0.45\linewidth}
\includegraphics[width=\linewidth,clip]{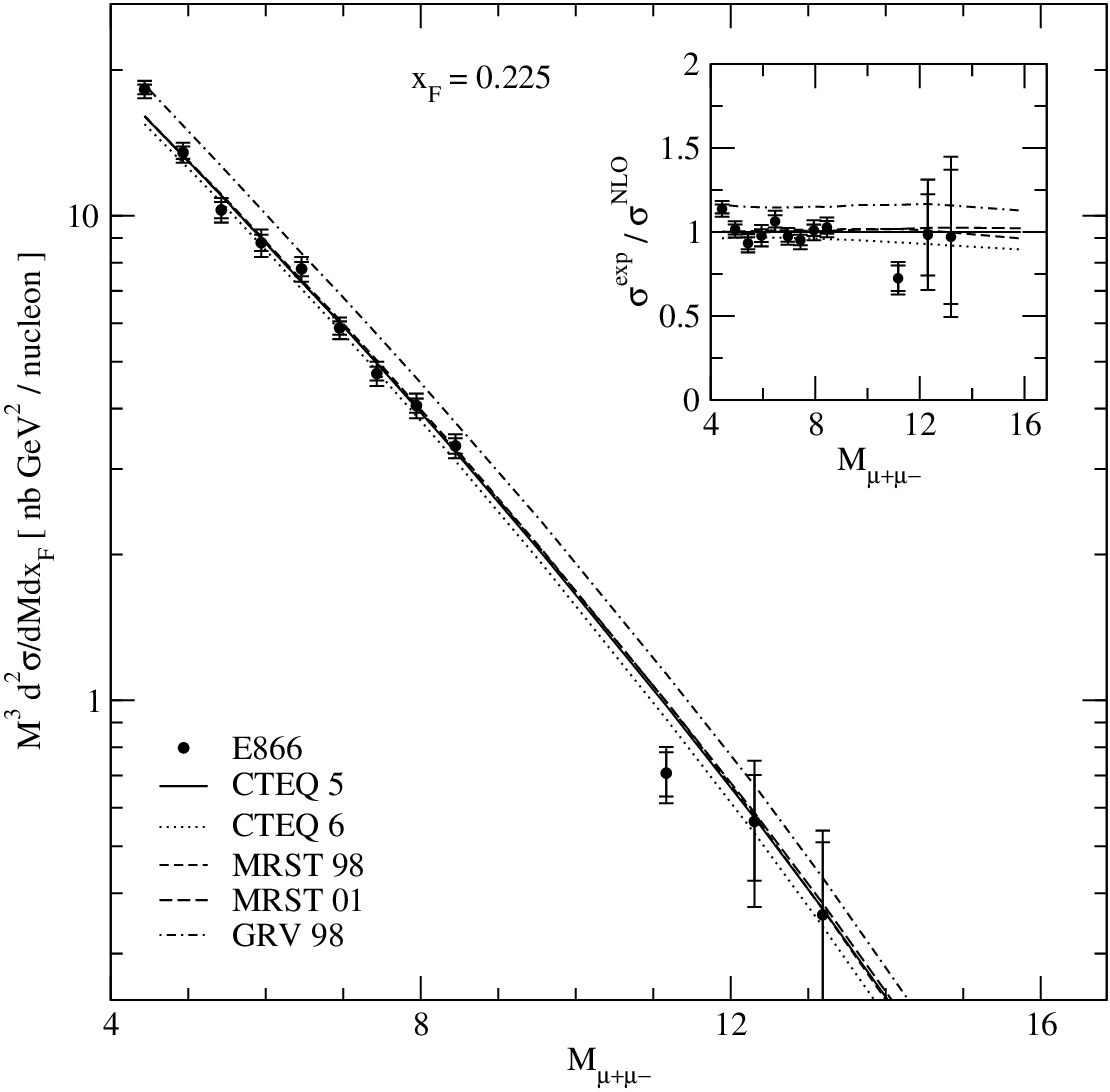}
\end{minipage}

\addtocontents{lof}{\  }
\addtocontents{lof}{\setlength{\baselineskip}{\singlespace}}
\begin{minipage}[tb]{\mycaplen}
\caption[Scaling form $M^3d^2\sigma/dMdx_F$ of the hydrogen and 
	deuterium cross sections for $0.2 \leq x_F < 0.25$.]{
	\setlength{\baselineskip}{\singlespace}
	Scaling form $M^3d^2\sigma/dMdx_F$ of the hydrogen (left) and 
	deuterium (right) cross sections for $0.2 \leq x_F < 0.25$.  The data are compared to 
	NLO calculations based on several different sets of parton distributions.
	The inset shows the ratio of the experimentally measured cross sections to
	NLO calculations based on the CTEQ 5 partons.
	}
\end{minipage}
\addtocontents{lof}{\setlength{\baselineskip}{\doublespace}}
\end{figure}

\begin{figure}\centering

\begin{minipage}[tb]{0.45\linewidth}
\includegraphics[width=\linewidth,clip]{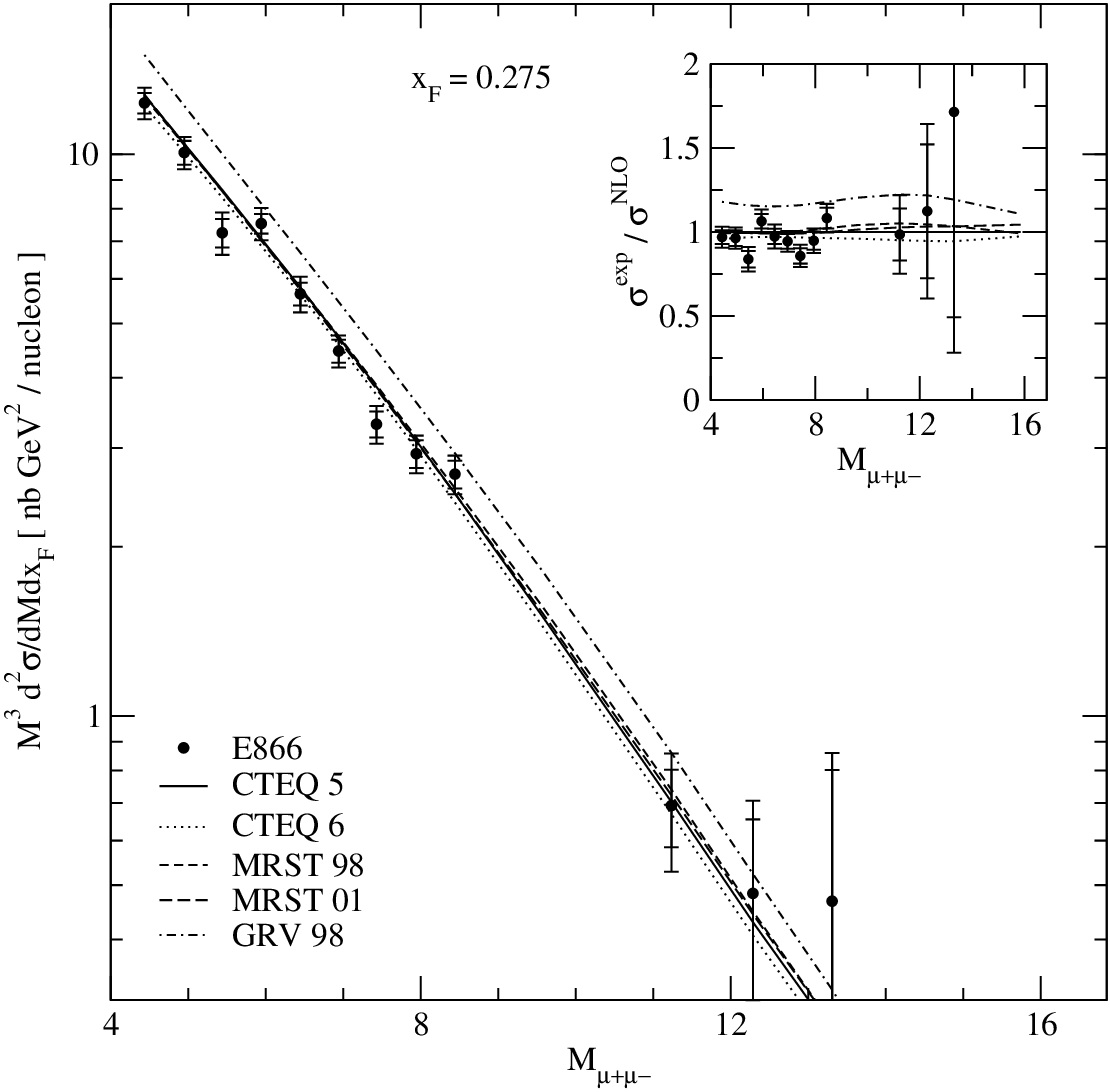}
\end{minipage}
\begin{minipage}[tb]{0.45\linewidth}
\includegraphics[width=\linewidth,clip]{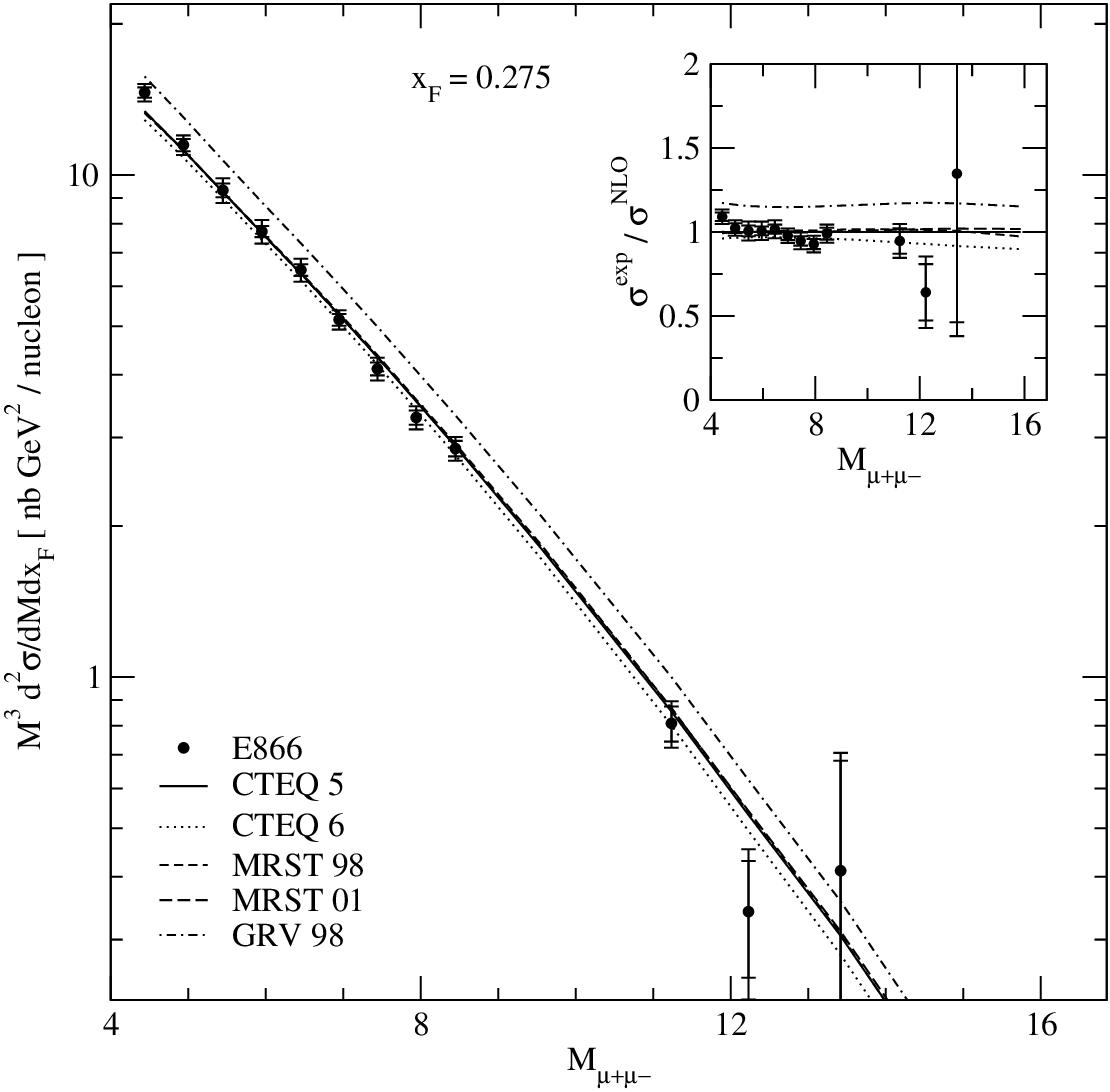}
\end{minipage}

\addtocontents{lof}{\  }
\addtocontents{lof}{\setlength{\baselineskip}{\singlespace}}
\begin{minipage}[tb]{\mycaplen}
\caption[Scaling form $M^3d^2\sigma/dMdx_F$ of the hydrogen and 
	deuterium cross sections for $0.25 \leq x_F < 0.3$.]{
	\setlength{\baselineskip}{\singlespace}
	Scaling form $M^3d^2\sigma/dMdx_F$ of the hydrogen (left) and 
	deuterium (right) cross sections for $0.25 \leq x_F < 0.3$.  The data are compared to 
	NLO calculations based on several different sets of parton distributions.
	The inset shows the ratio of the experimentally measured cross sections to
	NLO calculations based on the CTEQ 5 partons.
	}
\end{minipage}
\addtocontents{lof}{\setlength{\baselineskip}{\doublespace}}
\end{figure}

\begin{figure}\centering

\begin{minipage}[tb]{0.45\linewidth}
\includegraphics[width=\linewidth,clip]{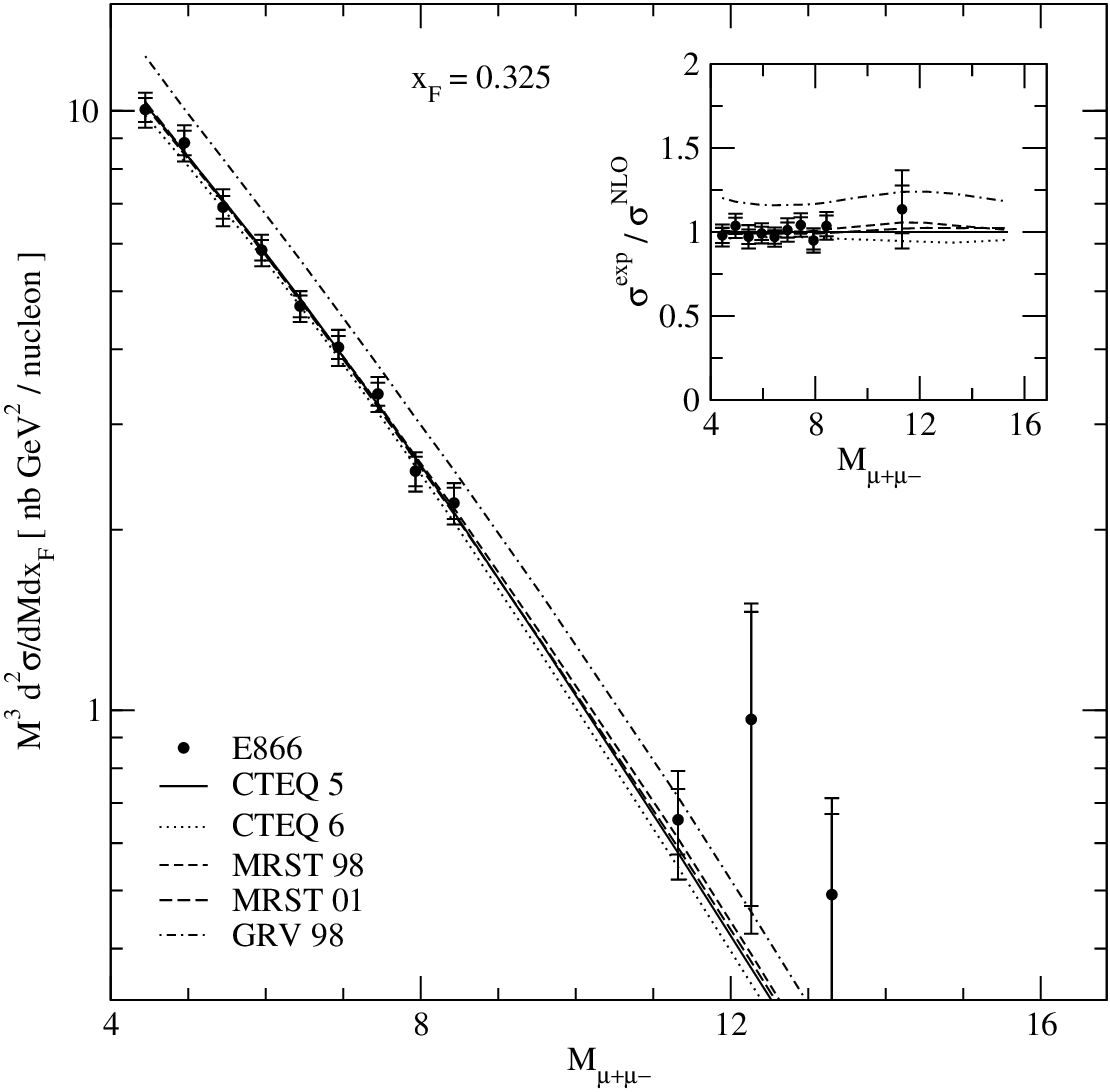}
\end{minipage}
\begin{minipage}[tb]{0.45\linewidth}
\includegraphics[width=\linewidth,clip]{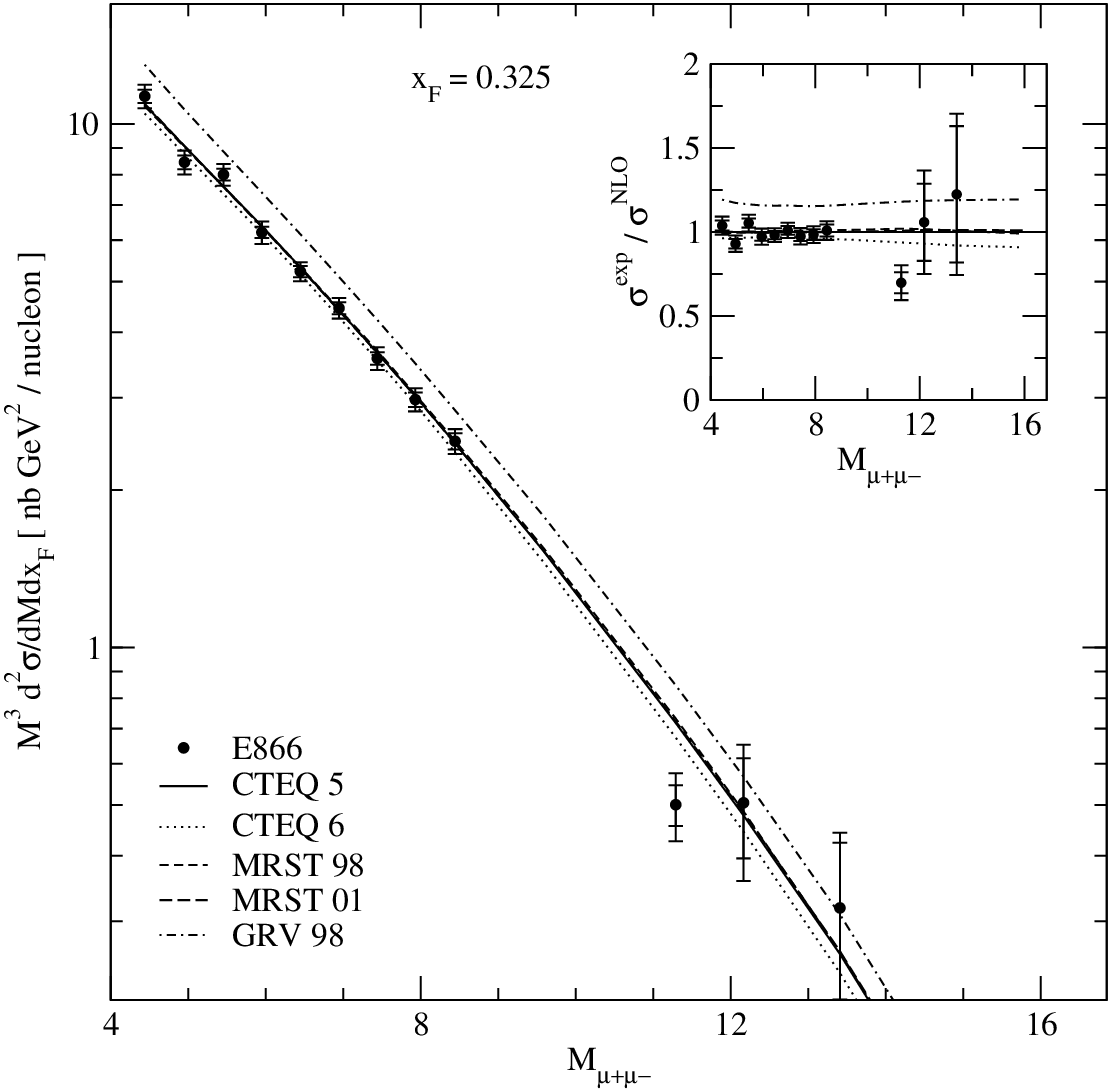}
\end{minipage}

\addtocontents{lof}{\  }
\addtocontents{lof}{\setlength{\baselineskip}{\singlespace}}
\begin{minipage}[tb]{\mycaplen}
\caption[Scaling form $M^3d^2\sigma/dMdx_F$ of the hydrogen and 
	deuterium cross sections for $0.3 \leq x_F < 0.35$.]{
	\setlength{\baselineskip}{\singlespace}
	Scaling form $M^3d^2\sigma/dMdx_F$ of the hydrogen (left) and 
	deuterium (right) cross sections for $0.3 \leq x_F < 0.35$.  The data are compared to 
	NLO calculations based on several different sets of parton distributions.
	The inset shows the ratio of the experimentally measured cross sections to
	NLO calculations based on the CTEQ 5 partons.
	}
\end{minipage}
\addtocontents{lof}{\setlength{\baselineskip}{\doublespace}}

\end{figure}

\begin{figure}\centering

\begin{minipage}[tb]{0.45\linewidth}
\includegraphics[width=\linewidth,clip]{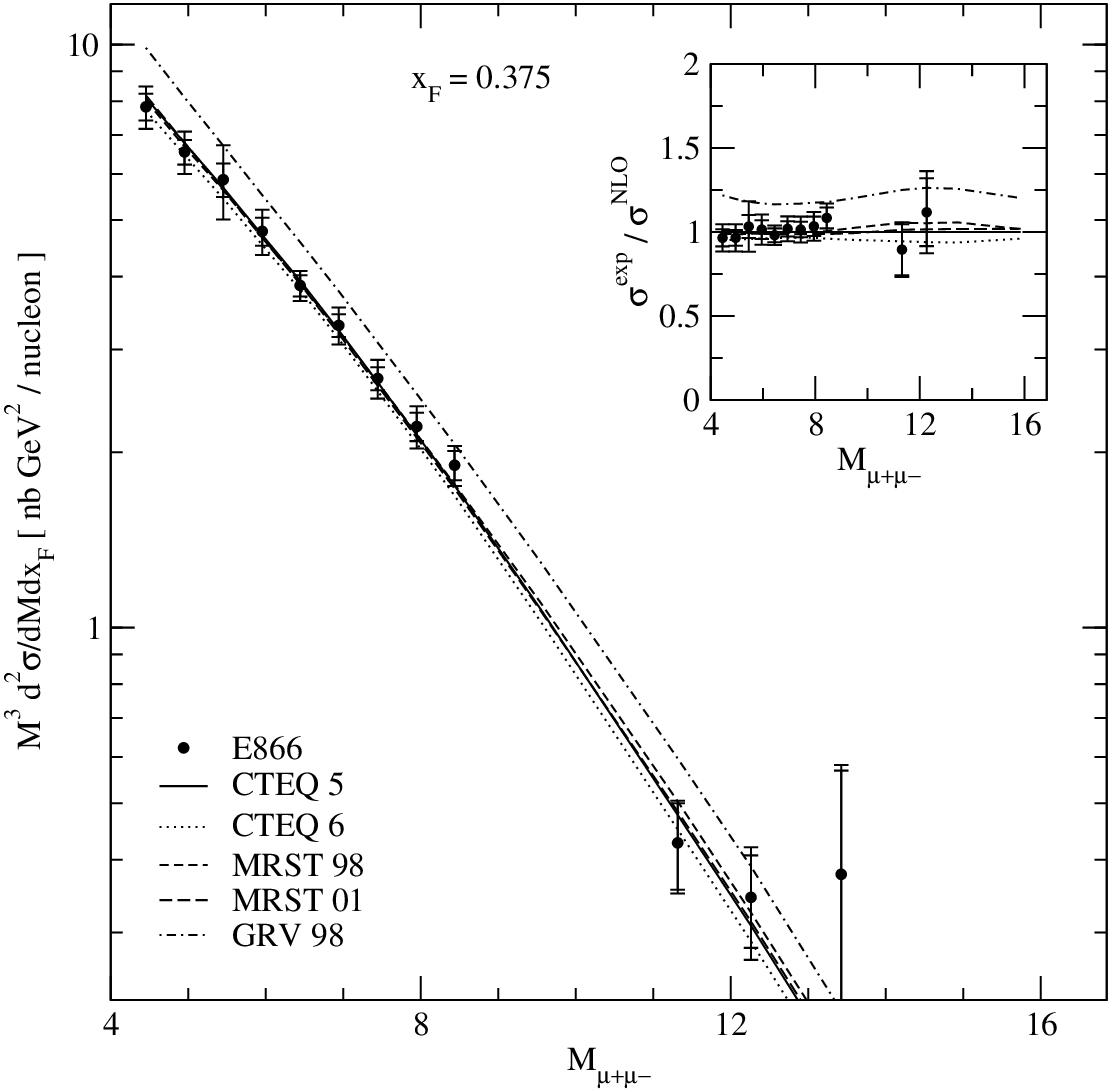}
\end{minipage}
\begin{minipage}[tb]{0.45\linewidth}
\includegraphics[width=\linewidth,clip]{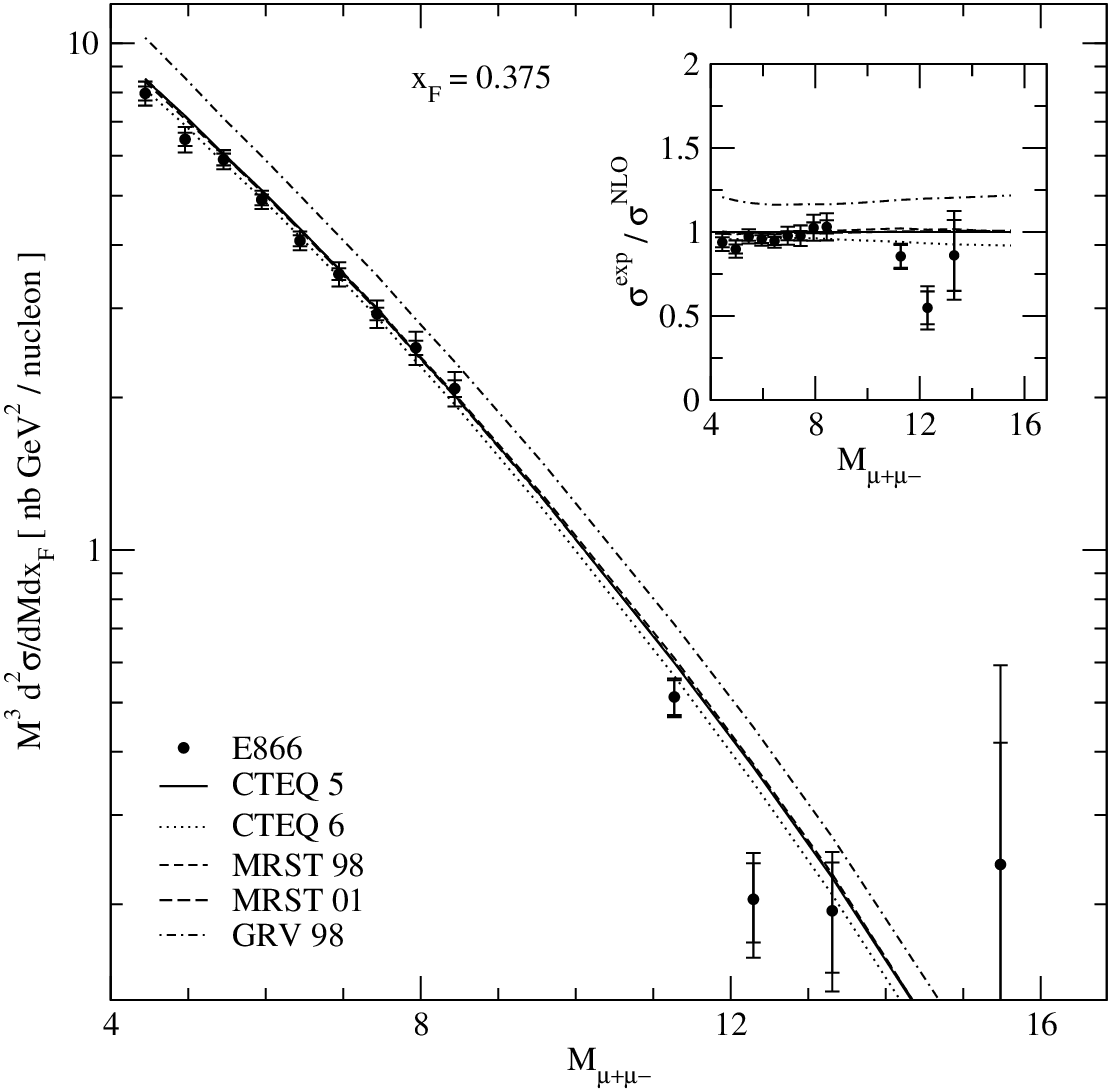}
\end{minipage}

\addtocontents{lof}{\  }
\addtocontents{lof}{\setlength{\baselineskip}{\singlespace}}

\begin{minipage}[tb]{\mycaplen}
\caption[Scaling form $M^3d^2\sigma/dMdx_F$ of the hydrogen and 
	deuterium cross sections for $0.35 \leq x_F < 0.4$.]{
	\setlength{\baselineskip}{\singlespace}
	Scaling form $M^3d^2\sigma/dMdx_F$ of the hydrogen (left) and 
	deuterium (right) cross sections for $0.35 \leq x_F < 0.4$.  The data are compared to 
	NLO calculations based on several different sets of parton distributions.
	The inset shows the ratio of the experimentally measured cross sections to
	NLO calculations based on the CTEQ 5 partons.
	}
\end{minipage}
\addtocontents{lof}{\setlength{\baselineskip}{\doublespace}}

\end{figure}

\begin{figure}\centering

\begin{minipage}[tb]{0.45\linewidth}
\includegraphics[width=\linewidth,clip]{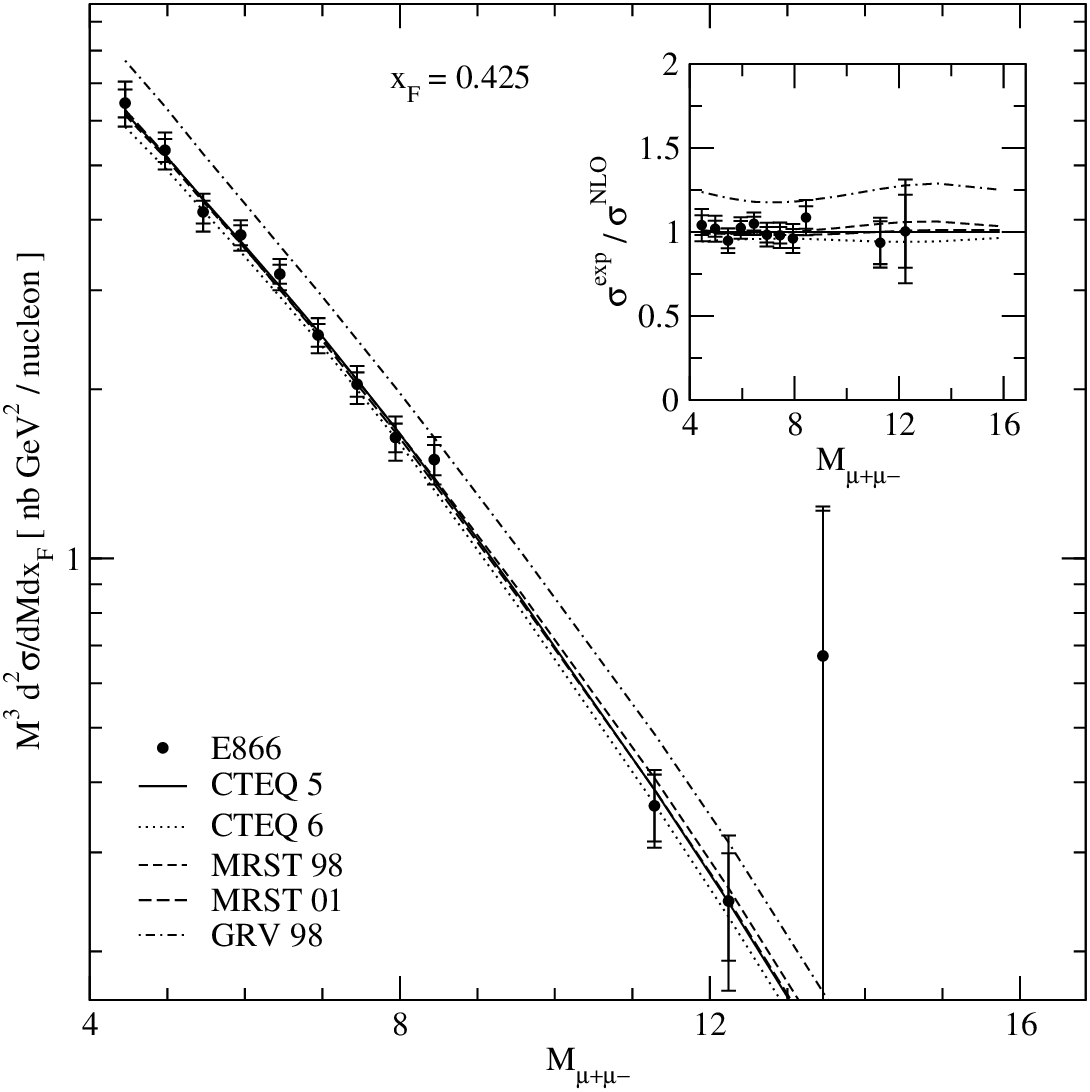}
\end{minipage}
\begin{minipage}[tb]{0.45\linewidth}
\includegraphics[width=\linewidth,clip]{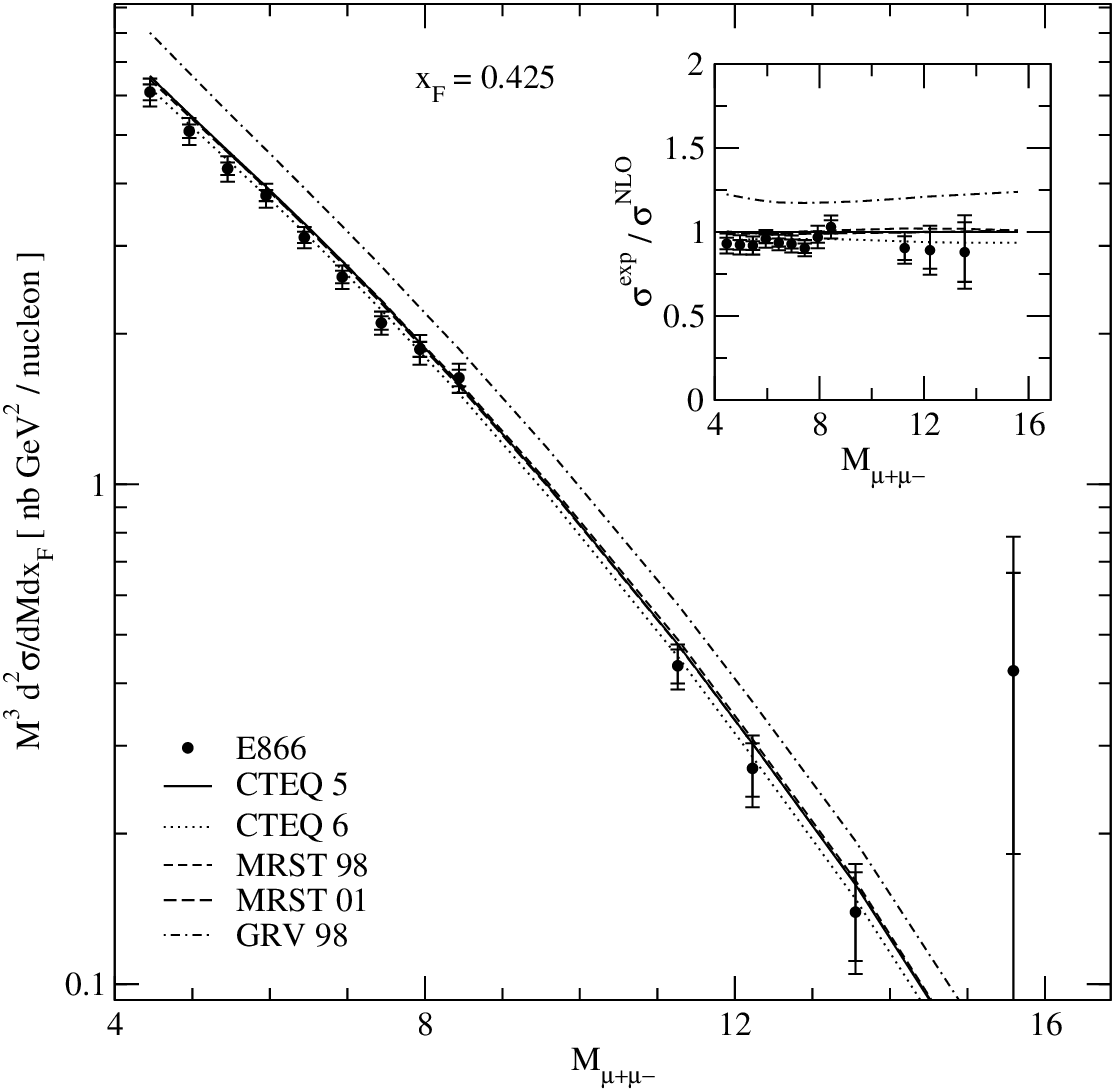}
\end{minipage}
\addtocontents{lof}{\  }
\addtocontents{lof}{\setlength{\baselineskip}{\singlespace}}

\begin{minipage}[tb]{\mycaplen}
\caption[Scaling form $M^3d^2\sigma/dMdx_F$ of the hydrogen and 
	deuterium cross sections for $0.4 \leq x_F < 0.45$.]{
	\setlength{\baselineskip}{\singlespace}
	Scaling form $M^3d^2\sigma/dMdx_F$ of the hydrogen (left) and 
	deuterium (right) cross sections for $0.4 \leq x_F < 0.45$.  The data are compared to 
	NLO calculations based on several different sets of parton distributions.
	The inset shows the ratio of the experimentally measured cross sections to
	NLO calculations based on the CTEQ 5 partons.
	}
\end{minipage}
\addtocontents{lof}{\setlength{\baselineskip}{\doublespace}}

\end{figure}

\begin{figure}\centering

\begin{minipage}[tb]{0.45\linewidth}
\includegraphics[width=\linewidth,clip]{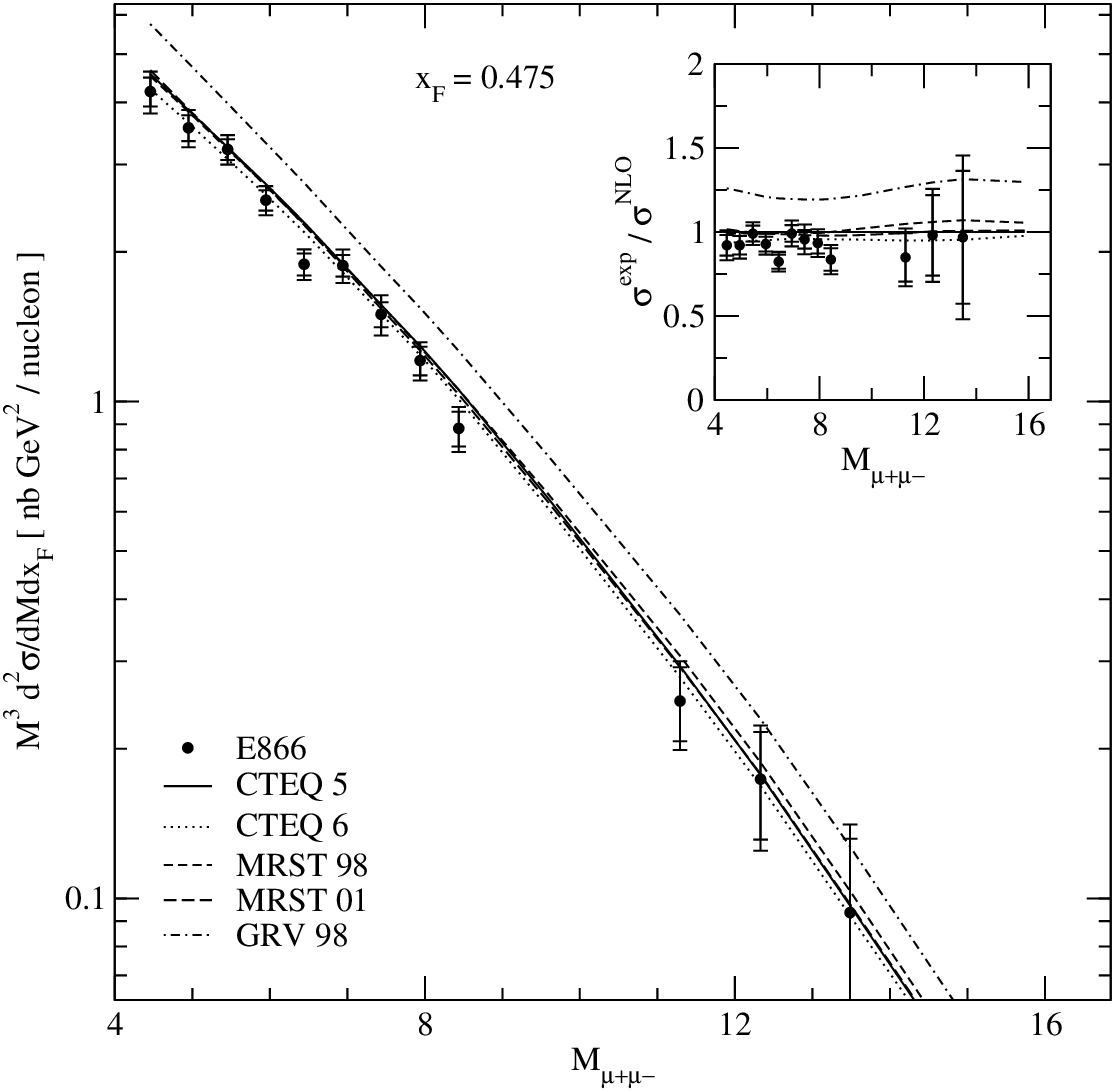}
\end{minipage}
\begin{minipage}[tb]{0.45\linewidth}
\includegraphics[width=\linewidth,clip]{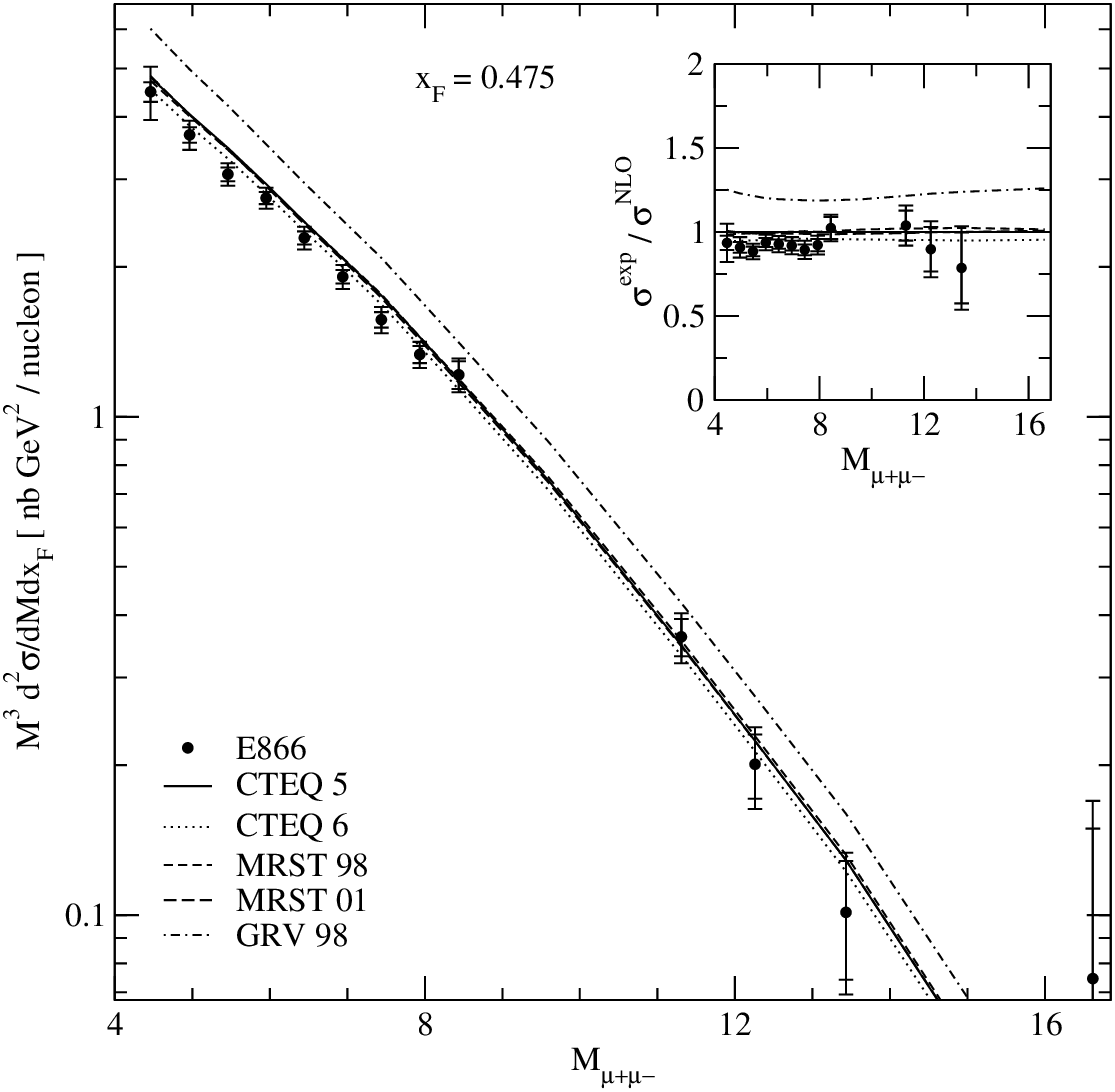}
\end{minipage}
\addtocontents{lof}{\  }
\addtocontents{lof}{\setlength{\baselineskip}{\singlespace}}
\begin{minipage}[tb]{\mycaplen}
\caption[Scaling form $M^3d^2\sigma/dMdx_F$ of the hydrogen and 
	deuterium cross sections for $0.45 \leq x_F < 0.5$.]{
	\setlength{\baselineskip}{\singlespace}
	Scaling form $M^3d^2\sigma/dMdx_F$ of the hydrogen (left) and 
	deuterium (right) cross sections for $0.45 \leq x_F < 0.5$.  The data are compared to 
	NLO calculations based on several different sets of parton distributions.
	The inset shows the ratio of the experimentally measured cross sections to
	NLO calculations based on the CTEQ 5 partons.
	}
\end{minipage}
\addtocontents{lof}{\setlength{\baselineskip}{\doublespace}}

\end{figure}

\begin{figure}\centering

\begin{minipage}[tb]{0.45\linewidth}
\includegraphics[width=\linewidth,clip]{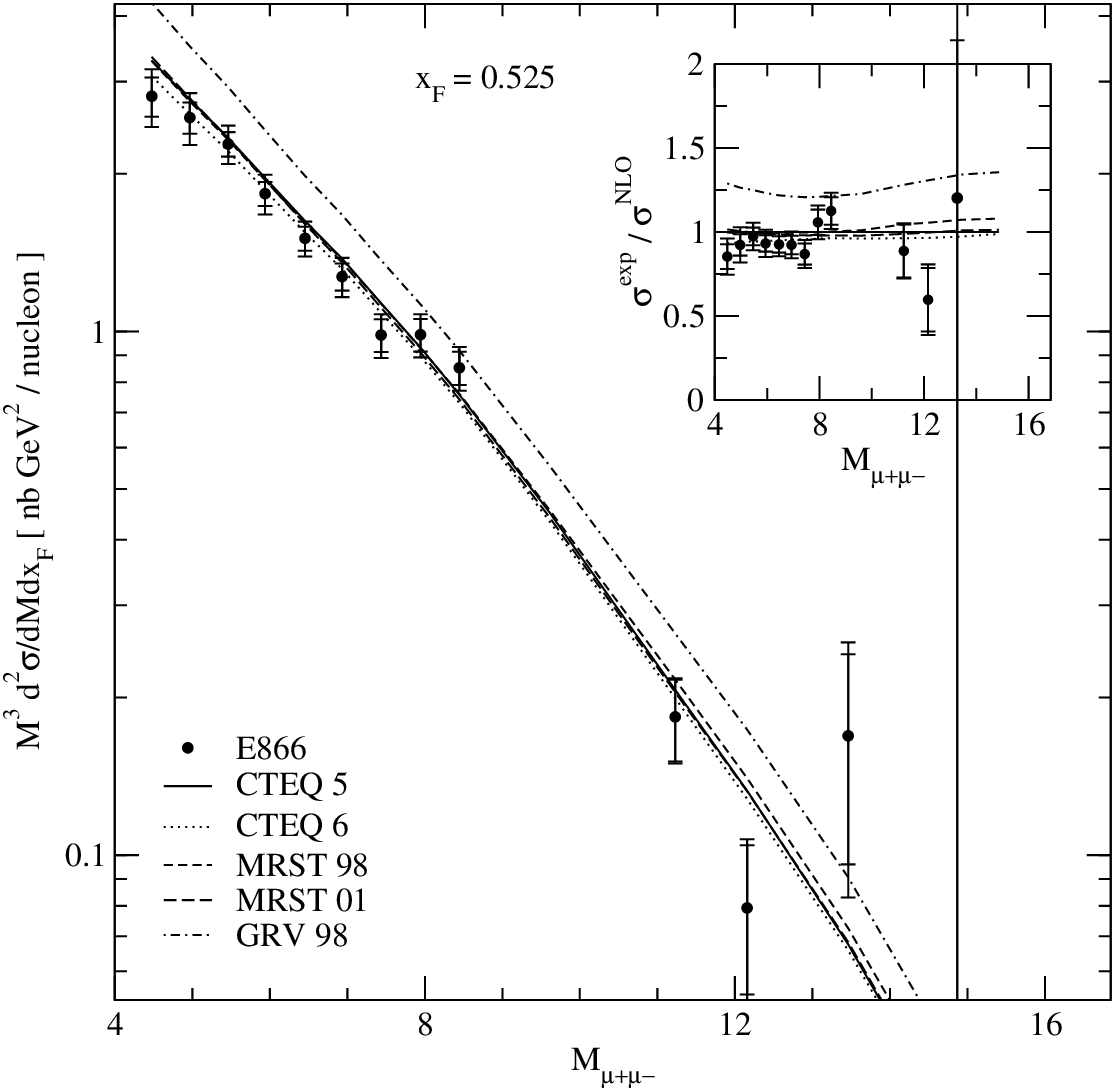}
\end{minipage}
\begin{minipage}[tb]{0.45\linewidth}
\includegraphics[width=\linewidth,clip]{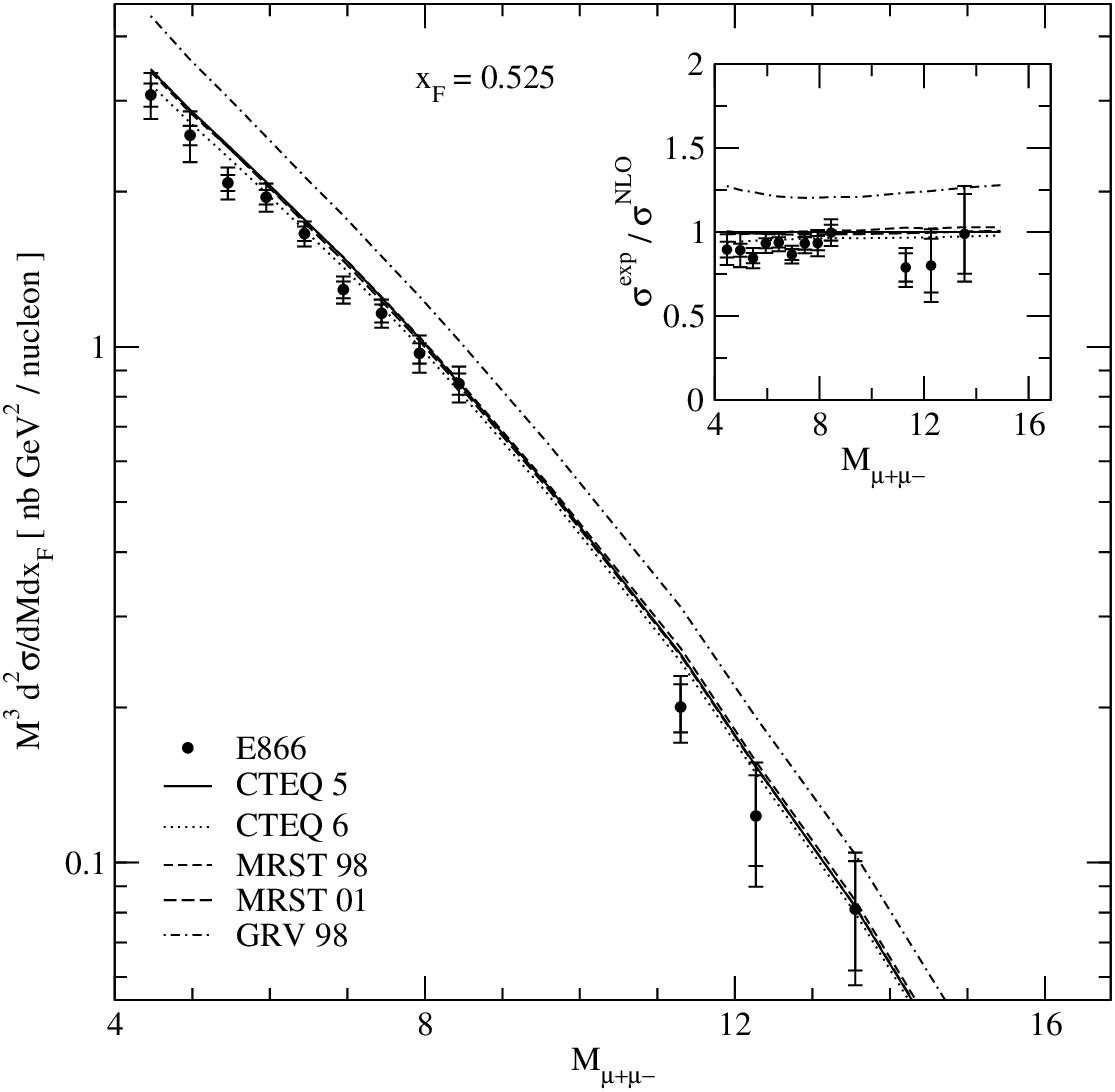}
\end{minipage}
\addtocontents{lof}{\  }
\addtocontents{lof}{\setlength{\baselineskip}{\singlespace}}
\begin{minipage}[tb]{\mycaplen}
\caption[Scaling form $M^3d^2\sigma/dMdx_F$ of the hydrogen (left) and 
	deuterium (right) cross sections for $0.5 \leq x_F < 0.55$.]{
	\setlength{\baselineskip}{\singlespace}
	Scaling form $M^3d^2\sigma/dMdx_F$ of the hydrogen (left) and 
	deuterium (right) cross sections for $0.5 \leq x_F < 0.55$.  The data are compared to 
	NLO calculations based on several different sets of parton distributions.
	The inset shows the ratio of the experimentally measured cross sections to
	NLO calculations based on the CTEQ 5 partons.
	}
\end{minipage}
\addtocontents{lof}{\setlength{\baselineskip}{\doublespace}}

\end{figure}

\begin{figure}\centering

\begin{minipage}[tb]{0.45\linewidth}
\includegraphics[width=\linewidth,clip]{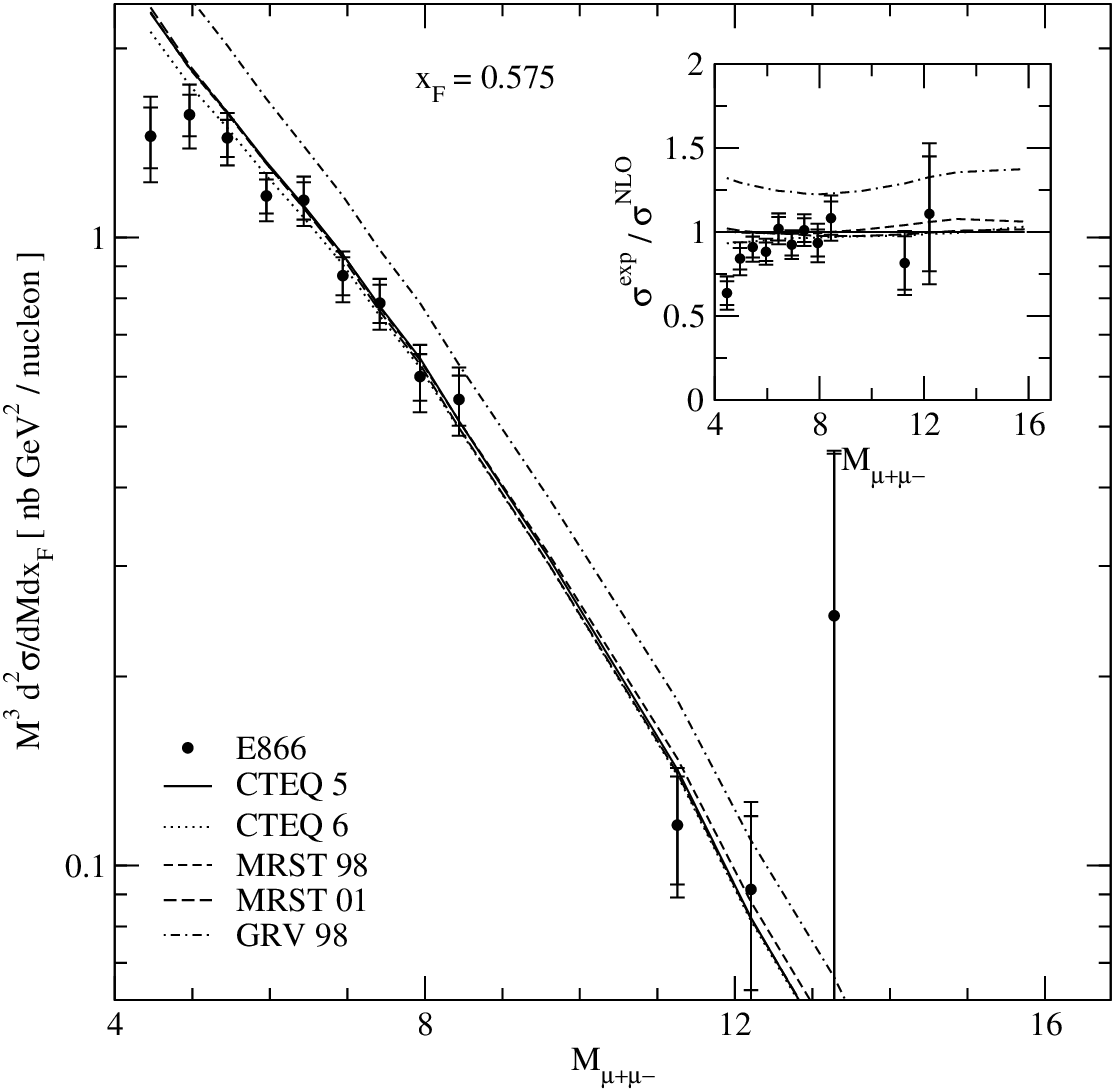}
\end{minipage}
\begin{minipage}[tb]{0.45\linewidth}
\includegraphics[width=\linewidth,clip]{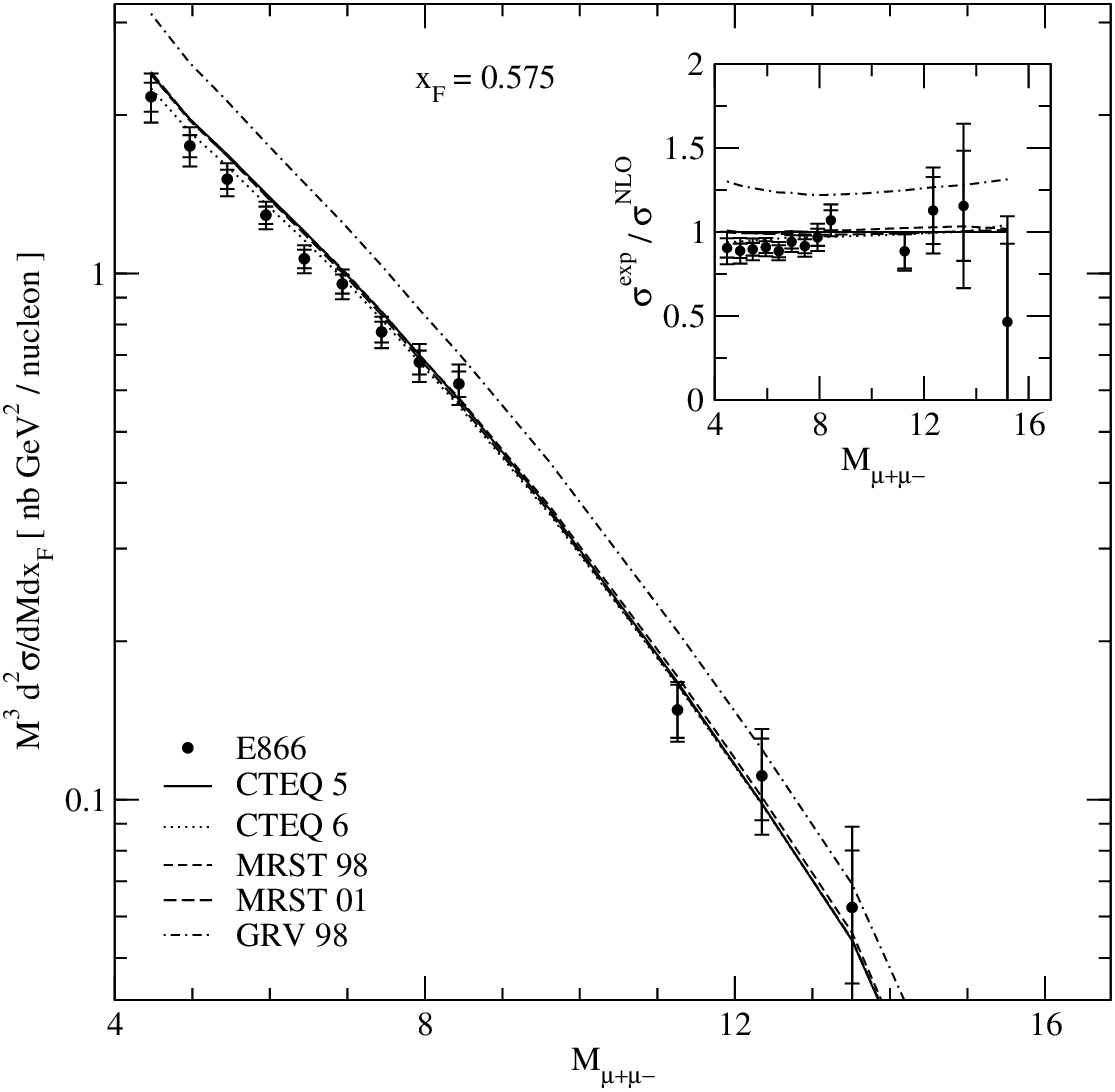}
\end{minipage}
\addtocontents{lof}{\  }
\addtocontents{lof}{\setlength{\baselineskip}{\singlespace}}

\begin{minipage}[tb]{\mycaplen}
\caption[Scaling form $M^3d^2\sigma/dMdx_F$ of the hydrogen (left) and 
	deuterium (right) cross sections for $0.55 \leq x_F < 0.6$.]{
	\setlength{\baselineskip}{\singlespace}
	Scaling form $M^3d^2\sigma/dMdx_F$ of the hydrogen (left) and 
	deuterium (right) cross sections for $0.55 \leq x_F < 0.6$.  The data are compared to 
	NLO calculations based on several different sets of parton distributions.
	The inset shows the ratio of the experimentally measured cross sections to
	NLO calculations based on the CTEQ 5 partons.
	}
\end{minipage}\addtocontents{lof}{\setlength{\baselineskip}{\doublespace}}

\end{figure}

\begin{figure}\centering

\begin{minipage}[tb]{0.45\linewidth}
\includegraphics[width=\linewidth,clip]{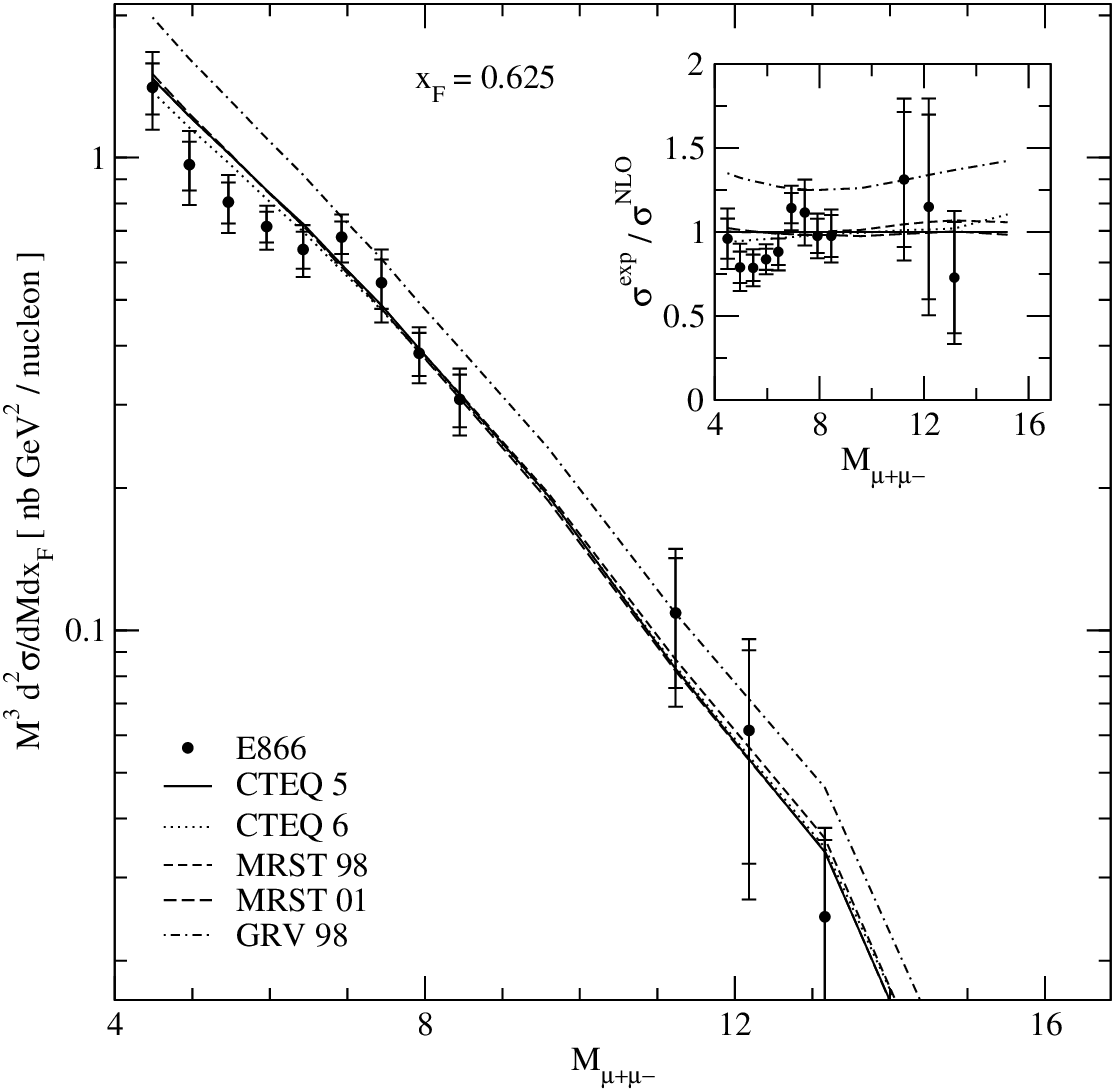}
\end{minipage}
\begin{minipage}[tb]{0.45\linewidth}
\includegraphics[width=\linewidth,clip]{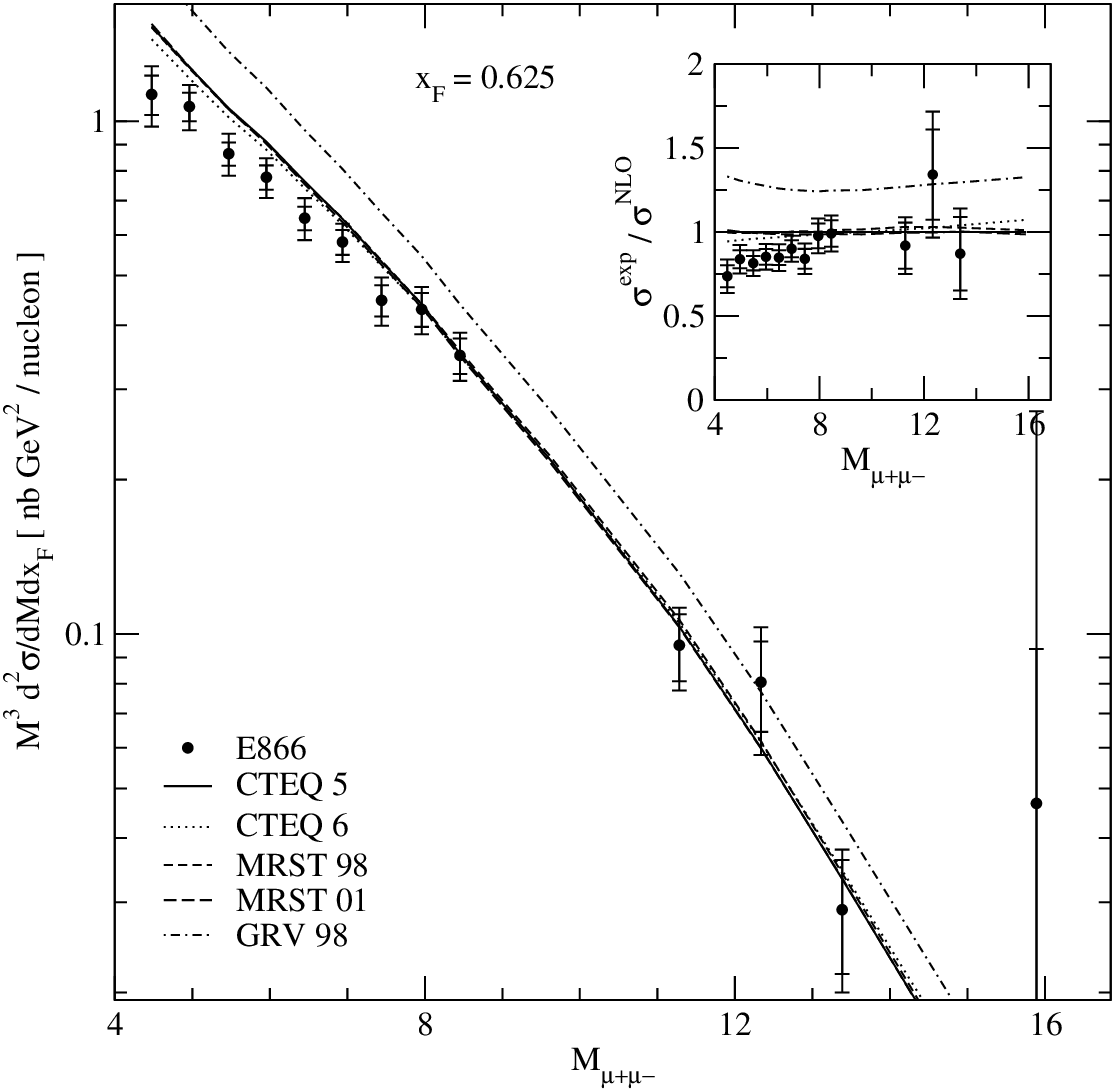}
\end{minipage}
\addtocontents{lof}{\  }
\addtocontents{lof}{\setlength{\baselineskip}{\singlespace}}

\begin{minipage}[tb]{\mycaplen}
\caption[Scaling form $M^3d^2\sigma/dMdx_F$ of the hydrogen and 
	deuterium cross sections for $0.6 \leq x_F < 0.65$.]{
	\setlength{\baselineskip}{\singlespace}
	Scaling form $M^3d^2\sigma/dMdx_F$ of the hydrogen (left) and 
	deuterium (right) cross sections for $0.6 \leq x_F < 0.65$.  The data are compared to 
	NLO calculations based on several different sets of parton distributions.
	The inset shows the ratio of the experimentally measured cross sections to
	NLO calculations based on the CTEQ 5 partons.
	}
\end{minipage}\addtocontents{lof}{\setlength{\baselineskip}{\doublespace}}

\end{figure}

\begin{figure}\centering

\begin{minipage}[tb]{0.45\linewidth}
\includegraphics[width=\linewidth,clip]{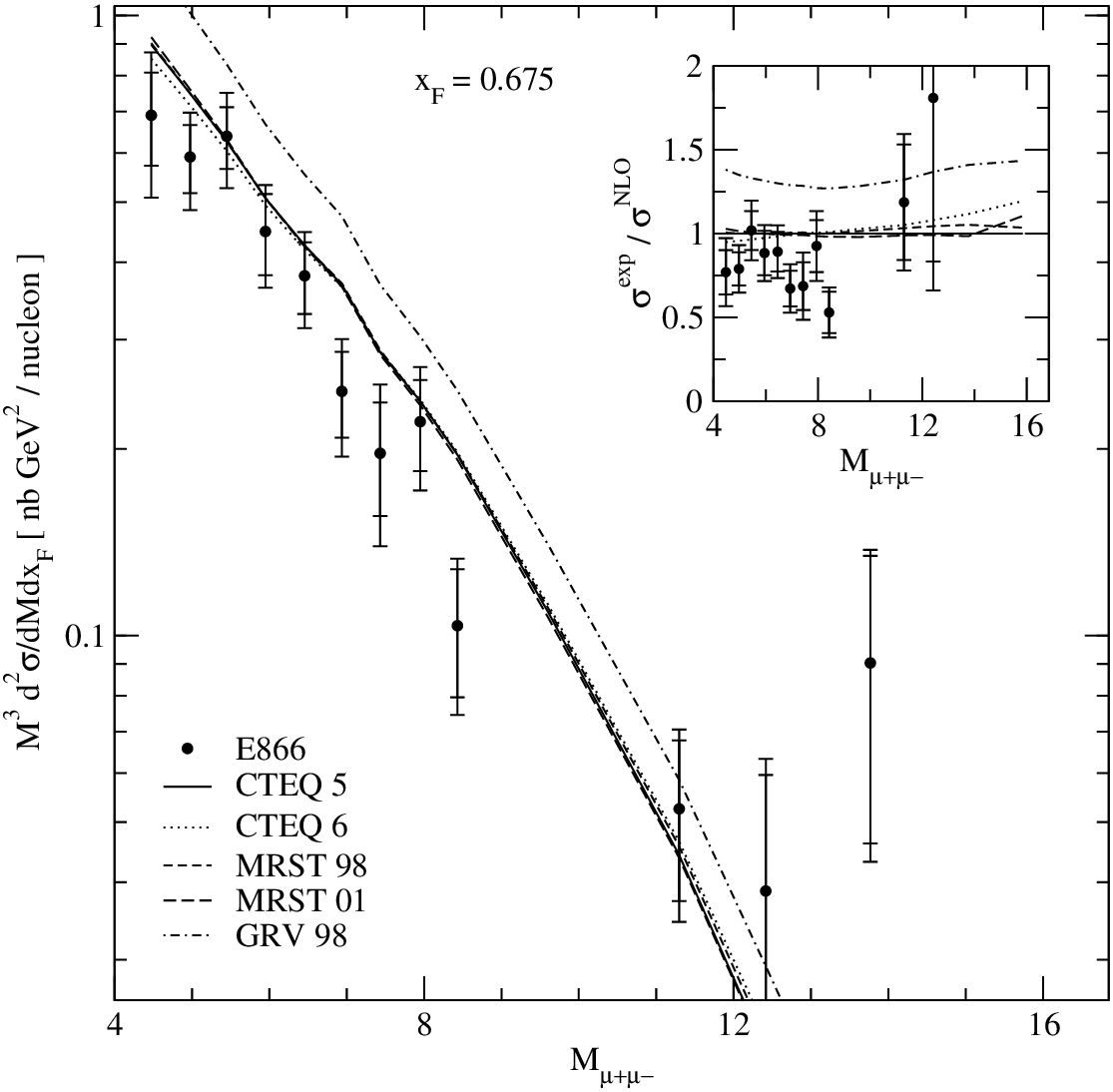}
\end{minipage}
\begin{minipage}[tb]{0.45\linewidth}
\includegraphics[width=\linewidth,clip]{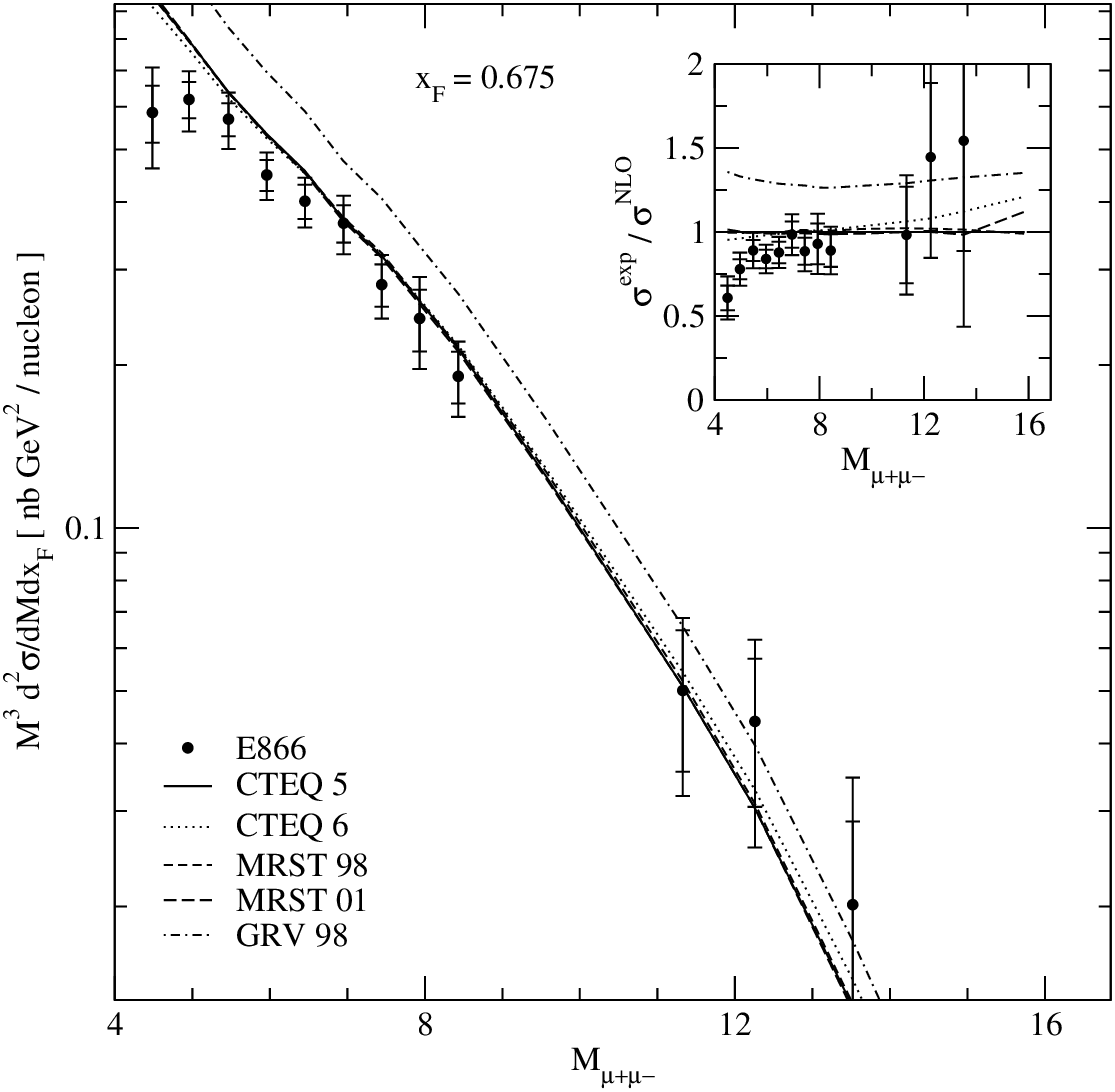}
\end{minipage}
\addtocontents{lof}{\  }
\addtocontents{lof}{\setlength{\baselineskip}{\singlespace}}

\begin{minipage}[tb]{\mycaplen}
\caption[Scaling form $M^3d^2\sigma/dMdx_F$ of the hydrogen and 
	deuterium cross sections for $0.65 \leq x_F < 0.7$.]{
	\setlength{\baselineskip}{\singlespace}
	Scaling form $M^3d^2\sigma/dMdx_F$ of the hydrogen (left) and 
	deuterium (right) cross sections for $0.65 \leq x_F < 0.7$.  The data are compared to 
	NLO calculations based on several different sets of parton distributions.
	The inset shows the ratio of the experimentally measured cross sections to
	NLO calculations based on the CTEQ 5 partons.}
\end{minipage}\addtocontents{lof}{\setlength{\baselineskip}{\doublespace}}

\end{figure}

\begin{figure}\centering

\begin{minipage}[tb]{0.45\linewidth}
\includegraphics[width=\linewidth,clip]{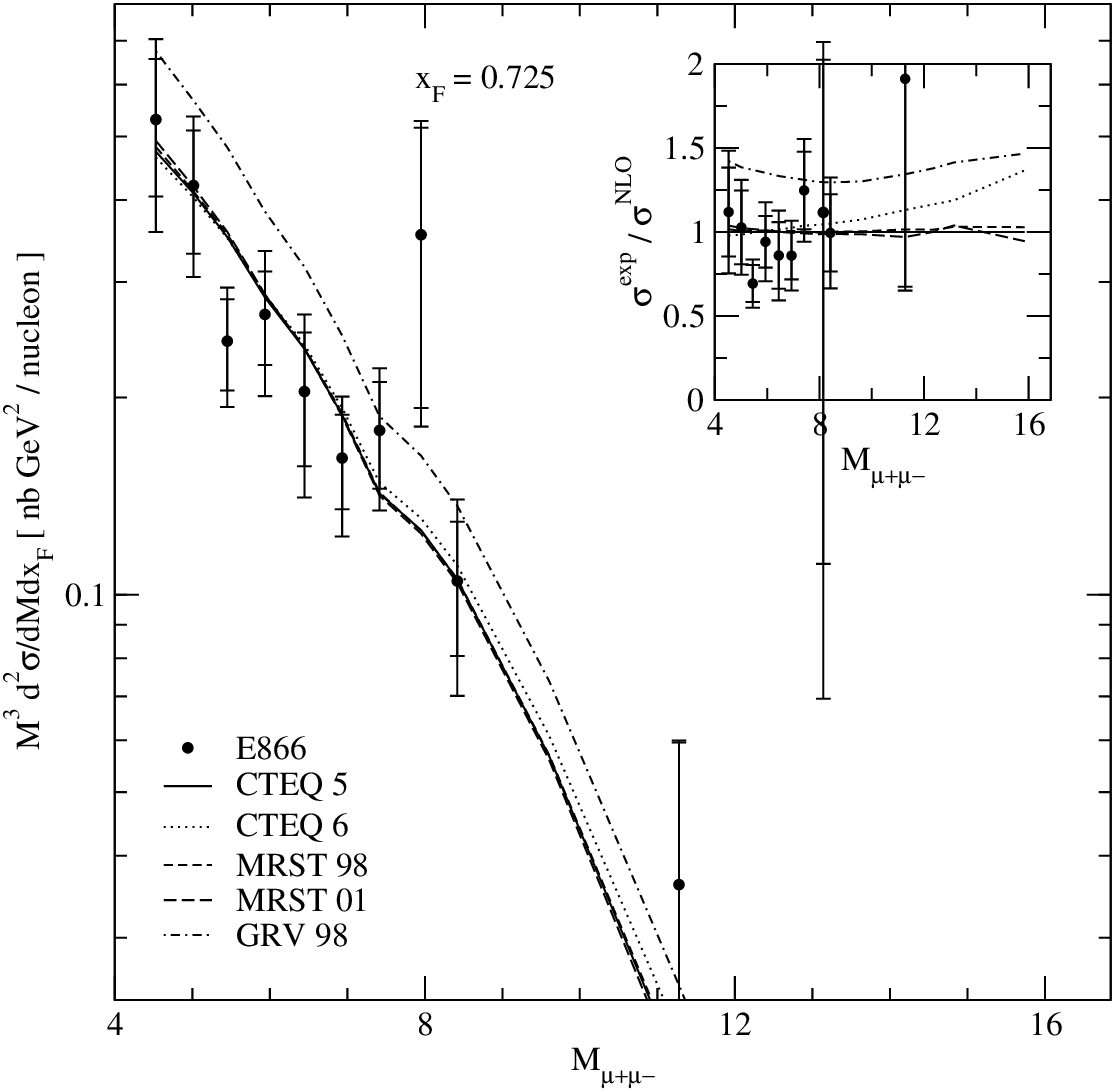}
\end{minipage}
\begin{minipage}[tb]{0.45\linewidth}
\includegraphics[width=\linewidth,clip]{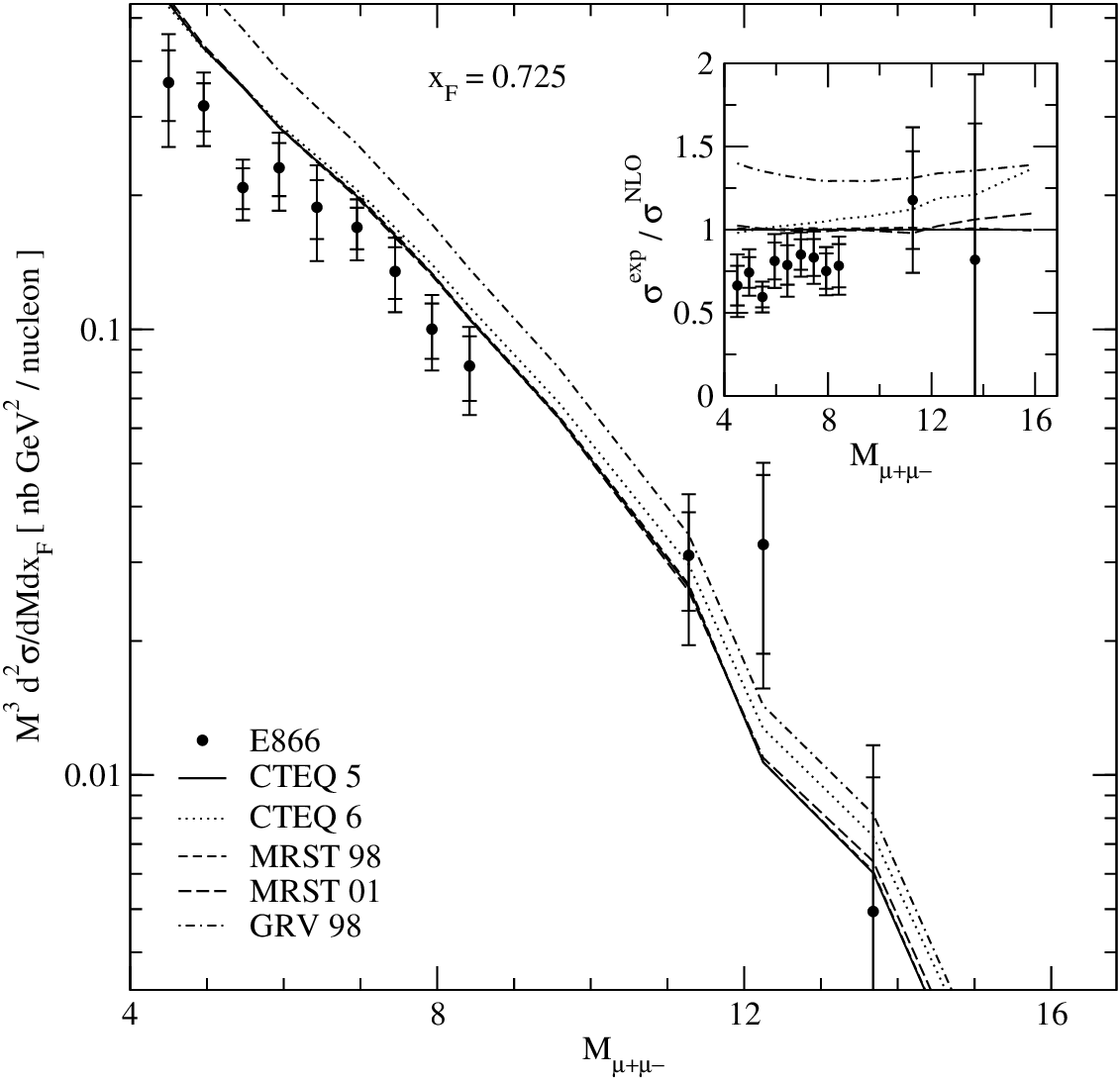}
\end{minipage}
\addtocontents{lof}{\  }
\addtocontents{lof}{\setlength{\baselineskip}{\singlespace}}

\begin{minipage}[tb]{\mycaplen}
\caption[Scaling form $M^3d^2\sigma/dMdx_F$ of the hydrogen  and 
	deuterium cross sections for $0.7 \leq x_F < 0.75$.]{
	\setlength{\baselineskip}{\singlespace}
	Scaling form $M^3d^2\sigma/dMdx_F$ of the hydrogen (left) and 
	deuterium (right) cross sections for $0.7 \leq x_F < 0.75$.  The data are compared to 
	NLO calculations based on several different sets of parton distributions.
	The inset shows the ratio of the experimentally measured cross sections to
	NLO calculations based on the CTEQ 5 partons.
	}
\end{minipage}\addtocontents{lof}{\setlength{\baselineskip}{\doublespace}}

\end{figure}

\begin{figure}\centering

\begin{minipage}[tb]{0.45\linewidth}
\includegraphics[width=\linewidth,clip]{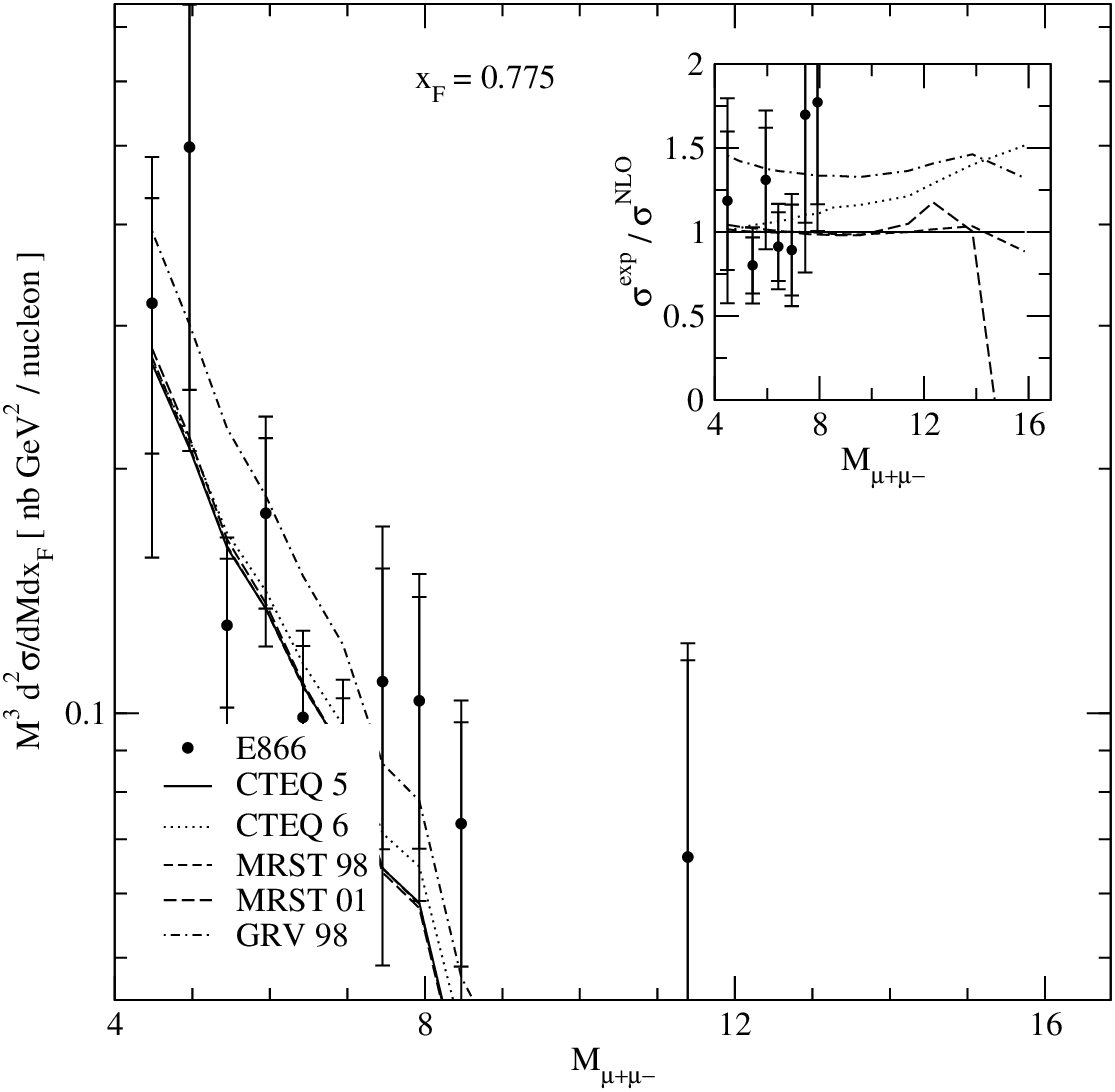}
\end{minipage}
\begin{minipage}[tb]{0.45\linewidth}
\includegraphics[width=\linewidth,clip]{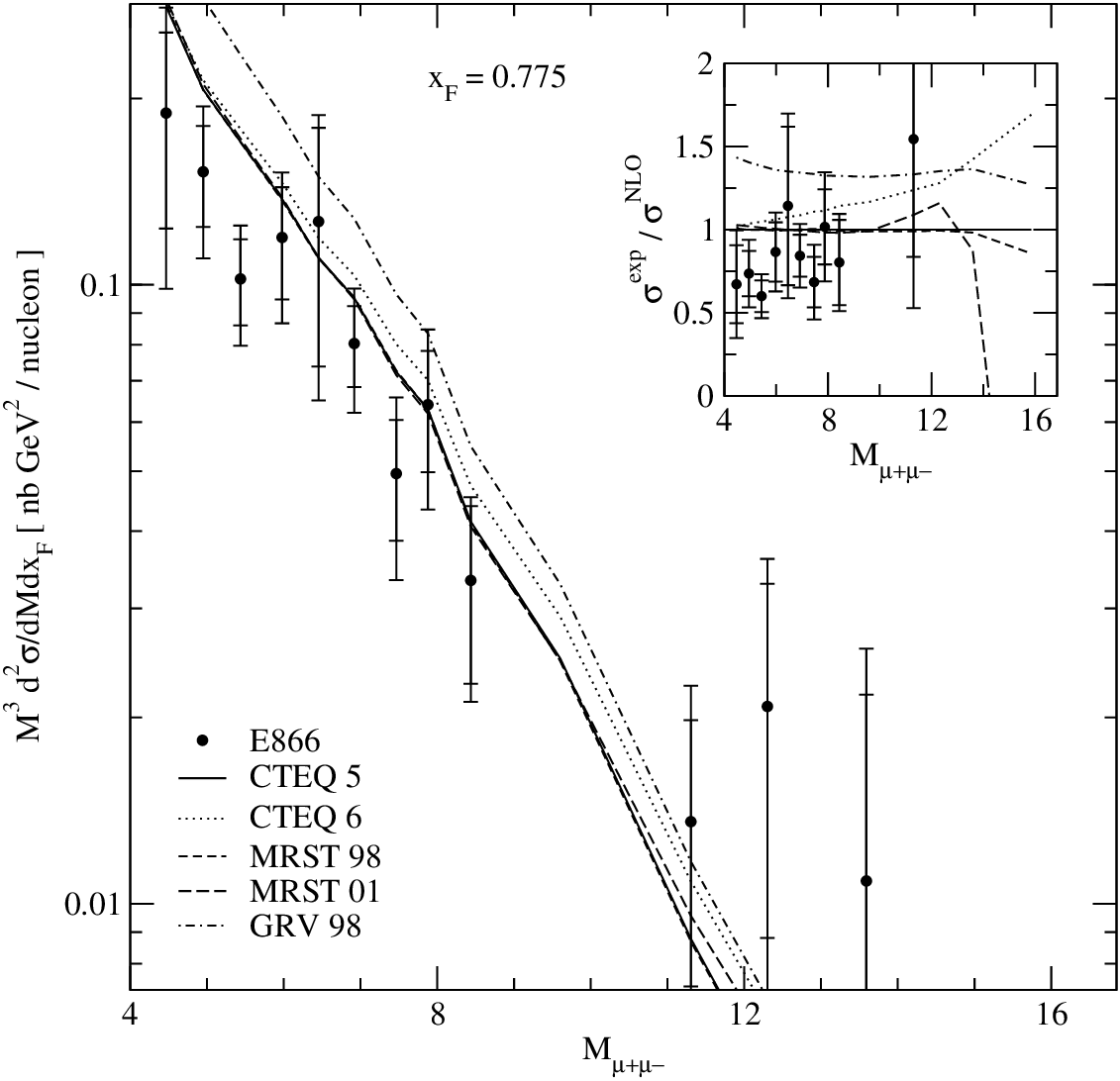}
\end{minipage}
\addtocontents{lof}{\  }
\addtocontents{lof}{\setlength{\baselineskip}{\singlespace}}

\begin{minipage}[tb]{\mycaplen}
\caption[Scaling form $M^3d^2\sigma/dMdx_F$ of the hydrogen and 
	deuterium cross sections for $0.75 \leq x_F < 0.8$.]{
	\setlength{\baselineskip}{\singlespace}
	\label{fig:d2sigma_last}Scaling form $M^3d^2\sigma/dMdx_F$ of the hydrogen (left) and 
	deuterium (right) cross sections for $0.75 \leq x_F < 0.8$.  The data are compared to 
	NLO calculations based on several different sets of parton distributions.
	The inset shows the ratio of the experimentally measured cross sections to
	NLO calculations based on the CTEQ 5 partons.
	}
\end{minipage}\addtocontents{lof}{\setlength{\baselineskip}{\doublespace}}

\end{figure}

\end{landscape}

\subsection{FNAL E866/NuSea Results -- $Ed^3\sigma/dp^3$}

The $p_T$-distributions are presented in terms of the invariant cross
section 
\begin{equation}
E\frac{d^3\sigma}{dp^3} = \frac{2E}{\pi\sqrt{s}} \frac{d^2\sigma}{dx_Fdp_T^2}
\end{equation}
\noindent where the $\phi$ dependence has been averaged over.
The cross sections are tabulated in tables \ref{table:d3sigma_first} through
\ref{table:d3sigma_last}, in $x_F$ bins, integrated over the specified
range in invariant mass.  The mean mass, $x_F$, $p_T$ and $E$ in each bin
are also tabulated.


\renewcommand{\spanner}[1]{
  \multicolumn{8}{|c|}{} \\ \nopagebreak
  \multicolumn{8}{|c|}{#1} \\ \nopagebreak
  \multicolumn{8}{|c|}{} \\ \nopagebreak
}

\renewcommand{\myheading}[0]{
  \hline 
  \multicolumn{8}{|c|}{} \\
   $p_T$ [GeV] & $<p_T>$ & $<M>$ & $<x_F>$ & $<E>$ &  $Ed^3\sigma/dp^3$  & Stat. Error. & Syst. Error \\
  \multicolumn{8}{|c|}{} \\
  \hline
  \multicolumn{8}{c}{} \\
}

\newcommand{\massbin}[3]{
    \hline \nopagebreak 
    \spanner{#1 $\leq M_{\mu^+\mu^-} <$ #2} \nopagebreak
    #3 \nopagebreak
    \multicolumn{8}{|c|}{}\\ \nopagebreak
    \hline \nopagebreak
  \multicolumn{8}{c}{}\\ \nopagebreak
}

\begin{landscape}

\addtocontents{lot}{\  }
\addtocontents{lot}{\setlength{\baselineskip}{\singlespace}}


\end{landscape}

\newpage
\def\nevents{175,000}
\def\ndevents{120,000}
\def\nhevents{55,000}

\def\Towell{Towell {\it et al.} \ }
\def\refTowell{\cite{bib:E866-Towell}}
\def\refHawker{\cite{bib:E866-Hawker}}

\section{DISCUSSION}

In the thirty years since the seminal experiment by Christenson {\it et al.}
\cite{bib:BNL-DY}, 
the Drell-Yan process for continuum dilepton production 
in hadronic interactions
remains an active area of experimental and theoretical investigation.
Several experiments have studied both pion- and proton-induced dileptons,
providing a wealth of data utilizing various beams, targets and center-of-mass
energies.  When these data are analyzed within the framework of perturbative
QCD along with data from other types of
hadronic interactions, they provide a powerful tool 
for extracting information about the parton distribution functions.

In the preceding chapters we have described the procedures used in and 
reported the results of a measurement of continuum diumon cross sections in 
800-GeV $pp$ and $pd$ interactions.  In this chapter, we compare our results
to existing dimuon data from $pA$ and $pd$ measurements performed by the CERN 
NA3 \cite{bib:CERN-NA3}, FNAL E605 \cite{bib:FNAL-E605} and FNAL E772 
\cite{bib:E772-Absolute} experiments.  Fermilab E866 has already published
\cite{bib:E866-Hawker,bib:E866-Towell} results on the deuterium-to-hydrogen
cross section ratio, and we will also demonstrate that our absolute 
measurements are consistent with those previous results.  We will also discuss
the level of agreement between our results and next-to-leading order 
calculations of the cross sections based on various sets of parton 
distribution fits.  We will highlight some important differences between data 
and theory, and discuss the likely impact our results will have on future fits 
to the parton distributions.

\subsection{Comparison with Other Experiments}

Following the work of Christenson {\it et al.}, several experiments 
measured Drell-Yan cross sections using various beams and targets.
The first high-statistics measurements to study the Drell-Yan process over
an extended range of kinematics examined dimuon production in $pA$
interactions.  CERN NA3 \cite{bib:CERN-NA3} studied $p$Pt at $\sqrt{s} = 27.4$ 
GeV, while Fermilab E605 \cite{bib:FNAL-E605} studied $p$Cu at 
$\sqrt{s} = 38.8$ GeV.  Less extensive measurements of the $pp$ and $pd$ cross 
sections in a narrow $x_F$ range centered around $x_F = 0$ were also carried 
out at CERN.  CERN R-209 \cite{bib:CERN-R-209} studied the $pp$ cross sections 
at $\sqrt{s} = 44$ and $62$ GeV, while CERN NA51 \cite{bib:NA51} studied 
the $pp$ and $pd$ cross sections\footnote{
\setlength{\baselineskip}{\singlespace}
CERN NA51 did not publish absolute measurements, only the asymmetry between 
$pp$ and $pd$ cross sections.} 
at $\sqrt{s} = 29$ GeV.  In the early 1990s, the $pd$ cross sections were
measured over a wide range of mass and $x_F$ by the Fermilab E772 collaboration
\cite{bib:E772-Absolute}.  They published results spanning the range 
$0.05 \leq x_F \leq 0.7$ based on an $83,000$ event dimuon sample.  
It is important to note that the E605 and E772 measurements were conducted
using essentially the same apparatus as E866.  Thus, these two experiments
provide important cross checks on the consistency of our results.

Figures \ref{fig:scaling_low} and \ref{fig:scaling_high} compare the E866 
measurements of the scaling form of the cross section $M^3d^2\sigma/dMdx_F$ 
in $pd$ interactions to the NA3, E605 and E772 measurements.  The data are 
plotted versus the scaling variable $\sqrt{\tau}$ to enable a direct 
comparison between the results measured at different $\sqrt{s}$.  The E866 
data, which are subject to an additional $\pm 6.5\%$ global normalization 
uncertainty, are in good agreement with these previous dimuon experiments 
over a broad range of mass and $x_F$.  The E605, E772 and E866 data typically 
agree to within less than the quoted normalization uncertainties\footnote{
\setlength{\baselineskip}{\singlespace}We
expect the systematic uncertainties in the normalization of these three 
experiments to be correlated at some level, since they used the same beam line 
monitors to calibrate the data, and some of the same calibration measurements
are common.} 
of the three experiments.
The only major area of disagreement is at larger values of $x_F$ and smaller 
masses, where the E866 measurement differs significantly from the E772 results.
In this kinematic region, the E772 measurement is subject to larger systematic
uncertainties ($\sim 20\%$) \cite{bib:E772-Absolute}.  The major differences
between E772 and E866 in this region are thought to be primarily 
due to uncertainties in the E772 magnetic field maps \cite{bib:PLM-priv-com}.

\begin{figure}
\centering
\includegraphics[width=0.9\linewidth,clip]{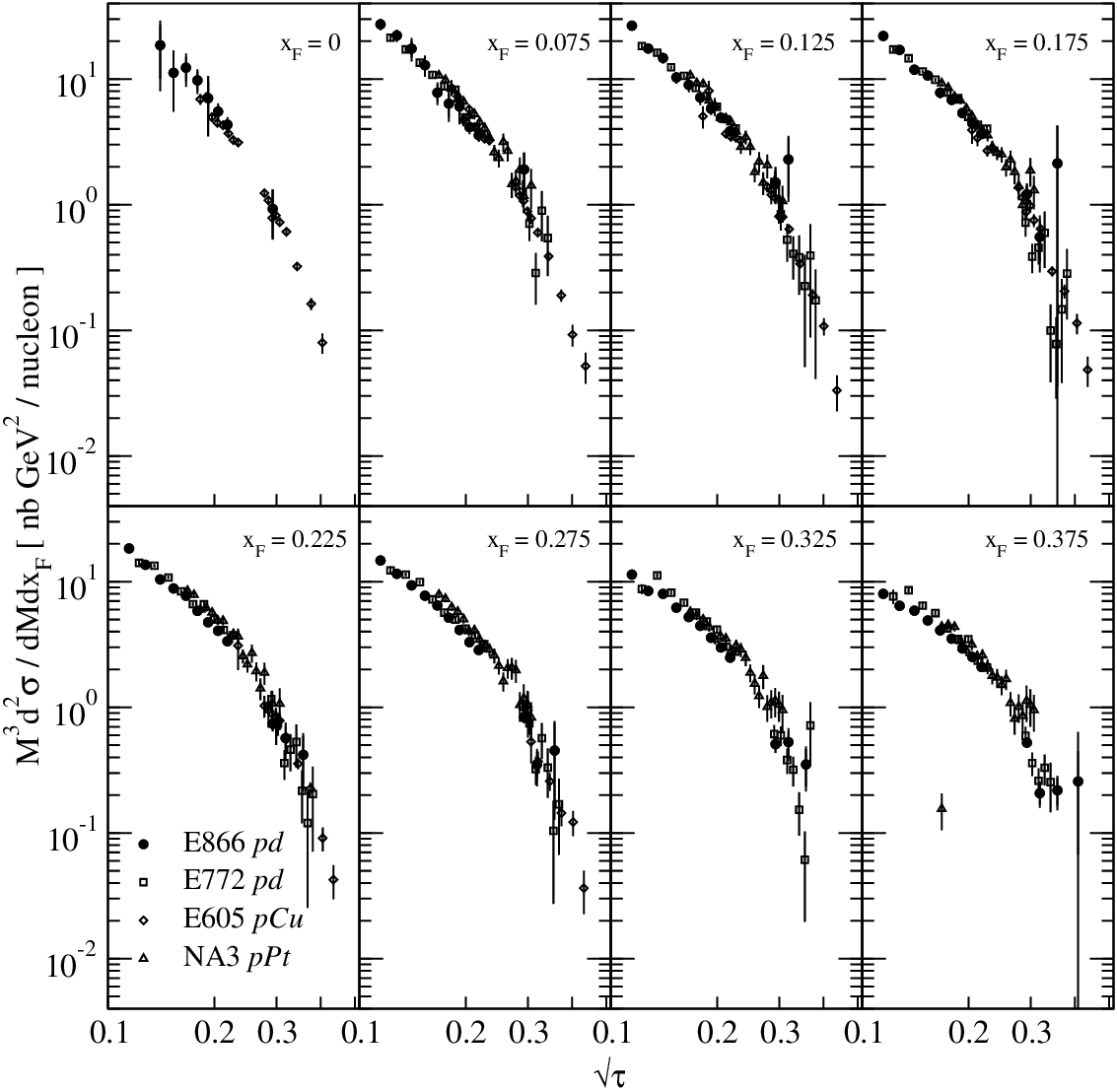}
\caption[Comparison between E866 and previous dimuon experiments.]{
	\setlength{\baselineskip}{\singlespace}
	\label{fig:scaling_low}Comparison between E866 and previous dimuon experiments.
	The scaling form $M^3d^2\sigma/dMdx_F$ is plotted versus $\sqrt{\tau}$ for 
	the E866 $pd$, E772 $pd$ \protect\cite{bib:E772-Absolute}, E605 $p$Cu
	\protect\cite{bib:FNAL-E605} and NA3 $p$Pt \protect\cite{bib:CERN-NA3} measurements.  
	Errors are statistical only, except for the E866 uncertainties which 
	are the linear sum of statistical and systematic uncertainties.  The 
	E866 data contain an additional $\pm 6.5\%$ systematic uncertainty in 
	the normalization.
	}
\end{figure}

\begin{figure}
\centering
\includegraphics[width=0.9\linewidth,clip]{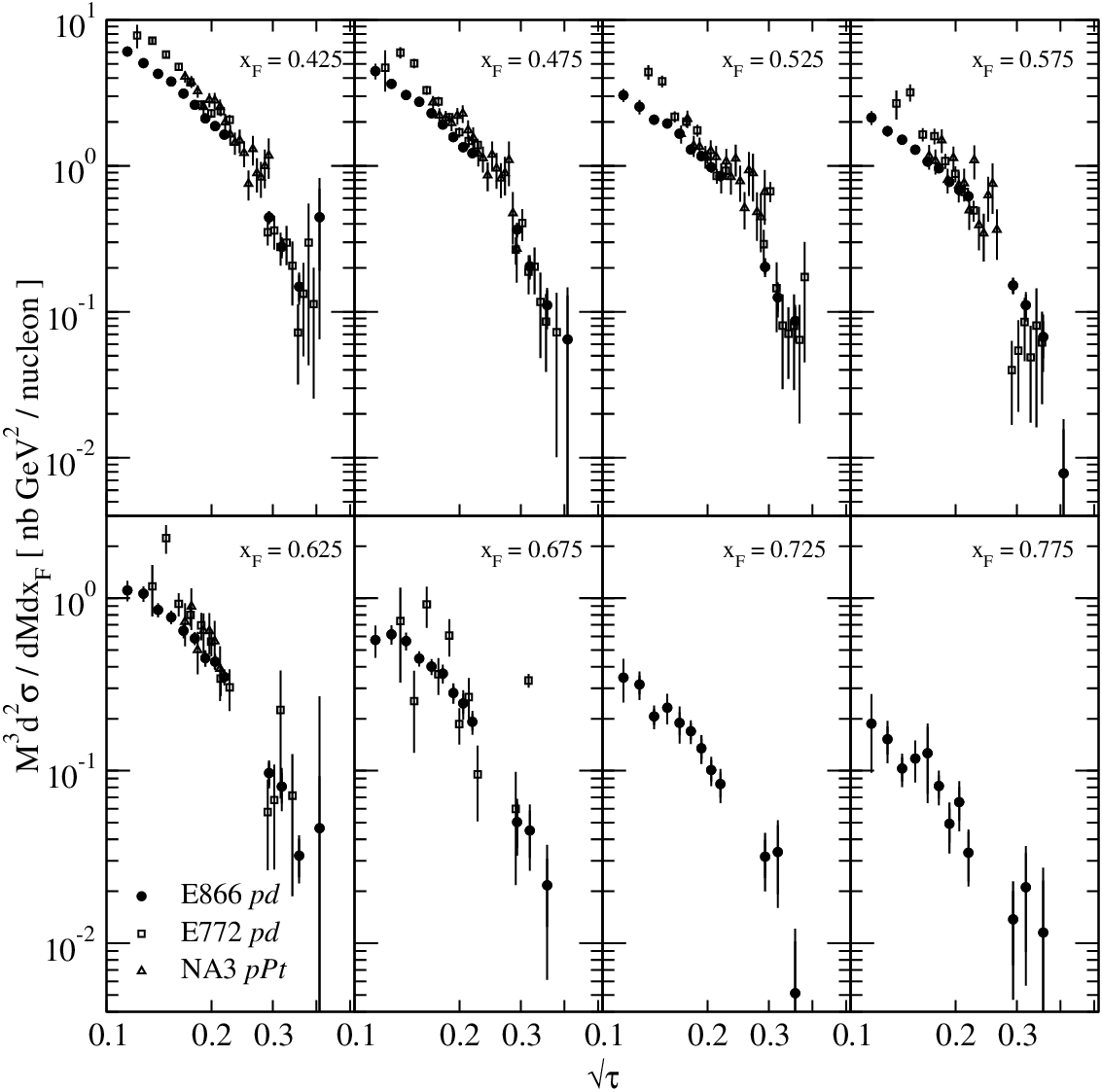}
\caption[Comparison between E866 and previous dimuon experiments.]{
	\setlength{\baselineskip}{\singlespace}
	\label{fig:scaling_high}Comparison between E866 and previous dimuon experiments.
	The scaling form $M^3d^2\sigma/dMdx_F$ is plotted versus $\sqrt{\tau}$ for 
	the E866 $pd$, E772 $pd$ \protect\cite{bib:E772-Absolute}, and NA3 $p$Pt
	\protect\cite{bib:CERN-NA3} measurements.  Errors are statistical only, 
	except for the E866 uncertainties which are the linear sum of the 
	statistical and systematic uncertainties.  The E866 data contain an 
	additional $\pm 6.5\%$ systematic uncertainty in the normalization.
	}
\end{figure}

\begin{figure}
  \includegraphics[clip,width=\linewidth]{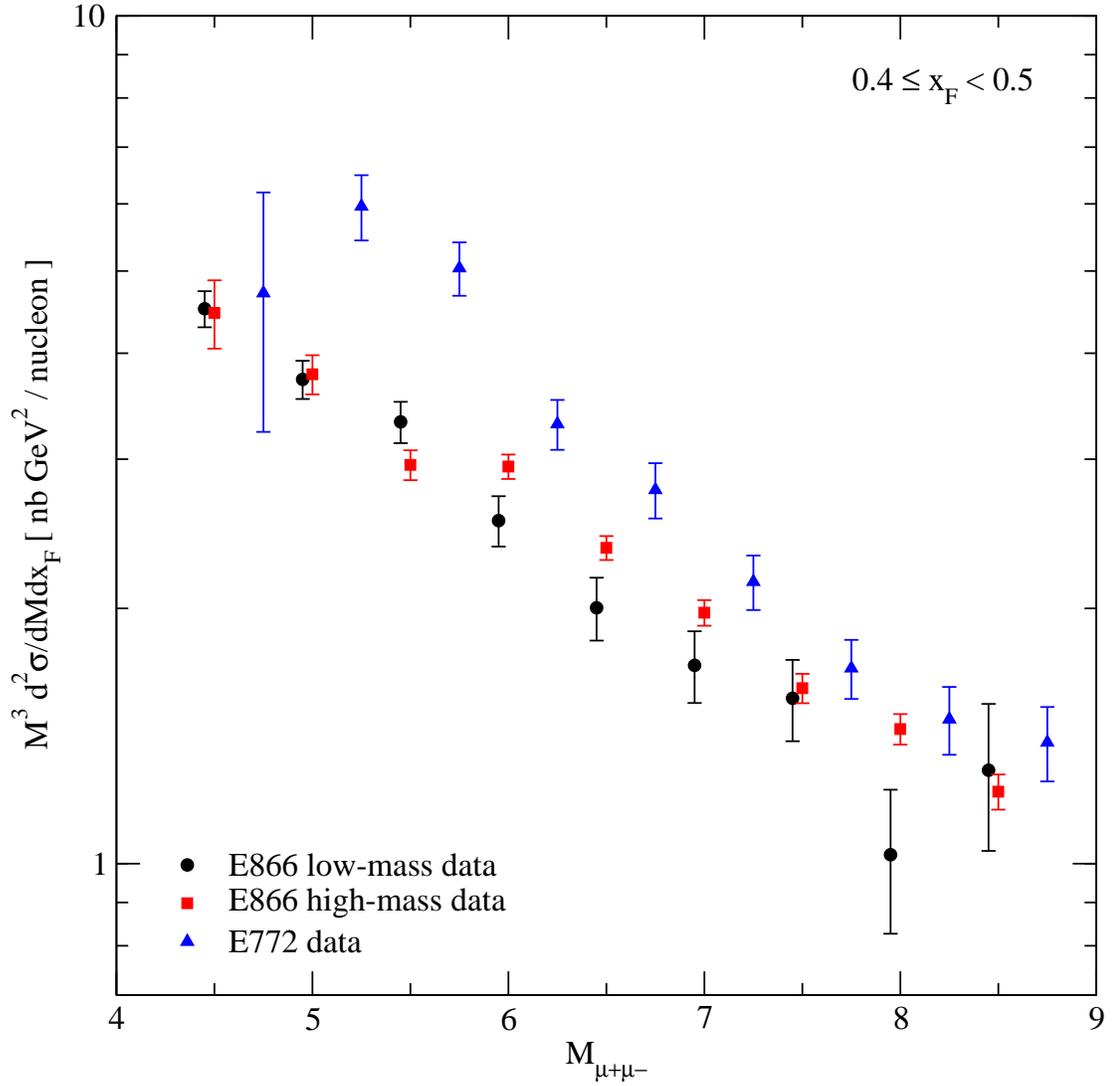}
  \caption[Detailed comparison between E866 and E772 at large $x_F$.]{
	\label{fig:e866ve772}\setlength{\baselineskip}{\singlespace}
  Detailed comparison between E866 and E772 in the range $0.4 \leq x_F < 0.5$.
  Deuterium results from the low- and high-mass E866 data sets are shown along with
  the E772 measurement.  The error bars represent statistical uncertainties only.
  The high-mass E866 data has been offset by 0.05 GeV.
  }
\end{figure}

Because the E772 and E866 measurements used essentially the same apparatus 
it is worth exploring this difference in a little more detail.  Figure 
\ref{fig:e866ve772} shows the results from the low-
and high-mass data settings used in E866, and the published E772 results.
They are shown  over the range $0.4\leq x_F< 0.5$ where the discrepancy is 
most pronounced.   The two E866 mass settings have very different
acceptances.  The acceptance in the high-mass data is plunging rapidly
with decreasing mass, while the low-mass acceptance is fairly large
in the same region.  The agreement between these data sets is quite
good, suggesting that the E866 data is generally free of the systematic
uncertainties present in the E772 data in this region.

\begin{figure}
\centering
\includegraphics[width=0.9\linewidth,clip]{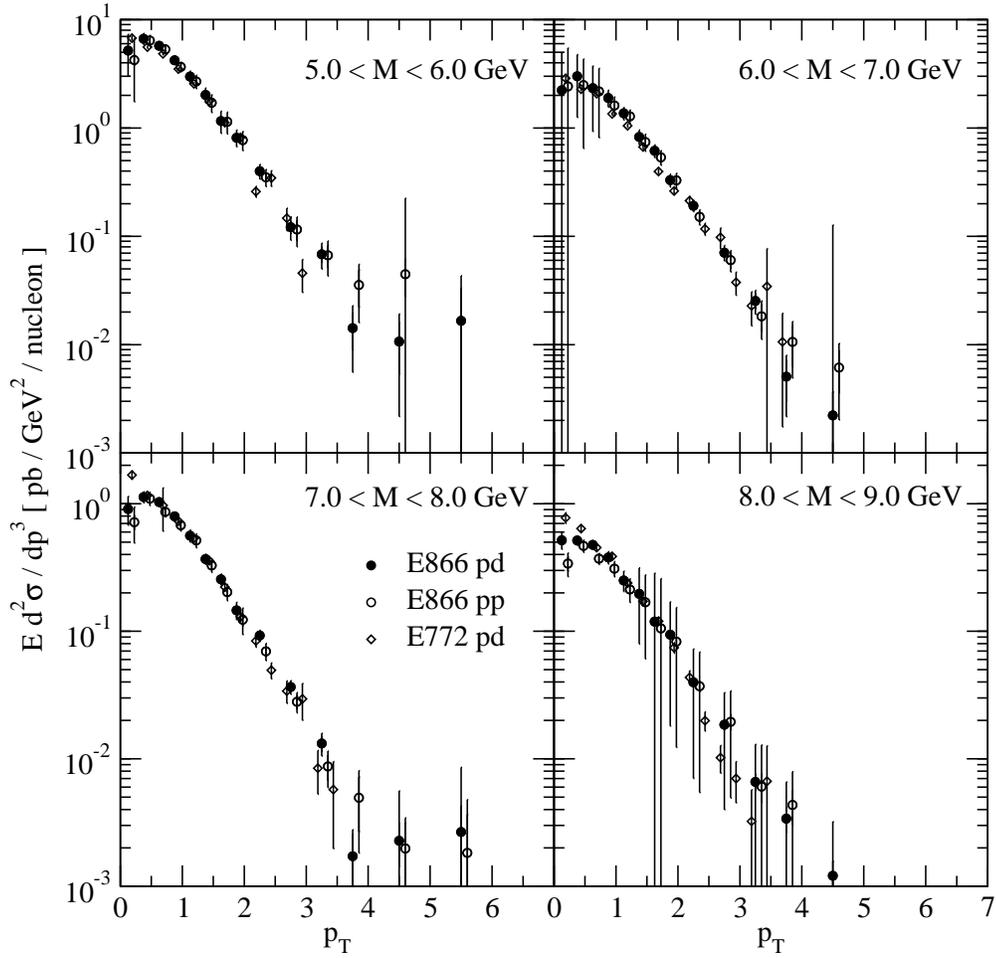}
\caption[Comparison of the E866 and E772 results for $E d^3\sigma/dp^3$]{
	\label{fig:d3sigma}
	\setlength{\baselineskip}{\singlespace}
        Comparison of the E866 and E772 results for $E d^3\sigma/dp^3$.
	Data are in the range $0 \leq x_F < 0.3$.
	Open (closed) circles are the E866 results for $pp$ ($pd$).
	Open diamonds represent the E772 results.  The E866 error bars are the
	linear sum of the statistical and systematic uncertainties.
	}
\end{figure}

Fermilab E772 has also published measurements of the triple-differential cross
section $E d^3\sigma/dp^3$, over a more restricted kinematic range 
($0.1 \leq x_F \leq 0.3$) than their double-differential measurements.
This range is below the $x_F$ range where the E772 and E866 measurements
disagree, and was imposed on E772 by correlations between their $x_F$ and $p_T$
acceptance \cite{bib:PLM-priv-com}, which was much less of a concern in the 
E866 data.  In the previous chapter we tabulated $E d^3\sigma/dp^3$ in various
$x_F$ bins covering our entire $x_F$ range.  To make a direct comparison 
between our results and those of E772, we recalculated our cross sections 
using the E772 binning.  These results are shown in figure \ref{fig:d3sigma} 
along with the E772 measurement.  In general, there is excellent agreement 
between the two experiments.  Only as $p_T\rightarrow 0$ GeV do the two 
experiments disagree.  The cross sections in this region are highly sensitive
to the precise alignment of the beam.  Uncertainties in either or both the
beam position and angle would lead to systematic shifts in the $p_T$ of
each event.  Thus, the differences between the two measurements may be
due to beam alignment issues in the two experiments.

\subsection{Cross Section Ratios}

We turn now to a comparison of our absolute measurements with the previously 
published cross section ratios.  The primary goal of Fermilab E866 was to measure the deuterium-to-hydrogen
cross section ratio as a function of $x_2$, which provides a clean 
probe of the Bj\"orken-$x$ dependence of the ratio of anti-down to 
anti-up quarks in the nucleon 
sea.\footnote{\setlength{\baselineskip}{\singlespace}Although not explicitly shown in chapter 1, the ratio of the
leading-order $pd$ and $pp$ cross sections can be written for large $x_F$ as 
$\frac{\sigma_{pd}}{\sigma_{pp}}(M,x_F;x_2) \approx 
\left\{ \frac{1 + \frac{1}{4}\frac{d}{u}(x_1)}{1 + \frac{1}{4}\frac{d}{u}(x_1)\frac{\bar{d}}{\bar{u}}(x_2)} \right\} \left( 1 + \frac{\bar{d}}{\bar{u}}(x_2) \right)$,
where $x_2$ is evaluated according to equation \ref{eqn:kinvert}.
}  Those results \cite{bib:E866-Hawker,bib:E866-Towell} have already been
published and incorporated into the various global analyses of the
parton distribution functions \cite{bib:CTEQ,bib:MRST,bib:GRV}.  

Although the previous analysis and this current work began with the same data 
sample, the tighter event selection criteria required for absolute 
normalization resulted in the elimination of much of the data above 
$x_2 \approx 0.15$.  Direct comparison between the results presented in chapter
5 and those of Towell {\it et al.} \cite{bib:E866-Towell} are complicated by 
this.  Further complicating matters, the results of Towell {\it et al.}
are differential only in a single variable ($x_2$),\footnote{
\setlength{\baselineskip}{\singlespace}Towell {\it et al.} did look at the 
cross section ratio versus other variables, but none which are directly 
comparable to our double-differential cross sections.} while our new 
results are differential in the mass and $x_F$ of the dimuon pair.

In figure \ref{fig:d2ratio}, we present the deuterium-to-hydrogen cross-section
ratio versus mass in four $x_F$ bins.  The ratios were calculated by
integrating the data in tables \ref{table:d3sigma_first} through 
\ref{table:d3sigma_last} over all $p_T$ and taking the ratio of deuterium
to hydrogen in each bin.  Next-to-leading order calculations of the cross
section ratio, based on the CTEQ 5 and CTEQ 6\footnote{
\setlength{\baselineskip}{\singlespace}The MRST 98, MRST 2001
and GRV 98 parton distributions give similar results.} parton distributions, 
are also shown in the figure.  Since the E866 cross section ratios were
used in the CTEQ 5 and CTEQ 6 global analyses, they provide a convenient
means to check the consistency between the two analyses.   
We also superimpose the previous cross section ratio results of Towell 
{\it et al.} on figure \ref{fig:d2ratio}.  Although those results are 
differential only in $x_2$, reference \cite{bib:E866-Towell} does quote a
mean mass and $x_F$ in each $x_2$ bin.  We therefore plot the cross section
ratios in each $x_2$ bin at the mean mass quoted in the reference, and only
in the $x_F$ bins appropriate to their mean $x_F$.  
Overall, the level of 
agreement between the data presented in chapter 5 and the cross section
ratios previously published by E866 is quite good.


To provide a more detailed and direct comparison with the previously 
published results, we calculated the absolute cross section 
$d^2\sigma/dx_1dx_2$ for hydrogen and deuterium using the same $x_2$ binning 
as reference \cite{bib:E866-Towell} and the same procedures as outlined in
chapter 4.  For this study, the event selection criteria were 
loosened to admit the large-$x_2$ events which were otherwise problematic
in the absolute cross sections analysis.\footnote{\setlength{\baselineskip}{\singlespace}Specifically, these events
tended to occupy regions of the spectrometer with either known efficiency 
problems or uncertainties in reconstructing the event kinematics.  These 
effects should almost completely cancel in ratio.  We did not include any
data from intermediate mass data sets 3 and 4, which account for approximately
$1/3$ of the total number of events used in reference \cite{bib:E866-Towell}.}
Figure \ref{fig:dhratio_vs_x2} shows the deuterium-to-hydrogen cross section
ratio versus $x_2$ as calculated in this analysis in three different
$x_1$ bins.  As with the previous analysis, there appears to be no significant
$x_1$ dependence on the cross section ratio.  Averaging the ratios over all 
$x_1$ we obtain a cross section ratio versus $x_2$ only, which is compared
with the results of Towell {\it et al.} in figure \ref{fig:dhratio_vs_x2}.
The only differences between the two analyses are almost completely due
to the larger event sample available to reference \cite{bib:E866-Towell}.

\begin{figure}
  \centering
  \includegraphics[width=0.9\linewidth]{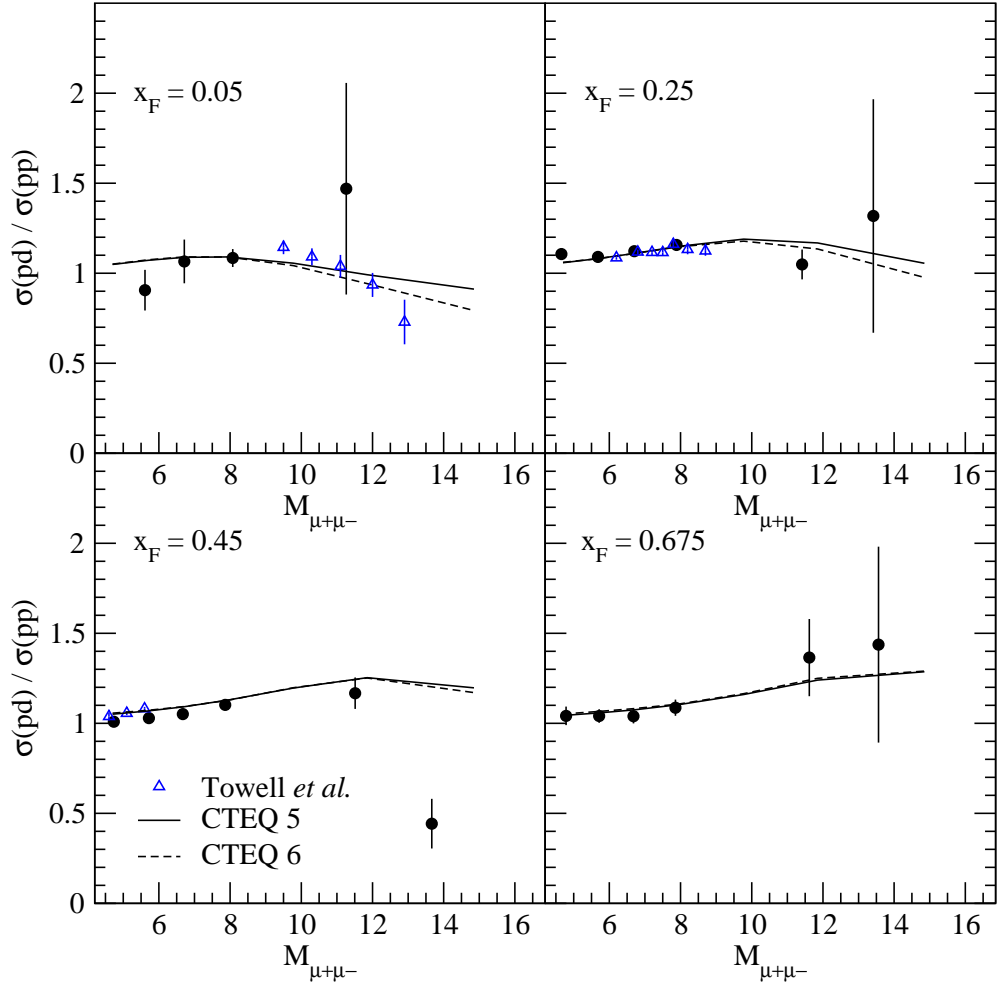}
  \caption[Ratio of $pd$ and $pp$ (per nucleon) cross sections.]{
	\label{fig:d2ratio}
	\setlength{\baselineskip}{\singlespace}
	Ratio of $pd$ and $pp$ (per nucleon) cross sections.
	The results of this analysis 
	(circles) are compared with the results (triangles) of the previous 
	analysis of \Towell \protect\cite{bib:E866-Towell}.  Although the results
 	of \Towell \ are differential only in $x_2$, the mean mass and $x_F$
	of the events in each $x_2$ bin was tabulated in the reference.
	We plot those results at those mean mass values, and only in $x_F$
	bins appropriate to the mean $x_F$.  NLO calculations of the ratio 
	based on the CTEQ 5 and CTEQ 6 parton distributions are also shown.
	}
\end{figure}

\begin{figure}
\centering
\includegraphics[width=0.9\linewidth,clip]{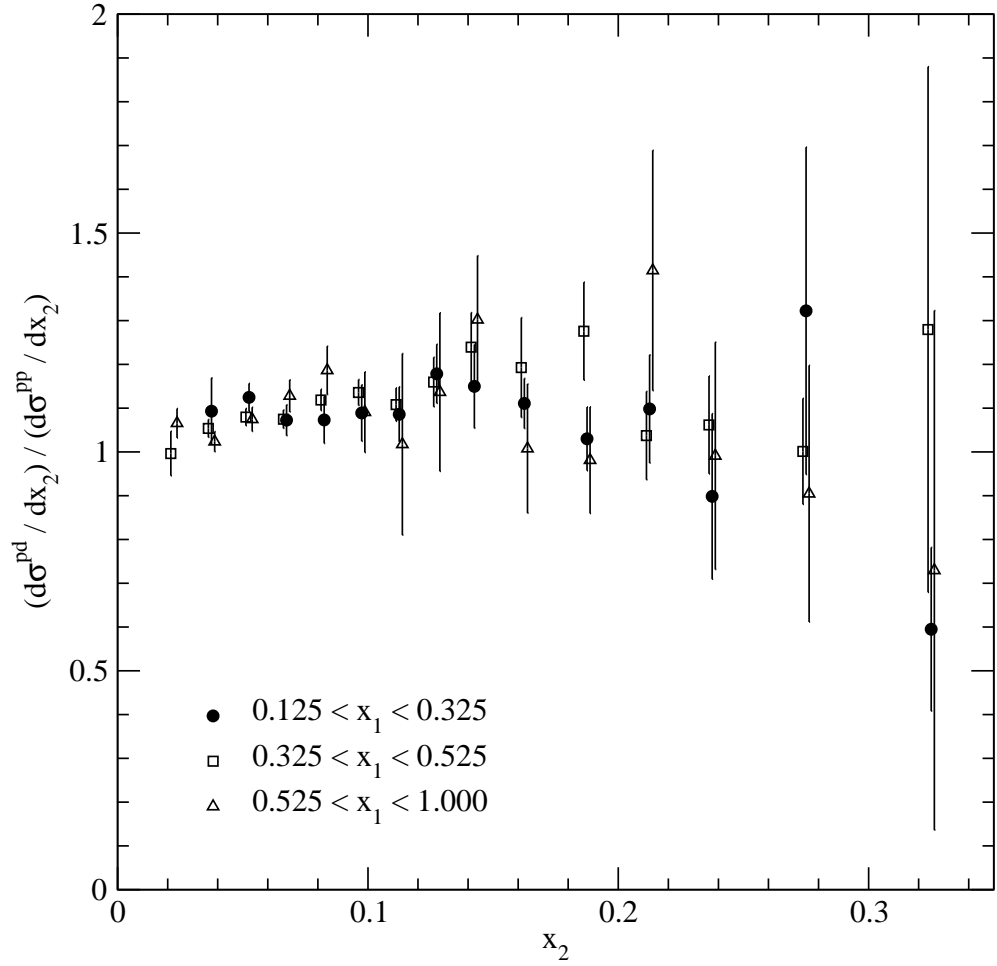}
\addtocontents{lof}{\  }
\addtocontents{lof}{\setlength{\baselineskip}{\singlespace}}
\caption[Ratio of $pd$ and $pp$ (per nucleon) cross sections plotted versus 
	$x_2$ in three $x_1$ bins.]{
	\label{fig:dhratio_x1_x2}
	\setlength{\baselineskip}{\singlespace}
	Ratio of $pd$ and $pp$ (per nucleon) cross sections plotted versus 
	$x_2$ in three $x_1$ bins.  The two larger $x_1$ bins are offset 
	slightly in the horizontal direction from the bin center for clarity.
	}
\addtocontents{lof}{\setlength{\baselineskip}{\doublespace}}
\end{figure}

\begin{figure}
\centering
\includegraphics[width=0.9\linewidth,clip]{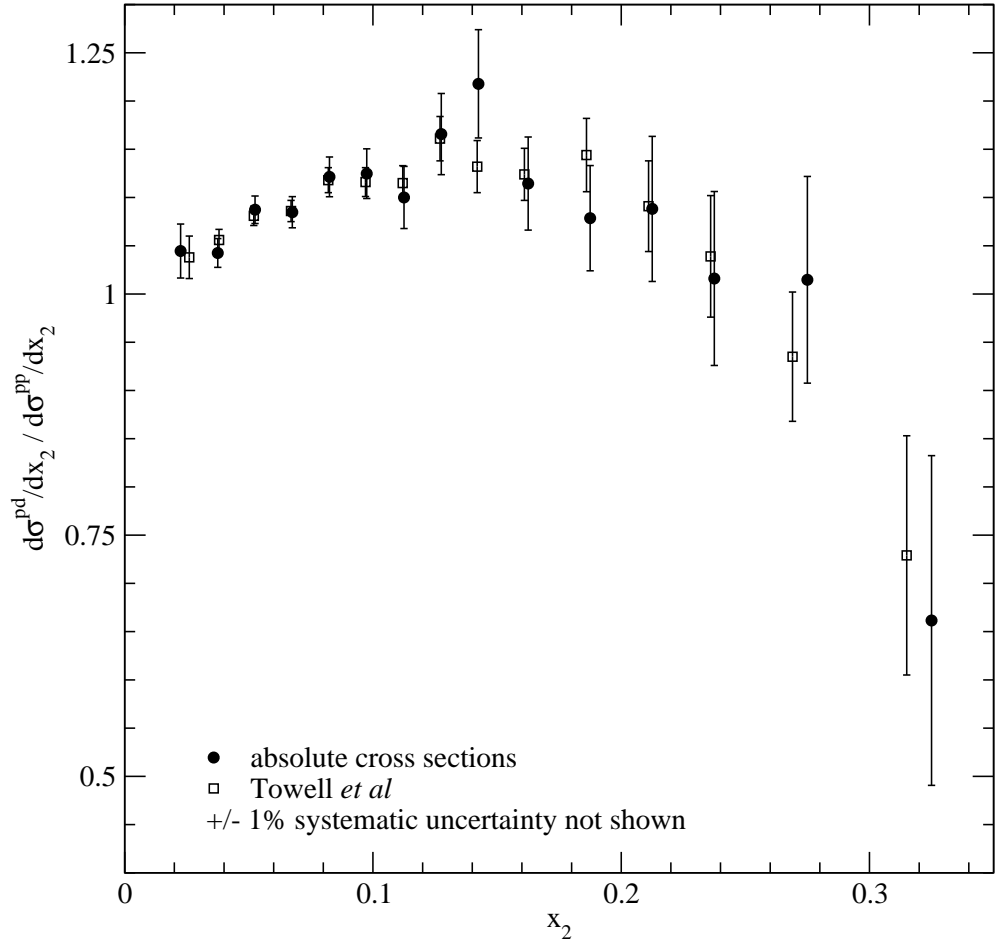}
\addtocontents{lof}{\  }
\addtocontents{lof}{\setlength{\baselineskip}{\singlespace}}
\caption[Ratio of $pd$ and $pp$ (per nucleon) cross sections plotted versus 
	$x_2$, at the center of each $x_2$ bin.]{
	\label{fig:dhratio_vs_x2}
	\setlength{\baselineskip}{\singlespace}
	Ratio of $pd$ and $pp$ (per nucleon) cross sections plotted versus 
	$x_2$, at the center of each $x_2$ bin.  The data are compared with 
	the previous E866 results, which are plotted at the mean $x_2$ in 
	each bin.  Both analyses are subject to an additional $\pm 1\%$ 
	systematic uncertainty in the relative normalization.
	}
\addtocontents{lof}{\setlength{\baselineskip}{\doublespace}}
\end{figure}

\subsection{Data versus Theory}

Detailed information on how our data will impact the parton distribution 
functions must inevitably await their incorporation into future global
analyses.  It is still useful to compare our results with theoretical
calculations of continuum dimuon cross sections.  Such comparisons will
demonstrate whether the existing sets of parton distributions are consistent
with our data in kinematic regions already constrained by previous
measurements (E605), and reveal where we expect our data to substantially
impact upon the parton distributions in the next round of fits.

As discussed back in chapter 1, continuum dimuon production is sensitive
to the gluon distributions through QCD processes at and beyond 
next-to-leading-order.  Dimuon production is dominated by gluon Compton 
scattering for $p_T > \frac{1}{2}M$ \cite{bib:Berger}, and the E866 results 
satisfy this kinematic condition over an extended range in mass and $x_F$.
Thus, in addition to information on the light-antiquark distributions, 
our data should provide a new and valuable source of information
about the gluon distributions.
Because this is still an area of ongoing theoretical 
investigation, in the following discussion we limit ourselves to a 
discussion of the quark and antiquark distributions.

\subsubsection{Scaling Form: $M^3d^2\sigma/dMdx_F$}

In the preceding chapter, the scaling form $M^3d^2\sigma/dMdx_F$ was tabulated
and plotted in mass and $x_F$ bins.  Figures \ref{fig:d2sigma_first} through
\ref{fig:d2sigma_last} plotted the $pp$ and $pd$ cross section in $x_F$ bins
versus the invariant mass of the dimuon pair.  In each $x_F$ bin we also 
compared the data to perturbative QCD calculations performed to
next-to-leading order.  These calculations were based on several of the
currently available and most commonly used sets of fits to the parton 
distributions. In general, they reproduce the shape and normalization of
the data well, save for the GRV 98 distributions which systematically
overestimate the data by $\sim 20\%$.  To better 
illustrate this, the theoretical calculations were fit to the data to
determine a ``$K^\prime$-factor'', which we define as the ratio of
the experimentally measured dimuon continuum to NLO theory.

\begin{table}[htb]
\addtocontents{lot}{\  }
\addtocontents{lot}{\setlength{\baselineskip}{\singlespace}}
  \caption[$K^\prime$-factors, where 
	$K^\prime = \frac{\sigma^{\text{exp}}}{\sigma^{\text{NLO}}}$,
	for 800-GeV $pp$ and $pd$ dimuon production.]{
	\label{table:kprime}
	\setlength{\baselineskip}{\singlespace}
	$K^\prime$-factors, where 
	$K^\prime = \frac{\sigma^{\text{exp}}}{\sigma^{\text{NLO}}}$,
	for 800-GeV $pp$ and $pd$ dimuon production.  The errors represent 
	the statistical and point-to-point systematic uncertainties in the 
	data.  The data are subject to an additional $\pm 6.5\%$ uncertainty 
	in the normalization.
	}
\addtocontents{lot}{\setlength{\baselineskip}{\doublespace}}
  \vspace{10pt}
  \centering
  \begin{tabular}{|l|cc|cc|}
  \hline
  \multicolumn{1}{|c|}{PDF} & $K^\prime_{pp}$ & $\chi^2 / ndf$ & $K^\prime_{pd}$ & $\chi^2 / ndf$ \\
  \hline\hline
  CTEQ5   & $0.9753 \pm 0.0037$ & $1.47$ & $0.9633 \pm 0.0017$ & $2.56$ \\
  CTEQ6   & $1.0150 \pm 0.0028$ & $1.44$ & $1.0003 \pm 0.0004$ & $2.62$ \\
  MRST 98 & $0.9732 \pm 0.0022$ & $1.43$ & $0.9604 \pm 0.0028$ & $2.42$ \\
  MRST 2001 & $0.9799 \pm 0.0034$ & $1.50$ & $0.9664 \pm 0.0078$ & $2.48$ \\ 
  GRV 98  & $0.8107 \pm 0.0043$ & $2.11$ & $0.8075 \pm 0.0002$ & $4.24$ \\
  \hline
  \end{tabular}
\end{table}

The $K^\prime$-factors for hydrogen and deuterium are shown in table
\ref{table:kprime}. All of the parton distributions studied, with the 
exception of the GRV 98 set, reproduce the normalization of the measured 
cross sections to better than our systematic uncertainties. 
This is not unexpected.  The valence distributions are well constrained
by DIS measurements, and the E605 Drell-Yan data (with which we agree)
are an important input into the fits, constraining the antiquark
sea over much of the range covered by E866 \cite{bib:CTEQ,bib:MRST,bib:GRV}.


It is interesting to note the improvement in the normalization of the CTEQ 6 
\cite{bib:CTEQ6}
based calculations relative to CTEQ 5 \cite{bib:CTEQ5}.  The main difference 
between the CTEQ 6
distributions and the previous CTEQ fit is a stiffer large-$x$ gluon, the
result of including Tevatron jet data in the newer analysis \cite{bib:CTEQ6}.  
The same data
has also been incorporated into the latest MRST fit \cite{bib:MRST01},
but without as significant a change in the gluon distributions.
Though there is improvement in the calculations based on the MRST 2001 
distributions over the previous MRST analysis, it is not as dramatic as 
that seen in the CTEQ distributions. 

The relatively poor quality of the fits shown in table \ref{table:kprime}
can be explained by behavior observed in the cross sections which was
not anticipated by the parton distribution functions.  The agreement between 
data and theory is quite good below $x_F \approx 0.3$, which is not surprising
given the degree of agreement with the E605 data previously discussed.
As $x_F$ increases beyond the range constrained by E605, the agreement between
the E866 data and the NLO calculations deteriorates, especially at low mass.  
While this kinematic range probes smaller values of $x_2$ than reached
in E605, it also probes values of $x_1$ (the momentum fraction of
the annihilating quark in the parton model) where the estimated uncertainties
on the valence quark distributions are about $\sim 20\%$.  Thus, we expect
our data to provide important constraints on both the valence and sea
quarks in the next round of fits.  

\subsubsection{$d^2\sigma/dx_1dx_2$}  

\begin{figure}
  \centering
  \includegraphics[clip,width=0.9\linewidth]{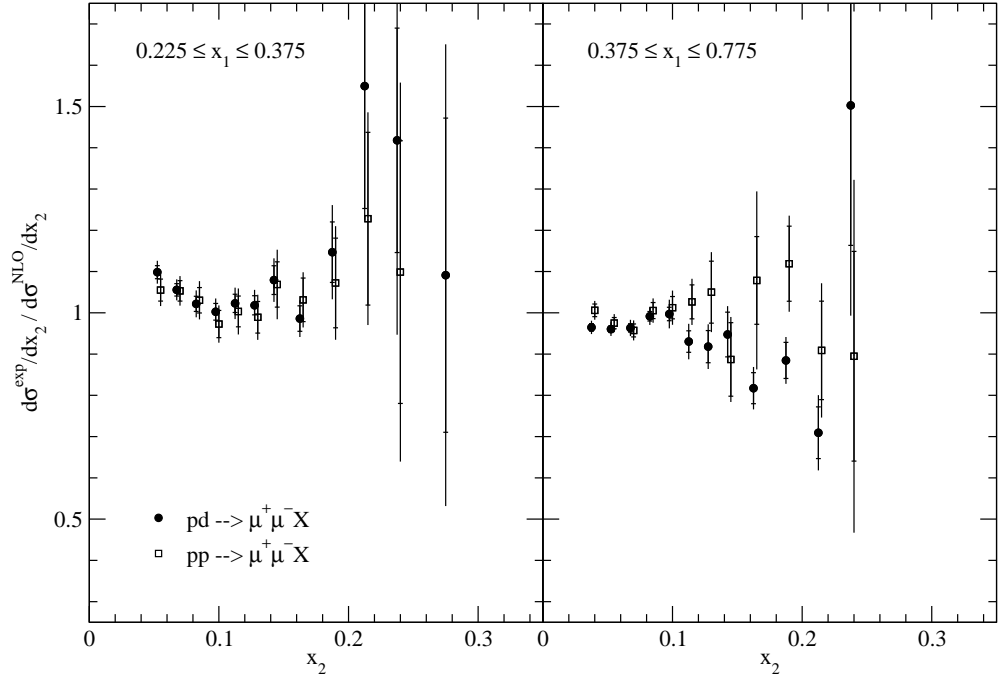}
\addtocontents{lof}{\  }
\addtocontents{lof}{\setlength{\baselineskip}{\singlespace}}
  \caption[Ratio of data to NLO calculations based on CTEQ 6 \protect\cite{bib:CTEQ6}
    versus $x_2$.]{
	\label{fig:dvthvsx2_binned}
	\setlength{\baselineskip}{\singlespace}
	Ratio of data to NLO calculations based on CTEQ 6 \protect\cite{bib:CTEQ6}
    versus $x_2$.  The left panel shows the range $0.225 \leq x_1 < 0.375$,
    and the right panel shows the range $0.375 \leq x_1 < 0.775$.}
\end{figure}

	To explore the impact of our data on future fits in more detail,
	we now examine the cross sections written in the form 
	$d^2\sigma/dx_1dx_2$.
	While higher-order QCD processes confuse the interpretation of $x_1$ 
	and $x_2$ here, one should recall that almost half of the cross 
	section is due to the leading-order $q\bar{q}$ annihilation for which 
	$x_1$ and $x_2$ have their usual meaning.  
	Writing the $pp$ and $pd$ cross sections in the limit of large $x_F$ 
	we have
	\begin{eqnarray}
	\begin{aligned}
	\frac{d^2\sigma_{pp}}{dx_1dx_2} & \propto \frac{4}{9} u(x_1) \bar{u}(x_2) + \frac{1}{9} d(x_1)\bar{d}(x_2) \\
	\frac{d^2\sigma_{pd}}{dx_1dx_2} & \propto \left\{ \frac{4 u(x_1) + d(x_1)}{9} \right\} \left( \bar{d}(x_2) + \bar{u}(x_2) \right).
	\end{aligned}
	\end{eqnarray}
	\noindent 
	%
	Thus, the shape of the cross sections relative to theoretical 
	calculations should help us understand the effect of our data on the 
	parton distributions.
	In the comparisons between data and theory which follow, the cross 
	sections of the form $d^2\sigma/dx_1dx_2$ were determined using the 
	same procedures outlined in previous chapters.  However, certain 
	limitations are imposed on this analysis by the Monte Carlo event 
	sample, which was generated over a limited range in mass and $x_F$.  
	Because of this, calculating the acceptance in the lowest-lying 
	$x_1$-$x_2$ bins is problematic -- the $x_1$-$x_2$ bins extend beyond 
	the kinematic limits of the Monte Carlo sample.  This mainly affects 
	the data below $x_2 = 0.03$, which partially overlaps the low-mass,
	large-$x_F$ data discussed in the previous section.
	To study the likely impact of our data on future fits to the antiquark
	distributions, we examine the behavior of data and theory as a function
	of $x_2$, the momentum fraction of the annihilating antiquark in the
	parton model.  
	After the cross sections differential in the momentum
	fractions of the interacting partons was determined, the ratio of those
	cross sections to next-to-leading-order calculations based on the 
	CTEQ 6 \cite{bib:CTEQ6} partons was calculated in each $x_1$-$x_2$ bin.
	Figure \ref{fig:dvthvsx2_binned} shows these ratios in two $x_1$ bins,
	each containing roughly the same number of events.  The difference in 
	shape (if any) between data and theory versus $x_2$ appears to be 
	consistent over the two ranges in $x_1$.

\addtocontents{lof}{\setlength{\baselineskip}{\doublespace}}

\begin{figure}
	  \centering
	  \includegraphics[clip,width=0.9\linewidth]{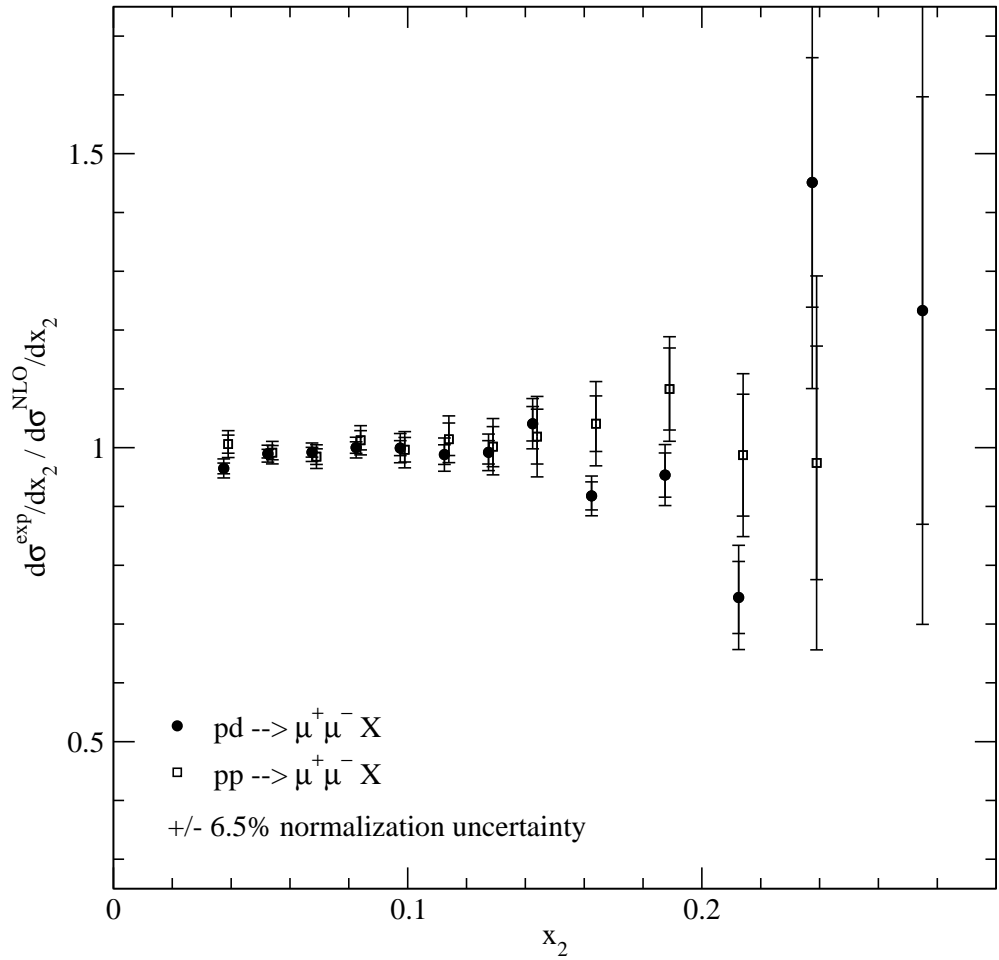}
\addtocontents{lof}{\  }
\addtocontents{lof}{\setlength{\baselineskip}{\singlespace}}
	  \caption[Ratio of data to NLO calculations based on CTEQ 6 \protect\cite{bib:CTEQ6}
	    versus $x_2$.]{
  		\setlength{\baselineskip}{\singlespace}
		\label{fig:dvthvsx2}Ratio of data to NLO calculations based on CTEQ 6 \protect\cite{bib:CTEQ6}
	    versus $x_2$.}
\addtocontents{lof}{\setlength{\baselineskip}{\doublespace}}
	\end{figure}

	In figure \ref{fig:dvthvsx2} we average together the ratios of data and
	theory over all $x_1$ bins to obtain the ratio of data to theory as a
	function of $x_2$ only.  The theoretical calculations clearly reproduce
	our data well, to within the quoted statistical and systematic errors
	in the data, over the $x_2$ range covered.  This indicates that the CTEQ 6
	fits to the light antiquarks are consistent with our data.  Since these
	fits are largely constrained by the E605 Drell-Yan measurements, this
	result was not unexpected.

\begin{figure}
	  \centering
	  \includegraphics[clip,width=0.9\linewidth]{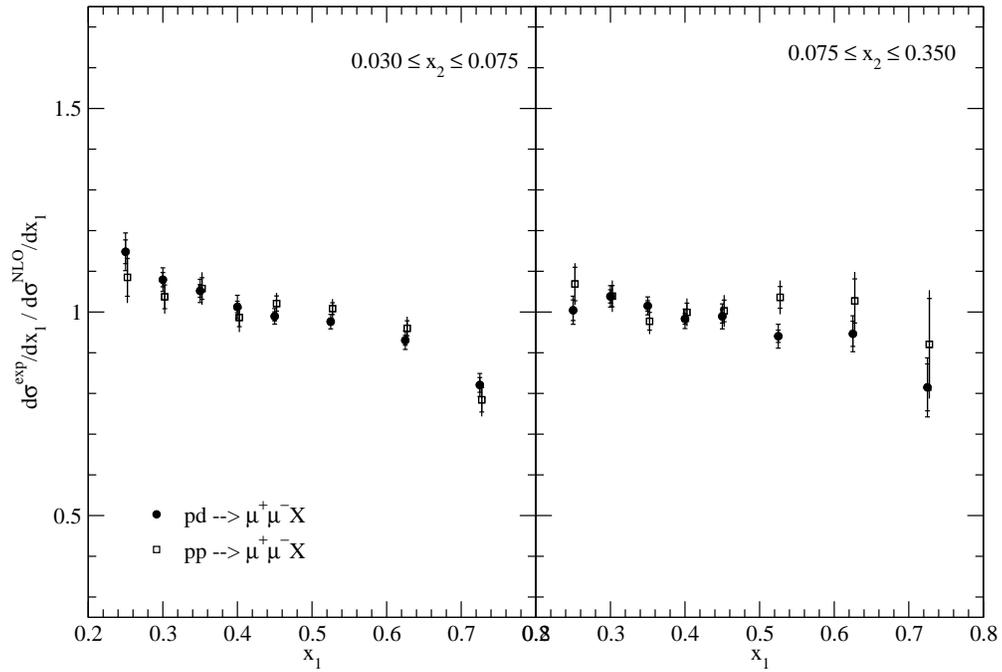}
\addtocontents{lof}{\  }
\addtocontents{lof}{\setlength{\baselineskip}{\singlespace}}
	  \caption[Ratio of data to NLO calculations based on CTEQ 6 \protect\cite{bib:CTEQ6}
	    versus $x_1$.]{
 	 	\setlength{\baselineskip}{\singlespace}
		\label{fig:dvthvsx1_binned}Ratio of data to NLO calculations based on CTEQ 6 \protect\cite{bib:CTEQ6}
	    versus $x_1$.  The left panel shows the range $0.225 \leq x_2 < 0.375$,
	    and the right panel shows the range $0.03 \leq x_2 < 0.350$.}
\addtocontents{lof}{\setlength{\baselineskip}{\doublespace}}
	\end{figure}

\begin{figure}
	  \centering
	  \includegraphics[clip,width=0.9\linewidth]{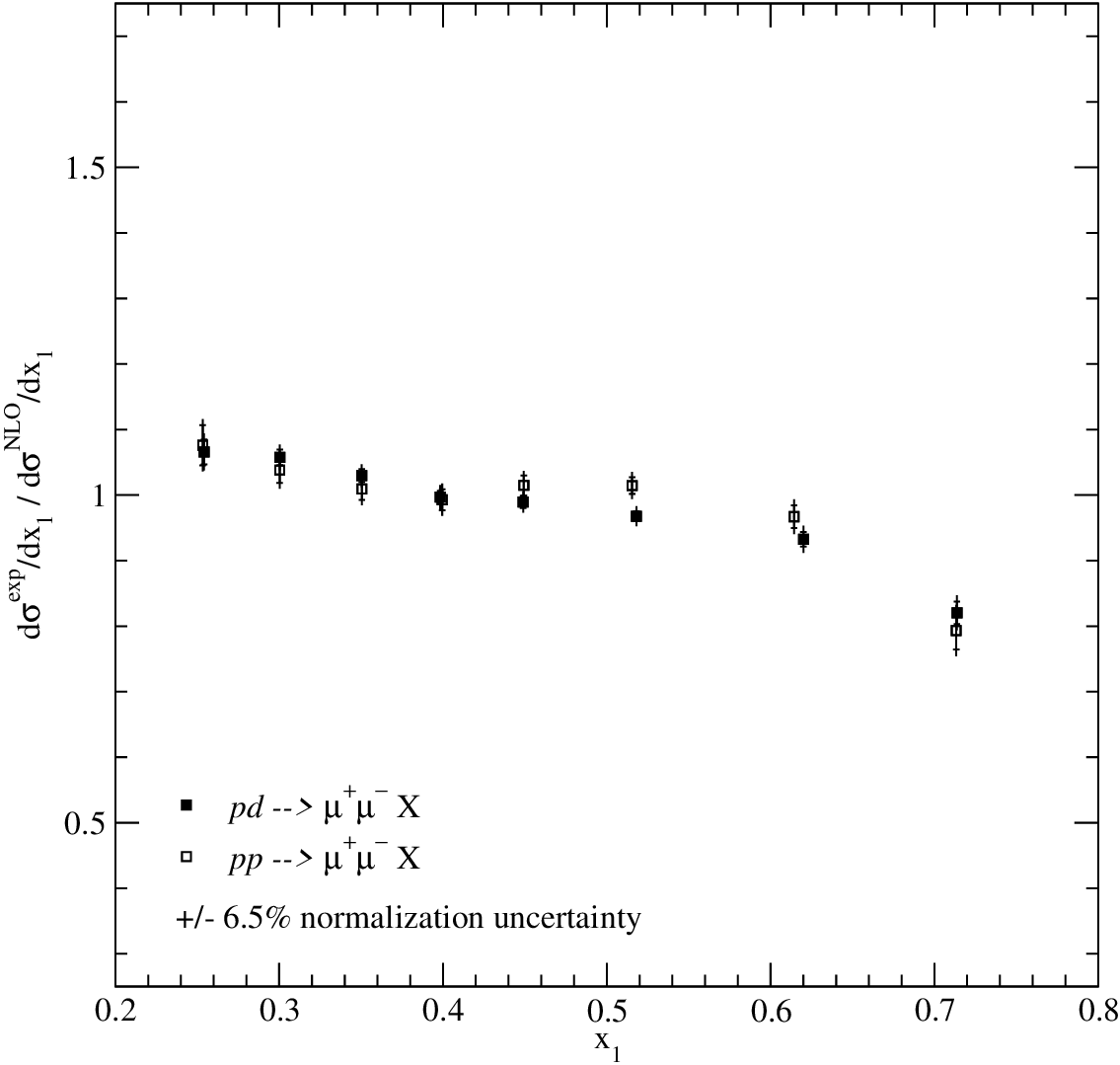}
\addtocontents{lof}{\  }
\addtocontents{lof}{\setlength{\baselineskip}{\singlespace}}
	  \caption[Ratio of data to NLO calculations based on CTEQ 6 \protect\cite{bib:CTEQ6}
	    versus $x_1$.]{
 			\setlength{\baselineskip}{\singlespace}
		\label{fig:dvthvsx1}Ratio of data to NLO calculations based on CTEQ 6 \protect\cite{bib:CTEQ6}
	    versus $x_1$.}
\addtocontents{lof}{\setlength{\baselineskip}{\doublespace}}
	\end{figure}

	The same analysis can be carried out to examine what our data imply for
	future fits to the valence quark distributions.  
	Figure \ref{fig:dvthvsx1_binned} shows data-divided-by-theory plotted 
	versus $x_1$, over two $x_2$ ranges each representing about half of the
	available statistics.
	There is an $x_1$ dependence which is consistent in both $x_2$ bins.
	Combining the results over all $x_2$ in figure \ref{fig:dvthvsx1}
	we see that at large $x_1$ the data falls about $20\%$ below the
	theoretical predictions.
	This seems to suggest that the valence quark distributions are smaller
	at large Bj\"orken-$x$ than 
	in current parton distributions,
	though not outside the estimated uncertainties of the fits 
	\cite{bib:CTEQ6}.  
	Parton distribution fits to DIS data at next-to-next-to-leading-order
	have also recently become available \cite{bib:MRSTNNLO}, and they
	also suggest a reduced valence distribution, relative to current
	NLO fits.

\subsection{Future Prospects}

Continuum dimuon production has been an active area of reseach for over
thirty years, and the outlook for future measurements remains bright.
Based on the cross section ratios published by E866 
\cite{bib:E866-Hawker,bib:E866-Towell}, Fermilab has accepted a proposal 
\cite{bib:E906-Proposal} for a new dimuon experiment, Fermilab Experiment
906, to be conducted using the 120-GeV proton beam from the Main Injector.
Like E866, the E906 measurements will examine the differences in continuum
dimuon production in $pp$ and $pd$ interactions.  Their primary goal is to 
measured $\frac{\bar{d}}{\bar{u}}(x)$ out to $x\approx 0.5$.  The extended
range in $x$ is made possible in large measure due to the seven-fold increase 
in cross section using 120-GeV protons relative to 800-GeV protons.

Flavor asymmetry is but one of the topics to be adressed with the new
experiment.  Absolute measurements of the hydrogen and deuterium cross
sections will also be possible, greatly extending the kinematic range
and statistical precision available to constrain the parton distributions.
Furthermore, these measurements may also be able to demonstrate an important 
property of QCD.  Logarithmic scaling violations have long been established 
in DIS measurements, but have not been conclusively shown using the Drell-Yan 
process.  The reasons for this include the relatively high $q^2$ probed in 
dimuon experiments, and the partial cancelation of the effect between the 
large-$x$ behavior of the valence quarks, and the low-$x$ behavior of the 
antiquarks \cite{bib:DY-SCALING-REVIEW,bib:E772-DY-POL}.  Given the 
statistical precision and overall understanding of systematic uncertainties 
attained with the E866 results, comparisons with the future measurements of 
E906 may well be able to demonstrate scaling violations in the Drell-Yan 
process.


\subsection{Conclusion}

	Fermilab Experiment 866 has performed an absolute measurement
	of dimuon cross sections in 800-GeV $pp$ and $pd$ interactions.
	These data represent the most extensive measurement of the Drell-Yan
	process to date, covering a wider range of kinematics with better
	statistical precision than has been achieved in previous
	measurements.
	The results are in good agreement with existing $pA$ and $pd$ 
	measurements, as well as NLO calculations based on the various sets 
	of parton distribution functions.
	The overall level of agreement between data and theory indicates 
	that the antiquark distributions are well understood over the
	kinematic range relevant to E866.
	However, differences between data and theory suggest that the 
	valence distributions are overestimated at large $x$ in current
	parameterizations.
	Given the statistical precision of the data, we can expect a major
	impact on future fits to the valence quarks.

\newpage


\addcontentsline{toc}{section}{REFERENCES}
\setlength{\baselineskip}{\singlespace}

\end{document}